\documentclass [aip, jmp, reprint, superscriptaddress, eqsecnum, onecolumn]{revtex4-1}
\usepackage {etex}

\usepackage [english]{babel}
\usepackage [T1]{fontenc}
\usepackage [utf8] {inputenc}
\usepackage {lmodern}
\usepackage {geometry}
\usepackage {amsfonts}
\usepackage {amsmath}	
\usepackage {amssymb}
\usepackage {amsxtra}
\usepackage [amsthm,thmmarks,amsmath]{ntheorem} 
\usepackage {wasysym}
\usepackage {mathtools}
\usepackage {calrsfs}
\usepackage {dsfont}
\usepackage {graphicx}
\usepackage {rotating}

\usepackage {setspace}
\usepackage {color}
\usepackage [all,cmtip]{xy}
\usepackage {hyperref}
\usepackage [nottoc,numbib]{tocbibind}
\usepackage {diagbox}
\usepackage {dcolumn}

\allowdisplaybreaks

\input diagxy

\DeclareFontFamily{OT1}{pzc}{}
\DeclareFontShape{OT1}{pzc}{m}{it}{<-> s * [1.30] pzcmi7t}{}
\DeclareMathAlphabet{\mathpzc}{OT1}{pzc}{m}{it}

\DeclareMathOperator{\R}{\mathbb R}
\DeclareMathOperator{\T}{\mathbb T}
\DeclareMathOperator{\C}{\mathbb C}
\DeclareMathOperator{\N}{\mathbb N}
\DeclareMathOperator{\Z}{\mathbb Z}
\DeclareMathOperator{\Q}{\mathbb Q}

\DeclareMathOperator{\uB}{\textup B}
\DeclareMathOperator{\uR}{\textup R}
\DeclareMathOperator{\uL}{\textup L}

\DeclareMathOperator{\uM}{\textup M}

\DeclareMathOperator{\PW}{\textup PW}
\DeclareMathOperator{\SW}{\textup SW}

\DeclareMathOperator{\fg}{\mathfrak g}
\DeclareMathOperator{\fk}{\mathfrak k}
\DeclareMathOperator{\ft}{\mathfrak t}
\DeclareMathOperator{\fb}{\mathfrak b}
\DeclareMathOperator{\fn}{\mathfrak n}
\DeclareMathOperator{\fA}{\mathfrak A}

\DeclareMathOperator{\fH}{\mathfrak H}

\DeclareMathOperator{\cB}{\mathcal B}

\DeclareMathOperator{\cD}{\mathcal D}

\DeclareMathOperator{\cS}{\mathcal S}
\DeclareMathOperator{\cF}{\mathcal F}

\DeclareMathOperator{\aut}{\textup{Aut}}
\DeclareMathOperator{\End}{\textup{End}}

\DeclareMathOperator{\img}{img}
\DeclareMathOperator{\loc}{loc}
\DeclareMathOperator{\supp}{supp}
\DeclareMathOperator{\ssupp}{sing\ supp}

\DeclareMathOperator{\limit}{lim}

\DeclareMathOperator{\vol}{vol}
\DeclareMathOperator{\id}{id}

\DeclareMathOperator{\rank}{rank}
\DeclareMathOperator{\Bohr}{Bohr}
\DeclareMathOperator{\disc}{disc}
\DeclareMathOperator{\Trig}{Trig}

\DeclareMathOperator{\tr}{tr}

\DeclarePairedDelimiter{\ceil}{\lceil}{\rceil}

\newtheoremstyle{breakdef}%
  {\item[\rlap{\vbox{\normalfont\bfseries\hbox{\llap{##2}\hskip\labelsep
          ##1:}\hbox{\\[0.1cm]}}}]}%
  {\item[\rlap{\vbox{\normalfont\bfseries\hbox{\llap{##2}\hskip\labelsep
          ##1 (##3):}\hbox{\\[0.1cm]}}}]}
\newtheoremstyle{breaksatz}%
  {\item[\rlap{\vbox{\normalfont\normalsize\bfseries\hbox{\llap{##2}\hskip\labelsep
          ##1:}\hbox{\\[0.1cm]}}}]}%
  {\item[\rlap{\vbox{\normalfont\normalsize\bfseries\hbox{\llap{##2}\hskip\labelsep
          ##1 (##3):}\hbox{\\[0.1cm]}}}]}
\newtheoremstyle{breaklem}%
  {\item[\rlap{\vbox{\normalfont\normalsize\bfseries\hbox{\llap{##2}\hskip\labelsep
          ##1:}\hbox{\\[0.1cm]}}}]}%
  {\item[\rlap{\vbox{\normalfont\normalsize\bfseries\hbox{\llap{##2}\hskip\labelsep
          ##1 (##3):}\hbox{\\[0.1cm]}}}]}
\newtheoremstyle{breakprop}%
  {\item[\rlap{\vbox{\normalfont\normalsize\bfseries\hbox{\llap{##2}\hskip\labelsep
          ##1:}\hbox{\\[0.1cm]}}}]}%
  {\item[\rlap{\vbox{\normalfont\normalsize\bfseries\hbox{\llap{##2}\hskip\labelsep
          ##1 (##3):}\hbox{\\[0.1cm]}}}]}
\newtheoremstyle{breakbem}%
  {\item[\rlap{\vbox{\hbox{\hskip\labelsep\normalfont\bfseries
          ##1 ##2:}\hbox{\\[0.1cm]}}}]}%
  {\item[\rlap{\vbox{\hbox{\hskip\labelsep\normalfont\bfseries
          ##1 ##2 (##3):}\hbox{\\[0.1cm]}}}]}
\newtheoremstyle{breakbsp}%
  {\item[\rlap{\vbox{\hbox{\hskip\labelsep\normalfont\bfseries
          ##1 ##2:}\hbox{\\[0.2cm]}}}]}%
  {\item[\rlap{\vbox{\hbox{\hskip\labelsep\normalfont\bfseries
          ##1 ##2 (##3):}\hbox{\\[0.2cm]}}}]}
\newtheoremstyle{breakkor}%
  {\item[\rlap{\vbox{\hbox{\hskip\labelsep\normalfont\bfseries
          ##1 ##2:}\hbox{\\[0.1cm]}}}]}%
  {\item[\rlap{\vbox{\hbox{\hskip\labelsep\normalfont\bfseries
          ##1 ##2 (##3):}\hbox{\\[0.1cm]}}}]}
\newtheoremstyle{proof}%
  {\item[\rlap{\vbox{\hbox{\hskip\labelsep\normalfont\bfseries
          \underline{##1:}}\hbox{\\[0.1cm]}}}]}%
  {\item[\rlap{\vbox{\hbox{\hskip\labelsep\normalfont\bfseries
          \underline{##1 (##3):}}\hbox{\\[0.1cm]}}}]}
\theoremstyle{breakkor} 
\newtheorem{Definition}{Definition}[section] 
\theoremstyle{breakkor}
\newtheorem{Theorem}[Definition]{Theorem}
\newtheorem{Conjecture}[Definition]{Conjecture}
\theoremstyle{breakkor}
\newtheorem{Lemma}[Definition]{Lemma}
\theoremstyle{breakkor}
\newtheorem{Proposition}[Definition]{Proposition}
\theoremstyle{breakkor}
\newtheorem{Corollary}[Definition]{Corollary}
\theoremstyle{breakkor}
\theorembodyfont{\normalfont}
\newtheorem{Remark}[Definition]{Remark}

\newtheorem{Examples}[Definition]{Examples}
\theoremstyle{proof}
\theoremsymbol{\fbox{}}
\newtheorem{Proof}{Proof}

\begin{document}
\title{Coherent states, quantum gravity and the Born-Oppenheimer approximation, II: Compact Lie Groups}
\author{Alexander Stottmeister}
\email{alexander.stottmeister@gravity.fau.de}
\author{Thomas Thiemann}
\email{thomas.thiemann@gravity.fau.de}
\affiliation{Institut für Quantengravitation, Lehrstuhl für Theoretische Physik III, Friedrich-Alexander-Universität Erlangen-Nürnberg, Staudtstraße 7/B2, D-91058 Erlangen, Germany}
\begin{abstract}
In this article, the second of three, we discuss and develop the basis of a Weyl quantisation for compact Lie groups aiming at loop quantum gravity-type models. This Weyl quantisation may serve as the main mathematical tool to implement the program of space adiabatic perturbation theory in such models. As we already argued in our first article, space adiabatic perturbation theory offers an ideal framework to overcome the obstacles that hinder the direct implementation of the conventional Born-Oppenheimer approach in the canonical formulation of loop quantum gravity. Additionally, we conjecture the existence of a new form of the Segal-Bargmann-Hall ``coherent state'' transform for compact Lie groups $G$, which we prove for $G=U(1)^{n}$ and support by numerical evidence for $G=SU(2)$. The reason for conjoining this conjecture with the main topic of this article originates in the observation, that the coherent state transform can be used as a basic building block of a coherent state quantisation (Berezin quantisation) for compact Lie groups $G$. But, as Weyl and Berezin quantisation for $\R^{2d}$ are intimately related by heat kernel evolution, it is natural to ask, whether a similar connection exists for compact Lie groups, as well. Moreover, since the formulation of space adiabatic perturbation theory requires a (deformation) quantisation as minimal input, we analyse the question to what extent the coherent state quantisation, defined by the Segal-Bargmann-Hall transform, can serve as basis of the former.
\end{abstract}
\maketitle
\tableofcontents
\section{Introduction}
\label{sec:intro} 
We have argued in our first article \cite{StottmeisterCoherentStatesQuantumI} that a realisation of the (time-dependent) Born-Oppenheimer approximation for multi-scale quantum dynamical systems, which are modelled by techniques used in loop quantum gravity, might be achieved along the lines of space adiabatic perturbation theory\cite{PanatiSpaceAdiabaticPerturbation}. This is, because space adiabatic perturbation theory avoids some technical limitations of the original Born-Oppenheimer approach, which in turn allows us to circumvent the so-called ``problem of  non-commutative fast-slow coupling'' (originally pointed out in the context of loop quantum gravity\cite{GieselBornOppenheimerDecomposition, StottmeisterCoherentStatesQuantumI}).
The main technical tool, necessary for a successful implementation of the ideas behind space adiabatic perturbation theory, is a Weyl quantisation associated with part of the multi-scale quantum system. More precisely, if the total quantum system is described as a coupled system decomposing into two sectors (for simplicity), one of which is called the slow sector, and the other one the fast sector (thinking of different relevant time scales), we will require the existence of a Weyl quantisation (in the sense of a real deformation quantisation) for the description of the slow subsystem. Then, if we further assume that the Weyl quantisation is scalable with a parameter $\varepsilon$ quantifying the separation of scales between the subsystems, we can introduce a systematic perturbation theory in the sense of Born and Oppenheimer by means of dequantising the slow sector and exploiting the induced $\varepsilon$-dependent $\star$-product in the resulting function spaces (in analogy with standard pseudo-differential operators and their symbolic calculus). \\
This said, it is the primary purpose of the present article to investigate the possibility of formulating a Weyl quantisation suitable for phase spaces of the type $T^{*}G$, $G$ a compact Lie group. The reason behind this objective is that such phase space serve as the main building block of loop quantum gravity-type models.\\[0.1cm]
Before we come to the main part of the article, which is composed of three sections, let us briefly outline its structure and content:\\[0.1cm]
Section \ref{sec:cst} introduces a new form of the Segal-Bargmann-Hall ``coherent state'' transform \cite{HallTheSegalBargmann} for compact Lie groups. This is motivated by the fact that unitary maps of this type provide the basis for the construction of a coherent state quantisation (also known as Berezin or Wick/Anti-Wick quantisation) of the co-tangent bundle, $T^{*}G$, of a compact Lie group $G$. But, before we enter into the discussion of the coherent state transform, we recall the definition of crossed product $C^{*}$-algebras, paying special attention to the transformation group $C^{*}$-algebra $C(G)\rtimes_{\uL}G$, as the latter is intimately connected with the Hilbert space $L^{2}(G)$, of which the coherent states are specific elements. Moreover, in foresight of section \ref{sec:wc}, introducing the $C^{*}$-algebra $C(G)\rtimes_{\uL}G$ is convenient, because it is fundamental to the construction of the Weyl quantisation for compact Lie groups, we are aiming at. A relation between coherent state quantisation and Weyl quantisation for compact Lie groups will be established subsection \ref{subsec:csquant} (with regard to their application to space adiabatic perturbation theory). For completeness, we provide the essential ingredients necessary to define the coherent states for compact Lie groups introduced by Hall \cite{HallTheSegalBargmann}, as well.\\
In section \ref{sec:wc}, we present possible ways to obtain Weyl (and Kohn-Nirenberg) quantisations for compact Lie groups. To this end, we follow the common philosophy \cite{RuzhanskyPseudoDifferentialOperators, TaylorNoncommutativeMicrolocalAnalysis} of constructing quantisations of functions (pseudo-differential operators) from left (or right) convolution kernels. Although, our constructions, which are based on results of Turunen and Rhuzhansky \cite{RuzhanskyPseudoDifferentialOperators} and Landsman \cite{LandsmanMathematicalTopicsBetween}, succeed to some extent, we are forced to deal with the dichotomy of choosing between local and global structures at various points. The latter can be traced back to the fact that the exponential map of a compact Lie group $G$, while still being onto, is no longer a diffeomorphims like in the case of $\R^{2d}$ (or nilpotent Lie groups in general). It appears, that the global formulas are easier to handle, when it comes to the composition of pseudo-differential operators (at least in the Kohn-Nirenberg formalism). But, it is the local setting, which is well adapted to deal with a semi-classical approximation of the commutation relations underlying the transformation group algebra $C(G)\rtimes_{\uL}G$, and allows for a more direct analogy with the original treatment of space adiabatic perturbation theory in \cite{PanatiSpaceAdiabaticPerturbation}. The main difficulty with the global formulas can be reduced to a lack of scale transformations compatible with the quantisation formulas and the algebraic structure of $C(G)\rtimes_{\uL}G$, i.e. we are missing a simple relation between pseudo-differential operators with different values of the quantisation parameter $\varepsilon$ (the adiabatic parameter in space adiabatic perturbation theory). Therefore, while asymptotic expansions of pseudo-differential operators are still conceivable in the global setting, a simple ordering in terms of powers of $\varepsilon$ is bound to fail. In subsection \ref{subsec:swt} and subsection \ref{subsec:u1bohr}, we further elaborate on this aspect: Namely, we relate the latter to the necessity of having a $\varepsilon$-scaleable Fourier transform on $G$. Then, we show that a partial solution can be achieved, if the so-called Stratonovich-Weyl transform for $G$ \cite{FigueroaMoyalQuantizationWith} is invoked, but, that a general solution seems to be obstructed by the rigidity of the representation theory of $G$ (integrality of highest weights). Only, in the case of $G=U(1)^{n}$, we are able to proceed further by, first, lifting everything to the universal covering group $\R^{n}$, and, second, passing to the Bohr compactification $\R^{n}_{\Bohr}$. The resulting theory of pseudo-differential operators connected with $\R_{\Bohr}$ is discussed and compared to the theory of almost-periodic pseudo-differential operators \cite{ShubinAlmostPeriodicFunctions, ShubinDifferentialAndPseudodifferential} in subsection \ref{subsec:u1bohr}, as well.\\
Finally, we present some concluding remarks and perspectives in section \ref{sec:con}.
\section{A new look at the Segal-Bargmann-Hall ``coherent state'' transform}
\label{sec:cst}
Although, this article is mainly concerned with the development of a Weyl quantisation for loop quantum gravity-type models, it was already noted in the preceding article \cite{StottmeisterCoherentStatesQuantumI} (cf. also \cite{FollandHarmonicAnalysisIn, FigueroaMoyalQuantizationWith}) that the (Stratonovich-)Weyl and Wick/Anti-Wick (de-)quantisations are closely related from a conceptual point of view (cf. \cite{RieffelQuantizationAndC} for a lucid overview), especially regarding their potential applicability in the context of space adiabatic perturbation theory (cf. \cite{PanatiSpaceAdiabaticPerturbation, TeufelAdiabaticPerturbationTheory}). In respect of their importance for Wick/Anti-Wick correspondences, we give a short review of the construction of coherent states on compact Lie Groups by Hall \cite{HallTheSegalBargmann, HallPhaseSpaceBounds} and conjecture a new version of the Segal-Bargmann-Hall transform (or resolution of unity) that is free from the ``dual'' heat kernel measure $\nu$ on the complexification $G_{\mathds{C}}$ Lie group $G$. We prove the conjecture in the case $G=U(1)$ and support it by some numerical evidence for $G=SU(2)$.\\
Since our constructions can be related to the so-called crossed product $C^{*}$-algebra $C(G)\rtimes_{\uL}G$, which will also be important for the other sections, we briefly recall its construction (cf. \cite{WilliamsCrossedProductsOf, BratteliOperatorAlgebrasAnd1}).
\subsection{Locally compact groups and crossed products}
\label{sec:cprd}
Given a locally compact group $G$, a $C^{*}$-algebra $\fA$ and a (strongly) continuous representation $\alpha:G\rightarrow\aut(\fA)$, one makes the
\begin{Definition}
\label{def:dynsys}
The triple $(\fA, G, \alpha)$ is called a ($C^{*}$-)\textup{dynamical system}. A \textup{covariant representation} of $(\fA, G, \alpha)$ is a triple $(\fH, \pi, U)$ consisting of a (non-degenerate) representation $\pi:\fA\rightarrow \cB(\fH)$ and (strongly) continuous unitary representation $U:G\rightarrow U(\fH)$, s.t.
\begin{align}
\label{eq:covrep}
\forall g\in G, A\in\fA:\ \pi(\alpha_{g}(A)) & = U_{g}\pi(A)U_{g}^{*}.
\end{align}
\end{Definition}
Let us illustrate this definition by giving a few
\begin{Examples}
\label{ex:covrep}
\begin{itemize}
	\item[1.] Clearly, $(\fA, \{e\}, \id)$ and $(\C, G, \id)$ are trivial examples of dynamical systems. The covariant representations of these correspond to (non-degenerate) representations of $\fA$ in the former, and to strongly continuous unitary representation of $G$ in the latter case.
	\item[2.] The left action $L$ of $G$ on itself gives rise to a continuous representation $\alpha_{\uL}:G\rightarrow\aut(C_{0}(G))$ on the $C^{*}$-algebra of continuous functions on $G$ vanishing at infinity by (cf. \cite{WilliamsCrossedProductsOf}, Lemma 2.5.)
\begin{align}
\label{eq:leftpullback}
\forall f\in C_{0}(G), g\in G:\ \alpha_{\uL}(g)(f) = L_{g^{-1}}^{*}f.
\end{align}
An important covariant representation of the dynamical system $(C_{0}(G), G, \alpha_{\uL})$ comes from the multiplier representation $\uM:C_{0}(M)\rightarrow \cB(L^{2}(G))$ and the left regular representation $\lambda:G\rightarrow U(L^{2}(G))$, where $L^{2}(G)$ is defined with respect to a (left) Haar measure. The compatibility of the pair $(M, \lambda)$ reflects the covariance condition \eqref{eq:covrep}:
\begin{align}
\label{eq:leftcovrep}
\forall f\in C_{0}(G), g\in G:\ \uM(\alpha_{\uL}(g)(f)) & = \lambda_{g}\uM(f)\lambda_{g}^{*}.
\end{align}
	\item[3.] In analogy with the preceding example, we consider the $C^{*}$-algebra $C_{b}(\R)$ of bounded continuous function on $\R$, and the (left) action $\tau$ of $\R$ on itself by translations. The triple $(C_{b}(\R), \R, (\tau^{-1})^{*})$ forms a dynamical system, and the triple $(L^{2}(\R), \uM, \lambda)$ is a covariant representation. Now, let us introduce the notation $U(x):=\uM(e^{ix(\ .\ )}),\ V(\xi):=\lambda_{-\xi},\ x,\xi\in\R$. This family of unitary operators in $L^{2}(\R)$ satisfies the canonical commutation relations in Weyl form:
\begin{align}
\label{eq:ccr}
V(\xi)U(x)V(\xi)^{*} & = \lambda_{-\xi}\uM(e^{ix(\ .\ )})\lambda_{\xi} \\ \nonumber
 & = \uM(\alpha_{\uL}(-\xi)(e^{ix(\ .\ )})) \\ \nonumber
 & = \uM(e^{ix((\ .\ )+\xi)}) \\ \nonumber
 & = e^{ix\xi}\uM(e^{ix(\ .\ )}) \\ \nonumber
 & = e^{ix\xi}U(x).
\end{align}
Thus, we obtain the Weyl algebra of 1-particle quantum systems as a dynamical system.
\end{itemize}
\end{Examples}
The importance of example 2 will become clear after the definition of the crossed product $C^{*}$-algebra $\fA\rtimes_{\alpha}G$ associated with a dynamical system, and the statement of theorem \ref{thm:svn}, which will be related to our conjecture of a new Segal-Bargmann-Hall transform for compact Lie groups. But, before we turn to the definition of the crossed product, we add an remark on dynamical systems with commutative $\fA$ (cf. \cite{WilliamsCrossedProductsOf}, especially Proposition 2.7.).
\begin{Remark}
\label{rem:abeliandynsys}
The Gelfand-Naimark theorem (cf. \cite{BratteliOperatorAlgebrasAnd1}) tells us that $\fA\cong C_{0}(X)$, where $X$ is the set of characters of $\fA$ with the locally compact Hausdorff weak$^{*}$-topology. Thus, the dynamical system $(\fA, G, \alpha)$ is isomorphic to the dynamical system $(C_{0}(X), G, \alpha)$. But, a dynamical system of the form $(C_{0}(X), G, \alpha)$ comes from a left $G$-space $(X,G)$, which is why these are called \textit{transformation group $C^{*}$-algebras}, and we may consider the fibration, $p:X\rightarrow G\setminus X$, of $X$ over the space of (left) orbits $G\setminus X$. Since every $G$-orbit $G\cdot x, x\in X$ can be identified with a quotient $G/H_{x}$, where $H_{x}$ is the stabiliser subgroup of $x\in X$, it is possible to associate dynamical systems $(C_{0}(G/H_{G\cdot x}), G, \alpha_{G/H_{G\cdot x}}),\ G\cdot x\in G\setminus X$ with sufficiently regular fibrations.\\
Interestingly, example 3, above, tells us that the associated $\R$-space is not $\R$ acting on itself by translations, but rather $(\beta\R,\R)$, i.e. an action of $\R$ on its Stone-\v{C}ech compactification $\beta\R$, since $C_{b}(\R)\cong C_{0}(\beta\R)$.
\end{Remark}
Coming to the discussion of the crossed product of a dynamical system $(\fA, G, \alpha)$, we indicate that it will be a $C^{*}$-algebra, denoted by $\fA\rtimes_{\alpha}G$, such that its (non-degenerate) representations correspond in a one-to-one fashion to covariant representations of the dynamical system. Furthermore, the crossed product is built in close analogy with the \textit{group $C^{*}$-algebra} $C^{*}(G)$, which turns out to be the special case $\C\rtimes_{\id}G$.
\begin{Definition}[cf. \cite{BratteliOperatorAlgebrasAnd1}, Definition 2.7.2.]
\label{def:l1alg}
Given a dynamical system $(\fA, G, \alpha)$, we denote by $dg$ and $\Delta$, respectively, a (left) Haar measure and its modular function on $G$. The completion of the pre-Banach *-algebra $C_{c}(G,\fA)$, equipped with
\begin{itemize}
	\item[1.] (multiplication, twisted convolution)
	\begin{align}
	\label{eq:l1mult}
	& (x\ast y)(g) := \int_{G}x(h)\alpha_{h}(y(h^{-1}g))dh,
	\end{align}
	\item[2.] (involution)
	\begin{align}
	\label{eq:l1inv}
	& x^{*}(g) := \Delta(g)^{-1}\alpha_{g}(x(g^{-1}))^{*},
	\end{align}
	\item[3.] (norm)
	\begin{align}
	\label{eq:l1norm}
	& ||x||_{1} := \int_{G}||x(h)||_{\fA}dh,\ \ \ x,y\in C_{c}(G,\fA),\ g\in G,
	\end{align}
\end{itemize}
is $L^{1}(G,\fA)$, the \textup{convolution Banach *-algebra} of $(\fA, G, \alpha)$.
\end{Definition}
Next, we need a $C^{*}$-norm on $L^{1}(G,\fA)$ to define $\fA\rtimes_{\alpha}G$.
\begin{Lemma}[cf. \cite{BratteliOperatorAlgebrasAnd1}, p. 138, and \cite{WilliamsCrossedProductsOf}, p. 52]
\label{lem:univnorm}
\begin{align}
\label{eq:univnorm}
||x|| & := \sup\{||\pi(x)||\ |\ \pi:L^{1}(G,\fA)\rightarrow \cB(\fH_{\pi}) - \textup{a Hilbert space representation}\}
\end{align}
defines a $C^{*}$-seminorm on $L^{1}(G,\fA)$, called the \textup{universal norm}. The universal norm $||\ .\ ||$ is dominated by $||\ .\ ||_{1}$. The completion of $L^{1}(G,\fA)$ w.r.t. $||\ .\ ||$ is a $C^{*}$-algebra.
\end{Lemma}
\begin{Definition}
\label{def:crossedprod}
The $C^{*}$-algebra $\overline{L^{1}(G,\fA)}^{||\ .\ ||}$ is called the \textup{($C^{*}$-)crossed product}, $\fA\rtimes_{\alpha}G$, of $(\fA, G, \alpha)$.
\end{Definition}
The one-to-one correspondence between covariant representations $(\fH,\pi,U)$ of $(\fA, G, \alpha)$ and (non-degenerate) representations $(\fH,\rho)$ of $\fA\rtimes_{\alpha}G$ is achieved via the \textit{integrated form} of the former (cf. \cite{WilliamsCrossedProductsOf}, Proposition 2.40.):
\begin{align}
\label{eq:intrep}
\rho(x) & := \int_{G}\pi(x(h))U_{h}dh,\ \ \ x\in C_{c}(G,\fA).
\end{align}
As we already mentioned above, example 2 is closely related to our conjectured new Segal-Bargmann-Hall transform for $L^{2}(G)$, where $G$ is a compact Lie group, which is why we state the following theorem concerning the properties of $C_{0}(G)\rtimes_{\uL}G$, and its representation coming from the covariant representation $(L^{2}(G),\uM,\lambda)$.
\begin{Theorem}[Stone-von Neumann, cf. \cite{WilliamsCrossedProductsOf}, Theorem 4.24.]
\label{thm:svn}
Given a locally compact group $G$, we have
\begin{align}
\label{eq:svn}
C_{0}(G)\rtimes_{\uL}G & \cong\mathcal{K}(L^{2}(G)).
\end{align}
Moreover, the integrated form of the covariant representation $(L^{2}(G), \uM, \lambda)$ of $(C_{0}(G), G, \alpha_{\uL})$ is a faithful irreducible representation of $C_{0}(G)\rtimes_{\uL}G$ onto $\mathcal{K}(L^{2}(G))$\footnote{Recall, that the defining representation is the unique, up to equivalence, irreducible representation of $\mathcal{K}(L^{2}(G))$ (cf. \cite{LandsmanMathematicalTopicsBetween}, Theorem I.2.2.6.).}.
\end{Theorem}

\subsection{Covariant coherent states}
\label{sec:covcs}
In this subsection, we recall the definition of the coherent states given by Hall \cite{HallTheSegalBargmann, HallPhaseSpaceBounds} for compact Lie groups. To this end, let $G$ denote an arbitrary compact Lie group ($\dim G = n$) with Lie algebra $\fg$. On $\fg$, we fix an $Ad$-invariant inner product $\langle\ ,\ \rangle_{\fg}$, e.g. the negative of the Killing form, and we normalise the Haar measure $dg$ on $G$ to coincide with the Riemannian volume measure coming from former. The inner product on $\fg$ gives rise to a Laplace-Beltrami operator $\Delta$ on $G$, which is a Casimir operator for $G$, and the associated heat equation,
\begin{align}
\label{eq:heateq}
\partial_{t}f_{t} & = \frac{1}{2}\Delta f_{t},
\end{align}
has a fundamental solution $\rho_{t},\ t>0$, the \textit{heat kernel}, at the identity in $G$. $\rho_{t}$ has a series expansion in terms of representation theoretical data of $G$:
\begin{align}
\label{eq:heatkernelseries}
\rho_{t}(g) & = \sum_{\pi\in\hat{G}}d_{\pi}e^{-\frac{t}{2}\lambda_{\pi}}\chi_{\pi}(g),
\end{align}
where $\hat{G}$ is the set of isomorphism classes of irreducible unitary representations of $G$, $d_{\pi}$ is the dimension, $\lambda_{\pi}$ the value of the Casimir operator $\Delta$ and $\chi_{\pi}$ the character of $\pi\in\hat{G}$. An important property of $\rho_{t}$ is, that it is a strictly positive $C^{\infty}$-class-function on $G$ for $t>0$.\\
The Lie group $G$ has a unique complexification $G_{\C}$\footnote{In view of the later sections of the article, we mention that the complexifications of $U(1)$ and $SU(2)$ are, respectively, $\C^{*}$ and $SL_{2}(\C)$.} (cf. \cite{BroeckerRepresentationsOfCompact, HallTheSegalBargmann}) with Lie algebra $\fg_{\C}$, which is the complexification of $\fg$. The inner product $\langle\ ,\ \rangle_{\fg}$ extends to a real-valued inner product on $\fg_{\C}$ via
\begin{align}
\label{eq:cxip}
\langle X_{1}+iY_{1},X_{2}+iY_{2}\rangle_{\fg_{\C}} & = \langle X_{1},X_{2}\rangle_{\fg} + \langle Y_{1},Y_{2}\rangle_{\fg},\ \ \ X_{1},X_{2},Y_{1},Y_{2}\in\fg,
\end{align}
giving rise to left-invariant Riemannian metric on $G_{\C}$. As for $G$, the Haar measure $dz$ on $G_{\C}$ is normalised w.r.t. to the Riemannian volume measure coming from $\langle\ ,\ \rangle_{\fg_{\C}}$. There is also a unique antiholomorphic antiautomorphism $\forall z\in G_{\C}:\ z\mapsto z^{*}$ extending the inversion $\forall g\in G:\ g\mapsto g^{-1}$ on $G$. It is related to the inversion on $G_{\C}$ via complex conjugation $\forall z\in G_{\C}:\ z^{*}=\bar{z}^{-1}$.\\
The coherent states for $G$ are constructed as result of the observation that $\rho_{t}$ admits a unique analytic continuation from $G$ to $G_{\C}$ (proved in \cite{HallTheSegalBargmann}). In terms of the series expansion, one makes the
\begin{Definition}
\label{def:cs}
The functions
\begin{align}
\label{eq:cs}
\Psi^{t}_{z}(g) & := \rho_{t}(g^{-1}z)= \sum_{\pi\in\hat{G}}d_{\pi}e^{-\frac{t}{2}\lambda_{\pi}}\chi_{\pi}(g^{-1}z),\ \ \ g\in G, z\in G_{\C}.
\end{align}
are called \textup{(covariant) coherent states} for $G$. Here, ``covariance'' refers to the behaviour of $\Psi^{t}_{z}$ under the (left) regular representation of $G$:
\begin{align}
\label{eq:cscov}
(\lambda_{h}\Psi^{t}_{z})(g) & = \Psi^{t}_{z}(h^{-1}g) \\ \nonumber
 & = \rho_{t}((h^{-1}g)^{-1}z) \\ \nonumber
 & = \rho_{t}(g^{-1}(hz)) \\ \nonumber
 & = \Psi^{t}_{hz}(g).
\end{align}
Clearly, this $G$-action on the set coherent states extends to a (simply) transitive, although non-unitary w.r.t. $L^{2}(G)$, $G_{\C}$-action.
\end{Definition}
Before we comment on further properties of these functions, we need to introduce some further notation (following closely \cite{HallPhaseSpaceBounds}). Namely, we need an analogue of $\rho_{t}$ on the $G_{\C}$-homogeneous space $G_{\C}/G$. The latter admits a unique $G_{\C}$-invariant Riemannian structure, which agrees at the identity coset with the restriction of $\langle\ ,\ \rangle_{\fg_{\C}}$ to $i\fg\subset\fg_{\C}$. The analogue of $\rho_{t}$ is the fundamental solution $\nu_{t}$ at the identity coset of the heat equation,
\begin{align}
\label{eq:dualheateq}
\partial_{t}f_{t} & = \frac{1}{4}\Delta f_{t},
\end{align}
on $G_{\C}/G$. $\nu_{t}$ can be identified with a left and right $G$-invariant function, also denoted $\nu_{t}$, on $G_{\C}$, which is normalised as
\begin{align}
\label{eq:dualnorm}
\int_{G_{\C}}\nu_{t}(z)\ dz & = \vol(G).
\end{align}
In addition to the structures introduced so far, we need some typical objects from the structure theory of compact Lie groups. That is, we fix a maximal torus $T\subset G$ with Lie algebra $\ft$, and real roots $R\subset\ft^{*}\cong\ft$,
\begin{align}
\label{eq:realroots}
\alpha\in R :\Leftrightarrow \alpha\neq0, \exists 0\neq X\in\fg_{\C}: \forall H\in\ft:\ [H,X]=2\pi i\alpha(H)X.
\end{align}
Furthermore, we pick a set of positive roots $R^{+}$ and denote by $\delta=\frac{1}{2}\sum_{\alpha\in R^{+}}\alpha$ half the sum of the positive roots. $W$ is the Weyl group of $T$, $C$ a fundamental Weyl chamber and $\Gamma\subset\ft$ the kernel of the exponential map restricted to $\ft$.\\
For $G_{\C}$, we invoke the (right) polar decomposition $z=ge^{iX}\in G_{\C},\ g\in G,\ X\in\fg$, which gives a diffeomorphism,
\begin{align}
\label{eq:rpolar}
\Phi:T^{*}G\cong G\times\fg\longrightarrow G_{\C},\ \Phi(g,X)=ge^{iX},
\end{align}
that turns the ``phase space'', $T^{*}G$, in a natural way into a K\"ahler manifold. The Haar measure $dz$ on $G_{\C}$ and the Liouville measure $dg\ dX$ on $T^{*}G\cong G\times\fg$ (by right translation), the latter being the product of the Haar measure on $G$ and the Lebesgue $dX$, normalised by means of $\langle\ ,\ \rangle$, on $\fg$, fit together in the following way (cf. \cite{HallPhaseSpaceBounds}, Lemma 5):
\begin{align}
\label{eq:liouville}
\forall f\in C_{c}(G_{\C}):\ \int_{G_{\C}}f(z)\ dz & = \int_{G}\int_{\fg}f(ge^{iX})\ dg\ \sigma(X)\ dX.
\end{align}
Here, $\sigma$ is the $Ad$-$G$-invariant function on $\fg$ determined by
\begin{align}
\label{eq:sigma}
\forall H\in\ft:\ \sigma(H) & = \prod_{\alpha\in R^{+}}\left(\frac{\sinh\alpha(H)}{\alpha(H)}\right)^{2}.
\end{align}
The explicit formula for the measure $\nu_{t}(z)\ dz$ under (right) polar decomposition is (cf. \cite{HallPhaseSpaceBounds}, Lemma 5):
\begin{align}
\label{eq:nupolar}
\nu_{t}(z)\ dz & = \frac{1}{(\pi t)^{\frac{n}{2}}}e^{-|\delta|^{2}_{\fg^{*}}t}e^{-\frac{1}{t}|X|^{2}_{\fg}}\eta(X)\ dg\ dX,
\end{align}
where $\eta$ is an analytic square root of $\sigma$:
\begin{align}
\label{eq:eta}
\forall H\in\ft:\ \eta(H) & = \prod_{\alpha\in R^{+}}\frac{\sinh\alpha(H)}{\alpha(H)}.
\end{align}
Finally, it is important to observe that every element $z\in G_{\C}$ has as decomposition of the form
\begin{align}
\label{eq:adjointdecomp}
z & =ge^{iH}h,\ \ \ g,h\in G, H\in\ft, 
\end{align}
because every element in $G$ is conjugate to some element in the maximal torus $T$.\\[0.1cm]
Let us now come back to the coherent states \eqref{eq:cs} and their properties. Firstly, they belong to the Hilbert space $L^{2}(G)$, which follows from the (analytically continued) heat kernel identity (cf. \cite{HallTheSegalBargmann}, Theorem 6):
\begin{align}
\label{eq:csprod}
(\Psi^{t}_{z},\Psi^{t}_{z'})_{L^{2}} & = \int_{G}\overline{\rho_{t}(g^{-1}z)}\rho_{t}(g^{-1}z')\ dg \\ \nonumber
 & = \rho_{2t}(z'^{-1}\bar{z})<\infty.
\end{align}
Moreover, one has an explicit formula for the norm of the coherent states (cf. \cite{HallPhaseSpaceBounds}, Eq. 8) derived from Urakawa's Poisson summation formula for the restriction of $\rho_{t}$ to the maximal torus $T$ \cite{UrakawaTheHeatEquation}:
\begin{align}
\label{eq:csnorm}
(\Psi^{t}_{z},\Psi^{t}_{z})_{L^{2}} & \ \ =\ \ \rho_{2t}(z^{-1}\bar{z}) = \rho_{2t}((z^{*}z)^{-1}) \\ \nonumber
 & \stackrel{\eqref{eq:adjointdecomp}}{=} \rho_{2t}(h^{-1}e^{-2iH}h) \\ \nonumber
 & \ \ =\ \ \rho_{2t}(e^{-2iH}), \\[0.25cm]
\label{eq:csnormexplicit}
\rho_{2t}(e^{-2iH}) & = \frac{1}{(4\pi t)^{\frac{n}{2}}}\ e^{|\delta|^{2}_{\fg^{*}}}\ e^{\frac{1}{t}|H|^{2}_{\fg}}\ \eta(H)^{-1} \\ \nonumber
 & \hspace{0.5cm}\times\sum_{\gamma_{0}\in\Gamma\cap\overline{C}}e^{i\delta(\gamma_{0})}\ e^{-\frac{1}{4t}|\gamma_{0}|^{2}_{\fg}}\ \frac{\sum_{\gamma\in W\cdot\gamma_{0}}e^{-\frac{i}{t}\langle\gamma,H\rangle_{\fg}}\prod_{\alpha\in R^{+}}\alpha\left(H+\frac{1}{2i}\gamma\right)}{\prod_{\alpha\in R^{+}}\alpha(H)}. 
\end{align}
Secondly, the coherent states provide a resolution of unity in $L^{2}(G)$, thus providing a unitary transformation, the (anti-)Segal-Bargmann-Hall transform,
\begin{align}
\label{eq:sbhtransform}
L^{2}(G) & \rightarrow\overline{\mathcal{H}}L^{2}(G_{\C},\nu_{t}), & \forall z\in G_{\C}:\Phi & \mapsto(\Psi^{t}_{z},\Phi)_{L^{2}},
\end{align}
which maps square integrable functions on $G$ isometrically onto antiholomorphic square integrable, w.r.t. $\nu_{t}(z)\ dz$, functions on $G_{\C}$:
\begin{align}
\label{eq:csres}
\forall \Phi_ {1},\Phi_{2}\in L^{2}(G):\ (\Phi_{1},\Phi_{2})_{L^{2}} & = \int_{G_{\C}}(\Phi_{1},\Psi^{t}_{z})_{L^{2}}(\Psi^{t}_{z},\Phi_{2})_{L^{2}}\ \nu_{t}(z)\ dz.
\end{align}
Alternatively, we will use the mnemonic (in the weak sense, using Dirac's notation):
\begin{align}
\label{eq:csresmnemonic}
\mathds{1} & =  \int_{G_{\C}}\ |\Psi^{t}_{z}\rangle\langle\Psi^{t}_{z}|\ \nu_{t}(z)\ dz.
\end{align}
\subsubsection{A new Segal-Bargmann-Hall ``coherent state'' transform}
\label{sec:newsbh}
Although, the coherent states \eqref{eq:cs} can be thought of as a generalisation of the \textit{standard coherent states} in $L^{2}(\R^{n})$ there is a certain asymmetry, already pointed out in \cite{HallCoherentStatesOn}, which results from the coherent states not being normalised, as would be standard in quantum physical treatments due to the need for a probabilistic interpretation of the ``overlap functions’’ $(\Phi_{1},\Phi_{2})_{L^{2}},\ \Phi_{1},\Phi_{2}\in L^{2}(G)$. Furthermore, the measure $\nu_{t}(z)\ dz$ is not proportional to $t$-scaled Liouville measure $(2\pi t)^{-n}\ dg\ dX$ as one would expect in relation to the ``correspondence principle’’. With regard to the resolution of unity \eqref{eq:csresmnemonic}, one has in the $\R^{n}$-case ($\R^{n}_{\C}=\C^{n},\ z=x+ip$):
\begin{align}
\label{eq:standardres}
\mathds{1} & =\ \ \ \ \int_{\C^{n}}\ |\Psi^{t}_{z}\rangle\langle\Psi^{t}_{z}|\ \nu_{t}(z)\ dz \\ \nonumber
 & = C_{t}\int_{\C^{n}}\ |\Psi^{t}_{z}\rangle\langle\Psi^{t}_{z}|\ (\langle\Psi^{t}_{z}|\Psi^{t}_{z}\rangle)^{-1}\ dz,
\end{align}
where $\Psi^{t}_{z}(x)=(2\pi t)^{-\frac{n}{2}}\ e^{-\frac{1}{2t}(z-x)^{2}},\ \nu_{t}(z)=(4\pi t)^{-\frac{n}{2}}\ e^{-\frac{1}{t}\Im(z)^{2}},\ dz=d^{n}x\ 2^{n}d^{n}p$ and $C_{t}=(4\pi t)^{-n}$. The equality between the first and second line in \eqref{eq:standardres} follows immediately, because
\begin{align}
\label{eq:csnormnu}
C_{t}((\Psi^{t}_{z},\Psi^{t}_{z})_{L^{2}})^{-1}=\nu_{t}(z).
\end{align}
But, the interesting point about \eqref{eq:standardres} is, that we already know that the first line carries over to arbitrary compact Lie groups, while the second line can be written down for arbitrary compact Lie groups as well, as it involves no additional structures. Moreover the constant $C_{t}$ may be compute from the norm of the coherent states $(\Psi^{t}_{z},\Psi^{t}_{z})_{L^{2}}$ as a function of $z\in\C^{n}$, explicitly it is obtained as an integral over the imaginary directions in $\C$:
\begin{align}
\label{eq:imint}
C_{t}^{-1} & = \int_{\R^{n}}\ ((\Psi^{t}_{z},\Psi^{t}_{z})_{L^{2}})^{-1}\ 2^{n}\ d^{n}\Im{z}.
\end{align}
Unfortunately, equation \eqref{eq:csnormnu} is not valid for arbitrary compact Lie groups, but, as we will argue below, the bounds on $(\Psi^{t}_{z},\Psi^{t}_{z})_{L^{2}}$ given in \cite{HallPhaseSpaceBounds} suffice to make sense out of an analogue of the second line of \eqref{eq:standardres}. Having made this observation, we come to the main conjecture of this section.
\begin{Conjecture}
\label{con:csres}
Given an arbitrary Lie group $G$, there exists a resolution of unity
\begin{align}
\label{eq:newcsres}
\mathds{1} & = C_{t}\int_{G}\int_{\fg}\ |\Psi^{t}_{\Phi(g,X)}\rangle\langle\Psi^{t}_{\Phi(g,X)}|\ (\langle\Psi^{t}_{\Phi(g,X)}|\Psi^{t}_{\Phi(g,X)}\rangle)^{-1}\ dg\ dX,\\ \nonumber
C_{t}^{-1} & = \vol(G)\int_{\fg}\ (\langle\Psi^{t}_{\Phi(g,X)}|\Psi^{t}_{\Phi(g,X)}\rangle)^{-1}\ dX \propto t^{-n}.
\end{align}
for small enough $t>0$. For commutative $G$ or $G=SU(2)$ \eqref{eq:newcsres} holds for all $t>0$.
\end{Conjecture}
We notice, that in contrast to \eqref{eq:csresmnemonic} the resolution of unity \eqref{eq:newcsres} lives on the phase space $T^{*}G\cong G\times\fg$, which is natural from a quantum physical perspective.\\[0.1cm]
A possible strategy for a proof could be provided by the fact, that $L^{2}(G)$ is an irreducible representation of the transformation group $C^{*}$-algebra $C(G)\rtimes_{\uL}G$ (see theorem \ref{thm:svn}). Thus, if the operator defined by the right hand side of the first line of \eqref{eq:newcsres} commuted with all representatives of $C(G)\rtimes_{\uL}G$, the conjecture would be (partly) proved by an appeal to Schur's lemma. \\[0.1cm]
Before, we argue in the favour of the conjecture, and prove it for $G=U(1)$, we obtain as a corollary an extension to (connected) Lie groups of compact type, i.e. those that admit an $Ad$-invariant inner product on their Lie algebras. The structure of these Lie groups is clarified by
\begin{Proposition}[cf. \cite{HallGeometricQuantizationAnd}, Proposition 2.2.]
\label{prop:cpttype}
Given a connected Lie group $K$ of compact type with $Ad$-invariant inner product $\langle\ ,\ \rangle$ on its Lie algebra $\fk$, there exists a compact connected Lie group $G$ and natural number $n\in\mathds{N}$, such that
\begin{align}
\label{eq:cpttypeiso}
K & \cong G\times\R^{n}
\end{align}
as Lie groups and the associated Lie algebra isomorphism $\fk\cong\fg\times\R^{n}$ is orthogonal.
\end{Proposition}
This implies
\begin{Corollary}
\label{cor:newcsrescpttype}
The resolution of unity \eqref{eq:newcsres} holds for arbitrary (connected) Lie groups $K\cong G\times\R^{n}$ of compact type, if the coherent states for $K$ are chosen as product states of the coherent states for $G$ and the standard coherent states for $\R^{n}$. The Liouville measure on $K$ is then the product measure of the Liouville measures on $G$ and $\R^{n}$, respectively.
\end{Corollary}
We conclude this section with an extended
\begin{Remark}[on Conjecture \ref{con:csres}]
\label{rem:csres}
From \eqref{eq:liouville}, \eqref{eq:nupolar}, \eqref{eq:csnorm} and \eqref{eq:csnormexplicit}, we have the following formula for the product of $(\Psi^{t}_{z},\Psi^{t}_{z})_{L^{2}}$ and $\nu_{t}(z)$:
\begin{align}
\label{eq:csnuprod}
(\Psi^{t}_{z},\Psi^{t}_{z})_{L^{2}}\ \nu_{t}(z) & \stackrel{z = g e^{iH}h}{=}(2\pi t)^{-n}\sigma(H)^{-1} \\ \nonumber
 &\hspace{0.5cm} \times\sum_{\gamma_{0}\in\Gamma\cap\overline{C}}e^{i\delta(\gamma_{0})}\ e^{-\frac{1}{4t}|\gamma_{0}|^{2}_{\fg}}\ \frac{\sum_{\gamma\in W\cdot\gamma_{0}}e^{-\frac{i}{t}\langle\gamma,H\rangle_{\fg}}\prod_{\alpha\in R^{+}}\alpha\left(H+\frac{1}{2i}\gamma\right)}{\prod_{\alpha\in R^{+}}\alpha(H)}.
\end{align}
Now, Hall shows in \cite{HallPhaseSpaceBounds}, Proposition 3, that the absolute value of sum over $\gamma\in W\cdot\gamma_{0}$ is bounded from above by an expression $P(|\gamma_{0}|_{\fg}/\sqrt{t}) \prod_{\alpha\in R^{+}}\alpha(H)$, where $P$ is a polynomial of degree equal to twice the number of positive roots $\#R^{+}$. This, immediately, leads to the conclusion that, for small enough $t>0$, we have constants $a_{t},\ b_{t}>0$ with $\lim_{t\rightarrow 0+}a_{t}=1=\lim_{t\rightarrow 0+}b_{t}$ exponentially fast\footnote{For $G$ commutative or $G=SU(2)$ Hall shows that $a_{t},\ b_{t}$ exist for all $t>0$.}, such that
\begin{align}
\label{eq:measureequiv}
(2\pi t)^{-n}b_{t}\leq (\Psi^{t}_{\Phi(g,X)},\Psi^{t}_{\Phi(g,X)})_{L^{2}}\ \nu_{t}(X)\ \sigma(X)\leq(2\pi t)^{-n}a_{t}.
\end{align}
This shows the equivalence of the measures $\nu_{t}(X)\sigma(X)\ dg\ dX$ and $((\Psi^{t}_{\Phi(g,X)},\Psi^{t}_{\Phi(g,X)})_{L^{2}})^{-1}\ dg\ dX$ on $G\times\fg$, and due to the finiteness and positivity of $a_{t}, b_{t}$, we know that the integrals in \eqref{eq:newcsres} makes sense. Namely, for all $\Phi_{1}\in L^{2}(G)$ we have:
\begin{align}
\label{eq:l2normequiv}
 &  & \frac{(2\pi t)^{n}}{a_{t}}||(\Psi^{t}_{(\ .\ )},\Phi_{1})_{L^{2}}||^{2}_{L^{2}(G_{\C},\nu_{t})} & \leq||(\Psi^{t}_{\Phi(\ .\ )},\Phi_{1})_{L^{2}}||^{2}_{L^{2}(G\times\fg,(||\Psi^{t}||^{2}_{L^{2}})^{-1})} \\ \nonumber
 &  &  & \leq\frac{(2\pi t)^{n}}{b_{t}}||(\Psi^{t}_{(\ .\ )},\Phi_{1})_{L^{2}}||^{2}_{L^{2}(G_{\C},\nu_{t})}\\[0.1cm] \nonumber
 & \Rightarrow\ & \ (\Psi^{t}_{(\ .\ )},\Phi_{1})_{L^{2}}\in \overline{\mathcal{H}}L^{2}(G_{\C},\nu_{t}) & \\ \nonumber
 & \Leftrightarrow\ & \ (\Psi^{t}_{\Phi(\ .\ )},\Phi_{1})_{L^{2}}\in \overline{\mathcal{H}}L^{2}(G\times\fg,(||\Psi^{t}_{\Phi(\ .\ )}||^{2}_{L^{2}})^{-1})
\end{align}
Next, we analyse the operator
\begin{align}
\label{eq:Aoperator}
A_{t} & := C_{t}\int_{G}\int_{\fg}\ |\Psi^{t}_{\Phi(g,X)}\rangle\langle\Psi^{t}_{\Phi(g,X)}|\ (\langle\Psi^{t}_{\Phi(g,X)}|\Psi^{t}_{\Phi(g,X)}\rangle)^{-1}\ dg\ dX
\end{align}
in some detail. To this end, we introduce the representative functions
\begin{align}
\label{eq:repfunc}
\forall\pi\in\hat{G},\ m,n=1,...,d_{\pi}:\ \langle\pi,m,n|g\rangle=\pi(g)_{mn},\ \ \ g\in G,
\end{align}
and find, because $\langle\pi,m,n|\Psi^{t}_{\Phi(g,X)}\rangle=e^{-\frac{t}{2}\lambda_{\pi}}\pi(ge^{iX})_{mn}$,
\begin{align}
\label{eq:repfuncexp}
 & \hspace{-0.75cm}\langle\pi,m,n|A_{t}|\pi',m',n'\rangle \\ \nonumber
 & = C_{t}\int_{G}\int_{\fg}\ \langle\pi,m,n|\Psi^{t}_{\Phi(g,X)}\rangle\langle\Psi^{t}_{\Phi(g,X)}|\pi',m',n'\rangle\ (\langle\Psi^{t}_{\Phi(g,X)}|\Psi^{t}_{\Phi(g,X)}\rangle)^{-1}\ dg\ dX \\ \nonumber
 & = C_{t}\int_{G}\int_{\fg}(\rho_{2t}(e^{-i2X}))^{-1}e^{-\frac{t}{2}(\lambda_{\pi}+\lambda_{\pi'})}\pi(ge^{iX})_{mn}\overline{\pi'(ge^{iX})_{m'n'}}\ dg\ dX \\ \nonumber
 & = C_{t}\int_{\fg} (\rho_{2t}(e^{-i2X}))^{-1}e^{-\frac{t}{2}(\lambda_{\pi}+\lambda_{\pi'})}\sum_{k,k'=1}^{d_{\pi},d_{\pi'}}\pi(e^{iX})_{kn}\overline{\pi'(e^{iX})_{k'n'}}\int_{G}\pi(g)_{mk}\pi'(g)_{m'k'}\ dg\ dX \\ \nonumber
 & = C_{t}\vol(G)\ d_{\pi}^{-1}\ \delta_{\pi,\pi'}\ \delta_{m,m'}\int_{\fg}(\rho_{2t}(e^{-i2X}))^{-1}e^{-t\lambda_{\pi}}\pi(e^{i2X})_{n'n}\ dX.
\end{align}
Defining an operator $A^{t}_{\pi}\in\End(V_{\pi})$ by the matrix elements
\begin{align}
\label{eq:Aoperatorred}
(A^{t}_{\pi})_{n'n} & = \int_{\fg}(\rho_{2t}(e^{-i2X}))^{-1}e^{-t\lambda_{\pi}}\pi(e^{i2X})_{n'n}\ dX,\ \ \ n,n'=1,...,d_{\pi},
\end{align}
we see that it commutes with $\pi(g), g\in G$, because $\rho_{2t}$ is an even class function and $dX$ is $Ad$-invariant:
\begin{align}
\label{eq:Aoperatorintertwine}
(\pi(g)A^{t}_{\pi})_{mn} & = \sum_{k=1}^{d_{\pi}}\pi(g)_{mk}(A^{t}_{\pi})_{kn} = \sum_{k=1}^{d_{\pi}}\pi(g)_{mk}\int_{\fg}(\rho_{2t}(e^{-i2X}))^{-1}e^{-t\lambda_{\pi}}\pi(e^{i2X})_{kn}\ dX \\ \nonumber
 & = \sum_{k=1}^{d_{\pi}}\int_{\fg}(\rho_{2t}(e^{-i2X}))^{-1}e^{-t\lambda_{\pi}}\pi(ge^{i2X}g^{-1})_{mk}\pi(g)_{kn}\ dX \\ \nonumber
 & = \sum_{k=1}^{d_{\pi}}\int_{\fg}(\rho_{2t}(e^{-i2X}))^{-1}e^{-t\lambda_{\pi}}\pi(e^{i2 Ad_{g}(X)})_{mk}\pi(g)_{kn}\ dX \\ \nonumber
 & = \sum_{k=1}^{d_{\pi}}\int_{\fg}(\rho_{2t}(e^{-i2 Ad_{g^{-1}}(X)}))^{-1}e^{-t\lambda_{\pi}}\pi(e^{i2X})_{mk}\pi(g)_{kn}\ dX \\ \nonumber
 & = \sum_{k=1}^{d_{\pi}}\int_{\fg}(\rho_{2t}(g^{-1}e^{-i2X}g))^{-1}e^{-t\lambda_{\pi}}\pi(e^{i2X})_{mk}\pi(g)_{kn}\ dX \\ \nonumber
 & = \sum_{k=1}^{d_{\pi}}\int_{\fg}(\rho_{2t}(e^{i2X}))^{-1}e^{-t\lambda_{\pi}}\pi(e^{i2X})_{mk}\pi(g)_{kn}\ dX \\ \nonumber
 & = \sum_{k=1}^{d_{\pi}}(A^{t}_{\pi})_{mk}\pi(g)_{kn} = (A^{t}_{\pi}\pi(g))_{mn}.
\end{align}
Thus, by Schur's lemma we have $A^{t}_{\pi}=d^{-1}_{\pi}\tr(A^{t}_{\pi})\mathds{1}$, and our conjecture is equivalent to the formula:
\begin{align}
\label{eq:traceformula}
C^{-1}_{t}d_{\pi}=\tr(A^{t}_{\pi}) & =\int_{\fg}(\rho_{2t}(e^{i2X}))^{-1}e^{-t\lambda_{\pi}}\chi_{\pi}(e^{i2X})\ dX = \int_{\fg}\frac{e^{-t\lambda_{\pi}}\chi_{\pi}(e^{i2X})}{\sum_{\pi'\in\hat{G}}d_{\pi'}e^{-t\lambda_{\pi'}}\chi_{\pi'}(e^{i2X})}\ dX \\ \nonumber
 & = \int_{\fg}\frac{\chi_{\pi}(e^{t\Delta}e^{i2X})}{\sum_{\pi'\in\hat{G}}d_{\pi'}\chi_{\pi'}(e^{t\Delta}e^{i2X})}\ dX = \int_{\fg}\frac{\chi_{\pi}(e^{t\Delta + i2X})}{\sum_{\pi'\in\hat{G}}d_{\pi'}\chi_{\pi'}(e^{\Delta + i2X})}\ dX \\ \nonumber
 & = \int_{\fg}\frac{\chi_{\pi}(e^{\frac{(t\tau + iX)^{2}}{t}+\frac{X^{2}}{t}})}{\sum_{\pi'\in\hat{G}}d_{\pi'}\chi_{\pi'}(e^{\frac{(t\tau + iX)^{2}}{t}+\frac{X^{2}}{t}})}\ dX = \int_{\fg}\frac{\chi_{\pi}(e^{\frac{(t\tau + iX)^{2}}{t}})}{\sum_{\pi'\in\hat{G}}d_{\pi'}\chi_{\pi'}(e^{\frac{(t\tau + iX)^{2}}{t}})}\ dX \\ \nonumber
 & = \int_{\fg}\frac{\tr_{V_{\pi}}(e^{\frac{(t\tau + iX)^{2}}{t}})}{\tr_{L^{2}(G)}(e^{\frac{(t\tau + iX)^{2}}{t}})}\ dX.
\end{align}
Here, we introduced the object $\tau=\sum_{i=1}^{n}\tau_{i}\otimes\tau_{i}$ for some orthonormal basis $\{\tau_{i}\}_{i=1,...,n}\subset\fg$, such that $\tau\cdot\tau=\sum_{i,j=1}^{n}\tau_{i}\tau_{j}\langle\tau_{i},\tau_{j}\rangle_{\fg}=\sum_{i=1}^{n}\tau_{i}^{2}=\Delta$ and $\tau\cdot X=\sum_{i=1}^{n}\tau_{i}X_{j}\langle\tau_{i},\tau_{j}\rangle_{\fg}$. From a physicist's point of view, the last line in \eqref{eq:traceformula} is especially attractive, because it resembles the average over $\fg$ of the Boltzmann-like distribution
\begin{align}
\label{eq:boltzmann}
p_{(V_{\pi}|L^{2}(G))}(X) & := \frac{\tr_{V_{\pi}}(e^{\frac{(t\tau + iX)^{2}}{t}})}{\tr_{L^{2}(G)}(e^{\frac{(t\tau + iX)^{2}}{t}})},\ \sum_{\pi\in\hat{G}}d_{\pi}p_{(V_{\pi}|L^{2}(G))}(X)=1\ \ \ X\in\fg.
\end{align}
Since characters, $\chi_{\pi},\ \pi\in\hat{G}$, and the (analytically continued) heat kernel $\rho_{t}$ are class functions, we may further simplify the expression \eqref{eq:traceformula} by Weyl's integration formula on $\fg$ (cf. \cite{DooleyHarmonicAnalysisAnd}):
\begin{align}
\label{eq:weylint}
\forall f\in C_{c}(\fg):\ \int_{\fg}f(X)\ dX & = \int_{C^{*}}\prod_{\alpha\in R^{+}}\alpha(H)^{2}\int_{G}f(Ad_{g}(H))\ dg\ dH \\ \nonumber
 & = \frac{1}{|W|} \int_{\ft}\prod_{\alpha\in R^{+}}\alpha(H)^{2}\int_{G}f(Ad_{g}(H))\ dg\ dH.
\end{align}
Here, $C^{*}\subset\ft$ corresponds to the Weyl chamber $C$ via $\langle\ ,\ \rangle$, $|W|$ is the number of elements in $W$, and $dH$ is a suitably normalised Lebesgue measure on $\ft$. Thus, we find for \eqref{eq:traceformula}:
\begin{align}
\label{eq:traceformulatorus}
C^{-1}_{t}d_{\pi}=\tr(A^{t}_{\pi}) & =\int_{\fg}(\rho_{2t}(e^{i2X}))^{-1}e^{-t\lambda_{\pi}}\chi_{\pi}(e^{i2X})\ dX \\ \nonumber
 & = \frac{\vol(G)}{|W|}\int_{\ft}\left(\prod_{\alpha\in R^{+}}\alpha(H)^{2}\right)(\rho_{2t}(e^{i2H}))^{-1}e^{-t\lambda_{\pi}}\chi_{\pi}(e^{i2H})\ dH.
\end{align}
To provide further evidence for our conjecture, we will prove it for $G=U(1)$, which also implies the conjecture for $G=U(1)^{n},\ n\in\mathds{N}$. Subsequently, we give numerical results that suggest the validity of the conjecture for $G=SU(2)$.\\[0.2cm]
If $G=U(1)$, we have $G_{\C}=\C^{*}$. The coherent state overlap function is found to be (cf. \cite{KowalskiCoherentStatesFor}):
\begin{align}
\label{eq:u1overlap}
\langle\Psi^{t}_{\xi}|\Psi^{t}_{\xi}\rangle & = \sum_{j\in\mathds{Z}}e^{-t j^{2}}e^{2jl} = \vartheta_{3}\left(\frac{il}{\pi}\right|\left.\frac{it}{\pi}\right) = \vartheta_{3}\left(\frac{l}{t}\right|\left.\frac{i\pi}{t}\right)\sqrt{\frac{\pi}{t}}e^{\frac{1}{t}l^{2}},
\end{align}
where we chose coordinates $\xi=e^{i(\varphi+il)},\ (\varphi,l)\in[0,2\pi)\times\R$ and $\vartheta_{3}$ denotes the third Jacobi theta function. The characters of the irreducible representations of $U(1)$ can be labeled by $j\in\mathds{Z}$, $\chi_{j}(\xi)=\xi^{l},\ d_{j}=1$, and the eigenvalues of the Casimir operator are $\lambda_{j}=j^{2}$. Putting everything together, \eqref{eq:traceformula} gives:
\begin{align}
\label{eq:u1traceformula}
C^{-1}_{t} & = \int_{\R}e^{-t j^{2}}e^{-2jl}\left(\vartheta_{3}\left(\frac{l}{t}\right|\left.\frac{i\pi}{t}\right)\sqrt{\frac{\pi}{t}}e^{\frac{1}{t}l^{2}}\right)^{-1}dl = \int_{\R}\sqrt{\frac{t}{\pi}}e^{-\frac{1}{t}(l+tj)^{2}}\left(\vartheta_{3}\left(\frac{l}{t}\right|\left.\frac{i\pi}{t}\right)\right)^{-1}dl \\ \nonumber
 & = \int_{\R}\sqrt{\frac{t}{\pi}}e^{-\frac{1}{t}l^{2}}\left(\vartheta_{3}\left(\frac{l}{t}\right|\left.\frac{i\pi}{t}\right)\right)^{-1}dl = t\int_{\R}\left(\sum_{n\in\mathds{Z}}e^{(l+i\frac{\pi n}{\sqrt{t}})^{2}}\right)^{-1}dl.
\end{align}
The last line follows from the $\mathds{Z}$-invariance of $\vartheta_{3}$ in the first argument. Clearly, this establishes the conjecture for $G=U(1)$. A Numerical evaluation of \eqref{eq:u1traceformula} indicates $C_{t}=\frac{1}{t}$. \\[0.2cm]
For $G=SU(2)$, the formulas become slightly more involved, although we end up with a \mbox{1-dimensional} integral as $\rank(SU(2))=1$. The characters of irreducible representations of $SU(2)$ are labeled by $n\in\mathds{N}$, $\chi_{n}(e^{2iH})=\frac{\sinh(2np)}{\sinh(2p)}$, for some suitable coordinate $p$ on $\ft\cong\R$. The dimension and the eigenvalue of the Casimir operator are $d_{n}=n$ and $\lambda_{n}=\frac{n^{2}-1}{4}$, respectively. The positive root is given by $\alpha(H)=p$. This allows for the computation of all objects involved in \eqref{eq:traceformulatorus}:
\begin{align}
\label{eq:su2traceformula}
C^{-1}_{t}d_{n} & = \frac{2\pi^{2}}{2}\int_{\R}p^{2}e^{-\frac{t}{4}(n^{2}-1)}\frac{\sinh(2np)}{\sinh(2p)}\left(\sum_{m\in\mathds{N}}m e^{-\frac{t}{4}(m^{2}-1)}\frac{\sinh(2mp)}{\sinh{2p}}\right)^{-1}\!\!\!dp \\ \nonumber
 & = \frac{\pi^{2}}{4}\int_{\R}p^{2}e^{-\frac{1}{t}(p-\frac{t}{2}n)^{2}}\!\left(\sum_{m\in\mathds{Z}}m e^{-\frac{1}{t}(p-\frac{t}{2}m)^{2}}\!\right)^{-1}\!\!\!\!\!\!dp \\ \nonumber
 & = \frac{\pi^{2}}{4}\int_{\R}\left(p+\frac{t}{2}n\right)^{2}\!\!e^{-\frac{1}{t}p^{2}}\!\left(\sum_{m\in\mathds{Z}}(m+n) e^{-\frac{1}{t}(p-\frac{t}{2}m)^{2}}\!\right)^{-1}\!\!\!\!\!\!dp \\ \nonumber
 & = \frac{\pi^{2}}{4}\int_{\R}p^{2}e^{-\frac{1}{t}(p-\frac{t}{2}n)^{2}}\left(\!e^{-\frac{1}{t}p^{2}}(\partial_{p}\vartheta_{3})\left(\frac{p}{2\pi i}\right|\left.\frac{it}{4\pi}\right)\!\right)^{-1}\!\!\!\!dp \\ \nonumber
 & = i\frac{\pi^{3}}{2}\int_{\R}p^{2}e^{-\frac{1}{t}(p-\frac{t}{2}n)^{2}}\left(\!e^{-\frac{1}{t}p^{2}}\vartheta'_{3}\left(\frac{p}{2\pi i}\right|\left.\frac{it}{4\pi}\right)\!\right)^{-1}\!\!\!\!dp.
\end{align}
Again, we need to prove an integral formula involving the Jacobi theta function $\vartheta_{3}$. Only this time, we have do deal with the derivative of $\vartheta_{3}$, which is why a simple shift $p\mapsto p+\frac{t}{2}n$ does not suffice to prove the formula. Nevertheless, a numerical evaluation of the integral (see table \ref{tab:su2int} and figures \ref{fig:su2int2d} \& \ref{fig:su2int3d})
\begin{align}
\label{eq:su2int}
I(t,n) & := 2i\int_{\R}p^{2}e^{-\frac{1}{t}(p-\frac{t}{2}n)^{2}}\left(e^{-\frac{1}{t}p^{2}}\vartheta'_{3}\left(\frac{p}{2\pi i}\right|\left.\frac{it}{4\pi}\right)\right)^{-1}\!\!\!dp
\end{align}
hints at the correctness of conjecture \ref{con:csres} for $G=SU(2)$, since the integral relation between $I(t,n)=n I(t,1)$ seems to be a rather strong requirement. Furthermore, table \ref{tab:su2int} indicates the relation $I(t,n)=t^{3}I(1,n)$, and thus $I(t,n) = \frac{t^{3}n}{8}\ \Rightarrow C_{t}=\frac{24}{(\pi t)^3}$.
\end{Remark}
\begin{table}[h]
\centering
\begin{ruledtabular}
\begin{tabular}{|d|d|d|d|d|d|}
	\text{\backslashbox{t}{n}} & 1 & 2 & 3 & 4 & 5 \\ \hline
	1 & 0.125 & 0.25 & 0.375 & 0.5 & 0.625 \\
	2 & 1 & 2 & 3 & 4 & 5 \\
	e & \sim2.51069 & \sim5.02138 & \sim7.53208 & \sim10.0428 & \sim12.5535 \\
	\pi & \sim3.87578 & \sim7.75157 & \sim11.6274 & \sim15.5031 & \sim19.3789 \\
	4 & 8 & 16 & 24 & 32 & 40 \\
\end{tabular}
\end{ruledtabular}
\caption{\label{tab:su2int}Numerical evaluation of $I(t,n)$.}
\end{table}
\begin{figure}[h]
\centering
	\includegraphics[width=\textwidth]{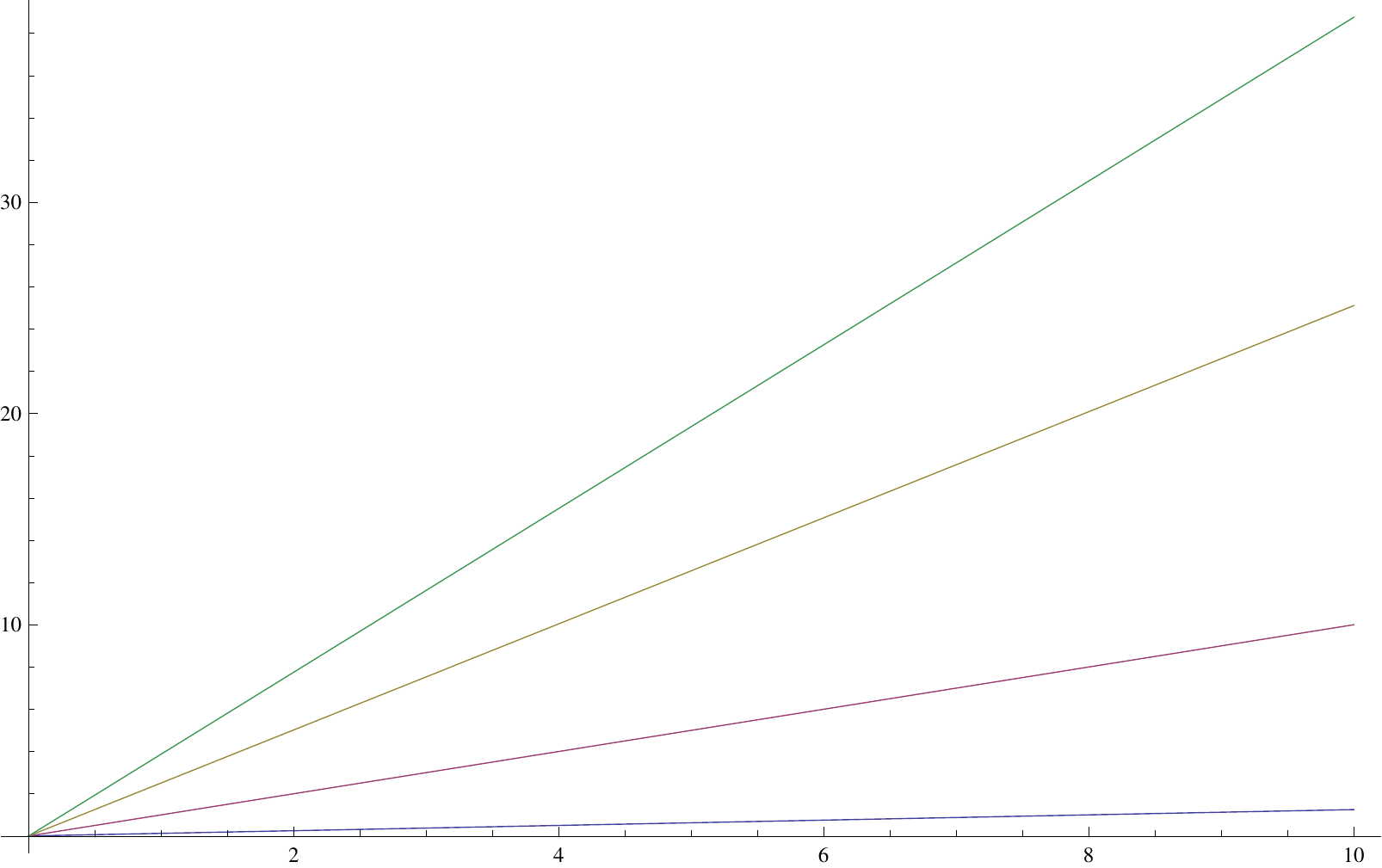}
	\caption{\label{fig:su2int2d}Plots of $I(t,n)$ ($t$ fixed): $I(1,n)$, $I(2,n)$, $I(e,n)$ and $I(\pi,n)$ (bottom to top).}
\end{figure}
\begin{figure}[h]
	\includegraphics[width=\textwidth]{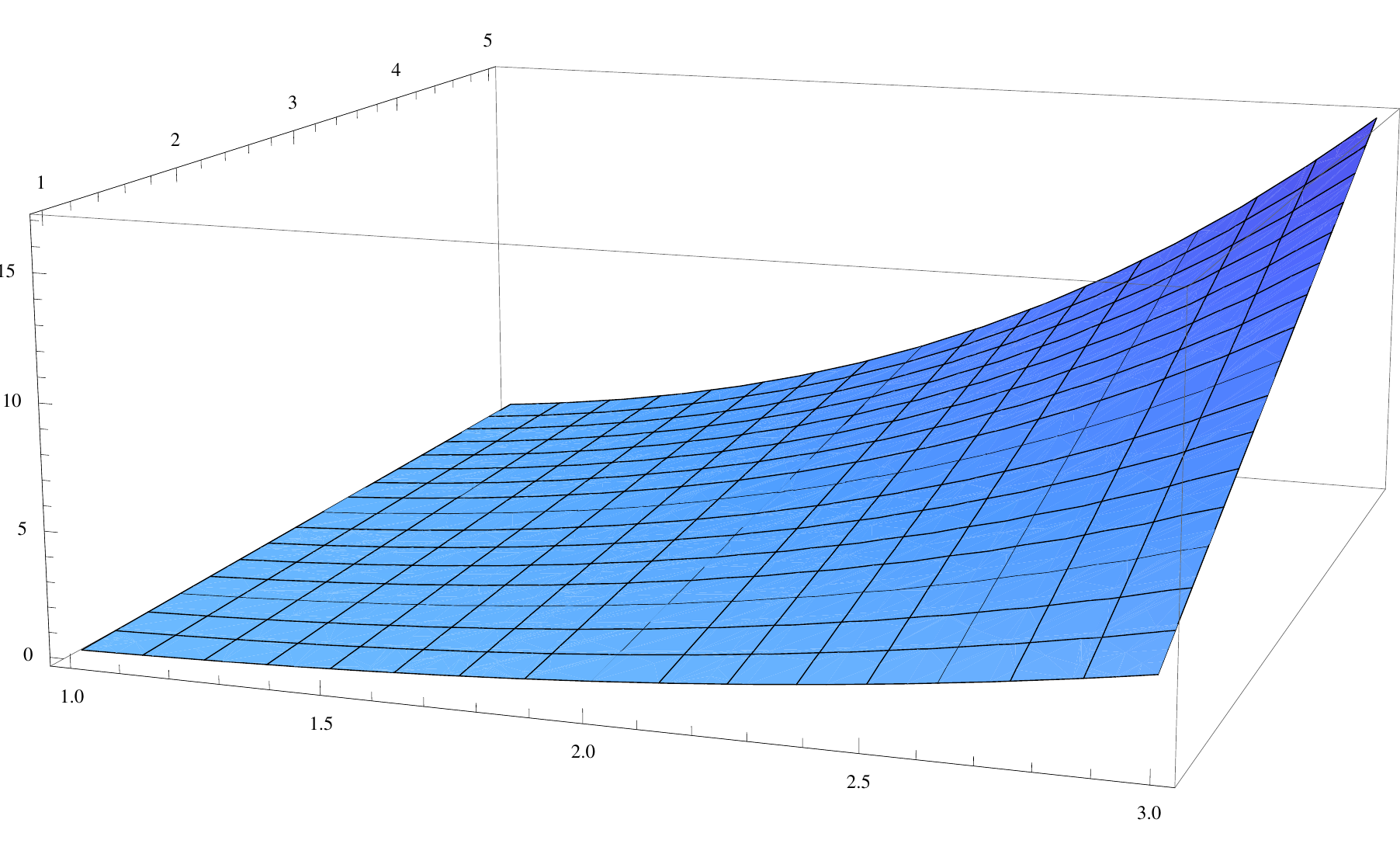}
	\caption{\label{fig:su2int3d}3D Plot of I(n,t).}
\end{figure}
\section{On Weyl  and Kohn-Nirenberg calculi for compact Lie groups}
\label{sec:wc}
In this section, we discuss local and global Weyl and Kohn-Nirenberg calculi for compact Lie groups $G$ ($\dim G = n$) to provide a (pseudo-differential) framework for the Born-Oppenheimer approximation \cite{MartinezOnAGeneral} or space-adiabatic perturbation theory \cite{PanatiSpaceAdiabaticPerturbation, TeufelAdiabaticPerturbationTheory} in loop quantum gravity (see also section II of the third article in this series\cite{StottmeisterCoherentStatesQuantumIII}). The need for local as well as global calculi is due to the fact that the exponential map is, while still onto, no longer a diffeomorphism for compact groups. Both, local and global calculi, are advantageous in certain situations: On the one hand, it is quite natural to handle the ``semi-classical limit’’ of the (quantum) commutation relations,
\begin{align}
\label{eq:quantcomm}
[f,f'] & =0, \\ \nonumber
[P_{X},f] & = - i\varepsilon\ R_{X}f, \\ \nonumber
[P_{X},P_{Y}] & = i\varepsilon\ P_{[X,Y]},
\end{align}
where $f,f'\in C^{\infty}(G),\ X,Y\in\fg$ and $R_{X}f=\frac{d}{dt}_{|t=0}L^{*}_{e^{tX}}f$, in local calculi via the Baker-Campbell-Hausdorff formula, on the other hand, the composition and computation of symbols of operators is simpler for the global calculi, and the class of admissible symbols is larger.\\
The local calculi are based on a generalised Weyl quantisation introduced by Landsman (cf. \cite{LandsmanStrictdeformationQuantization, LandsmanMathematicalTopicsBetween}) based on (strict) Rieffel deformations (cf. \cite{RieffelLieGroupConvolution, RieffelDeformationQuantizationAnd}), while the global calculi are closely related to the Kohn-Nirenberg calculus of Ruzahnsky and Turunen (cf. \cite{RuzhanskyPseudoDifferentialOperators, RuzhanskyGlobalQuantizationOf}). In contrast to those existing accounts on (pseudo-differential) quantisation on compact Lie groups, we arrive at the calculi from the perspective of dequantisation of the transformation group $C^{*}$-algebra $C(G)\rtimes_{\uL}G\cong\mathcal{K}(L^{2}(G))$, which is a natural quantum algebra over a single edge of loop quantum gravity (see section II of the third article in this series\cite{StottmeisterCoherentStatesQuantumIII}). Furthermore, it is important to note, from the point of view of applications, that the Weyl calculi are favored over the Kohn-Nirenberg calculi, because the former are real, i.e. hermitean/self-adjoint operators tend to have hermitean/self-adjoint symbols (cf. \cite{StottmeisterCoherentStatesQuantumI}).\\
We start, in subsection \ref{subsubsec:globalcalc}, with the definition of the global calculi. In subsection \ref{subsubsec:pwsymbols} we introduce the local calculi and the ``Paley-Wiener-Schwartz’’ symbol spaces $S^{K,m}_{\rho,\delta}$. For the latter, we prove a completeness result w.r.t. asymptotic expansions by adopting the method of kernel cut-off operators from the calculus of Volterra-Mellin operators (cf. \cite{KrainerVolterraFamiliesOf, KrainerTheCalculusOf}). For the global calculi we provide a reformulation in terms of the Stratonovich-Weyl transform of Figueroa, Gracia-Bond{\'i}a and V{\'a}rilly (cf. \cite{FigueroaMoyalQuantizationWith, VarillyTheMoyalRepresentation}) in subsection \ref{subsec:swt}, which additionally gives rise to a scaled $\varepsilon$-scaled integral transform on $G$. As a byproduct, we prove strictness of the Stratonovich-Weyl quantisation.\\
Following this, we comment on the relation of the calculi to coherent state quantisation in subsection \ref{subsec:csquant}. We conclude the section with a closer look a the special case $G=U(1)$ and a possible extension of the global calculi to $G=\R_{\textup{Bohr}}$, where we will argue that the global calculus is suitable for dealing with symbols that are not analytic in the momenta.
\subsection{Pseudo-Differential Operators on Compact Lie groups}
\label{sec:compact}
Before we state the definitions for the types of pseudo-differential operators on compact Lie groups that we intend to discuss, we start with a (informal) motivation, i.e. we refrain from defining the function spaces on which the following formulae will be well-defined: \\[0.1cm]
For quantum mechanics on $\R^{n}$ the commutation relations \eqref{eq:quantcomm} correspond to the standard commutation relations for position and momentum,
\begin{align}
\label{eq:canonicalcomm}
[Q_{i},Q_{j}] & = 0 = [P_{i},P_{j}] \\ \nonumber
[P_{i},Q_{j}] & = -i\varepsilon\ \delta_{i,j},\ \ \ \forall i,j=1,...,n,
\end{align}
which are often presented in their Weyl form obtained by considering the (formal) exponentials $W(x,\xi):=e^{i(x\cdot Q+\xi\cdot P)},$ $x,\xi\in\R^{n},$ (\textit{Weyl elements}). In the Schrödinger representation on $L^{2}(\R^{n})$ the action of the Weyl elements is defined to be:
\begin{align}
\label{eq:stweyl}
(W(x,\xi)\Psi)(q) & = e^{\frac{i\varepsilon}{2}x\cdot\xi}e^{i\xi\cdot q}\Psi(q+\varepsilon\xi),\ \ \ \Psi\in L^{2}(\R^{n}).
\end{align}
Moreover, the Weyl elements provide a means of quantising functions $\sigma$ on phase space $T^{*}\R^{n}\cong\R^{n}\times\R^{n}$ to operators $A_{\sigma}$ on $L^{2}(\R^{n})$ by Fourier transformation, i.e.:
\begin{align}
\label{eq:stweylquant}
A_ {\sigma} & := \frac{1}{(2\pi)^{n}}\int_{\R^{2n}}dx\ d\xi\ \mathcal{F}[\sigma](x,\xi)W(x,\xi) = \int_{\R^{2n}}dq\ dp\ \sigma(q,p)\hat{W}(q,p) = (\hat{W},\sigma)_{L^{2}(\R^{2n})},
\end{align}
where we introduce the \textit{Fourier-Weyl elements} $\hat{W}(q,p) := \frac{1}{(2\pi)^{2n}}\int_{\R^{2n}}dx\ d\xi\ e^{-i(x\cdot q+\xi\cdot p)}W(x,\xi)$, $\hat{W}(q,p)^{*}=\hat{W}(q,p)$, and $\mathcal{F}[\sigma](x,\xi)=\frac{1}{(2\pi)^{n}}\int_{\R^{2n}}dq\ dp\ \sigma(q,p)e^{-i(x\cdot q+\xi\cdot p)}$ is the Fourier transform of $\sigma$. Combining \eqref{eq:stweyl} and \eqref{eq:stweylquant}, we find
\begin{align}
\label{eq:stweylpseudo}
(A_{\sigma}\Psi)(q) & = \frac{1}{(2\pi\varepsilon)^{n}}\int_{\R^{2n}}dx\ d\xi\ \sigma(\tfrac{1}{2}(q+x),\xi)e^{\frac{i}{\varepsilon}\xi\cdot(q-x)}\Psi(x).
\end{align}
Thus, $A_{\sigma}$ is pseudo-differential operator on $L^{2}(\R^{n})$ with \textit{amplitude} oder \textit{symbol} $\sigma$. In addition to the quantisation formula \eqref{eq:stweylquant}, the Fourier-Weyl elements also give rise to a useful formula for the dequantisation of an operator $A$ on $L^{2}(\R^{n})$, i.e. finding a symbol $\sigma_{A}$ s.t. $A_{\sigma_{A}}=A$, because of the (distributional) orthogonality relations
\begin{align}
\label{eq:stweyltrace}
\left(\frac{\varepsilon}{2\pi}\right)^{n}(W(x,\xi),W(x',\xi'))_{HS} & = \left(\frac{\varepsilon}{2\pi}\right)^{n}\tr(W(x,\xi)^{*}W(x',\xi')) = \delta^{(n)}(x-x')\delta^{(n)}(\xi-\xi') \\ \nonumber
(2\pi\varepsilon)^{n}(\hat{W}(q,p),\hat{W}(q',p'))_{HS} & = (2\pi\varepsilon)^{n}\tr(\hat{W}(q,p)^{*}\hat{W}(q',p')) = \delta^{(n)}(q-q')\delta^{(n)}(p-p').
\end{align}
Applying \eqref{eq:stweyltrace} to the product $A_{\rho}=A_{\sigma}A_{\tau}$ of two operators $A_{\sigma}$ and $A_{\tau}$, we find the well-known formula for the \textit{twisted or Moyal product} $\rho=\sigma\star_{\varepsilon}\tau$ of the symbols $\sigma$ and $\tau$, and its asymptotic expansion (cf. \cite{FollandHarmonicAnalysisIn}):
\begin{align}
\label{eq:sttwistprod}
\rho(q,p) & = \frac{1}{(2\pi)^{2n}}\!\int_{\R^{2n}}\!\!\!\!\ dq'\ \!dp'\ \!e^{i(q\cdot q'+p\cdot p')}\!\int_{\R^{2n}}\!\!\!\!\ dq''\ \!dp''\ \!\mathcal{F}[\sigma](q'',p'')\mathcal{F}[\tau](q'-q'',p'-p'')e^{\frac{i\varepsilon}{2}(q'\cdot p'' - p'\cdot q'')} \\ \nonumber
 & \sim \exp\left(-\frac{i\varepsilon}{2}(\nabla_{p}\cdot\nabla_{x}-\nabla_{q}\cdot\nabla_{\xi})\right)_{|(q,p)=(x,\xi)}\sigma(q,p)\tau(x,\xi).
\end{align}
Clearly, to arrive at \eqref{eq:sttwistprod} we need to evaluate the \textit{tri-kernel} $\tr(\hat{W}(q,p)^{*}\hat{W}(q',p')\hat{W}(q'',p''))$, which is possible, because linear combinations of (Fourier-)Weyl elements are closed under products. \\[0.1cm]
Thus, we conclude that the structures required for a theory of pseudo-differential operators adapted to quantum systems defined by the commutation relations \eqref{eq:quantcomm} or their exponential form (which we recognize as covariant representation of $(C(G),G,\alpha_{\uL})$\footnote{Strictly speaking, we suppressed the ``classicality parameter’’ $\varepsilon$. We will comment on this issue below.}$^{,}$\footnote{We incorporate a minus sign in the definition of the group translations, $U_{g}=L^{*}_{g^{-1}}$, because we intend to work with right invariant vector fields on $G$ while maintaining a left action.}),
\begin{align}
\label{eq:quantcommexp}
[f,f'] & = 0, \\ \nonumber
U_{g}fU_{g}^{*} & = L^{*}_{g^{-1}}f \\ \nonumber
U_{g}U_{g'} & = U_{gg'},\ \ \ \ f,f'\in C(G),\ g,g'\in G,
\end{align}
are a set of (Fourier-)Weyl elements, closed under product if possible, and a notion of Fourier transform. Moreover, to obtain a practical calculus for the application of the Born-Oppenheimer scheme and space-adiabatic perturbation theory we should require the existence of a formula similar to \eqref{eq:sttwistprod} for symbols of operator products, which admits a suitable asymptotic expansion.\\[0.1cm]
Let us also add a short comment on the choice of Weyl elements \eqref{eq:stweyl}: Namely, we could make the alternative definitions
\begin{align}
\label{eq:stkn}
 W_{R}(x,\xi) & := e^{i x\cdot Q}e^{i\xi\cdot P}, & W_{L}(x,\xi) & := e^{i\xi\cdot P}e^{i x\cdot Q},
\end{align}
which lead to the Kohn-Nirenberg pseudo-differential operators $A^{L}_{\sigma}$, $A^{R}_{\sigma}$ associated with a symbol $\sigma$ (cf. \cite{FollandHarmonicAnalysisIn}), which are standard in treatments of pseudo-differential operators on manifolds (cf. \cite{HoermanderTheAnalysisOf3, RuzhanskyPseudoDifferentialOperators}):
\begin{align}
\label{eq:stknpseudoR}
(A^{R}_{\sigma}\Psi)(q) & = \frac{1}{(2\pi\varepsilon)^{n}}\int_{\R^{2n}}dx\ d\xi\ \sigma(q,\xi)e^{\frac{i}{\varepsilon}\xi\cdot(q-x)}\Psi(x), \\
\label{eq:stknpseudoL}
(A^{L}_{\sigma}\Psi)(q) & = \frac{1}{(2\pi\varepsilon)^{n}}\int_{\R^{2n}}dx\ d\xi\ \sigma(x,\xi)e^{\frac{i}{\varepsilon}\xi\cdot(q-x)}\Psi(x).
\end{align}
Although, the Kohn-Nirenberg pseudo-differential operators generalise in a much more straightforward manner to manifolds, because \eqref{eq:stknpseudoR} and \eqref{eq:stknpseudoL} can be localised in the position variable $q$\cite{RieffelQuantizationAndC}, and thus to general (compact) Lie groups, they are disadvantageous w.r.t. to the Born-Oppenheimer scheme and space-adiabatic perturbation theory, because they do not provide a real quantisation in contrast to \eqref{eq:stweylpseudo}, i.e. $\big(A^{R/L}_{\sigma}\big)^{*}\neq A^{R/L}_{\sigma^{*}}$, due to the asymmetric treatment of $Q$ and $P$.
\subsubsection{A global calculus}
\label{subsubsec:globalcalc}
Having identified the ingredients necessary for the definition of pseudo-differential operators, we will now explain how these are realized for a compact Lie group $G$ in a global fashion. With regard to the notation, we stick to section \ref{sec:cst}. As noted above, the commutation relations \eqref{eq:quantcomm} correspond in exponential form \eqref{eq:quantcommexp} to a covariant representation of $(C(G),G,\alpha_{\uL})$, and thus to the transformation group $C^{*}$-algebra $C(G)\rtimes_{\uL}G$. In the (faithful) integrated representation $\rho_{\uL}$ on $L^{2}(G)$ a function $F\in C(G\times G)\subset C(G)\rtimes_{\uL}G$ acts according to
\begin{align}
\label{eq:leftintegrated}
(\rho_{\uL}(F)\Psi)(g) & = \int_{G}dh\ F(h,g)(U_{h}\Psi)(g) = \int_{G}dh\ F(h,g)\Psi(h^{-1}g) \\ \nonumber
 & = \int_{G}dh\ F(gh^{-1},g)\Psi(h),\ \ \ \Psi\in L^{2}(G).
\end{align}
Clearly, \eqref{eq:leftintegrated} still to makes sense as a continuous operator $\rho_{\uL}(F):C^{\infty}(G)\rightarrow\cD'(G)$ for \mbox{$F\in\cD'(G\times G)\cong\cD'(G)\hat{\otimes}\cD'(G)$}, and if we restrict $F$ to be in $\cD'(G,C^{\infty}(G))\cong\cD'(G)\hat{\otimes}\ \!C^{\infty}(G)$ it defines a continuous operator on $C^{\infty}(G)$\footnote{We endow $C^{\infty}(G)=\cD(G)$ and $\cD'(G)$ with their usual nuclear topologies. $\hat{\otimes}$ denotes the complete tensor product of nuclear locally convex spaces. Clearly, the adjective ``nuclear’’ is a misnomer for ``nucular’’, but the former is by now well established.}. Therefore, it is possible to make
\begin{Definition}
\label{def:knelments}
For $F^{R}(\pi,m,n;h):=\pi(\ .\ )_{mn}\delta_{h},\ F^{L}(\pi,m,n;h):=\pi(h^{-1}\ .\ )_{mn}\delta_{h} \in\cD'(G,C^{\infty}(G))$, where $\pi\in\hat{G}$, $m,n=1,...,d_{\pi}$ is a unitary irreducible representation, we obtain the operators
\begin{align}
\label{eq:knelements}
\left(\rho_{\uL}\left(F^{R}(\pi,m,n;h)\right)\Psi\right)(g) & = \pi(g)_{mn}\Psi(h^{-1}g), \\ \nonumber
\left(\rho_{\uL}\left(F^{L}(\pi,m,n;h)\right)\Psi\right)(g) & = \pi(h^{-1}g)_{mn}\Psi(h^{-1}g),\ \ \ \Psi\in C^{\infty}(G),
\end{align}
which extend to operators on $L^{2}(G)$ by continuity.
\end{Definition}
In view of \eqref{eq:stkn}, these operators generalise the Weyl elements used in the definition of the Kohn-Nirenberg pseudo-differential operators \eqref{eq:stknpseudoR} and \eqref{eq:stknpseudoL}. In the following, we will restrict attention to the (right) elements $F^{R}(\pi,m,n;h)$, although all statements hold in slightly modified form for $ F^{L}(\pi,m,n;h)$, as well. The analogy with \eqref{eq:stknpseudoR} and \eqref{eq:stknpseudoL} becomes even clearer, when we consider the Weyl elements in context of the (global) Fourier transform between $G$ and its (unitary) dual $\hat{G}$\footnote{Note, that we sometimes abuse notation and explicitly use a representative of an isomorphism class $\pi\in\hat{G}$, but the appearance of traces on the representation space in the right places makes all formulae independent of the representative.}, and its inverse,
\begin{align}
\label{eq:globalft}
\mathcal{F}[\Psi](\pi) = \hat{\Psi}(\pi) & := \int_{G}dg\ \Psi(g)\pi(g),\ \ \ \Psi\in L^{2}(G), \\ \nonumber
\mathcal{F}^{-1}[\Phi](g) = \check{\Phi}(g) & := \sum_{\pi\in\hat{G}}d_{\pi}\tr(\pi(g)^{*}\Phi(\pi)),\ \ \ \Phi\in L^{2}(\hat{G}):
\end{align}
We define the Fourier-Weyl elements
\begin{align}
\label{eq:knfourier}
\hat{F}^{R}(g;\pi,m,n) & := \sum_{\pi'\in\hat{G}}d_{\pi'}\sum_{m',n'=1}^{d_{\pi'}}\overline{\pi'(g)_{m'n'}}\int_{G}dh\ \overline{\pi(h)_{mn}}F^{R}(\pi',m',n';h) \\ \nonumber
 & = \delta_{g}\overline{\pi(\ .\ )_{mn}}\in C^{\infty}(G,\cD'(G)).
\end{align}
Then, for $\sigma\in\hat{\cD'}(\hat{G},C^{\infty}(G))=\mathcal{F}_{1}(\cD'(G,C^{\infty}(G)))$, we have
\begin{align}
\label{eq:knquant}
F^{R}_{\sigma}(h,g) & = \sum_{\pi\in\hat{G}}d_{\pi}\sum_{m,n=1}^{d_{\pi}}\int_{G}dg'\ \sigma(\pi,g')_{mn}\hat{F}^{R}(g';\pi,m,n)(h,g) \\ \nonumber
 & = \sum_{\pi\in\hat{G}}d_{\pi}\tr(\pi(h)^{*}\sigma(\pi,g)) = \check{\sigma}^{1}(h,g),
\end{align}
which is analogous to \eqref{eq:stweylquant}, and continuous to be well-defined in the distributional sense for $\sigma\in\mathcal{F}_{1}(\cD'(G\times G))$. The representation of $\check{\sigma}$ on $C^{\infty}(G)$ via $\rho_{\uL}$ leads to the pseudo-differential operators of Ruzhansky and Turunen \cite{RuzhanskyPseudoDifferentialOperators} (up to the fact that those authors employ the right convolution algebra $C(G)\rtimes_{\uR}G$):
\begin{align}
\label{eq:knpseudo}
\left(\rho_{\uL}\left(F^{R}_{\sigma}\right)\Psi\right)(g) & = \int_{G}dh\ F^{R}_{\sigma}(gh^{-1},g)\Psi(h) \\ \nonumber
 & = \int_{G}\check{\sigma}(gh^{-1},g)\Psi(h) = \sum_{\pi\in\hat{G}}d_{\pi}\tr(\pi(g)^{*}\sigma(\pi,g)\hat{\Psi}(\pi)).
\end{align}
As a consequence of the Peter-Weyl theorem and the fact that (left) convolution is well-defined on $\cD'(G)$ the (Fourier-)Weyl elements satisfy (distributional) orthogonality relations:
\begin{align}
\label{eq:kntrace}
\tr\left(\rho_{\uL}\left(F^{R}(\pi,m,n;h)^{*}\right)\rho_{\uL}\left(F^{R}(\pi',m',n';h')\right)\right) & = d_{\pi}^{-1}\delta_{\pi,\pi'}\delta_{m,m'}\delta_{n,n'}\delta_{h}(h'), \\ \nonumber
\tr\left(\rho_{\uL}\left(F^{R}(g;\pi,m,n)^{*}\right)\rho_{\uL}\left(F^{R}(g';\pi',m',n')\right)\right) & = \delta_{g'}(g)d_{\pi}^{-1}\delta_{\pi,\pi'}\delta_{m,m'}\delta_{n,n'},
\end{align}
where the adjoint $(\ .\ )^{*}$ has to be taken the sense of \eqref{eq:l1inv}. Furthermore, the linear span of Weyl elements is closed under products due to complete reducibility of unitary representations of $G$. in fact, we have:
\begin{align}
\label{eq:knprod}
& \left(F^{R}(\pi,m,n;h)\ast_{\uL}F^{R}(\pi',m',n';h')\right)(g,g') \\ \nonumber
& = \pi(g')_{mn}\pi'(h^{-1}g')\delta_{hh'}(g) = \sum_{k'=1}^{d_{\pi'}}\pi'(h^{-1})_{m'k'}\pi(g')_{mn}\pi'(g')_{k'n'}\delta_{hh'}(g) \\ \nonumber
& = \sum_{k'=1}^{d_{\pi'}}\pi'(h^{-1})_{m'k'}\sum_{\pi''\in\hat{G}}\sum_{s=0}^{N_{\pi,\pi'}^{\pi''}}\sum_{M,N=1}^{d_{\pi''}}C(\pi,m;\pi',k'|\pi'',M;s)\overline{C(\pi,n;\pi',n'|\pi'',N;s)}\pi''(g')_{MN}\delta_{hh'}(g) \\ \nonumber
& = \sum_{k'=1}^{d_{\pi'}}\pi'(h^{-1})_{m'k'}\sum_{\pi''\in\hat{G}}\sum_{s=1}^{N_{\pi,\pi'}^{\pi''}}\sum_{M,N=1}^{d_{\pi''}}C(\pi,m;\pi',k'|\pi'',M;s)\overline{C(\pi,n;\pi',n'|\pi'',N;s)} \\[-0.5cm] \nonumber
 & \hspace{9cm}\times F^{R}(\pi'',M,N;hh')(g,g'),
\end{align}
where we denote by the $C$'s the Clebsch-Gordan coefficients associated with the decomposition
\begin{align}
\label{eq:repdecomp}
\pi\otimes\pi'\cong\bigoplus_{\pi''\in\hat{G}}\bigoplus_{s=1}^{N^{\pi''}_{\pi,\pi'}}\pi''\ \ \textup{with multiplicities}\ N^{\pi''}_{\pi,\pi'}\in\mathds{N}_{0}.
\end{align}
Dequantisation of a continuous operator $A$ on $C^{\infty}(G)$ takes the form:
\begin{align}
\label{eq:kndequant}
\sigma_{A}(\pi,g)_{mn} & := \tr\left(\rho_{\uL}\left(\hat{F}^{R}(g;\pi,m,n)^{*}\right)A\right),
\end{align}
which agrees with the formula of Ruzhansky and Turunen \cite{RuzhanskyPseudoDifferentialOperators} for $A=\rho_{\uL}(F),\ F\in\cD'(G,C^{\infty}(G))$:
\begin{align}
\label{eq:kndequantequiv}
\sigma_{F}(\pi,g)_{mn} & = \tr\left(\rho_{\uL}\left(\hat{F}^{R}(g;\pi,m,n)^{*}\right)\rho_{\uL}(F)\right) = \int_{G}dh\ F(h,g)\pi(h)_{mn} = \hat{F}(\pi,g)_{mn}.
\end{align}
By the Schwartz kernel theorem (cf. \cite{AmannVectorValuedDistributions, RuzhanskyPseudoDifferentialOperators}) this covers already the general case of continuous operators on $C^{\infty}(G)$.\\[0.1cm]
Before we try to deform the Kohn-Nirenberg quantisation, defined so far, to obtain a Weyl quantisation, i.e. a real quantisation, we collect some of the properties of the former in a
\begin{Proposition}
\label{prop:knquant}
Given continuous operators $A,B:C^{\infty}(G)\rightarrow\cD'(G)$ with $F_{A},F_{B}\in\cD'(G\times G)$, the symbols \mbox{$\sigma_{A},\sigma_{B}\in\mathcal{F}_{1}(\cD'(G\times G))$} satisfy:
\begin{itemize}
	\item[1.] The map $A\mapsto\sigma_{A}$ is a linear homeomorphism between $L(C^{\infty}(G),\cD'(G))$ and $\hat{\cD'}(\hat{G}\times G)$ with $\sigma_{\mathds{1}}(\pi,g)=\pi(e)$.
	\item[2.] $\sigma_{A^{*}}(\pi,g) = \sigma_{F^{*}_{A}}(\pi,g) = \sigma_{\alpha_{\uL}^{-1}(F_{A})}(\pi,g)^{*}$, where $(\alpha_{\uL}^{-1}(F))(h,g)=F(h,hg)$, \mbox{$F\in\cD'(G\times G)$.}
	\item[3.] $\sigma_{U_{h}AU^{*}_{h}}(\pi,g) = \sigma_{(\alpha_{h^{-1}}\times L_{h^{-1}})^{*}F_{A}}(\pi,g) = \pi(h)\sigma_{A}(\pi,h^{-1}g)\pi(h)^{*}$, where $\alpha_{h}(g)=hgh^{-1}$.
	\item[4.] If $U^{R}:G\rightarrow U(L^{2}(G))$ denotes the right regular representation, i.e. $(U^{R}_{h}\Psi)(g)=\Psi(gh)$ for $\Psi\in L^{2}(G),\ h,g\in G$, we have: $\sigma_{U^{R}_{h}AU^{R*}_{h}}(\pi,g)=\sigma_{(\id\times R_{h})^{*}F_{A}}(\pi,g)=\sigma_{A}(\pi,gh)$.
	\item[5.] $(A,B)_{HS} = \tr(A^{*}B) = (F_{A},F_{B})_{L^{2}(G\times G)} = (\sigma_{A},\sigma_{B})_{L^{2}(\hat{G}\times G)}$, if $A,B$ are Hilbert-Schmidt operators. Moreover, the maps $HS(L^{2}(G))\ni A\mapsto F_{A}\in L^{2}(G\times G)$  and $HS(L^{2}(G))\ni A\mapsto \sigma_{A}\in L^{2}(\hat{G}\times G)$ are unitary.
	\item[6.] $\sigma_{AB}(\pi,g) = \int_{G}dh\ F_{A}(h,g)\pi(h)\sigma_{B}(\pi,h^{-1}g)$, if $A,B:C^{\infty}(G)\rightarrow C^{\infty}(G)$.
	\item[7.] The symbol $\sigma_{A}$ can be computed by
	\begin{align}
	\label{eq:knsymbolformula}
	\sigma_{A}(\pi,g) & = (\pi A \pi^{*})(g) = \pi(g)\int_{G}dh\ F_{A}(gh^{-1},g)\pi(h)^{*}.
	\end{align}
\end{itemize}
\begin{Proof}
\begin{itemize}
	\item[1.] The linearity of $A\mapsto\sigma_{A}$ and $\sigma_{\mathds{1}}(\pi,g)=\pi(e)$ are evident from the definition. The homeomorphism property follows from the Schwartz kernel theorem and the fact that the (partial) Fourier transform sets up a homeomorphism between $\cD'(G\times G)$ and $\hat{\cD'}(\hat{G}\times G)$.
	\item[2.] For $\Psi\in C^{\infty}(G)$, we have
\begin{align}
\label{eq:knadjoint}
(\sigma_{A^{*}}(\pi),\Psi) & = \int_{G}dg\ \int_{G}dh\ F_{A^{*}}(h,g)\pi(h)\Psi(g) = \int_{G}dg\ \int_{G}dh\ F^{*}_{A}(h,g)\pi(h)\Psi(g) \\ \nonumber
 & = \int_{G}dg\ \int_{G}dh\ \overline{F_{A}(h^{-1},h^{-1}g)}\pi(h)\Psi(g) = \int_{G}dg\ \int_{G}dh\ \overline{F_{A}(h,hg)}\pi(h)^{*}\Psi(g) \\ \nonumber
 & = \int_{G}dg\ \left(\int_{G}dh\ F_{A}(h,hg)\pi(h)\right)^{*}\Psi(g) \\ \nonumber
 & = \int_{G}dg\ \left(\int_{G}dh\ \alpha^{-1}_{\uL}(F_{A})(h,g)\pi(h)\right)^{*}\Psi(g).
\end{align}
	\item[3.]Referring to the definition of $A=\rho_{\uL}(F_{A})$, we find:
\begin{align}
\label{eq:kncovariance1}
U_{h}AU^{*}_{h} & = U_{h}\rho_{\uL}(F_{A})U^{*}_{h} =\rho_{\uL}((\alpha_{h^{-1}}\times L_{h^{-1}})^{*}F_{A}).
\end{align}
Therefore, we have:
\begin{align}
\label{eq:kncovariance2}
\sigma_{U_{h}AU^{*}_{h}}(\pi,g) & = \sigma_{(\alpha_{h^{-1}}\times L_{h^{-1}})^{*}F_{A}}(\pi,g) = \int_{G}dh'\ (\alpha_{h^{-1}}\times L_{h^{-1}})^{*}F_{A}(h',g)\pi(h') \\ \nonumber
 & = \int_{G}dh' F_{A}(h',h^{-1}g)\pi(\alpha_{h}(h')) = \pi(h)\int_{G}dh'\ F_{A}(h',h^{-1}g)\pi(h')\pi(h)^{*} \\ \nonumber
 & = \pi(h)\sigma_{A}(\pi,h^{-1}g)\pi(h)^{*}.
\end{align}
	\item[4.] This follows along the same lines as 3.
	\item[5.] It is well known (cf. \cite{ReedMethodsOfModern1}), that $A,B$ are Hilbert-Schmidt operators if and only if $F_{A},F_{B}\in L^{2}(G\times G)$, which gives the second equality and the unitarity of $A\mapsto F_{A}$. The third equality and the unitarity of $A\mapsto\sigma_{A}$ follow, because the (partial) Fourier transform is a unitary map from $L^{2}(G\times G)$ to $L^{2}(\hat{G}\times G)$.
	\item[6.] By assumption the product $AB$ is well-defined and gives rise to a (left) convolution kernel $F_{AB}$, which is found from
\begin{align}
\label{eq:knprod1}
AB & = \rho_{\uL}(F_{A})\rho_{\uL}(F_{B}) = \rho_{\uL}(F_{A}\ast_{\uL}F_{B}).
\end{align}
The formula for $\sigma_{AB}$ follows from direct computation:
\begin{align}
\label{eq:knprod2}
\sigma_{AB}(\pi,g) & = \int_{G}dh\ (F_{A}\ast_{\uL}F_{B})(h,g)\pi(h) \\ \nonumber
 & = \int_{G}dh\ \int_{G}dh' F_{A}(h',g)F_{B}(h'^{-1}h,h'^{-1}g)\pi(h) \\ \nonumber
 & = \int_{G}dh\ \int_{G}dh' F_{A}(h',g)\pi(h')F_{B}(h'^{-1}h,h'^{-1}g)\pi(h'^{-1}h) \\ \nonumber
 & = \int_{G}dh'\  F_{A}(h',g)\pi(h')\int_{G}dh\ F_{B}(h,h'^{-1}g)\pi(h) \\ \nonumber
 & = \int_{G}dh\ F_{A}(h,g)\pi(h)\sigma_{B}(\pi,h^{-1}g).
\end{align}
	\item[7.] Employing \eqref{eq:kndequantequiv}, we find:
	\begin{align}
	\label{eq:knsymbolformulaproof}
	\hat{F}_{A}(\pi,g) & = \int_{G}dh\ F_{A}(h,g)\pi(h) = \pi(g)\int_{G}dh\ F_{A}(h,g)\pi(g^{-1}h) \\ \nonumber
	 & = \pi(g)\int_{G}dh\ F_{A}(gh,g)\pi(h) \\ \nonumber
	 & = \pi(g)\int_{G}dh\ F_{A}(gh^{-1},h)\pi(h)^{*} = (\pi A \pi^{*})(g).
	\end{align}
\end{itemize}
\end{Proof}
\end{Proposition}
The last property is extremely useful in the actual computation of symbols $\sigma_{A}$ of operators \mbox{$A:C^{\infty}(G)\rightarrow\cD'(G)$}, as it does not require the computation of the (left) convolution kernel $F_{A}$.  To see how this works, we compute the symbols of $P_{X}=-i\varepsilon R_{X},\ X\in\fg,$ and $f\in C^{\infty}(G)$:
\begin{align}
\label{eq:knelementarysymbols}
\sigma_{P_{X}}(\pi,g) & = (\pi P_{X}\pi^{*}) = \pi(g) (P_{X}\pi^{*})(g) = -i\varepsilon\pi(g)\frac{d}{dt}_{|t=0}\pi(e^{tX}g)^{*} \\ \nonumber
 & = -i\varepsilon\pi(g)\frac{d}{dt}_{|t=0}\pi(g)^{*}\pi(e^{-tX}) = i\varepsilon d\pi(X), \\ \nonumber
\sigma_{f}(\pi,g) & = (\pi f\pi^{*})(g) = \pi(g)f(g)\pi(g)^{*} = f(g)\mathds{1}_{V_{\pi}}.
\end{align}
Combining this with the sixth property, gives rise to the commutation relations \eqref{eq:quantcomm}:
\begin{align}
\label{eq:knelementarycomm}
\sigma_{ff'}(\pi,g) & = \pi(g)(ff'\pi^{*})(g) = f(g)f'(g)\mathds{1}_{V_{\pi}} = \sigma_{f}(\pi,g)\sigma_{f'}(\pi,g), \\[0.1cm] \nonumber
 \sigma_{P_{X}f} & = \pi(g)(P_{X}f\pi^{*})(g) = \pi(g)(P_{X}\pi^{*})(g)f(g) - i\varepsilon(R_{X}f)(g)\mathds{1}_{V_{\pi}} \\ \nonumber
 & = \sigma_{P_{X}}(\pi,g)\sigma_{f}(\pi,g) - i\varepsilon\sigma_{R_{X}f}(\pi,g), \\[0.1cm] \nonumber
 \sigma_{P_{X}P_{Y}}(\pi,g) & = \pi(g)(P_{X}P_{Y}\pi^{*})(g) = i\varepsilon\pi(g)(P_{X}\pi^{*})(g)d\pi(Y)=-\varepsilon^{2}d\pi(X)d\pi(Y) = \sigma_{P_{X}}\sigma_{P_{Y}}.
\end{align}
The second property of the above proposition gives a measure to which extent the quantisation $\hat{\cD'}(\hat{G}\times G)\ni\sigma\mapsto\rho_{\uL}(F^{R}_{\sigma})\in L(C^{\infty}(G),\cD'(G))$ fails to be real. Interestingly, there is a simple way to cure this, if we were allowed to take square roots in $G$, which is indeed possible for compact Lie groups by means of the exponential map $\exp:\fg\rightarrow G$, as the latter is onto (cf. \cite{BroeckerRepresentationsOfCompact}). Moreover, we can define $\sqrt{g},\ \forall g\in G$, s.t. $\sqrt{g^{-1}}=\sqrt{g}^{-1}$, but there is also a price to pay: Namely, $\sqrt{\ .\ }:G\rightarrow G$ is in general not a smooth homomorphism, but only a measurable map.
\begin{Definition}
\label{def:weylquant}
For $\sigma\in\hat{\cD'}(\hat{G}\times G)$, s.t. $\alpha^{\frac{1}{2}}_{\uL}\left(F^{R}_{\sigma}\right)(h,g)=F^{R}_{\sigma}(h,\sqrt{h}^{-1}g)$ is in $\cD'(G\times G)$, e.g. if $\sqrt{\ .\ }$ is smooth on $\ssupp_{1}(F^{R}_{\sigma})$, we define the Weyl quantisation $F^{W}_{\sigma}$ of $\sigma$ by
\begin{align}
\label{eq:weylquant}
F^{W}_{\sigma}(h,g) & := \alpha^{\frac{1}{2}}_{\uL}\left(F^{R}_{\sigma}\right)(h,g).
\end{align}
The Weyl elements $F^{W}(\pi,m,n;h)$ are the Weyl quantisations of the symbols
\begin{align}
\label{eq:weylelementssym}
(\sigma_{(\pi,m,n;h)}(\pi',h'))_{m'n'} & :=d^{-1}_{\pi}\delta_{\pi,\pi'}\delta_{m',m}\delta_{n,n'}\delta_{h}(h'),
\end{align}
where $\pi,\pi'\in\hat{G},\ m,n=1,...,d_{\pi}$, \mbox{$m',n'=1,...,d_{\pi'}$,} $h,h'\in G$. Explicitly, we have
\begin{align}
\label{eq:weylelements}
F^{W}(\pi,m,n;h)(h',g) & = \pi(\sqrt{h}^{-1}g)_{mn}\delta_{h}(h') \\ \nonumber
 & = \sum_{k=1}^{d_{\pi}}\pi(\sqrt{h}^{-1})_{mk}\pi(g)_{kn}\delta_{h}(h') \\ \nonumber
 & = \sum_{k=1}^{d_{\pi}}\pi(\sqrt{h}^{-1})_{mk}F^{R}(\pi,k,n;h)(h',g).
\end{align}
\end{Definition}
The following lemma shows that the Weyl quantisation is real, and that the Weyl element satisfies appropriate orthogonality relations.
\begin{Lemma}
\label{lem:weylquant}
For $\sigma\in\hat{\cD'}(\hat{G}\times G)$ as in definition \ref{def:weylquant}, we have
\begin{align}
\label{eq:weyladjoint}
\rho_{\uL}\left(F^{W}_{\sigma}\right)^{*} & = \rho_{\uL}\left(F^{W*}_{\sigma}\right) = \rho_{\uL}\left(F^{W}_{\sigma^{*}}\right).
\end{align}
Moreover, the Weyl elements satisfy
\begin{align}
\label{eq:weyltrace}
\tr\left(\rho_{\uL}\left(F^{W}(\pi,m,n;h)^{*}\right)\rho_{\uL}\left(F^{W}(\pi',m',n';h')\right)\right) & = d_{\pi}^{-1}\delta_{\pi,\pi'}\delta_{m,m'}\delta_{n,n'}\delta_{h}(h').
\end{align}
\begin{Proof}
The first statement is a consequence of the adjointness property (2.) in proposition \ref{prop:knquant}:
\begin{align}
\label{eq:weyladjointcalc}
F^{W*}_{\sigma}(h,g) & =\overline{ F^{W}_{\sigma}(h^{-1},h^{-1}g)} =\overline{ F^{R}_{\sigma}(h^{-1},\sqrt{h}h^{-1},g)} = \overline{F^{R}_{\sigma}(h^{-1},\sqrt{h}^{-1}g)} \\ \nonumber
 & = \sum_{\pi\in\hat{G}}d_{\pi}\overline{\tr\left(\pi(h)\sigma(\pi,\sqrt{h}^{-1}g)\right)} \\ \nonumber
  & = \sum_{\pi\in\hat{G}}d_{\pi}\tr\left(\pi(h)^{*}\sigma(\pi,\sqrt{h}^{-1}g)^{*}\right) \\ \nonumber
  & = F^{R}_{\sigma^{*}}(h,\sqrt{h}^{-1}g) = F^{W}_{\sigma^{*}}(h,g).
\end{align}
The second statement follows from the orthogonality relations \eqref{eq:kntrace} and \eqref{eq:weylelements}:
\begin{align}
\label{eq:weyltracecalc}
 & \tr\left(\rho_{\uL}\left(F^{W}(\pi,m,n;h)^{*}\right)\rho_{\uL}\left(F^{W}(\pi',m',n';h')\right)\right) \\ \nonumber
 &\ \ \ \ =\sum_{k=1}^{d_{\pi}}\sum_{k'=1}^{d_{\pi'}}\overline{\pi(\sqrt{h}^{-1})_{mk}}\pi'(\sqrt{h'}^{-1})_{m'k'}\tr\left(\rho_{\uL}\left(F^{R}(\pi,k,n;h)^{*}\right)\rho_{\uL}\left(F^{R}(\pi',k',n';h')\right)\right) \\ \nonumber
 &\ \ \ \ =  d_{\pi}^{-1}\delta_{\pi,\pi'}\delta_{n,n'}\delta_{h}(h')\sum_{k=1}^{d_{\pi}}\pi(\sqrt{h})_{km}\pi(\sqrt{h}^{-1})_{m'k} = d_{\pi}^{-1}\delta_{\pi,\pi'}\delta_{m,m'}\delta_{n,n'}\delta_{h}(h'),
\end{align}
where the last line makes sense because $\sqrt{\ .\ }:G\rightarrow G$ admits a unique pointwise evaluation.
\end{Proof}
\end{Lemma}
\begin{Remark}
\label{rem:fourierweyl}
Unfortunately, the definition of Fourier-Weyl elements seems to be problematic, as can be seen from a (formal) calculation:
\begin{align}
\label{eq:fourierweyl}
\hat{F}^{W}(g;\pi,m,n) (h,g')& = \sum_{\pi'\in\hat{G}}d_{\pi'}\sum_{m',n'=1}^{d_{\pi'}}\overline{\pi'(g)_{m'n'}}\int_{G}dh'\ \overline{\pi(h')_{mn}}F^{W}(\pi',m',n';h')(h,g') \\ \nonumber
 & = \overline{\pi(h)_{mn}}\delta_{\sqrt{h}g}(g')
\end{align}
From the last line, we infer that the definition of $\hat{F}^{W}(g;\pi,m,n)$ would require the composition of $\delta$ with $\sqrt{\ .\ }$, which is not necessarily well-defined, because $\sqrt{\ .\ }$ is in general not even continuous.
\end{Remark}
 Still, we can define a dequantisation map, if we restrict ourselves to operators \mbox{$A:C^{\infty}(G)\rightarrow\cD'(G)$,} s.t. $\alpha^{-\frac{1}{2}}_{\uL}\left(F_{A}\right)(h,g) = F_{A}(h,\sqrt{h}g)$ is in $\cD'(G\times G)$, in analogy with definition \ref{def:weylquant}.
\begin{Definition}
\label{def:weyldequant}
The Weyl symbol of an operator $A:C^{\infty}\rightarrow\cD'(G)$, s.t. $\alpha^{-\frac{1}{2}}_{\uL}\left(F_{A}\right)(h,g) = F_{A}(h,\sqrt{h}g)$ is in $\cD'(G\times G)$, is
\begin{align}
\label{eq:weylsymbol}
\sigma^{W}_{A}(\pi,g) & := \mathcal{F}_{1}\left[\alpha^{-\frac{1}{2}}_{\uL}\left(F_{A}\right)\right](\pi,g) = \int_{G}dh\ \alpha^{-\frac{1}{2}}_{\uL}\left(F_{A}\right)(h,g)\pi(h).
\end{align} 
\end{Definition}
From the definitions it is obvious, that we have
\begin{Corollary}
\label{cor:weylquant}
The Weyl quantisation $\sigma\mapsto F^{W}_{\sigma}$ and the Weyl symbol map $A\mapsto\sigma^{W}_{A}$ are inverse to each other.
\end{Corollary}
In accordance with proposition \ref{prop:knquant}, we collect some properties of the Weyl (de)quantisation, although, at this point, we refrain from specifying the sets of operators or symbols for which the procedure is well-defined any further. We will cure this in the following subsection, where we define the local calculi.
\begin{Proposition}
\label{prop:weylquant}
Given continuous operators $A,B:C^{\infty}(G)\rightarrow\cD'(G)$, s.t. the Weyl symbols $\sigma^{W}_{A},\sigma^{W}_{B}\in\hat{\cD'}(\hat{G}\times G)$ are well-defined, then we have
\begin{itemize}
	\item[1.] $\sigma^{W}_{U_{h}AU^{*}_{h}}(\pi,g) = \pi(h)\sigma^{W}_{(\id\times L_{h^{-1}})^{*}F_{A}}(\pi,g)\pi(h)^{*}$.
	\item[2.] $\sigma^{W}_{U^{R}_{h}AU^{R*}_{h}}(\pi,g) = \sigma^{W}_{A}(\pi,gh)$.
	\item[3.] If the product $AB$ and its Weyl symbol $\sigma^{W}_{AB}$ are well-defined, we have
\begin{align}
\label{eq:weylprod}
\sigma^{W}_{AB}(\pi,g) & = \!\!\!\!\!\sum_{\pi',\pi''\in\hat{G}}\!\!\!\!d_{\pi'}d_{\pi''}\!\!\int_{G}\!\!dg'\!\!\int_{G}\!\!dg''\pi(g')\pi(g'')\tr(\pi'(g')^{*}\sigma^{W}_{A}(\pi'\!,\!\sqrt{g'^{-1}}\sqrt{g'g''}g)) \\[-0.5cm] \nonumber
&\hspace{5cm} \times\tr(\pi''(g'')^{*}\sigma_{B}(\pi''\!,\!\sqrt{g''^{-1}}g'^{-1}\!\sqrt{g'g''}g)).
\end{align} 
\end{itemize}
\begin{Proof}
The properties 1.-3. are proved in a completely analogous way as the corresponding properties in proposition \ref{prop:knquant}.
\end{Proof}
\end{Proposition}
\begin{Remark}
\label{rem:weylelementarysymbols}
The Weyl symbols of the elementary operators appearing in \eqref{eq:quantcomm} equal their Kohn-Nirenberg symbols, since we have $F_{P_{X}}(h,g)=-i\varepsilon(R_{X}\delta_{e})(h),\ X\in\fg,$ and $F_{f}(h,g)=\delta_{e}(h)f(g),\ f\in C^{\infty}(G)$:
\begin{align}
\label{eq:weylelementarysymbols}
&\ &\ \alpha^{-\frac{1}{2}}_{\uL}(F_{f})(h,g) & =\delta_{e}(h)f(\sqrt{h}g)=\delta_{e}(h)f(g) \\ \nonumber
&\Rightarrow\ & \ \sigma^{W}_{f}(\pi,g) & = f(g)\mathds{1}_{V_{\pi}} \\ \nonumber
&\ &\ \alpha^{-\frac{1}{2}}_{\uL}(F_{P_{X}})(h,g) & =-i\varepsilon(R_{X}\delta_{e})(h) \\ \nonumber
&\Rightarrow\ &\ \sigma^{W}_{P_{X}}(\pi,g) & = i\varepsilon d\pi(X),
\end{align}
where there first line makes sense, because $\sqrt{\ .\ }:G\rightarrow G$ is uniquely defined everywhere, although it is not continuous.
\end{Remark}
\begin{Remark}
\label{rem:weylprod}
If we choose $G=\R^{n}$, and therefore $\hat{G}=\R^{n}$, and $\sigma_{A},\sigma_{B}\in\mathcal{S}(\R^{2n})$, we can make still make sense out of the product formula \eqref{eq:weylprod}, and a simple calculation shows that it is equivalent to the twisted product \eqref{eq:sttwistprod}:
\begin{align}
\label{eq:weylprodequiv}
\sigma^{W}_{AB}(p,x) &\ \ \ \ =\ \ \ \ \ \ \ \ \int_{\R^{2n}}\!\!\!\!\frac{dp'\ dx'}{(2\pi)^{n}}\int_{\R^{2n}}\!\!\!\!\frac{dp''\ dx''}{(2\pi)^{n}}e^{ip\cdot(x'+x'')}e^{-ip'\cdot x'}e^{-ip''\cdot x''}\sigma^{W}_{A}(p',\tfrac{1}{2}x''+x) \\[-0.25cm] \nonumber
 & \hspace{9cm}\times\sigma^{W}_{B}(p'',-\tfrac{1}{2}x'+x) \\ \nonumber
 &\!\!\!\!\!\!\!\!\stackrel{\substack{x'\ \mapsto\ -2(x'-x) \\ x''\ \mapsto\ \ \ 2(x''-x)}}{=} \int_{\R^{2n}}\!\!\!\!\frac{dp'\ dx'}{(4\pi)^{n}}\int_{\R^{2n}}\!\!\!\!\frac{dp''\ dx''}{(4\pi)^{n}}e^{-2i(p-p')\cdot x'}e^{2i(p-p'')\cdot x''}e^{2i(p-p')\cdot x}e^{-2i(p-p'')\cdot x} \\[-0.25cm] \nonumber
 & \hspace{9cm}\times\sigma^{W}_{A}(p',x'')\sigma^{W}_{B}(p'',x') \\ \nonumber
 &\ \ \ \ =\ \ \ \ \ \ \ \ \int_{\R^{2n}}\!\!\!\!\frac{dp'\ dx'}{(4\pi)^{n}}\int_{\R^{2n}}\!\!\!\!\frac{dp''\ dx''}{(4\pi)^{n}}e^{2i((p-p')\cdot(x-x')-(p-p'')\cdot(x-x''))}\sigma^{W}_{A}(p',x'')\sigma^{W}_{B}(p'',x'),
\end{align}
where we recognize the last line as an alternative formula for the twisted product \eqref{eq:sttwistprod} (cf. \cite{FollandHarmonicAnalysisIn}).
\end{Remark}
So far, we have not dealt with the question of the existence of an asymptotic expansion for the symbol $\sigma^{(W)}_{AB}$ of an operator product $AB:C^{\infty}(G)\rightarrow C^{\infty}(G)$. This will be done in the next subsection, where we introduce an $\varepsilon$-dependent expansion of $\sigma^{(W)}_{AB}$ in the local calculi. In contrast, Ruzhansky and Turunen \cite{RuzhanskyPseudoDifferentialOperators} define a global symbolic calculus, but the $\varepsilon$-dependence remains rather opaque in their setting. We will come back to this point in subsections \ref{subsec:u1bohr}.
\subsubsection{A local calculus of Paley-Wiener-Schwartz symbols}
\label{subsubsec:pwsymbols}
Following Rieffel and Landsman \cite{RieffelLieGroupConvolution, LandsmanStrictdeformationQuantization, LandsmanMathematicalTopicsBetween}, we localise the quantisations discussed in the previous subsection in the sense, that we pass from global symbols $\sigma\in\hat{\cD'}(\hat{G}\times G)$, living on $G$ and its unitary dual $\hat{G}$, to local symbols $\sigma\in\cD'(T^{*}G)$, living on $T^{*}G\cong G\times\fg^{*}$ (by right translation), via the exponential map $\exp:\fg\rightarrow G$. More precisely, this works as follows: $\exp$ defines a diffeomorphism between an open neighbourhood $U\subset\fg$ of $0\in\fg$ (possibly $Ad$-invariant) and an open neighbourhood $V\subset G$ of $e\in G$\footnote{For $G=U(1),SU(2)$ we can choose the maximal neighbourhood $V=G\setminus\{-1\}$.}. Additionally, we define Fourier transform and its inverse between functions on $\fg$ and $\fg^{*}$:
\begin{align}
\label{eq:localfourier}
\mathcal{F}[F](\theta) = \hat{F}(\theta) & := \int_{\fg}dX\ e^{-i\theta(X)}F(X),\ F\in L^{1}(\fg),\theta\in\fg^{*} \\ \nonumber
\mathcal{F}^{-1}[\sigma](X) = \check{\sigma}(X) & :=\int_{\fg^{*}}\frac{d\theta}{(2\pi)^{n}}e^{i\theta(X)}\sigma(\theta),\ \sigma\in L^{1}(\fg^{*}),X\in\fg,
\end{align}
where we fix the normalisation of the Lebesque measures $dX$ and $d\theta$ via the exponential map and the Haar measure $dg$ on $G$ (cp. section \ref{sec:cst}):
\begin{align}
\label{eq:haarmeasurenorm}
\int_{V}dg\ f(g) & = \int_{U}dX\ j(X)^{2}f(\exp(X)),\ f\in C_{c}(V).
\end{align}
Here, $j$ is the analytic function $j(H)=\prod_{\alpha\in R^{+}}\frac{\sin(\alpha(H))}{\alpha(H)},\ H\in\ft\subset\fg,$ in the notation of section \ref{sec:cst} (see especially \eqref{eq:liouville} and \eqref{eq:sigma}). Now, we are in a position to make the
\begin{Definition}
\label{def:localtoglobal}
Given a function $\sigma\in C^{\infty}_{\PW,U}(\fg^{*})\hat{\otimes}\ \!C^{\infty}(G)$ \footnote{The subscript $\PW$ stands for ``Paley-Wiener’’, because of the Paley-Wiener-Schwartz theorem (cf. \cite{HoermanderTheAnalysisOf1}), which characterises the image of $\cD(\R^{n})$ (and even $\mathcal{E}'(\R^{n})$) under the Fourier transform.} s.t. $\check{\sigma}^{1}$, the inverse Fourier transform of $\sigma$ in the first variable, is in $\cD(\fg)\hat{\otimes}\ \!C^{\infty}(G)$ with $\supp_{1}(\check{\sigma}^{1})\subset U$, we define $F_{\sigma}\in C^{\infty}(G)\hat{\otimes}\ \!C^{\infty}(G)$ by:
\begin{align}
\label{eq:localtoglobal}
F_{\sigma}(h,g) & := \check{\sigma}^{1}(X_{h},g) = \int_{\fg^{*}}\frac{d\theta}{(2\pi)^{n}}e^{i\theta(X_{h})}\sigma(\theta,g)\ \textup{for}\ X_{h}:=\exp^{-1}(h),
\end{align}
which is well-defined due to the support properties of $\check{\sigma}^{1}$. We also define an $\varepsilon$-scaled version of \eqref{eq:localtoglobal}:
\begin{align}
\label{eq:epsilonlocaltoglobal}
\forall\epsilon\in(0,1]:\ F^{\varepsilon}_{\sigma}(h,g) & := \varepsilon^{-n}\check{\sigma}^{1}(\varepsilon^{-1}X_{h},g) = \int_{\fg^{*}}\frac{d\theta}{(2\pi\varepsilon)^{n}}e^{\frac{i}{\varepsilon}\theta(X_{h})}\sigma(\theta,g).
\end{align}
We call $F^{(\varepsilon)}_{\sigma}$ the \textup{($\varepsilon$-scaled) Kohn-Nirenberg quantisation} of $\sigma$, as it defines a compact operator on $L^{2}(G)$ via the left integrated representation $\rho_{\uL}:C(G)\rtimes_{\uL}G\rightarrow\mathcal{K}(L^{2}(G))$. Clearly, the $\varepsilon$-scaled quantisation extends to those $\sigma$, s.t. $\supp_{1}(\check{\sigma}^{1})\subset\varepsilon^{-1}U=:U_{\varepsilon}$. \\[0.1cm]
By standard distributional reasoning, the quantisation extends to distributions $\sigma\in\hat{\mathcal{E}'}_{U}(\fg^{*})\hat{\otimes}\cD'(G)$, i.e. $\check{\sigma}^{1}\in\mathcal{E}'_{U}(\fg)\hat{\otimes}\cD'(G)$, where $\mathcal{E}'_{U}(\fg)$ is the space of compactly supported distributions in $U\subset\fg$.\\[0.1cm]
Furthermore, if we define the (smooth) square root $\sqrt{\ .\ }:V\subset G\rightarrow V\subset G$ by $\sqrt{g}=\exp(\tfrac{1}{2}X_{g})$, we can deform the Kohn-Nirenberg quantisation to a Weyl quantisation in analogy with the previous subsection:
\begin{align}
\label{eq:localweylquant}
F^{W,\varepsilon}_{\sigma}(h,g) & := \alpha^{\frac{1}{2}}_{\uL}(F^{\varepsilon}_{\sigma})(h,g) = F^{\varepsilon}_{\sigma}(h,\sqrt{h^{-1}}g).
\end{align}
The Weyl quantisation will, in general, not be well-defined for $\sigma\in\hat{\mathcal{E}'}_{U}(\fg^{*})\hat{\otimes}\cD'(G)$, but surely for\\ $\sigma\in\hat{\mathcal{E}'}_{U}(\fg^{*})\hat{\otimes}\ \!C^{\infty}(G)$.
\end{Definition}
\begin{Corollary}
\label{cor:localtoglobal}
The Kohn-Nirenberg and Weyl quantisation have following adjointness (in the sense of \eqref{eq:l1inv}) and covariance (w.r.t. to $G$) properties (cp. proposition \ref{prop:knquant} \& \ref{prop:weylquant}):
\begin{itemize}
	\item[1.] $\left(F^{(\varepsilon)}_{\sigma}\right)^{*} = \alpha_{\uL}(F^{(\varepsilon)}_{\overline{\sigma}})$ and $\left(F^{W,(\varepsilon)}_{\sigma}\right)^{*} = F^{W,(\varepsilon)}_{\overline{\sigma}}$,
	\item[2.] $U_{h}F^{(\varepsilon)}_{\sigma}U^{*}_{h} = \alpha_{\uL}(h)\left(F^{(\varepsilon)}_{\left(Ad^{*}_{h^{-1}}\right)^{*}\sigma}\right)$ \& $U_{h}F^{W,(\varepsilon)}_{\sigma}U^{*}_{h} = \alpha_{\uL}\!\left(h\sqrt{h^{-1}(\!\ .\!\ )h}\sqrt{(\!\ .\!\ )^{-1}}\right)\!\left(F^{W,(\varepsilon)}_{\left(Ad^{*}_{h^{-1}}\right)^{*}\sigma}\right)$,
	\item[3.] $U^{R}_{h}F^{(\varepsilon)}_{\sigma}U^{R*}_{h} =\alpha_{\uR}(h)\left(F^{(\varepsilon)}_{\sigma}\right) = F^{(\varepsilon)}_{\left(R^{*}_{g^{-1}}\right)^{*}\sigma}$ \& $U^{R}_{h}F^{W,(\varepsilon)}_{\sigma}U^{R*}_{h} = \alpha_{\uR}(h)\left(F^{W,(\varepsilon)}_{\sigma}\right) = F^{W,(\varepsilon)}_{\left(R^{*}_{g^{-1}}\right)^{*}\sigma}$,
\end{itemize}
where $\sigma\in\hat{\mathcal{E}'}_{U}(\fg^{*})\hat{\otimes}\cD'(G)$, $h\in G$, $\alpha_{\uR}$ is the right regular representation of $G$, $R^{*}$ is the pullback (right) action on $T^{*}G$ and $Ad^{*}:G\rightarrow GL(\fg^{*})$ denotes the coadjoint action.
\begin{Proof}
We prove the statements for $\sigma\in C^{\infty}_{\PW,U}(\fg^{*})\hat{\otimes}\ \!C^{\infty}(G)$, which implies them by standard distributional reasoning for $\sigma\in\hat{\mathcal{E}'}_{U}(\fg^{*})\hat{\otimes}\cD'(G)$ or $\hat{\mathcal{E}'}_{U}(\fg^{*})\hat{\otimes}\ \!C^{\infty}(G)$.
\begin{itemize}
	\item[1.]
\begin{align}
\label{eq:localadjoint}
\left(F^{\varepsilon}_{\sigma}\right)^{*}(h,g) & = \overline{F^{\varepsilon}_{\sigma}(h^{-1},h^{-1}g)} = \overline{\varepsilon^{-n}\check{\sigma}^{1}(\varepsilon^{-1}X_{h^{-1}},h^{-1}g)} \\ \nonumber
 & = \overline{\varepsilon^{-n}\check{\sigma}^{1}(-\varepsilon^{-1}X_{h},h^{-1}g)} = \varepsilon^{-n}\check{\overline{\sigma}}^{1}(\varepsilon^{-1}X_{h},h^{-1}g) =  \alpha_{\uL}(F^{\varepsilon}_{\overline{\sigma}})(h,g).
\end{align}
Here, the third equality follows from $h^{-1}=\exp(X_{h})^{-1}=\exp(-X_{h})$, while the fourth equality follows from the interplay of the inverse Fourier transform and complex conjugation. By the same reasoning, we find:
\begin{align}
\label{eq:localweyladjoint}
\left(F^{W,\varepsilon}_{\sigma}\right)^{*}(h,g) & = \overline{F^{\varepsilon}_{\sigma}(h^{-1},\sqrt{h}h^{-1}g)} = \overline{\varepsilon^{-n}\check{\sigma}^{1}(\varepsilon^{-1}X_{h^{-1}},\sqrt{h^{-1}}g)} \\ \nonumber
 & = \overline{\varepsilon^{-n}\check{\sigma}^{1}(-\varepsilon^{-1}X_{h},\sqrt{h^{-1}}g)} = \varepsilon^{-n}\check{\overline{\sigma}}^{1}(\varepsilon^{-1}X_{h},\sqrt{h^{-1}}g) \\ \nonumber
 & =  F^{W,\varepsilon}_{\overline{\sigma}}(h,g).
\end{align}
	\item[2.] 
\begin{align}
\label{eq:localleftcov}
(U_{h}F^{\varepsilon}_{\sigma}U_{h}^{*})(h',g) & = F^{\varepsilon}_{\sigma}(h^{-1}h'h,h^{-1}g) = \varepsilon^{-n}\check{\sigma}^{1}(\varepsilon^{-1}X_{h^{-1}h'h},h^{-1}g) \\ \nonumber
 & = \varepsilon^{-n}\check{\sigma}^{1}(\varepsilon^{-1}Ad_{h^{-1}}(X_{h'}),h^{-1}g) \\ \nonumber
 & = \varepsilon^{-n}\mathcal{F}^{-1}_{1}[(Ad^{*}_{h^{-1}})^{*}\sigma](\varepsilon^{-1}X_{h'},h^{-1}g) \\ \nonumber
 & = \alpha_{\uL}(h)\left(F^{\varepsilon}_{\left(Ad^{*}_{h^{-1}}\right)^{*}\sigma}\right)(h',g).
\end{align}
The third equality follows from the definition of the adjoint action $Ad:G\rightarrow GL(\fg)$, and the fourth equality follows from the $Ad$-invariance of the Lebesgue measure $d\theta$ and the definition of the coadjoint action. Analogously, we find:
\begin{align}
\label{eq:localweylleftcov}
(U_{h}F^{W,\varepsilon}_{\sigma}U_{h}^{*})(h',g) & = F^{\varepsilon}_{\sigma}(h^{-1}h'h,\sqrt{(h^{-1}h'h)^{-1}}h^{-1}g) \\ \nonumber
 & = \varepsilon^{-n}\check{\sigma}^{1}(\varepsilon^{-1}X_{h^{-1}h'h},\sqrt{h'^{-1}}\sqrt{h'}\sqrt{(h^{-1}h'h)^{-1}}h^{-1}g) \\ \nonumber
 & = \varepsilon^{-n}\check{\sigma}^{1}(\varepsilon^{-1}Ad_{h^{-1}}(X_{h'}),\sqrt{h'^{-1}}\sqrt{h'}\sqrt{(h^{-1}h'h)^{-1}}h^{-1}g) \\ \nonumber
 & = \varepsilon^{-n}\mathcal{F}^{-1}_{1}[(Ad^{*}_{h^{-1}})^{*}\sigma](\varepsilon^{-1}X_{h'},\sqrt{h'^{-1}}\sqrt{h'}\sqrt{(h^{-1}h'h)^{-1}}h^{-1}g) \\ \nonumber
 & = \alpha_{\uL}(h\sqrt{h^{-1}h'h}\sqrt{h'^{-1}})\left(F^{W,\varepsilon}_{\left(Ad^{*}_{h^{-1}}\right)^{*}\sigma}\right)(h',g).
\end{align}
	\item[3.] By commutativity of left and right action, it follows:
\begin{align}
\label{eq:localrightcov}
(U_{h}F^{\varepsilon}_{\sigma}U_{h}^{*})(h',g) & = F^{\varepsilon}_{\sigma}(h',gh) = \varepsilon^{-n}\check{\sigma}^{1}(\varepsilon^{-1}X_{h'},gh) = F^{\varepsilon}_{\left(R^{*}_{g^{-1}}\right)^{*}\sigma},
\end{align}
\begin{align}
\label{eq:localweylrightcov}
(U_{h}F^{W,\varepsilon}_{\sigma}U_{h}^{*})(h',g) & = F^{\varepsilon}_{\sigma}(h',\sqrt{h'^{-1}}gh) = \varepsilon^{-n}\check{\sigma}^{1}(\varepsilon^{-1}X_{h'},\sqrt{h'^{-1}}gh) = F^{W,\varepsilon}_{\left(R^{*}_{g^{-1}}\right)^{*}\sigma}.
\end{align}
\end{itemize}
\end{Proof}
\end{Corollary}
Our definition of the Weyl quantisation is indeed equivalent to the one given by Landsman in terms of a ``geodesic midpoint construction’’ (cf. \cite{LandsmanMathematicalTopicsBetween}, Definition II.3.4.4.)
\begin{Lemma}
\label{lem:localweylequiv}
The operator defined by the Weyl quantisation $F^{W,\varepsilon}_{\sigma}$ of $\sigma\in C^{\infty}_{\PW,U}(\fg^{*})\hat{\otimes}\ \!C^{\infty}(G)$ is equivalent to the operator defined by the Weyl kernel
\begin{align}
\label{eq:localweylkernel}
K^{W,(\varepsilon)}_{\sigma}(h,g) & = \varepsilon^{-n}\check{\sigma}^{1}(\varepsilon^{-1}\nu_{\delta}^{-1}(h,g)),
\end{align}
where $\nu_{\delta}^{-1}:V\times V\rightarrow TV\cong V\times\fg$ maps $(h,g)$ to the tangent vector at the midpoint of the geodesic from $h$ to $g$ (w.r.t. an invariant metric on $G$).
\begin{Proof}
By definition the operator corresponding to the Weyl quantisation $F^{W,\varepsilon}_{\sigma}$ is:
\begin{align}
\label{eq:localweylconvolution}
\forall\Psi\in L^{2}(G):\ \left(\rho_{\uL}\left(F^{W,\varepsilon}_{\sigma}\right)\Psi\right)(g) & = \int_{G}dh\ F^{W,\varepsilon}_{\sigma}(h,g)\Psi(h^{-1}g) = \int_{G}dh\ F^{W,\varepsilon}_{\sigma}(gh^{-1},g)\Psi(h).
\end{align}
Thus, the kernel of $\rho_{\uL}\left(F^{W,\varepsilon}_{\sigma}\right)$ is
\begin{align}
\label{eq:localweylkernelcomputation}
K_{\rho_{\uL}\left(F^{W,\varepsilon}_{\sigma}\right)}(h,g) & = F^{W,\varepsilon}_{\sigma}(gh^{-1},g) \\ \nonumber
 & = \varepsilon^{-n}\check{\sigma}^{1}(\varepsilon^{-1}X_{gh^{-1}},\sqrt{gh^{-1}}^{-1}g) \\ \nonumber
 & = \varepsilon^{-n}\check{\sigma}^{1}(\varepsilon^{-1}X_{gh^{-1}},\exp(-\tfrac{1}{2}X_{gh^{-1}})g).
\end{align}
But, $(\exp(-\tfrac{1}{2}X_{gh^{-1}})g,X_{gh^{-1}})$ is exactly the point in $V\times\fg$ corresponding to the tangent vector at the midpoint of the geodesic $\gamma_{h\rightarrow g}:[0,1]\rightarrow V,\ \gamma_{h\rightarrow g}(t)=\exp(tX_{gh^{-1}})h,$ under right translation, because $\exp(\tfrac{1}{2}X_{gh^{-1}})h = \exp((1-\tfrac{1}{2})X_{gh^{-1}})h = \exp(-\tfrac{1}{2}X_{gh^{-1}})\exp(X_{gh^{-1}})h = \exp(-\tfrac{1}{2}X_{gh^{-1}})g$. We conclude:
\begin{align}
\label{eq:localweylequiv}
K_{\rho_{\uL}\left(F^{W,\varepsilon}_{\sigma}\right)}(h,g) & = \varepsilon^{-n}\check{\sigma}^{1}(\varepsilon^{-1}\nu_{\delta}^{-1}(h,g)) = K^{W,(\varepsilon)}_{\sigma}(h,g).
\end{align}
\end{Proof}
\end{Lemma}
Therefore, we have the following theorem proven by Landsman in the context of Riemannian manifolds.
\begin{Theorem}[cf. \cite{LandsmanMathematicalTopicsBetween}, Theorem II.3.5.1. \& Theorem III.2.8.1]
\label{thm:localweylquant}
The composition of $\rho_{\uL}$ and the Weyl quantisation
\begin{align}
\label{eq:strictlocalweylquant}
Q^{W}_{\varepsilon}:=\rho_{\uL}\circ F^{W,\varepsilon}_{(\ .\ )}:C^{\infty}_{\PW,U}(\fg^{*})\hat{\otimes}\ \!C^{\infty}(G)\rightarrow\mathcal{K}(L^{2}(G))
\end{align}
is a nondegenerate strict quantization of $C^{\infty}_{\PW,U}(\fg^{*})\hat{\otimes}\ \!C^{\infty}(G)\subset C_{0}(G\times\fg^{*})\cong C_{0}(T^{*}G)$ on $\varepsilon\in(0,1]$, i.e.
we have for all $\sigma,\tau\in C^{\infty}_{\PW,U}(\fg^{*},\R)\hat{\otimes}\ \!C^{\infty}(G,\R)$:
\begin{itemize}
	\item[1.] (nondegeneracy): \\[0.1cm] $\forall\epsilon\in(0,1]:\ Q^{W}_{\varepsilon}(\sigma)=0 \Leftrightarrow \sigma=0$.
	\item[2.] (Rieffel's condition): \\[0.1cm] $\varepsilon\mapsto||Q^{W}_{\varepsilon}(\sigma)||$ is continuous on $(0,1]$, especially \mbox{$\limit_{\varepsilon\rightarrow0}||Q^{W}_{\varepsilon}(\sigma)|| = ||\sigma||_{\infty}$.}
	\item[3.] (von Neumann's condition): \\[0.1cm] $\limit_{\varepsilon\rightarrow0}\left|\left|\tfrac{1}{2}\left(Q^{W}_{\varepsilon}(\sigma)Q^{W}_{\varepsilon}(\tau)+Q^{W}_{\varepsilon}(\tau)Q^{W}_{\varepsilon}(\sigma)\right) - Q^{W}_{\varepsilon}(\sigma\tau)\right|\right|=0$.
	\item[4.] (Dirac's condition): \\[0.1cm] $\limit_{\varepsilon\rightarrow0}\left|\left|\tfrac{i}{\varepsilon}\left[Q^{W}_{\varepsilon}(\sigma),Q^{W}_{\varepsilon}(\tau)\right] - Q^{W}_{\varepsilon}\left(\{\sigma,\tau\}_{T^{*}G}\right)\right|\right| = 0$.
\end{itemize}
Here, $\{\ ,\ \}_{T^{*}G}$ is the canonical Poisson structure on $T^{*}G$, which takes the following for on $G\times\fg^{*}$ by right translation (cp. \cite{LandsmanMathematicalTopicsBetween}, Proposition III.1.4.1):
\begin{align}
\label{eq:canonicalpoisson}
\forall \sigma,\tau\in C^{\infty}(T^{*}G):\ \{\sigma,\tau\}_{T^{*}G} & = \langle\partial_{\theta}\sigma, R\tau\rangle - \langle R\sigma, \partial_{\theta}\tau\rangle + \{\sigma,\tau\}_{-}, 
\end{align}
where $(\langle X, Rf\rangle)(g) = (R_{X}f)(g) = \frac{d}{dt}_{|t=0}f(\exp(tX)g)$ is the right differential on $G$, and \\ $\{f,f'\}_{-}(\theta)=-\theta([\partial_{\theta}f,\partial_{\theta}f'])$ is the (minus) Lie-Poisson structure on $\fg^{*}$.
\end{Theorem}
\begin{Remark}
\label{rem:weylsmoothing}
For $\sigma\in C^{\infty}_{\PW,U}(\fg^{*})\hat{\otimes}\ \!C^{\infty}(G)$, we have $K_{\rho_{\uL}(F^{(W),\varepsilon}_{\sigma})}\in C^{\infty}(G)\hat{\otimes}\ \!C^{\infty}(G)\cong C^{\infty}(G\times G)$ by definition. Therefore, $Q^{(W)}_{\varepsilon}(\sigma)$ is not only a compact operator on $L^{2}(G)$, but preserves $C^{\infty}(G)$ and extends to a smoothing operator from $\cD'(G)$ to $C^{\infty}(G)$ by the properties of convolution (cf. \cite{HoermanderTheAnalysisOf1}, Theorem 4.1.1.).
\end{Remark}
Before we introduce the Paley-Wiener-Schwartz symbol spaces $S^{K,m}_{\PW,\rho,\delta}\subset\hat{\mathcal{E}'}_{U}(\fg^{*})\hat{\otimes}\ \!C^{\infty}(G)$, we discuss the quantization of symbols that are polynomial in the momentum variables $\theta\in\fg^{*}$. To make this precise, we recall that the left and right pullback actions of $G$ on $T^{*}G$ are strongly Hamiltonian and compatible, i.e. we have bi-equivariant Poisson momentum maps (cf. \cite{LandsmanMathematicalTopicsBetween}, section III.1.4):
\begin{align}
\label{eq:momentummaps}
 J^{L^{*}_{(\ .\ )^{-1}}}: & T^{*}(G)\rightarrow\fg^{*}, &  & \\ \nonumber
 J^{L^{*}_{(\ .\ )^{-1}}}(\theta,g) & = \theta, & J^{L^{*}_{(\ .\ )^{-1}}}(L^{*}_{h^{-1}}(\theta,g)) & = Ad^{*}_{h}\left(J^{L^{*}_{(\ .\ )^{-1}}}(\theta,g)\right),\\ \nonumber
\{J^{L^{*}_{(\ .\ )^{-1}}}_{X},J^{L^{*}_{(\ .\ )^{-1}}}_{Y}\}_{T^{*}G} & = -J^{L^{*}_{(\ .\ )^{-1}}}_{[X,Y]}, & \ J^{L^{*}_{(\ .\ )^{-1}}}(R^{*}_{h^{-1}}(\theta,g)) & = J^{L^{*}_{(\ .\ )^{-1}}}(\theta,g), \\[0.25cm] \nonumber
 J^{R^{*}_{(\ .\ )^{-1}}}: & T^{*}(G)\rightarrow\fg^{*}, &  & \\ \nonumber
 J^{R^{*}_{(\ .\ )^{-1}}}(\theta,g) & = -Ad^{*}_{g^{-1}}(\theta), & J^{R^{*}_{(\ .\ )^{-1}}}(R^{*}_{h^{-1}}(\theta,g)) & = Ad^{*}_{h}\left(J^{R^{*}_{(\ .\ )^{-1}}}(\theta,g)\right), \\ \nonumber
\{J^{R^{*}_{(\ .\ )^{-1}}}_{X},J^{R^{*}_{(\ .\ )^{-1}}}_{Y}\}_{T^{*}G} & = -J^{R^{*}_{(\ .\ )^{-1}}}_{[X,Y]}, & \ J^{R^{*}_{(\ .\ )^{-1}}}(L^{*}_{h^{-1}}(\theta,g)) & = J^{R^{*}_{(\ .\ )^{-1}}}(\theta,g),
\end{align}
\begin{align}
\{J^{L^{*}_{(\ .\ )^{-1}}}_{X},J^{R^{*}_{(\ .\ )^{-1}}}_{Y}\}_{T^{*}G} = 0,\ \ \ \ X,Y\in\fg
\end{align}
where we identified $T^{*}G\cong G\times\fg^{*}$ by right translation, as above, and defined $J^{\bullet}_{X}(\theta,g):=J^{\bullet}(\theta,g)(X),\ X\in\fg$. This allows us, to make the notion of polynomial symbols precise:
\begin{Definition}
\label{def:localpoly}
Given a (smooth) function $\sigma$ on $T^{*}G\cong G\times\fg^{*}$ with values in multilinear maps from $\R\oplus\bigoplus_{n=1}^{N}\fg^{\oplus n}$ to $\R$, $N\in\mathds{N}_{0}$, of the form
\begin{align}
\label{eq:localpoly}
\sigma(\theta,g)(\oplus X) & = f_{0}(g)+\sum_{n=1}^{N}\sum_{i_{1},...,i_{n}}f_{i_{1}...i_{n}}(g)J^{L^{*}_{(\ .\ )^{-1}}}_{X_{i_{1}}}(\theta,g)...J^{L^{*}_{(\ .\ )^{-1}}}_{X_{i_{n}}}(\theta,g) \\ \nonumber
 & =  f_{0}(g)+\sum_{n=1}^{N}\sum_{i_{1},...,i_{n}}f_{i_{1}...i_{n}}(g)\theta(X_{i_{1}})\theta(X_{i_{n}})
\end{align}
for $\oplus X\in\bigoplus_{n=1}^{N}\fg^{\oplus n},\ \{f_{0},f_{i_{1}...i_{n}}\}_{i_{1},...,i_{n}}\subset\cD'(G)$, we call each $\sigma(\oplus X)\in\hat{\mathcal{E}'}_{U}(\fg^{*})\hat{\otimes}\cD'(G)$ a \textup{polynomial symbol of degree $\leq N$}. If $\sigma(\oplus X)\in\hat{\mathcal{E}'}_{U}(\fg^{*})\hat{\otimes}\ \!C^{\infty}(G)$, we say that $\sigma$ is smooth.
\end{Definition}
The Paley-Wiener-Schwartz theorem \cite{HoermanderTheAnalysisOf1} allows us to characterise the quantisation of polynomial symbols:
\begin{Corollary}
\label{cor:localpoly}
For $N\in\mathds{N}_{0}$: $\sigma$ is a polynomial symbol of degree $\leq N$ $\Leftrightarrow$ $\check{\sigma}^{1}$ is a distribution of order $\leq N$ in $\fg$ with values in $\cD'(G)$ and $\supp_{1}(\check{\sigma}^{1})\subset\{0\}$ $\Leftrightarrow$ $F^{(\varepsilon)}_{\sigma}$ is a distribution of order $\leq N$ in $G$ with values in $\cD'(G)$ and $\supp_{1}(F^{(\varepsilon)}_{\sigma})\subset\{e\}$.
\end{Corollary}
Among the polynomial symbols, we find the special cases $\sigma_{f}(\theta,g):=f(g),\ f\in C^{\infty}(G),$ and $\sigma_{X}(\theta,g):=J^{L^{*}_{(\ .\ )^{-1}}}_{X}(\theta,g)$, $X\in\fg,$. After a moments reflection, we see that the quantisation of these symbols gives rise to the commutation relations \eqref{eq:quantcomm}:
\begin{align}
\label{eq:localpositionmomentum}
\forall\Psi\in C^{\infty}(G):\ \left(Q^{(W)}_{\varepsilon}(\sigma_{f})\Psi\right)(g) & = f(g)\Psi(g), & \left(Q^{(W)}_{\varepsilon}(\sigma_{X})\Psi\right)(g) & = -i\varepsilon(R_{X}\Psi)(G), 
\end{align}
\begin{align}
\label{eq:localquantcomm}
 & Q^{(W)}_{\varepsilon}(\{\sigma_{f},\sigma_{f'}\}_{T^{*}G}) = \tfrac{i}{\varepsilon}[Q^{(W)}_{\varepsilon}(\sigma_{f}),Q^{(W)}_{\varepsilon}(\sigma_{f'})] = 0,\\ \nonumber
 & Q^{(W)}_{\varepsilon}(\{\sigma_{X},\sigma_{f}\}_{T^{*}G}) = \tfrac{i}{\varepsilon}[Q^{(W)}_{\varepsilon}(\sigma_{X}),Q^{(W)}_{\varepsilon}(\sigma_{f})] = R_{X}f, \\ \nonumber
 & Q^{(W)}_{\varepsilon}(\{\sigma_{X},\sigma_{Y}\}_{T^{*}G}) = \tfrac{i}{\varepsilon}[Q^{(W)}_{\varepsilon}(\sigma_{X}),Q^{(W)}_{\varepsilon}(\sigma_{Y})] = i\varepsilon R_{[X,Y]}.
\end{align}
Let us come to the definition of the Paley-Wiener-Schwartz symbol spaces $S^{K,m}_{\PW,\rho,\delta}$, which are analogous to the (classical) symbol spaces $S^{m}_{\rho,\delta}$ in the theory of pseudo-differential operators and Weyl quantisation on $\R^{n}$ (cf. \cite{HoermanderTheAnalysisOf1, HoermanderTheAnalysisOf3, FollandHarmonicAnalysisIn}). The main obstacle to such a definition is the fact that the exponential is no longer a diffeomorphism, which is why we need to deal with compactly supported instead of tempered distributions in $\fg$, and thus by the Paley-Wiener-Schwartz theorem with entire analytic functions on $\fg^{*}$ by means of the Fourier transform. Concerning the development of an asymptotic calculus, the analyticity requirement forbids the use of $0$-excision functions, which are standard in pseudo-differential calculus. Luckily, this alleged shortcoming can be dealt with by the method of kernel cut-off operator from the theory of Volterra-Mellin pseudo-differential operators (cf. \cite{KrainerVolterraFamiliesOf, KrainerTheCalculusOf}).
\begin{Definition}
\label{def:pwsymbol}
Let $K\sqsubset\fg$ be a convex compact subset and $m\in\R,\ 0\leq\delta\leq\rho\leq1$. A function $\sigma\in C^{\infty}(\fg_{\C}^{*},C^{\infty}(G))$ belongs to the space of \textup{Paley-Wiener-Schwartz symbols} $S^{K,m}_{\PW,\rho,\delta}$ if the following conditions are satisfied:
\begin{itemize}
	\item[1.] $\sigma:\fg_{\C}^{*}\rightarrow C^{\infty}(G)$ is (weakly) holomorphic\footnote{Since $C^{\infty}(G)$ is a Fr\'echet space in its natural topology, there is no need to distinguish between weak and strong holomorphicity (cf. \cite{RudinFunctionalAnalysis}, Theorem 3.31).},
	\item[2.] $\forall\alpha,\beta\in\N^{n}_{0}:\exists C_{\alpha\beta}>0:\forall\theta\in\fg_{\C}^{*}:\ \sup_{g\in G}|(R^{\alpha}\partial^{\beta}_{\theta}\sigma)(\theta,g)|\leq C_{\alpha\beta}\langle\theta\rangle^{m-|\beta|\rho+|\alpha|\delta}e^{H_{K}(\Im(\theta))}$,
\end{itemize}
where $\langle\theta\rangle:=(1+|\theta|_{\fg_{\C}^{*}}^{2})^{\frac{1}{2}}$ is the \textup{standard regularized distance}, $H_{K}(\Im(\theta)):=\sup_{X\in K}\Im(\theta)(X)$ is the \textup{supporting function of $K$}, and we use the standard multi index notation (w.r.t. a fixed ordered basis $\{\tau_{i}\}_{i=1}^{n}\subset\fg$ and its dual in $\fg^{*}$). Clearly, the definition is independent of the ordering of the right multi-differentials $R^{\alpha}=R^{\alpha_{1}}_{1}...R^{\alpha_{n}}_{n}$, because the commutator $[R_i,R_j]=f^{k}_{ij}R_{k}$ reduces the order and $\delta\geq0$.\\
$S^{K,m}_{\PW,\rho}(\fg^{*}_{\C})$ denotes the analogue of $S^{K,m}_{\PW,\rho,\delta}$ with $C^{\infty}(G)$ replaced by $\C$.
\end{Definition}
Clearly, smooth polynomial symbols of degree $\leq N$ belong to $S^{K,N}_{\PW,1,0}$ for all $K\sqsubset\fg$. The following relations among the symbol spaces are immediate consequences of the definition:
\begin{align}
\label{eq:symbolrelations}
S^{K,m}_{\PW,\rho,\delta}\subset S^{K,m}_{\PW,\rho',\delta}, &\ \rho\geq\rho', & S^{K,m}_{\PW,\rho,\delta}\subset S^{K,m}_{\PW,\rho,\delta'}, &\ \delta\leq\delta' \\ \nonumber
S^{K,m}_{\PW,\rho,\delta}\subset S^{K',m}_{\PW,\rho,\delta}, &\ K\subset K', & S^{K,m}_{\PW,\rho,\delta}\subset S^{K,m'}_{\PW,\rho,\delta}, &\ m\leq m'.
\end{align}
This suggests the definition of the following spaces:
\begin{align}
\label{eq:pwsymbolunion}
S^{K,\infty}_{\PW,\rho,\delta}:=\bigcup_{m\in\R}S^{K,m}_{\PW,\rho,\delta},&\ \ \ S^{K,-\infty}_{\PW}:=\bigcap_{m\in\R}S^{K,m}_{\PW,\rho,\delta}.
\end{align}
The Kohn-Nirenberg and Weyl quantisation of the restriction $\sigma_{|\fg^{*}}$ of $\sigma\in S^{K,m}_{\PW,\rho,\delta}$ to $\fg^{*}\subset\fg^{*}_{\C}$ define continuous operators on $C^{\infty}(G)$. In the following, we will abuse notation and suppress the restriction index.
\begin{Corollary}
\label{cor:pwsymbol}
Given $\sigma\in S^{K,m}_{\PW,\rho,\delta}$, there exists $\varepsilon\in(0,1]$ s.t. $K\subset U_{\varepsilon}$. Then $F^{W,\varepsilon}_{\sigma}$ defines a continuous operator on $C^{\infty}(G)$ via $\rho_{\uL}$. Moreover, if $\sigma\in S^{K,-\infty}_{\PW,\rho,\delta}$, then $\rho_{\uL}\left(F^{W,\varepsilon}_{\sigma}\right)$ is a smoothing operator from $\cD'(G)$ to $C^{\infty}(G)$.
\begin{Proof}
By the Paley-Wiener-Schwartz theorem, $\check{\sigma}^{1}$ is a distribution of order $\leq N$, for $N\in\N_{0}$ s.t. $m\leq N$, in $\fg$ with values in $C^{\infty}(G)$ and $\supp_{1}(\check{\sigma}^{1})\subset K$, which implies the first statement. The second statement follows from remark \ref{rem:weylsmoothing}.
\end{Proof}
\end{Corollary}
The optimal constants $C_{\alpha\beta}>0$ in definition \ref{def:pwsymbol} turn the symbol spaces $S^{K,m}_{\PW,\rho,\delta}$ into Fr\'echet spaces.
\begin{Proposition}
\label{prop:pwsfrechet}
Fix a convex compact subset $K\sqsubset\fg$ and $m\in\R,\ 0\leq\delta\leq\rho\leq1$. The countable family of (ordered) seminorms
\begin{align}
\label{eq:pwsnorms}
||\sigma||^{(K,m,\rho,\delta)}_{k} & := \sup_{\substack{\alpha,\beta\in\N^{n}_{0} \\ |\alpha|+|\beta|\leq k}}\sup_{(g,\theta)\in G\times\fg_{\C}^{*}}\langle\theta\rangle^{-m+|\beta|\rho-|\alpha|\delta}e^{-H_{K}(\Im(\theta))}|(R^{\alpha}\partial^{\beta}_{\theta}\sigma)(\theta,g)|,
\end{align}
$k\in\N_{0}, \sigma\in S^{K,m}_{\PW,\rho,\delta}$, defines a Fr\'echet space topology on $S^{K,m}_{\PW,\rho,\delta}$.
\begin{Proof}
We show that $S^{K,m}_{\PW,\rho,\delta}$ is Hausdorff and complete in the locally convex topology defined by the seminorms. Since metrisability follows, because the family of seminorms is countable, we may conclude that the spaces are Fr\'echet.
\begin{itemize}
	\item[1.] Assuming $||\sigma||^{(K,m,\rho,\delta)}_{k}=0$ for some $k\in\N_{0}$, we have $||\sigma||^{(K,m,\rho,\delta)}_{0}=0$, implying $u\equiv 0$. Therefore, $S^{K,m}_{\PW,\rho,\delta}$ is separated, and thus Hausdorff.
	\item[2.] Let us assume that $\{\sigma_{i}\}_{i=1}^{\infty}\subset S^{K,m}_{\PW,\rho,\delta}$ is a Cauchy sequence, i.e. $\{\sigma_{i}\}_{i=1}^{\infty}$ is Cauchy for all seminorms $||\ .\ ||^{(K,m,\rho,\delta)}_{k},\ k\in\N_{0}$. By the definition of the seminorms, we know that the convergence of $\{R^{\alpha}\sigma_{i}\}_{i=1}^{\infty}$ is uniform on compact sets in $G\times\fg^{*}_{\C}$ for all $\alpha\in\N^{n}_{0}$, which implies the existence of limit $\sigma\in C^{\infty}(\fg^{*}_{\C},C^{\infty}(G))$ in this sense. Similarly, we have compact convergence of $\{\Lambda(R^{\alpha}\sigma_{i})\}_{i=1}^{\infty}$ for all $\Lambda\in\mathcal{E}'_{\beta}(G),\ \alpha\in\N^{n}_{0}$, where $\mathcal{E}'_{\beta}(G)$ is the dual of $C^{\infty}(G)$ with its strong topology (cf. \cite{RobertsonTopologicalVectorSpaces}), which implies the existence (holomorphic) limits $\sigma^{\Lambda,\alpha}\in\mathcal{O}(\fg^{*}_{\C})$ with linear dependence on $\Lambda\in\mathcal{E}'_{\beta}(G)$. Next, we show that the maps $\Lambda\mapsto\sigma^{\Lambda,\alpha}(\theta),\ \theta\in\fg^{*}_{\C},$ are bounded, which allows us to conclude that there exists $\sigma^{\alpha}(\theta)\in C^{\infty}(G)$ s.t. $\sigma^{\Lambda,\alpha}(\theta)=\Lambda(\sigma^{\alpha}(\theta))$, because $\mathcal{E}'_{\beta}(G)$ is Fr\'echet-Montel, and thus reflexive:
\begin{align}
\label{eq:symbolproofboundedness1}
\forall\theta\in\fg^{*}_{\C}:\forall\epsilon>0\exists i_{0}\in\N:\forall i\geq i_{0}:\ |\sigma^{\Lambda,\alpha}(\theta)| & \leq |\sigma^{\Lambda,\alpha}(\theta)-\sigma^{\Lambda,\alpha}_{i}(\theta)| + |\sigma^{\Lambda,\alpha}_{i}(\theta)| \leq \epsilon + |\sigma^{\Lambda,\alpha}_{i}(\theta)| \\ \nonumber
 & \leq \epsilon + \sup_{\substack{\sigma_{k}(\theta)\in\{\sigma_{j}(\theta)\}_{j=1}^{\infty} \\ k\in\N}}|\Lambda((R^{\alpha}\sigma_{k})(\theta))|.
\end{align}
As $\{(R^{\alpha}\sigma_{j})(\theta)\}_{j=1}^{\infty}\subset C^{\infty}(G)$ is (weakly) bounded, the boundedness of the maps $\Lambda\mapsto\sigma^{\Lambda,\alpha}(\theta),\ \theta\in\fg^{*}_{\C},$ follows:
\begin{align}
\label{eq:symbolproofboundedness2}
\forall\Lambda\in\mathcal{E}'_{\beta}(G):\exists k_{\Lambda}\in\N_{0}, C_{\Lambda,\alpha}>0:\hspace{1cm} & \\ \nonumber
 \!\!\sup_{\substack{\sigma_{i}(\theta)\in\{\sigma_{j}(\theta)\}_{j=1}^{\infty} \\ i\in\N}}\!\!\!\!\!\!|\Lambda(\sigma^{\alpha}_{i}(\theta))| & \leq C_{\Lambda,\alpha} \sup_{\substack{i\in\N}}\sup_{\substack{ g\in G \\ |\gamma|\leq k_{\Lambda}}}|(R^{\alpha+\gamma}\sigma_{i})(\theta,g)| \\ \nonumber
 & \leq C_{\Lambda,\alpha} \langle\theta\rangle^{m+(k_{\Lambda}+|\alpha|)\delta}e^{H_{K}(\Im(\theta))} \sup_{\substack{i\in\N}}||\sigma_{i}||^{(K,m,\rho,\delta)}_{k_{\Lambda}+|\alpha|} \\ \nonumber
 & \leq \langle\theta\rangle^{m+k_{\Lambda}\delta}e^{H_{K}(\Im(\theta))}M_{k_{\Lambda}+|\alpha|}.
\end{align}
By construction, the map $\theta\mapsto\sigma(\theta)$ is weakly holomorphic, and $\sigma^{\alpha}(\theta)=(R^{\alpha}\sigma)(\theta)$, because the above implies:
\begin{align}
\label{eq:rightderlimit1}
\forall\alpha\in\N^{n}_{0},\theta\in\fg^{*}_{\C}:\forall\epsilon>0:\exists i_{0}&\in\N: \forall i,j\geq i_{0}: \sup_{g\in G}|(R^{\alpha}\sigma_{i})(\theta,g)-(R^{\alpha}\sigma)(\theta,g)| & <\epsilon
\end{align}
\begin{align}
\label{eq:rightderlimit2}
& \Rightarrow & \forall\Lambda\in\mathcal{E}'_{\beta}(G),\alpha\in\N^{n}_{0},\theta\in\fg^{*}_{\C}: & \forall\epsilon>0:\exists i_{0}\in\N:\forall i\geq i_{0}:  \\ \nonumber
& & |\Lambda(\sigma^{\alpha}(\theta))-\Lambda((R^{\alpha}\sigma)(\theta))| & \leq |\Lambda(\sigma^{\alpha}(\theta))-\Lambda(\sigma^{\alpha}_{i}(\theta))| + |\Lambda((R^{\alpha}\sigma_{i})(\theta))-\Lambda((R^{\alpha}\sigma)(\theta))| \\ \nonumber
 & & & \leq |\Lambda(\sigma^{\alpha}(\theta))-\Lambda(\sigma^{\alpha}_{i}(\theta))| \\ \nonumber
& & &\hspace{0.5cm} + C_{\Lambda,\alpha}\sup_{\substack{g\in G \\ |\gamma|\leq k_{\Lambda}}}|(R^{\alpha}\sigma_{i})(\theta,g)-(R^{\alpha}\sigma)(\theta,g)| \\ \nonumber
 & & & < \epsilon
\end{align}
Furthermore, we know from the assumptions that the sequences
\begin{align}
\label{eq:tausequences}
\{\tau^{K,m,\rho,\delta}_{\alpha,\beta,i}= \langle\ .\ \rangle^{-m+\rho|\beta|-\delta|\alpha|}e^{-H_{K}(\Im(\ .\ ))}(R^{\alpha}\partial^{\beta}_{\theta}\sigma_{i})\}_{i=1}^{\infty}\subset C_{b}(\fg^{*}_{\C}\times G)
\end{align}
uniformly converge to limits $\tau^{K,m,\rho,\delta}_{\alpha,\beta}\in C_{b}(\fg^{*}_{\C}\times G)$. It remains to be concluded that:
\begin{align}
\label{eq:tausigmaequal}
\tau^{K,m,\rho,\delta}_{\alpha,\beta}(\theta,g) = \langle\theta\rangle^{-m+\rho|\beta|-\delta|\alpha|}e^{-H_{K}(\Im(\theta))}(R^{\alpha}\partial^{\beta}_{\theta}\sigma)(\theta,g),
\end{align}
which follows from the convergence properties established so far:
\begin{align}
\label{eq:tausigmaconv}
\forall\alpha,\beta\in\N^{n}_{0}, & (g,\theta)\in G\times\fg^{*}_{\C}: \forall\epsilon>0:\exists i_{0}\in\N:\forall i\geq i_{0}: \\ \nonumber
 & |\tau^{K,m,\rho,\delta}_{\alpha,\beta}(\theta,g) - \langle\theta\rangle^{-m+\rho|\beta|-\delta|\alpha|}e^{-H_{K}(\Im(\theta))}(R^{\alpha}\partial^{\beta}_{\theta}\sigma)(\theta,g)| \\ \nonumber
 & \leq |\tau^{K,m,\rho,\delta}_{\alpha,\beta}(\theta,g)-\langle\theta\rangle^{-m+\rho|\beta|-\delta|\alpha|}e^{-H_{K}(\Im(\theta))}(R^{\alpha}\partial^{\beta}_{\theta}\sigma_{i})(\theta,g)| \\ \nonumber
 &\hspace{0.5cm} + \langle\theta\rangle^{-m+\rho|\beta|-\delta|\alpha|}e^{-H_{K}(\Im(\theta))}|(R^{\alpha}\partial^{\beta}_{\theta}\sigma_{i})(\theta,g)-(R^{\alpha}\partial^{\beta}_{\theta}\sigma)(\theta,g)| \\ \nonumber
 & < \epsilon
\end{align}
\end{itemize}
\end{Proof}
\end{Proposition}
\begin{Remark}
\label{rem:symbolunion}
By the preceding proposition, we can give $S^{K,\infty}_{\PW,\rho,\delta}$ a strict inductive limit topology (cf. \cite{AmannVectorValuedDistributions}), which makes it an LF-space.
\end{Remark}
\begin{Lemma}
\label{lem:pwsmult}
For $\sigma\in S^{K,m}_{\PW,\rho,\delta},\ \tau\in S^{K',m'}_{\PW,\rho',\delta'}$ s.t. $\max(\delta,\delta')\leq\min(\rho,\rho')$, we have continuous maps:
\begin{itemize}
	\item[1.] $\forall\alpha,\beta\in\N^{n}_{0}:\ \sigma\mapsto R^{\alpha}\partial^{\beta}_{\theta}\sigma\in S^{K,m-|\beta|\rho+|\alpha|\delta}_{\PW,\rho,\delta}$.
	\item[2.] $(\sigma,\tau)\mapsto\sigma\tau\in S^{K+K',m+m'}_{\PW,\min(\rho,\rho'),\max(\delta,\delta')}$.
\end{itemize}
This implies that the Poisson bracket \eqref{eq:canonicalpoisson} defines a bilinear operation
\begin{align}
\label{eq:poissonbilinear}
\{\ ,\ \}:S^{K,m}_{\PW,\rho,\delta}\times S^{K',m'}_{\PW,\rho,\delta}\rightarrow S^{K+K',m+m'-\min(\rho-\delta,2\rho-1)}_{\PW,\rho,\delta}.
\end{align}
\begin{Proof}
\begin{itemize}
	\item[1.] From the commutation relations $[R_i,R_j]=f^{k}_{ij}R_{k}$ and $\delta\geq0$, we conclude:
\begin{align}
\label{eq:pwsdiff}
\sup_{g\in G}|(R^{\gamma}\partial^{\epsilon}_{\theta}(R^{\alpha}\partial^{\beta}_{\theta}\sigma))(\theta,g)| & \leq \sup_{g\in G}|(R^{\alpha+\gamma}\partial^{\beta+\epsilon}_{\theta}\sigma)(\theta,g)| \\ \nonumber
 &\hspace{0.5cm} + \sum_{\substack{\epsilon\in\N^{n}_{0} \\ |\zeta|<|\alpha|+|\gamma|}}\sup_{g\in G}|(R^{\zeta}\partial^{\beta+\epsilon}_{\theta}\sigma)(\theta,g)| \\ \nonumber
 & \leq \langle\theta\rangle^{m-|\beta+\epsilon|\rho}e^{H_{K}(\Im(\theta))}\bigg(\!C_{(\alpha+\gamma)(\beta+\epsilon)}\langle\theta\rangle^{|\alpha+\gamma|\delta}+\!\!\!\!\!\!\!\!\sum_{\substack{\epsilon\in\N^{n}_{0} \\ |\zeta|<|\alpha|+|\gamma|}}\!\!\!\!\!\!\!\!C_{\zeta(\beta+\epsilon)}\langle\theta\rangle^{|\zeta|\delta}\!\bigg) \\ \nonumber
 & \leq C'_{(\alpha+\gamma)(\beta+\epsilon)}\langle\theta\rangle^{m-|\beta+\epsilon|\rho+|\alpha+\gamma|\delta}e^{H_{K}(\Im(\theta))}.
\end{align}
	\item[2.] Using the Leibniz formula and the fact that $H_{K}+H_{K'}=H_{K+K'}$, we find:
\begin{align}
\label{eq:pwsmult}
 & \sup_{g\in G}|(R^{\alpha}\partial^{\beta}_{\theta}(\sigma\tau))(\theta,g)| \\ \nonumber
 &\ \leq \sum_{\substack{\gamma\in\N^{n}_{0} \\ \gamma\leq\alpha}}\binom{\alpha}{\gamma}\sum_{\substack{\epsilon\in\N^{n}_{0} \\ \epsilon\leq\beta}}\binom{\beta}{\epsilon}\sup_{g\in G}|(R^{\gamma}\partial^{\epsilon}_{\theta}\sigma)(\theta,g)| \sup_{g\in G}|(R^{\alpha-\gamma}\partial^{\beta-\epsilon}_{\theta}\tau)(\theta,g)| \\ \nonumber
 &\ \leq \sum_{\substack{\gamma\in\N^{n}_{0} \\ \gamma\leq\alpha}}\binom{\alpha}{\gamma}\sum_{\substack{\epsilon\in\N^{n}_{0} \\ \epsilon\leq\beta}}\binom{\beta}{\epsilon}C_{\gamma\epsilon}\langle\theta\rangle^{m-|\epsilon|\rho+|\gamma|\delta}e^{H_{K}(\Im(\theta))}C'_{(\alpha-\gamma)(\beta-\epsilon)}\langle\theta\rangle^{m'-|\beta-\epsilon|\rho'+|\alpha-\gamma|\delta'}e^{H_{K'}(\Im(\theta))} \\ \nonumber
 &\ \ \leq C''_{\alpha\beta}e^{H_{K}(\Im(\theta))+H_{K'}(\Im(\theta))}\langle\theta\rangle^{m+m'}\sum_{\substack{\gamma\in\N^{n}_{0} \\ \gamma\leq\alpha}}\binom{\alpha}{\gamma}\langle\theta\rangle^{|\gamma|\delta+|\alpha-\gamma|\delta'}\sum_{\substack{\epsilon\in\N^{n}_{0} \\ \epsilon\leq\beta}}\binom{\beta}{\epsilon}\langle\theta\rangle^{|\epsilon|\rho-|\beta-\epsilon|\rho'} \\ \nonumber
 &\ \leq C'''_{\alpha\beta}\langle\theta\rangle^{m+m'-|\beta|\min(\rho,\rho')+|\alpha|\max(\delta,\delta')}e^{H_{K+K'}(\Im(\theta))}.
\end{align}
\end{itemize}
\end{Proof}
\end{Lemma}
As a preparation for the main theorem of this subsection, we define the kernel cut-off operator.
\begin{Definition}[cp. \cite{KrainerVolterraFamiliesOf}, Definition 3.6]
\label{def:kernelcutoff}
For $\varphi\in C^{\infty}(\fg)$, we define the \textup{kernel cut-off operator} by
\begin{align}
\label{eq:kernelcutoff}
(H(\varphi)\sigma)(\theta,g) & := \check{\sigma}^{1}(e^{-i\theta(\ .\ )}\varphi) = \int_{\fg}dX\ e^{-i\theta(X)}\varphi(X)\int_{\fg^{*}}\frac{d\theta'}{(2\pi)^{n}}e^{i\theta'(X)}\sigma(\theta,g)
\end{align}
for $\sigma\in S^{K,m}_{\PW,\rho,\delta}, (g,\theta)\in G\times\fg^{*}_{\C}$.
\end{Definition}
\begin{Remark}
\label{rem:kernelcutoff}
If $\theta\in\fg^{*}\subset\fg^{*}_{\C}$, we have
\begin{align}
\label{eq:kernelcutoffredef}
(H(\varphi)\sigma)(\theta,g) & = \check{\sigma}^{1}(e^{-i\theta(\ .\ )}\varphi) = \check{\sigma}^{1}(e^{-i\theta(\ .\ )}\varphi\chi) \\ \nonumber
 & = \int_{\fg}dX\ e^{-i\theta(X)}\varphi(X)\chi(X)\int_{\fg^{*}}\frac{d\theta'}{(2\pi)^{n}}e^{i\theta'(X)}\sigma(\theta,g) \\ \nonumber
 & = \int_{\fg}dX\ \varphi(X)\chi(X)\int_{\fg^{*}}\frac{d\theta'}{(2\pi)^{n}}e^{-i\theta'(X)}\sigma(\theta-\theta',g),
\end{align}
for some cut-off function $\chi\in C^{\infty}_{c}(\fg)$ about $\supp_{1}(\check{\sigma}^{1})$, i.e. $\chi\equiv1$ on some relatively compact neighbourhood $U$ of $\supp_{1}(\check{\sigma}^{1})$ and $\chi\equiv0$ on $\fg\setminus U'$ for some relatively compact neighbourhood $U'\supset U$. Clearly, both sides define holomorphic functions of $\theta$, that are equal on $\fg^{*}\subset\fg^{*}_{\C}$, and thus are equal on $\fg^{*}_{\C}$. The holomorphicity of the last expression can be concluded from
\begin{align}
\label{eq:kernelcutoffhol}
\int_{\fg}dX\ \varphi(X)\chi(X)\int_{\fg^{*}}\frac{d\theta'}{(2\pi)^{n}}e^{-i\theta'(X)}\sigma(\theta-\theta',g) & = \int_{\fg^{*}}\frac{d\theta'}{(2\pi)^{n}}e^{-i\theta'(X)}\sigma(\theta-\theta',g)\mathcal{F}[\phi\chi](\theta')
\end{align}
and the observation that differentiation under the integral is permitted, which follows from \mbox{$\sigma(\theta-(\!\ .\!\ ),g)\mathcal{F}[\phi\chi]\in C^{\infty}_{\PW}(\fg^{*})$.} But, the last line in \eqref{eq:kernelcutoffredef} is independent of $\chi$ due to the support properties of $\check{\sigma}^{1}$, which gives us
\begin{align}
\label{eq:kernelcutoffalt}
(H(\varphi)\sigma)(\theta,g) & = \int_{\fg}dX\ \varphi(X)\int_{\fg^{*}}\frac{d\theta'}{(2\pi)^{n}}e^{-i\theta'(X)}\sigma(\theta-\theta',g).
\end{align}
\end{Remark}
Let us establish some important properties of the kernel cut-off operator.
\begin{Theorem}[cp. \cite{KrainerVolterraFamiliesOf}, Theorem 3.7.]
\label{thm:kernelcutoff}
The kernel cut-off operator $H:C^{\infty}_{b}(\fg)\times S^{K,m}_{\PW,\rho,\delta}\rightarrow S^{K,m}_{\PW,\rho,\delta}$ continuous. If $\rho>0$ we have the asymptotic expansion
\begin{align}
\label{eq:kernelcutoffasymptotic}
H(\varphi)\sigma & \sim \sum_{\alpha\in\N^{n}_{0}}\frac{(-1)^{|\alpha|}}{\alpha !}((-i\partial_{X})^{\alpha}\varphi)(0)\partial^{\alpha}_{\theta}\sigma
\end{align}
in $S^{K,m}_{\PW,\rho,\delta}$.
\begin{Proof}
Since $S^{K,m}_{\PW,\rho,\delta}$ is a Fr\'echet space it suffices to prove that the $||\ .\ ||^{(K,m,\rho,\delta)}_{k}$-seminorms of $H(\varphi)\sigma$ are bounded by the $||\ .\||^{(K,m,\rho,\delta)}_{k}$-seminorms of $\sigma$ and the $||\ .\ ||_{\infty,k}$-seminorms of $\varphi$. By standard regularization techniques for oscillatory integrals, we have, for large enough $M\in\N_{0}$ and all $\alpha,\beta\in\N^{n}_{0}$:
\begin{align}
\label{eq:kernelcutoffestimate}
 & |(R^{\alpha}\partial^{\beta}_{\theta}H(\varphi)\sigma)(\theta,g)| \\ \nonumber
 & = |(H(\varphi)(R^{\alpha}\partial^{\beta}_{\theta}\sigma))(\theta,g)| \\ \nonumber
 & =  \int_{\fg}\frac{dX}{\langle X\rangle^{2n}}\left(\left(1-\Delta_{X}\right)^{M}\varphi\right)(X) \int_{\fg^{*}}\frac{d\theta'}{(2\pi)^{n}}e^{-i\theta'(X)}\left(\left(1-\Delta_{\theta'}\right)^{n}\langle\theta'\rangle^{-2M}(R^{\alpha}\partial^{\beta}_{\theta}\sigma)(\theta-\theta',g)\right) .
\end{align}
The contribution of $\varphi$ to the integral can be estimated by:
\begin{align}
\label{eq:phiestimate}
\left|\left(\left(1-\Delta_{X}\right)^{M}\varphi\right)(X)\right| & \leq \sum^{M}_{m=0}\binom{M}{m}\left|\left(\Delta_{X}^{m}\varphi\right)(X)\right| \\ \nonumber
 & \leq 2^{M}||\varphi||_{\infty,2 M}.
\end{align}
Applying the Leibniz rule and the estimates
\begin{align}
\label{eq:sigmaestimate}
|\partial^{\gamma}_{\theta'}\langle\theta'\rangle^{-2M}|\ \ \ \ \ \ &\ \ \ \ \ \ \leq C_{\gamma,M}\langle\theta'\rangle^{-2M-|\gamma|}, \\[0.25cm] \nonumber
|(R^{\alpha}\partial^{\beta+\gamma}_{\theta}\sigma)(\theta-\theta',g)| &\ \ \ \ \ \ \leq\ \ \ \ \ \ C_{\alpha,\beta+\gamma}e^{H_{K}(\Im)(\theta-\theta')}\langle\theta-\theta'\rangle^{m-|\beta+\gamma|\rho+|\alpha|\delta} \\ \nonumber
 & \underset{\substack{\theta'\in\fg^{*} \\ \textup{Peetre's ineq.}}}{\leq} C'_{\alpha,\beta+\gamma}e^{H_{K}(\Im)(\theta)}\langle\theta\rangle^{m-|\beta+\gamma|\rho+|\alpha|\delta}\langle\theta'\rangle^{|m-|\beta+\gamma|\rho+|\alpha|\delta|},
\end{align}
we get:
\begin{align}
& \left|\left(\left(1-\Delta_{\theta'}\right)^{n}\langle\theta'\rangle^{-2M}(R^{\alpha}\partial^{\beta}_{\theta}\sigma)(\theta-\theta',g)\right)\right| \\ \nonumber
\leq &\ C''_{\alpha,\beta,n,M}e^{H_{K}(\Im)(\theta)}\langle\theta\rangle^{m-|\beta|\rho+|\alpha|\delta}\langle\theta'\rangle^{-2M+|m|+(2n+|\beta|)\rho+|\alpha|\delta}||\varphi||_{\infty,2 M}||\sigma||^{(K,m,\rho,\delta)}_{|\alpha|+|\beta|+2n},
\end{align}
and thus the seminorm estimate:
\begin{align}
\label{eq:kernelcutoffseminorm}
 & ||H(\varphi)\sigma||^{(K,m,\rho,\delta)}_{k} \\ \nonumber
\leq &\ C'''_{k,n,M}\underbrace{\left(\int_{\fg}\frac{dX}{\langle X\rangle^{2n}}\right)}_{=:C_{2n}<\infty}\underbrace{\left(\int_{\fg^{*}}\frac{d\theta'}{(2\pi)^{n}}\langle\theta'\rangle^{-2M+|m|+(2n+k)\rho+k\delta}\right)}_{<\infty\ \textup{for large}\ M}||\varphi||_{\infty,2 M}||\sigma||^{(K,m,\rho,\delta)}_{k+2n}.
\end{align}
To obtain the asymptotic expansion, we consider the Taylor expansion of $\varphi$ at $X=0$ of order $N-1$:
\begin{align}
\label{eq:varphitaylor}
\varphi(X) & = \sum_{|\alpha|\leq N-1}\frac{1}{\alpha !}(\partial^{\alpha}_{X}\varphi)(0)X^{\alpha} + \frac{1}{(N-1)!}\sum_{|\alpha|=N}\binom{N}{\alpha}X^{\alpha}\underbrace{\int^{1}_{0}ds\ (1-s)^{N-1}(\partial^{\alpha}_{X}\varphi)(sX)}_{=:\varphi^{\alpha}_{(N)}(X)\in C^{\infty}_{b}(\fg)}.
\end{align}
Plugging this expression into the kernel cut-off operator and integrating by parts, we find:
\begin{align}
\label{eq:taylorkernelcutoff}
(H(\varphi)\sigma)(\theta,g) & = \sum_{|\alpha|\leq N-1}\frac{1}{\alpha !}(\partial^{\alpha}_{X}\varphi)(0)\int_{\fg}dX\ \int_{\fg^{*}}\frac{d\theta'}{(2\pi)^{n}}e^{-i\theta'(X)}\sigma(\theta-\theta',g)X^{\alpha} \\ \nonumber
&\ \ \ + \frac{1}{(N-1)!}\sum_{|\alpha|=N}\binom{N}{\alpha}\int_{\fg}dX\ \int_{\fg^{*}}\frac{d\theta'}{(2\pi)^{n}}e^{-i\theta'(X)}\sigma(\theta-\theta',g)X^{\alpha} \\[-0.5cm] \nonumber
&\hspace{6cm}\times\underbrace{\int^{1}_{0}ds\ (1-s)^{N-1}(\partial^{\alpha}_{X}\varphi)(sX)}_{=:\varphi^{\alpha}_{(N)}(X)\in C^{\infty}_{b}(\fg)} \\ \nonumber
& = \sum_{|\alpha|\leq N-1}\frac{(-1)^{|\alpha|}}{\alpha !}(-i\partial^{\alpha}_{X}\varphi)(0)\int_{\fg}dX\ \int_{\fg^{*}}\frac{d\theta'}{(2\pi)^{n}}e^{-i\theta'(X)}(\partial^{\alpha}_{\theta}\sigma)(\theta-\theta',g) \\ \nonumber
&\ \ \ + \frac{(-i)^{N}}{(N-1)!}\sum_{|\alpha|=N}\binom{N}{\alpha}\int_{\fg}dX\ \int_{\fg^{*}}\frac{d\theta'}{(2\pi)^{n}}e^{-i\theta'(X)}(\partial^{\alpha}_{\theta}\sigma)(\theta-\theta',g) \\[-0.5cm] \nonumber
&\hspace{6cm}\times\int^{1}_{0}ds\ (1-s)^{N-1}(\partial^{\alpha}_{X}\varphi)(sX) \\ \nonumber
& = \sum_{|\alpha|\leq N-1}\frac{(-1)^{|\alpha|}}{\alpha !}(-i\partial^{\alpha}_{X}\varphi)(0)\!\!\!\!\!\!\underbrace{(\partial^{\alpha}_{\theta}\sigma)}_{\in S^{K,m-|\alpha|\rho}_{\PW,\rho,\delta}}\!\!\!\!\!(\theta,g) \\ \nonumber
&\ \ \ + \frac{(-i)^{N}}{(N-1)!}\sum_{|\alpha|=N}\binom{N}{\alpha}\underbrace{(H(\varphi^{\alpha}_{(N)})((\partial^{\alpha}_{\theta}\sigma)))}_{\in S^{K,m-N\rho}_{\PW,\rho,\delta}}(\theta,g),
\end{align}
where the statement in the last line follows from lemma \ref{lem:pwsmult} and the continuity property of $H$, which was shown before. The result follows by the definition of asymptotic expansions, i.e.: \\[0.1cm]
Given $\{m_{k}\}_{k=1}^{\infty}\subset\R$, s.t. $\limit_{k\rightarrow\infty}m_{k}=-\infty$ and $m:=\max_{k\in\N}m_{k}$, and $\sigma_{k}\in S^{K,m_{k}}_{\PW,\rho,\delta}$, $\sigma\in S^{K,m}_{\PW,\rho,\delta}$, we say that $\sum_{k=1}^{\infty}a_{k}$ is asymptotic to $a$, $a\sim\sum_{k=1}^{\infty}a_{k}$, if
\begin{align}
\label{eq:asymptoticdef}
\forall M\in\R:\exists k_{0}\in\N:\forall k'\geq k_{0}:\ &\ a-\sum_{k=1}^{k'}a_{k}\in S^{K,M}_{\PW,\rho,\delta}.
\end{align}
$a$ is unique up to $S^{K,-\infty}_{\PW,\rho,\delta}$ (smoothing symbols).
\end{Proof}
\end{Theorem}
The preceding theorem implies the important
\begin{Corollary}[cp. \cite{KrainerVolterraFamiliesOf}, Corollary 3.8.]
\label{cor:cutoffkernel}
Given a cut-off function $\varphi\in C^{\infty}_{c}(\fg)$ around $X=0$, we have continuous operator
\begin{align}
\label{eq:kernelcutoffoperator}
\id-H(\varphi):S^{K,m}_{\PW,\rho,\delta}\longrightarrow S^{K,-\infty}_{\PW,\rho,\delta}.
\end{align}
\begin{Proof}
The Taylor expansion of $1-\varphi$ at $X=0$ vanishes to infinite order, and $\id-H(\varphi) = H(1-\varphi)$.
\end{Proof}
\end{Corollary}
Now, we can state the main theorem of this section (cp. \cite{KrainerVolterraFamiliesOf}, Theorem 3.16.).
\begin{Theorem}[Asymptotic completeness of the Paley-Wiener-Schwartz symbols]
\label{thm:asymptoticcompletness}
The symbol spaces $S^{K,m}_{\PW,\rho,\delta}$ are asymptotically complete in the following sense:\\[0.1cm]
Given $\{m_{k}\}_{k=1}^{\infty}\subset\R$, s.t. $\limit_{k\rightarrow\infty}m_{k}=-\infty$ and $m:=\max_{k\in\N}m_{k}$, and $\sigma_{k}\in C^{\infty}(G,S^{K,m_{k}}_{\PW,\rho,\delta})$, there exists $\sigma\in C^{\infty}(G,S^{K,m}_{\PW,\rho,\delta})$ s.t. $a\sim\sum_{k=1}^{\infty}a_{k}$. We call $a$ a \textup{resummation} of $\sum_{k=1}^{\infty}a_{k}$.
\end{Theorem}
Before we prove the theorem, we need some lemmata, following the idea of proof for Volterra symbols by Krainer \cite{KrainerVolterraFamiliesOf}.
\begin{Lemma}
\label{lem:auxestimate1}
Given $\beta\in\N^{n}_{0}$, $\varphi\in C^{\infty}_{c}(\fg)$ and $\sigma\in S^{K,-(2(n+1)+|\beta|)}_{\PW,\rho}(\fg^{*}_{\C})$, we have:
\begin{align}
\label{eq:auxestimate1}
\sup_{\theta\in\fg^{*}_{\C}}e^{-2r|\Im(\theta)|}|(H((-i\partial_{X})^{\beta}\varphi_{c})(e^{ir\tau_{n}(\ .\ )}\sigma))(\theta)| 
 & \leq k_{n}(\varphi,\beta)\frac{1}{c^{n+1}}||e^{ir\tau_{n}(\ .\ )}\sigma||^{(2r,-(2(n+1)+|\beta|),\rho,\delta)}_{0},
\end{align}
for some constant $k_{n}(\varphi,\beta)>0$, where $c\in[0,\infty)$, $\varphi_{c}(X)=\varphi(cX)$, $r:=\inf\{r'\in[0,\infty)\ |\ K\subset B_{r'}(0)=\{X\in\fg\ |\ |X|\leq r'\}\}$ and $e^{ir\tau_{n}(\ .\ )}(\theta):=e^{ir\theta(\tau_{n})}$ with $\tau_{n}$ the nth basis vector of $\fg$.
\begin{Proof}
First, observe that multiplication of $\sigma$ with $e^{ir\tau_{n}(\ .\ )}$ shifts the support of $\check{\sigma}$, $\mathcal{F}^{-1}[e^{ir\tau_{n}(\ .\ )}\sigma]=\check{\sigma}(\ .\ +r\tau_{n})=:\check{\sigma}_{r}$, s.t. $0\in\fg$ is not an interior point of $\supp(\mathcal{F}^{-1}[e^{ir\tau_{n}(\ .\ )}\sigma])\subset K-r\tau_{n}\subset B_{2r}(0)$. Second, we have the equivalent estimates:
\begin{align}
\label{eq:auxestimate1equiv}
\sup_{\substack{\theta\in\fg^{*}_{\C} \\ |\alpha|=|\beta|+2(n+1) }}\!\!\!\!\!\!e^{-2r|\Im(\theta)|}|\theta^{\alpha}(e^{ir\tau_{n}(\ .\ )}\sigma)(\theta)| < \infty\Leftrightarrow\sup_{\theta\in\fg^{*}_{\C}}e^{-2r|\Im(\theta)|}\langle\theta\rangle^{|\beta|+2(n+1)}|(e^{ir\tau_{n}(\ .\ )}\sigma)(\theta)| < \infty.
\end{align}
Now, the assumptions imply that:
\begin{itemize}
	\item[1.] $\mathcal{F}^{-1}[e^{ir\tau_{n}(\ .\ )}\sigma]_{|\fg_{+}}\equiv 0$, where $\fg_{+}:=\{X\in\fg\ |\ \theta_{n}(X)>0\}$ ($\theta_{n}$ is dual to $\tau_{n}$).
	\item[2.] $\mathcal{F}^{-1}[e^{ir\tau_{n}(\ .\ )}\sigma]\in C^{|\beta|+n+1}$, since
\begin{align}
\label{eq:auxestimateder}
(\partial^{\alpha}_{X}\mathcal{F}^{-1}[e^{ir\tau_{n}(\ .\ )}\sigma])(X) & = i^{|\alpha|}\int_{\fg^{*}}\frac{d\theta}{(2\pi)^{n}}\theta^{\alpha}e^{i\theta(X)}e^{ir\theta(\tau_{n})}\sigma(\theta) \\ \nonumber
 & = i^{|\alpha|}\int_{\fg^{*}}\frac{d\theta}{(2\pi)^{n}}\langle\theta\rangle^{-(n+1)}\left(\langle\theta\rangle^{n+1}\theta^{\alpha}e^{i\theta(X)}e^{ir\theta(\tau_{n})}\sigma(\theta)\right)
\end{align}
is finite for $|\alpha|\leq\beta+n+1$ by \eqref{eq:auxestimate1equiv}.
\end{itemize}
The Taylor expansion of $\mathcal{F}^{-1}[e^{ir\tau_{n}(\ .\ )}\sigma]$ at $X=0$ up to order $|\beta|+n$ reads:
\begin{align}
\label{eq:auxestimatetaylor}
\mathcal{F}^{-1}[e^{ir\tau_{n}(\ .\ )}\sigma](X) & = \sum_{|\alpha|=|\beta|+n+1}\!\!\!\!R_{\alpha}(0)X^{\alpha}, & |R_{\alpha}(0)| & \leq\frac{1}{\alpha!}\sup_{\substack{X\in K-r\tau_{n} \\ |\gamma|=|\alpha| }}|\partial^{\gamma}_{X}\mathcal{F}^{-1}[e^{ir\tau_{n}(\ .\ )}\sigma](X)| \\ \nonumber
 & & & =\frac{1}{\alpha!}\sup_{\substack{X\in K \\ |\gamma|=|\alpha| }}|(\partial^{\gamma}_{X}\check{\sigma})(X)|
\end{align}
since $\partial^{\alpha}_{X}\mathcal{F}^{-1}[e^{ir\tau_{n}(\ .\ )}\sigma]$ vanishes at $X=0$ for all $|\alpha|<|\beta|+n+1$.
Now, we come to the proof of \eqref{eq:auxestimate1}:
\begin{align}
 & \sup_{\theta\in\fg^{*}_{\C}}e^{-2r|\Im(\theta)|}|(H((-i\partial_{X})^{\beta}\varphi_{c})(e^{ir\tau_{n}(\ .\ )}\sigma))(\theta)| \\ \nonumber
 & = \sup_{\theta\in\fg^{*}_{\C}}e^{-2r|\Im(\theta)|}|\int_{\fg}dX\ e^{-i\theta(X)}((-i\partial_{X})^{\beta}\varphi_{c})(X)\check{\sigma}_{r}(X)| \\ \nonumber
 & \leq \sup_{\theta\in\fg^{*}_{\C}}e^{-2r|\Im(\theta)|}\int_{\fg}dX\ e^{|\Im(\theta)||X|}|((-i\partial_{X})^{\beta}\varphi_{c})(X)\check{\sigma}_{r}(X)| \\ \nonumber
 & \leq \int_{\fg}dX\ |((-i\partial_{X})^{\beta}\varphi_{c})(X)\check{\sigma}_{r}(X)| = \int_{\fg}\frac{dX}{\langle X\rangle^{\frac{n+1}{2}}}\ \langle X\rangle^{\frac{n+1}{2}}|((-i\partial_{X})^{\beta}\varphi_{c})(X)\check{\sigma}_{r}(X)| \\ \nonumber
 & \leq \underbrace{\left(\int_{\fg}\frac{dX}{\langle X\rangle^{n+1}}\right)^{\frac{1}{2}}}_{=:C_{n+1}<\infty}\left(\int_{\fg}dX\ \langle X\rangle^{n+1}|((-i\partial_{X})^{\beta}\varphi_{c})(X)\check{\sigma}_{r}(X)|^{2}\right)^{\frac{1}{2}} \\ \nonumber
 & \leq C_{n+1}^{\frac{1}{2}}\bigg(\int_{\fg}dX\langle X\rangle^{n+1}|((-i\partial_{X})^{\beta}\varphi_{c})(X)|^{2}\bigg(\sum_{|\alpha|=|\beta|+n+1}\!\!\!\!\!\!\!\!\!|X^{\alpha}||R_{\alpha}(0)|\bigg)^{2}\bigg)^{\frac{1}{2}} \\ \nonumber
 & \leq C_{n+1}^{\frac{1}{2}}\sup_{\substack{X\in K \\ |\gamma|=|\beta|+n+1 }}|(\partial^{\gamma}_{X}\check{\sigma})(X)|\bigg(\int_{\fg}dX\langle X\rangle^{n+1}|((-i\partial_{X})^{\beta}\varphi_{c})(X)\!\!\!\!\sum_{|\alpha|=|\beta|+n+1}\!\!\!\!\!\!\!\!\! \frac{1}{\alpha!}X^{\alpha}|^{2}\bigg)^{\frac{1}{2}} \\
\nonumber
& \leq C_{n+1}^{\frac{1}{2}}\frac{C_{n+1}}{(2\pi)^{n}}\bigg(\sup_{\theta\in\fg^{*}_{\C}}e^{-2r|\Im(\theta)|}\langle\theta\rangle^{|\beta|+2(n+1)}|(e^{ir\tau_{n}(\ .\ )}\sigma)(\theta)|\bigg) \\ \nonumber
&\hspace{4cm}\times\bigg(\int_{\fg}dX\langle X\rangle^{n+1}|((-i\partial_{X})^{\beta}\varphi_{c})(X)\!\!\!\!\sum_{|\alpha|=|\beta|+n+1}\!\!\!\!\!\!\!\!\! \frac{1}{\alpha!}X^{\alpha}|^{2}\bigg)^{\frac{1}{2}} \\ \nonumber
& = C_{n+1}^{\frac{1}{2}}\frac{C_{n+1}}{(2\pi)^{n}}||e^{ir\tau_{n}(\ .\ )}\sigma||^{(2r,-(2(n+1)+|\beta|),\rho,\delta)}_{0} \\ \nonumber
&\hspace{4cm}\times\bigg(c^{-(3n+2)}\int_{\fg}dX\underbrace{\langle c^{-1}X\rangle^{n+1}}_{\leq\langle X\rangle^{n+1}}|((-i\partial_{X})^{\beta}\varphi)(X)\!\!\!\!\sum_{|\alpha|=|\beta|+n+1}\!\!\!\!\!\!\!\!\! \frac{1}{\alpha!}X^{\alpha}|^{2}\bigg)^{\frac{1}{2}} \\ \nonumber
& \leq C_{n+1}^{\frac{1}{2}}\frac{C_{n+1}}{(2\pi)^{n}}||e^{ir\tau_{n}(\ .\ )}\sigma||^{(2r,-(2(n+1)+|\beta|),\rho,\delta)}_{0} \\ \nonumber
&\hspace{4cm}\times\bigg(c^{-2(n+1)}\int_{\fg}dX\langle X\rangle^{n+1}|((-i\partial_{X})^{\beta}\varphi)(X)\!\!\!\!\sum_{|\alpha|=|\beta|+n+1}\!\!\!\!\!\!\!\!\! \frac{1}{\alpha!}X^{\alpha}|^{2}\bigg)^{\frac{1}{2}} \\ \nonumber
& = \underbrace{C_{n+1}^{\frac{1}{2}}\frac{C_{n+1}}{(2\pi)^{n}}\bigg(\int_{\fg}dX\langle X\rangle^{n+1}|((-i\partial_{X})^{\beta}\varphi)(X)\!\!\!\!\sum_{|\alpha|=|\beta|+n+1}\!\!\!\!\!\!\!\!\! \frac{1}{\alpha!}X^{\alpha}|^{2}\bigg)^{\frac{1}{2}}}_{=:k_{n}(\varphi,\beta)} \\ \nonumber
&\hspace{6cm}\times c^{-(n+1)}||e^{ir\tau_{n}(\ .\ )}\sigma||^{(2r,-(2(n+1)+|\beta|),\rho,\delta)}_{0}
\end{align}
\end{Proof}
\end{Lemma}
\begin{Lemma}
\label{lem:auxestimate2}
Let $N\in\N_{0}$, $\{M_{\alpha\beta}\}_{|\alpha|+|\beta|\leq N}\subset\N$, $\varphi\in C^{\infty}_{c}(\fg)$ and $\sigma\in\mathcal{O}(\fg^{*}_{\C},C^{\infty}(G))$, s.t. 
\begin{align}
\label{eq:auxestimate2cond}
\sup_{\substack{ (g,\theta)\in G\times\fg^{*}_{\C} \\ |\alpha|+|\beta|\leq N }}e^{-r|\Im(\theta)|}\langle\theta\rangle^{2\ceil*{\frac{M_{\alpha\beta}}{2}}+2(n+1)}|(R^{\alpha}\partial^{\beta}_{\theta}\sigma)(\theta,g)| & < \infty,
\end{align}
then we have:
\begin{align}
\label{eq:auxestimate2}
\ &\ \sup_{\substack{ (g,\theta)\in G\times\fg^{*}_{\C} \\ |\alpha|+|\beta|\leq N }}e^{-2r|\Im(\theta)|}\langle\theta\rangle^{M_{\alpha\beta}}|(R^{\alpha}\partial^{\beta}_{\theta}H(\varphi_{c})(e^{ir\tau_{n}(\ .\ )}\sigma))(\theta,g)| \\ \nonumber
\leq\ &\ \tilde{k}_{n}(\varphi,N,\{M_{\alpha\beta}\})\frac{1}{c^{n+1}}\sup_{\substack{ (g,\theta)\in G\times\fg^{*}_{\C} \\ |\alpha|+|\beta|\leq N }}e^{-2r|\Im(\theta)|}\langle\theta\rangle^{2\ceil*{\frac{M_{\alpha\beta}}{2}}+2(n+1)}|(R^{\alpha}\partial^{\beta}_{\theta}(e^{ir\tau_{n}(\ .\ )}\sigma))(\theta,g)|
\end{align}
for some constant $\tilde{k}_{n}(\varphi,M,N)>0$, where the notation of lemma \ref{lem:auxestimate1} is employed.
\begin{Proof}
The assumptions imply:
\begin{align}
\label{eq:auxestimate2equiv}
\sup_{\substack{ (g,\theta)\in G\times\fg^{*}_{\C} \\ |\alpha|+|\beta|\leq N }}e^{-2r|\Im(\theta)|}\langle\theta\rangle^{2\ceil*{\frac{M_{\alpha\beta}}{2}}+2(n+1)}|(R^{\alpha}\partial^{\beta}_{\theta}(e^{ir\tau_{n}(\ .\ )}\sigma))(\theta,g)| & < \infty,
\end{align}
as multiplication of $\sigma$ with $e^{ir\tau_{n}(\ .\ )}$ shifts the support of $\check{\sigma}^{1}$, and modifies the decay properties of $\sigma$ in the imaginary directions of $\fg^{*}_{\C}$ according to the Paley-Wiener-Schwartz theorem. Next, we observe, that the following properties hold, due to the definition of $H$:
\begin{itemize}
	\item[1.] $R^{\alpha}\partial^{\beta}_{\theta}H(\varphi_{c})(e^{ir\tau_{n}(\ .\ )}\sigma) = H(\varphi_{c})(R^{\alpha}\partial^{\beta}_{\theta}(e^{ir\tau_{n}(\ .\ )}\sigma))$.
	\item[2.] $\theta^{\alpha}(H(\varphi_{c})(e^{ir\tau_{n}(\ .\ )}\sigma))(\theta,g) = \sum_{\beta\leq\alpha}\binom{\alpha}{\beta}H((-i\partial_{X})^{\beta}\varphi_{c})((\ .\ )^{\alpha-\beta}e^{ir\tau_{n}(\ .\ )}\sigma)(\theta,g)$.
	\item[3.] $\langle\theta\rangle^{M_{\alpha\beta}}\leq\sum_{k=0}^{\ceil*{\frac{M_{\alpha\beta}}{2}}}\binom{\ceil*{\frac{M_{\alpha\beta}}{2}}}{k}\sum_{|\gamma|=k}\binom{k}{\gamma}\theta^{2\gamma}$.
\end{itemize}
Combining these properties with lemma \ref{lem:auxestimate1}, we find:
\begin{align}
\ &\hspace{0.4cm} \sup_{\substack{ (g,\theta)\in G\times\fg^{*}_{\C} \\ |\alpha|+|\beta|\leq N }}e^{-2r|\Im(\theta)|}\langle\theta\rangle^{M_{\alpha\beta}}|(R^{\alpha}\partial^{\beta}_{\theta}H(\varphi_{c})(e^{ir\tau_{n}(\ .\ )}\sigma))(\theta,g)| \\\nonumber
\ & \leq \sup_{\substack{ (g,\theta)\in G\times\fg^{*}_{\C} \\ |\alpha|+|\beta|\leq N }}e^{-2r|\Im(\theta)|}\sum_{k=0}^{\ceil*{\frac{M_{\alpha\beta}}{2}}}\binom{\ceil*{\frac{M_{\alpha\beta}}{2}}}{k}\sum_{|\gamma|=k}\binom{k}{\gamma}|\theta^{2\gamma}(R^{\alpha}\partial^{\beta}_{\theta}H(\varphi_{c})(e^{ir\tau_{n}(\ .\ )}\sigma))(\theta,g)| \\ \nonumber
\ & \leq \sup_{\substack{ (g,\theta)\in G\times\fg^{*}_{\C} \\ |\alpha|+|\beta|\leq N }}e^{-2r|\Im(\theta)|}\sum_{k=0}^{\ceil*{\frac{M_{\alpha\beta}}{2}}}\binom{\ceil*{\frac{M_{\alpha\beta}}{2}}}{k}\sum_{|\gamma|=k}\binom{k}{\gamma}\sum_{\zeta\leq2\gamma}\binom{2\gamma}{\zeta} \\ \nonumber
\ & \hspace{4cm}\times|H((-i\partial_{X})^{\zeta}\varphi_{c})((\ .\ )^{2\gamma-\zeta}R^{\alpha}\partial^{\beta}_{\theta}(e^{ir\tau_{n}(\ .\ )}\sigma))(\theta,g)| \\ \nonumber
\ & \leq \sup_{|\alpha|+|\beta|\leq N}e^{-2r|\Im(\theta)|}\sum_{k=0}^{\ceil*{\frac{M_{\alpha\beta}}{2}}}\binom{\ceil*{\frac{M_{\alpha\beta}}{2}}}{k}\sum_{|\gamma|=k}\binom{k}{\gamma}\sum_{\zeta\leq2\gamma}\binom{2\gamma}{\zeta} \\ \nonumber
\ &\hspace{4cm}\times\sup_{(g,\theta)\in G\times\fg^{*}_{\C}}|H((-i\partial_{X})^{\zeta}\varphi_{c})((\ .\ )^{2\gamma-\zeta}R^{\alpha}\partial^{\beta}_{\theta}(e^{ir\tau_{n}(\ \!.\ \!)}\sigma))(\theta,g)| \\ \nonumber
\ &\!\!\!\! \underset{\textup{\eqref{eq:auxestimate1}}}{\leq}\sup_{|\alpha|+|\beta|\leq N}\sum_{k=0}^{\ceil*{\frac{M_{\alpha\beta}}{2}}}\binom{\ceil*{\frac{M_{\alpha\beta}}{2}}}{k}\sum_{|\gamma|=k}\binom{k}{\gamma}\sum_{\zeta\leq2\gamma}\binom{2\gamma}{\zeta}\frac{k(\varphi,\zeta)}{c^{n+1}} \\ \nonumber
\ &\hspace{4cm}\times\sup_{(g,\theta)\in G\times\fg^{*}_{\C}}e^{-2r|\Im(\theta)|}\langle\theta\rangle^{|\zeta|+2(n+1)}|(\theta^{2\gamma-\zeta}R^{\alpha}\partial^{\beta}_{\theta}(e^{ir\tau_{n}(\ .\ )}\sigma))(\theta,g)| \\ \nonumber
\ & \leq \sup_{|\alpha|+|\beta|\leq N}\!\!\!\underbrace{\sum_{k=0}^{\ceil*{\frac{M_{\alpha\beta}}{2}}}\!\!\binom{\ceil*{\frac{M_{\alpha\beta}}{2}}}{k}\!\!\sum_{|\gamma|=k}\!\!\binom{k}{\gamma}\!\!\sum_{\zeta\leq2\gamma}\!\!\binom{2\gamma}{\zeta}\frac{k(\varphi,\zeta)}{c^{n+1}}}_{\leq\frac{1}{c^{n+1}}\tilde{k}(\varphi,N,\{M_{\alpha\beta}\})}\\ \nonumber
\ &\hspace{4cm}\times\sup_{(g,\theta)\in G\times\fg^{*}_{\C}}\!\!\!\!\!\!e^{-2r|\Im(\theta)|}\langle\theta\rangle^{2\ceil*{\frac{M_{\alpha\beta}}{2}}+2(n+1)}|(R^{\alpha}\partial^{\beta}_{\theta}(e^{ir\tau_{n}(\ \!.\ \!)}\sigma))(\theta,g)|,
\end{align}
which concludes the proof.
\end{Proof}
\end{Lemma}
With the help of the preceding lemmata, we can prove a crucial convergence result for the symbol spaces (cp. \cite{KrainerVolterraFamiliesOf}, Proposition 3.14.).
\begin{Proposition}
\label{prop:convergenceprop}
Given $\{m_{k}\}_{k=1}^{\infty}\subset\R$, s.t. $m_{k}\geq m_{k+1}\underset{k\rightarrow\infty}{\longrightarrow}-\infty$, and a countable system of bounded sets $\{S_{k_{j}}\}_{j\in\N}\subset S^{K,m_{k}}_{\PW,\rho,\delta}$ for every $k\in\N$, then there exists a sequence $\{c_{i}\}_{i=1}^{\infty}\subset[1,\infty)$, with $c_{i}< c_{i+1}\underset{i\rightarrow\infty}{\longrightarrow}\infty$, s.t. 
\begin{align}
\label{eq:convergentsum}
\sum_{i=k}^{\infty}\sup_{\sigma\in S_{i_{j}}}p(H(\varphi_{d_{i}})(e^{ir\tau_{n}(\ .\ )}\sigma)) & < \infty,
\end{align}
for all $j,k\in\N$, all continuous seminorms $p$ on $S^{B_{2r}(0),m_{k}}_{\PW,\rho,\delta}$ and all sequences $\{d_{i}\}_{i=1}^{\infty}\subset[1,\infty)$ with $\forall i\in\N: d_{i}\geq c_{i}$. Here, we use again the notation of lemma \ref{lem:auxestimate1}.
\begin{Proof}
Without loss of generality, we may assume that $\{m_{k}\}_{k=1}^{\infty}\subset\R_{-}$ and $S_{k_{j}}\subset S_{k_{j+1}}$ for all $j,k\in\N$. For all $l\in\N$, we define (ordered) seminorms
\begin{align}
\label{eq:convergenceseminorms}
q^{2r,\rho,\delta}_{l}(\sigma) & := \sup_{\substack{ (g,\theta)\in G\times\fg^{*}_{\C} \\ |\alpha|+|\beta|\leq l }}e^{-2r|\Im(\theta)|}\langle\theta\rangle^{-m_{l}+|\beta|\rho-|\alpha|\delta}|(R^{\alpha}\partial^{\beta}_{\theta}\sigma)(\theta,g)|,\ q^{2r,\rho,\delta}_{l}\leq q^{2r,\rho,\delta}_{l+1}.
\end{align}
Using the preceding lemmata, we find for suitable $\sigma$ and $c\in[1,\infty)$:
\begin{align}
\label{eq:lemmaestimate}
q^{2r,\rho,\delta}_{l}(H(\varphi_{c})(e^{ir\tau_{n}(\ .\ )}\sigma)) & = \sup_{\substack{ (g,\theta)\in G\times\fg^{*}_{\C} \\ |\alpha|+|\beta|\leq l }}e^{-2r|\Im(\theta)|}\langle\theta\rangle^{-m_{l}+|\beta|\rho-|\alpha|\delta}|(R^{\alpha}\partial^{\beta}_{\theta}H(\varphi_{c})(e^{ir\tau_{n}(\ .\ )}\sigma))(\theta,g)| \\ \nonumber
 & \leq \frac{1}{c^{n+1}}\tilde{k}(\varphi,l,\{M_{\alpha\beta}\})\\ \nonumber
\ &\hspace{0.5cm}\times\sup_{\substack{ (g,\theta)\in G\times\fg^{*}_{\C} \\ |\alpha|+|\beta|\leq l }}e^{-2r|\Im(\theta)|}\langle\theta\rangle^{2\ceil*{\frac{M_{\alpha\beta}}{2}}+2(n+1)}|(R^{\alpha}\partial^{\beta}_{\theta}(e^{ir\tau_{n}(\ .\ )}\sigma))(\theta,g)|,
\end{align}
where $M_{\alpha\beta}:=\ceil*{-m_{l}+|\beta|\rho-|\alpha|\delta}$. By assumption on the sequence $\{m_{k}\}_{k=1}^{\infty}$, we can find $i_{0}\in\N,\ i_{0}\geq l$, s.t. $2\ceil*{\frac{M_{\alpha\beta}}{2}}+2(n+1)\leq -m_{i_{0}}+|\beta|\rho-|\alpha|\delta$, for $|\alpha|+|\beta|\leq l$, i.e. we need $m_{i_{0}}+2(n+2)+1\leq m_{l}$. Thus, we get the estimate:
\begin{align}
\label{eq:centralestimate}
\forall l\in\N:\exists i_{0}\in\N:\forall i\geq i_{0}: &\ q^{2r,\rho,\delta}_{l}(H(\varphi_{c})(e^{ir\tau_{n}(\!\ .\!\ )}\sigma))\leq\frac{\tilde{k}(\varphi,l,\{M_{\alpha\beta}\})}{c^{n+1}}q^{2r,\rho,\delta}_{i}(e^{ir\tau_{n}(\!\ .\!\ )}\sigma),
\end{align}
for $\sigma\in S^{K,m_{i}}_{\PW,\rho,\delta}$. The existence of the sequence $\{c_{i}\}_{i=1}^{\infty}\subset[1,\infty)$ with the prescribed properties follows by induction. Following Krainer \cite{KrainerVolterraFamiliesOf}, we construct sequences $\{c_{l_{i}}\}_{i=1}^{\infty}\subset[1,\infty)$ for $l\in\N$, and take $\{c:=c_{i_{i}}\}_{i=1}^{\infty}$:
\begin{itemize}
	\item[1.] Let $l=1$: By \eqref{eq:centralestimate}, we can find a sequence $\{c_{1_{i}}\}_{i=1}^{\infty}\subset[1,\infty)$, $c_{i}< c_{i+1}\underset{i\rightarrow\infty}{\longrightarrow}\infty$, s.t. for all $i\in\N$ with $m_{i}+2(n+2)+1\leq m_{1}$
\begin{align}
\label{eq:sequenceinductionstart}
\sup_{\sigma\in S_{i_{1}}}q^{2r,\rho,\delta}_{1}(H(\varphi_{d_{i}})(e^{ir\tau_{n}(\ .\ )}\sigma)) & < 2^{-i}
\end{align}
holds for all $\{d_{i}\}_{i=1}^{\infty}\subset[1,\infty)$ with $\forall i\in\N: d_{i}\geq c_{1_{i}}$.
	\item[2.] Let $\{c_{l_{i}}\}_{i=1}^{\infty}\subset[1,\infty)$ be constructed: By \eqref{eq:centralestimate}, we find a subsequence $\{c_{(l+1)_{i}}\}_{i=1}^{\infty}\subset\{c_{l_{i}}\}_{i=1}^{\infty}$, s.t. for all $i\in\N$ with $m_{i}+2(n+2)+1\leq m_{l+1}$
\begin{align}
\label{eq:sequenceinductionstep}
\sup_{\sigma\in S_{i_{l+1}}}q^{2r,\rho,\delta}_{l+1}(H(\varphi_{d_{i}})(e^{ir\tau_{n}(\ .\ )}\sigma)) & < 2^{-i}
\end{align}
holds for all $\{d_{i}\}_{i=1}^{\infty}\subset[1,\infty)$ with $\forall i\in\N: d_{i}\geq c_{(l+1)_{i}}$.
	\item[3.] By construction, the diagonal sequence $\{c:=c_{i_{i}}\}_{i=1}^{\infty}\subset[1,\infty)$, has the property $c_{i}\geq c_{l_{i}}$ for $i\geq l$ and  $c_{i}< c_{i+1}\underset{i\rightarrow\infty}{\longrightarrow}\infty$.
	\item[4.] Let $j,k\in\N$ and $p$ be a continuous seminorm on $S^{B_{2r}(0),m_{k}}_{\PW,\rho,\delta}$, then there exist $l_{0}$ s.t. the restriction of $p$ to $S^{B_{2r}(0),m_{i}}_{\PW,\rho,\delta}$ is dominated by $q^{2r,\rho,\delta}_{l_{0}}$ with a constant independent of $i$ and $S_{i_{j}}\subset S_{i_{l_{0}}}$ for almost all $i\in\N$. These assertions follow from the inclusion properties of the symbol spaces (see \eqref{eq:symbolrelations}). Employing the continuity of $H$ (see theorem \ref{thm:kernelcutoff}), we conclude that the series \eqref{eq:convergentsum} is indeed convergent for given data $j,k\in\N$ and $p$.
\end{itemize}
\end{Proof}
\end{Proposition}
\begin{Proof}[of Theorem \ref{thm:asymptoticcompletness}]
Without loss of generality, we may assume $m_{k}\geq m_{k+1}\underset{k\rightarrow\infty}{\longrightarrow}-\infty$. For $j,k\in\N$, we define $S_{k_{j}}:=\{(R_{1}^{\alpha}\sigma_{k})(g)\ |\ g\in G,\ |\alpha|\leq j\}\subset S^{K,m_{k}}_{\PW,\rho,\delta}$. Here, $R_{1}$ is the right differential in the first group variable. Since $\sigma_{k}\in C^{\infty}(G,S^{K,m_{k}}_{\PW,\rho,\delta})$, we know that the sets $S_{k_{j}}$ are bounded in $S^{K,m_{k}}_{\PW,\rho,\delta}$. Now, we choose a cut-off function $\varphi\in C^{\infty}_{c}(\fg)$ around $X=0$, and apply proposition \ref{prop:convergenceprop} to obtain a sequence $\{c_{i}\}_{i=1}^{\infty}\subset[1,\infty)$, s.t.
\begin{align}
\label{eq:asymptoticconvergence}
\sum_{i=k}^{\infty}\sup\left\{p\left((H(\varphi_{c_{i}})(e^{ir\tau_{n}(\ .\ )}R^{\alpha}\sigma))(g)\right)\ |\ g\in G,\ |\alpha|\leq j\right\} & < \infty
\end{align}
for all continuous seminorms $p$ on $S^{B_{2r}(0),m_{k}}_{\PW,\rho,\delta}$. Therefore, the sum
\begin{align}
\label{eq:asymptoticresum}
a^{(r)} & :=\sum_{i=1}^{\infty}H(\varphi_{c_{i}})(e^{ir\tau_{n}(\ .\ )}\sigma)
\end{align}
is unconditionally convergent in $C^{\infty}(G,S^{K-r\tau_{n},m}_{\PW,\rho,\delta})$. Now, we define:
\begin{align}
\label{eq:asymptoticresumshift}
a & := e^{-ir\tau_{n}(\ .\ )}a^{(r)},
\end{align}
which is in $C^{\infty}(G,S^{K,m}_{\PW,\rho,\delta})$. It follows that the $a\sim\sum_{k=1}^{\infty}a_{k}$:
\begin{align}
\label{eq:asymptoticity}
a - \sum_{i=1}^{k}a_{i} & =  e^{-ir\tau_{n}(\ .\ )}\left(a^{r}-\sum_{i=1}^{k} e^{ir\tau_{n}(\ .\ )}a_{i}\right) \\ \nonumber
 & =  e^{-ir\tau_{n}(\ .\ )}\sum_{i=k+1}^{\infty}H(\varphi_{c_{i}})(e^{ir\tau_{n}(\ .\ )}\sigma) + e^{-ir\tau_{n}(\ .\ )}\sum_{i=1}^{k}(\id-H(\varphi_{c_{i}}))(e^{ir\tau_{n}(\ .\ )}\sigma) \\ \nonumber
 & = \underbrace{e^{-ir\tau_{n}(\ .\ )}\sum_{i=k+1}^{\infty}H(\varphi_{c_{i}})(e^{ir\tau_{n}(\ .\ )}\sigma)}_{\in C^{\infty}\left(G,S^{K,m_{k+1}}_{\PW,\rho,\delta}\right)} + \underbrace{e^{-ir\tau_{n}(\ .\ )}\sum_{i=1}^{k}H(1-\varphi_{c_{i}})(e^{ir\tau_{n}(\ .\ )}\sigma)}_{\in C^{\infty}\left(G,S^{K,-\infty}_{\PW,\rho,\delta}\right)\ \textup{Cor.\ \ref{cor:cutoffkernel}} }.
\end{align}
\end{Proof}
We conclude the subsection by showing that the operator product of two Weyl quantisations $F^{W,\varepsilon}_{\sigma},\ F^{W,\varepsilon}_{\tau}$ of Paley-Wiener-Schwartz symbols $\sigma\in S^{K,m}_{\PW,\rho,\delta}, \tau\in S^{K',m'}_{\PW,\rho,\delta}$, assuming that $\varepsilon\in(0,1]$ is small enough, has a formal expansion in $\varepsilon$ that qualifies as an asymptotic series for certain values of $0\leq\delta\leq\rho\leq1$. Moreover the series has finitely many non-vanishing terms if $\sigma$ and $\tau$ are polynomial. Unfortunately, we can (so far) not establish that this series is asymptotic to the dequantisation of the operator product, because the image of the Weyl quantisation is not obviously closed under products\footnote{This would amount to the fact that the Weyl quantisation is a strict deformation quantisation.}. To this end, we recall that the group product in $V\subset G$ can be pulled back to $U\subset\fg$, and the Dynkin-Baker-Campbell-Hausdorff formula tells us, that we may write
\begin{align}
\label{eq:bchformula}
X_{hg} = \exp^{-1}(hg) & = \exp^{-1}(X_{h})\ast\exp^{-1}(X_{g}) = X_{h}+X_{g}+\sum_{k=1}^{\infty}P_{k}(X_{h},X_{g}) \\ \nonumber
& = X_{h}+X_{g}+\frac{1}{2}[X_{h},X_{g}]+\frac{1}{12}([X_{h},[X_{h},X_{g}]]+[X_{g},[X_{g},X_{h}]]) + \textup{higher\ orders},
\end{align}
for sufficiently small $X_{h},X_{g}\in U$ (cf. \cite{BlanesOnTheConvergence} for convergence properties of \eqref{eq:bchformula}). Here, the $P_{k},\ k\in\N,$ are Lie-Polynomials. The $\varepsilon$-scaled version of the product, $\exp(\varepsilon(X_{h}\ast_{\varepsilon}X_{g})) = \exp(\varepsilon X_{h})\exp(\varepsilon X_{g})$, is:
\begin{align}
\label{eq:epsilonbchformula}
\varepsilon^{-1}((\varepsilon X_{h})\ast(\varepsilon X_{g}))& = X_{h}\ast_{\varepsilon}X_{\varepsilon} = X_{h}+X_{g}+\sum_{k=1}^{\infty}\varepsilon^{k}P_{k}(X_{h},X_{g}) \\ \nonumber
& = X_{h}+X_{g}+\frac{\varepsilon}{2}[X_{h},X_{g}]+\frac{\varepsilon^{2}}{12}([X_{h},[X_{h},X_{g}]]+[X_{g},[X_{g},X_{h}]]) + O(\varepsilon^{3}).
\end{align}
If we apply this formula to the twisted convolution \eqref{eq:l1mult} of the Weyl quantisations of $F^{W,\varepsilon}_{\sigma},\ F^{W,\varepsilon}_{\tau}$, we get for $\varepsilon$ small enough:
\begin{align}
\label{eq:twistedepsilonproduct}
 & \left(F^{W,\varepsilon}_{\sigma}\ast F^{W,\varepsilon}_{\tau}\right)(h,g) \\ \nonumber
 & = \int_{G}dh'\ F^{W,\varepsilon}_{\sigma}(h',g)F^{W,\varepsilon}_{\tau}(h'^{-1}h,h'^{-1}g) \\ \nonumber
 & = \int_{G}dh'\  F^{\varepsilon}_{\sigma}(h',\sqrt{h'^{-1}}g)F^{\varepsilon}_{\tau}(h'^{-1}h,\sqrt{h^{-1}h'}h'^{-1}g) \\ \nonumber
 & = \varepsilon^{-2n}\int_{G}dh'\ \check{\sigma}^{1}\big(\varepsilon^{-1}X_{h'},\exp(-\tfrac{1}{2}X_{h'})g\big)\check{\tau}^{1}\big(\varepsilon^{-1}X_{h'^{-1}h},\exp(-\tfrac{1}{2}X_{h'^{-1}h})\exp(-X_{h'})g\big) \\ \nonumber
 & = \varepsilon^{-2n}\int_{G}dh'\ \check{\sigma}^{1}\big(\varepsilon^{-1}X_{h'},\exp(-\tfrac{1}{2}X_{h'})g\big)\\[-0.25cm] \nonumber
 & \hspace{2cm}\times\check{\tau}^{1}\big(\varepsilon^{-1}((-X_{h'})\ast X_{h}),\exp(-\tfrac{1}{2}((-X_{h'})\ast X_{h}))\exp(-X_{h'})g\big) \\ \nonumber
 & = \varepsilon^{-2n}\int_{\fg}dX_{h'}\ j(X_{h'})^{2}\ \check{\sigma}^{1}\big(\varepsilon^{-1}X_{h'},\exp(-\tfrac{1}{2}X_{h'})g\big)\\[-0.25cm] \nonumber
 & \hspace{2cm}\times\check{\tau}^{1}\big(\varepsilon^{-1}((-X_{h'})\ast X_{h}),\exp(-\tfrac{1}{2}((-X_{h'})\ast X_{h}))\exp(-X_{h'})g\big) \\ \nonumber
 & = \varepsilon^{-n}\int_{\fg}dX_{h'}\ j(\varepsilon X_{h'})^{2}\ \check{\sigma}^{1}\big(X_{h'},\exp(-\tfrac{\varepsilon}{2}X_{h'})g\big)\\[-0.25cm] \nonumber
 & \hspace{2cm}\times\check{\tau}^{1}\big(\varepsilon^{-1}((-\varepsilon X_{h'})\ast X_{h}),\exp(-\tfrac{1}{2}((-\varepsilon X_{h'})\ast X_{h}))\exp(-\varepsilon X_{h'})g\big) \\ \nonumber
 & = \varepsilon^{-n}\int_{\fg}dX_{h'}\ j(\varepsilon X_{h'})^{2}\ \check{\sigma}^{1}\big(X_{h'},\exp(-\tfrac{\varepsilon}{2}X_{h'})g\big) \\[-0.25cm] \nonumber
 & \hspace{2cm}\times\check{\tau}^{1}\big(((-X_{h'})\ast_{\varepsilon}(\varepsilon^{-1}X_{h})),\exp(-\tfrac{\varepsilon}{2}((-X_{h'})\ast_{\varepsilon}(\varepsilon^{-1}X_{h})))\exp(-\varepsilon X_{h'})g\big) \\ \nonumber
 & = \varepsilon^{-n}\int_{\fg}dX_{h'}\ j(\varepsilon X_{h'})^{2}\ \check{\sigma}^{1}\big(X_{h'},\exp(-\tfrac{\varepsilon}{2}X_{h'})\exp(\tfrac{\varepsilon}{2}(\varepsilon^{-1}X_{h}))\exp(-\tfrac{1}{2}(X_{h}))g\big) \\[-0.25cm] \nonumber
 & \hspace{2cm}\times\check{\tau}^{1}\big(((-X_{h'})\ast_{\varepsilon}(\varepsilon^{-1}X_{h})),\\ \nonumber
 & \hspace{3cm}\exp(-\tfrac{\varepsilon}{2}((-X_{h'})\ast_{\varepsilon}(\varepsilon^{-1}X_{h})))\exp(-\varepsilon X_{h'})\exp(\tfrac{\varepsilon}{2}(\varepsilon^{-1}X_{h}))\exp(-\tfrac{1}{2}(X_{h}))g\big) \\ \nonumber
 & = \varepsilon^{-n}\int_{\fg}dX_{h'}\ j(\varepsilon X_{h'})^{2}\ \check{\sigma}^{1}\big(X_{h'},\exp(\varepsilon((-\tfrac{1}{2}X_{h'})\ast_{\varepsilon}(\tfrac{1}{2}(\varepsilon^{-1}X_{h}))))\exp(-\tfrac{1}{2}(X_{h}))g\big) \\[-0.25cm] \nonumber
 & \hspace{2cm}\times\check{\tau}^{1}\big(((-X_{h'})\!\ast_{\varepsilon}\!(\varepsilon^{-1}X_{h})), \\ \nonumber
 & \hspace{3cm}\exp(\varepsilon((-\tfrac{1}{2}((-X_{h'})\!\ast_{\varepsilon}\!(\varepsilon^{-1}X_{h})))\ast_{\varepsilon}(((-X_{h'})\ast_{\varepsilon}(\tfrac{1}{2}(\varepsilon^{-1}X_{h}))))))\exp(-\tfrac{1}{2}(X_{h}))g\big),
\end{align}
where we switched integration from $G$ to $\fg$ by means of the exponential map (see \eqref{eq:haarmeasurenorm}), which is permitted due to the support properties of $\check{\sigma}^{1}$ and $\check{\tau}^{1}$, changed integration variables $X_{h'}\mapsto\varepsilon^{-1}X_{h'}$, and successively replaced the group product in $V$ by its $\varepsilon$-scaled pullback $\ast_{\epsilon}$ in $U$.\\[0.1cm]
Now, we would like to write $\left(F^{W,\varepsilon}_{\sigma}\ast F^{W,\varepsilon}_{\tau}\right)(h,g) = F^{W,\epsilon}_{\rho}(h,g)$ for some $\rho\in S^{K+K',m+m'}_{\PW,\rho,\delta}$, which would define the twisted product of symbols $\rho = \sigma\star_{\varepsilon}\tau$. The shift in the support $(K,K')\mapsto K+K'$ is to be expected, because of the relation of $\star_{\epsilon}$ to the convolution of $\check{\sigma}^{1}$ and $\check{\tau}^{1}$. Unfortunately, as already mentioned above, the last line in \eqref{eq:twistedepsilonproduct}, is not obviously of the form $ F^{W,\epsilon}_{\rho}(h,g)$. Nevertheless, if we apply \eqref{eq:epsilonbchformula}, we can deduce an asymptotic series in $S^{K+K',m+m'}_{\PW,\rho,\delta}$ by successive integration by parts. To this end, we explicitly write out the inverse Fourier transforms in \eqref{eq:twistedepsilonproduct}:
\begin{align}
\label{eq:twistedepsilonproductexplicit}
 & \left(F^{W,\varepsilon}_{\sigma}\ast F^{W,\varepsilon}_{\tau}\right)(h,g) \\ \nonumber
 & = \varepsilon^{-n}\int_{\fg}dX_{h'}\ j(\varepsilon X_{h'})^{2}\int_{\fg^{*}}\frac{d\theta}{(2\pi)^{n}}e^{i\theta(X_{h'})}\sigma(\theta,\exp(\varepsilon((-\tfrac{1}{2}X_{h'})\ast_{\varepsilon}(\tfrac{1}{2}(\varepsilon^{-1}X_{h}))))\exp(-\tfrac{1}{2}(X_{h}))g) \\ \nonumber
 &\hspace{0.5cm}\times\int_{\fg^{*}}\frac{d\theta'}{(2\pi)^{n}}e^{i\theta'(((-X_{h'})\ast_{\varepsilon}(\varepsilon^{-1}X_{h})))} \\[-0.25cm] \nonumber
 & \hspace{1.5cm}\times\tau(\theta',\exp(\varepsilon((-\tfrac{1}{2}((-X_{h'})\ast_{\varepsilon}(\varepsilon^{-1}X_{h})))\ast_{\varepsilon}(((-X_{h'})\ast_{\varepsilon}(\tfrac{1}{2}(\varepsilon^{-1}X_{h}))))))\exp(-\tfrac{1}{2}(X_{h}))g).
\end{align}
From the Dynkin-Baker-Campbell-Hausdorff formula \eqref{eq:bchformula}, and the fact that the $P_{k},\ k\in\N,$ are Lie-Polynomials, we infer that
\begin{align}
\label{eq:bchshift}
(-X_{h'})\ast_{\varepsilon}(\varepsilon^{-1}X_{h}) & = -(X_{h'}-\varepsilon^{-1}X_{h})+\sum_{k=1}^{\infty}\varepsilon^{k}P'_{k}(X_{h'},X_{h'}-\varepsilon^{-1}X_{h}),
\end{align}
for some (new) Lie-Polynomials $P'_{k},\ k\in\N$. \\ Therefore, we can achieve the following rewriting of $e^{i\theta'(((-X_{h'})\ast_{\varepsilon}(\varepsilon^{-1}X_{h})))}$:
\begin{align}
\label{eq:bchdifferential}
e^{i\theta'(((-X_{h'})\ast_{\varepsilon}(\varepsilon^{-1}X_{h})))} & = e^{i\theta'(((-X_{h'})\ast_{\varepsilon}(\varepsilon^{-1}X_{h}))+(X_{h'}-\varepsilon^{-1}X_{h}))}e^{-i\theta'(X_{h'}-\varepsilon^{-1}X_{h})} \\ \nonumber
 & = \prod_{k=1}^{\infty}e^{i\varepsilon^{k}\theta'(P'_{k}(X_{h'},X_{h'}-\varepsilon^{-1}X_{h}))}e^{-i\theta'(X_{h'}-\varepsilon^{-1}X_{h})} \\ \nonumber
 & = \prod_{k=1}^{\infty}\varepsilon^{k}Q_{k}(\varepsilon,X_{h'},\theta',\partial_{\theta'})e^{-i\theta'(X_{h'}-\varepsilon^{-1}X_{h})},
\end{align}
where $Q_{k}(\varepsilon,X_{h'},\theta',\partial_{\theta'}),\ k\in\N,$ are differential operators of infinite order. If we continue by performing Taylor expansions of $j(\varepsilon X_{h'})^2$, $\sigma(\theta,\exp(\varepsilon((-\tfrac{1}{2}X_{h'})\ast_{\varepsilon}(\tfrac{1}{2}(\varepsilon^{-1}X_{h}))))\exp(-\tfrac{1}{2}(X_{h}))g)$ and $\tau(\theta',\exp(\varepsilon((-\tfrac{1}{2}((-X_{h'})\ast_{\varepsilon}(\varepsilon^{-1}X_{h})))\ast_{\varepsilon}(((-X_{h'})\ast_{\varepsilon}(\tfrac{1}{2}(\varepsilon^{-1}X_{h}))))))\exp(-\tfrac{1}{2}(X_{h}))g)$ in $\varepsilon$, trade all $X_{h'}$- and $(X_{h'}-\varepsilon^{-1}X_{h})$-dependence by differentiation of $e^{i\theta(X_{h'})}$ and $e^{-i\theta'(X_{h'}-\varepsilon^{-1}X_{h})}$ for $\theta$- and $\theta'$-differentials, and finally perform repeated integrations by parts, culminating in an expression where the $dX_{h'}$ integration gives an oscillatory integral representation of $\delta^{(n)}(\theta-\theta')$, we will get an infinite sum in orders of $\varepsilon$
\begin{align}
\label{eq:asymptoticproductsum}
\left(F^{W,\varepsilon}_{\sigma}\ast_{\uL} F^{W,\varepsilon}_{\tau}\right)\ &\ \overset{?}{\sim}\ \sum_{k=0}^{\infty}\varepsilon^{k}C^{W,\varepsilon}_{k}(\sigma,\tau)
\end{align}
consisting of expressions of the form $C^{W,\varepsilon}_{k}(\sigma,\tau)$, where each order $\varepsilon^k$ can be written as products of differentials of $\sigma$ and $\tau$ of total order $\leq k$ each. Due to the fact, that $j\neq1$ for general compact Lie groups $G$, the differential operations on $\sigma$ and $\tau$ will not solely be determined by the Dynkin-Baker-Campbell-Hausdorff formula, and thus by the Poisson bracket \eqref{eq:canonicalpoisson}, but also be affected by differential operators determined by the structure of the (positive) roots of $G$. Nevertheless, it follows from the Taylor series of $j$ that those additional differential operators will lead to lower order terms in $S^{K+K',m+m'}_{\PW,\rho,\delta}$, than those coming from the Poisson bracket, if $\rho>\delta$. Taking a closer look at the formula for the Poisson bracket \eqref{eq:canonicalpoisson} and \eqref{eq:poissonbilinear}, we realise that the condition $\rho>\delta$, known from the $\R^{n}$-case, has to be supplemented by $\rho>\frac{1}{2}$ for the expression \eqref{eq:asymptoticproductsum} to qualify as an asymptotic sum. We note that, by theorem \ref{thm:asymptoticcompletness}, there exists a resummation of the right hand side of \eqref{eq:asymptoticproductsum}, which is unique up to smoothing symbols in $S^{K+K',-\infty}_{\PW,\rho,\delta}$. Thus, from a practical point of view it would suffice to show that the difference between the operators defined by the left hand side and the resummation of the right hand side is sufficiently ``small’’ in a sense to be made precise, if asymptoticity in $S^{K+K',m+m'}_{\PW,\rho,\delta}$ were to fail. \\[0.1cm]
Finally, let us be a bit more explicit, and give the expression for \eqref{eq:asymptoticproductsum} up to order $\varepsilon$. Additionally, we provide the expression up to order $\varepsilon^{2}$ before partial integration:
\begin{align}
\label{eq:secondorderproduct}
 & j(\varepsilon X_{h'})^{2}e^{i\theta'(((-X_{h'})\ast_{\varepsilon}(\varepsilon^{-1}X_{h})))}\sigma(\theta,\exp(\varepsilon((-\tfrac{1}{2}X_{h'})\!\ast_{\varepsilon}\!(\tfrac{1}{2}(\varepsilon^{-1}X_{h}))))\exp(-\tfrac{1}{2}(X_{h}))g) \\ \nonumber
 &\ \times\tau(\theta',\exp(\varepsilon((-\tfrac{1}{2}((-X_{h'})\ast_{\varepsilon}(\varepsilon^{-1}X_{h})))\ast_{\varepsilon}(((-X_{h'})\ast_{\varepsilon}(\tfrac{1}{2}(\varepsilon^{-1}X_{h}))))))\exp(-\tfrac{1}{2}(X_{h}))g) \\[0.25cm] \nonumber
\sim_{O(\varepsilon^{3})}\ & \ e^{-i\theta'(X_{h'}-\varepsilon^{-1}X_{h})}\Bigg(\sigma(\theta,\sqrt{h^{-1}}g)\tau(\theta',\sqrt{h^{-1}}g) \\ \nonumber
\ & \ + \frac{\varepsilon}{2}\bigg(i\theta'([X_{h'},X_{h'}-\varepsilon^{-1}X_{h}])\sigma(\theta,\sqrt{h^{-1}}g)\tau(\theta',\sqrt{h^{-1}}g)  \\ \nonumber
\ & \ \ -(R_{X_{h'}-\varepsilon^{-1}X_{h}}\sigma)(\theta,\sqrt{h^{-1}}g)\tau(\theta',\sqrt{h^{-1}}g)-\sigma(\theta,\sqrt{h^{-1}}g)(R_{X_{h'}}\tau)(\theta',\sqrt{h^{-1}}g)\bigg) \\ \nonumber
\ & \ +\frac{\varepsilon^{2}}{4}\bigg(\Big(-\frac{4}{3}\sum_{\alpha\in R^{+}}\alpha(X_{h'})^{2}-\frac{1}{2}(\theta'([X_{h'},X_{h'}-\varepsilon^{-1}X_{h}]))^{2} \\ \nonumber
\ & \ -\frac{1}{3}i\theta'(2[X_{h'},X_{h'}-\varepsilon^{-1}X_{h}]_{(2)}+[X_{h'}-\varepsilon^{-1}X_{h},X_{h'}]_{(2)})\Big)\sigma(\theta,\sqrt{h^{-1}}g)\tau(\theta',\sqrt{h^{-1}}g) \\ \nonumber
\ & \ +\frac{1}{2}\Big(\sigma(\theta,\sqrt{h^{-1}}g)\big((R^{2}_{X_{h'}}\tau)(\theta',\sqrt{h^{-1}}g)+(R_{[X_{h'},X_{h'}-\varepsilon^{-1}X_{h}]}\tau)(\theta',\sqrt{h^{-1}}g)\big) \\ \nonumber
\ & \ +\big((R^{2}_{X_{h'}-\varepsilon^{-1}X_{h}}\sigma)(\theta,\sqrt{h^{-1}}g)+(R_{[X_{h'},X_{h'}-\varepsilon^{-1}X_{h}]}\sigma)(\theta,\sqrt{h^{-1}}g)\big)\tau(\theta',\sqrt{h^{-1}}g)\Big) \\ \nonumber
\ & \ + (R_{X_{h'}-\varepsilon^{-1}X_{h}}\sigma)(\theta,\sqrt{h^{-1}}g)(R_{X_{h'}}\tau)(\theta',\sqrt{h^{-1}}g) \\ \nonumber
\ & \ -i\theta'([X_{h'},X_{h'}\!-\!\varepsilon^{-1}X_{h}]) \\[-0.25cm] \nonumber
\ & \ \hspace{0.5cm}\times\Big(\!(R_{X_{h'}-\varepsilon^{-1}X_{h}}\sigma)(\theta,\sqrt{h^{-1}}g)\tau(\theta',\sqrt{h^{-1}}g)\!+\!\sigma(\theta,\sqrt{h^{-1}}g)(R_{X_{h'}}\tau)(\theta',\sqrt{h^{-1}}g)\!\Big)\!\bigg)\!\Bigg)\!.
\end{align}
Integrating this expression at order $\varepsilon$, we have:
\begin{align}
\label{eq:firstorderproduct}
\left(F^{W,\varepsilon}_{\sigma}\ast F^{W,\varepsilon}_{\tau}\right)\ &\ \sim_{O(\varepsilon^{2})}\ F^{W,\varepsilon}_{\sigma\tau} - \frac{i\varepsilon}{2}F^{W,\varepsilon}_{\{\sigma,\tau\}_{T^{*}G}}, \\ \nonumber
\sigma\star_{\varepsilon}\tau\ &\ \sim_{O(\varepsilon^{2})}\ \sigma\tau - \frac{i\varepsilon}{2}\{\sigma,\tau\}_{T^{*}G}
\end{align}
as expected from theorem \ref{thm:localweylquant}.
\subsection{Scaled Fourier transforms, the Stratonovich-Weyl transform and coadjoint orbits}
\label{subsec:swt}
The global and local definitions of pseudo-differential operators we introduced in the previous subsection appear to be intimately tied to the natural representation $L^{2}(G)$ of the transformation group $C^{*}$-algebra $C(G)\rtimes_{\alpha_{\uL}}G$, and its regularity properties w.r.t. the group translations $U_{g},\ g\in G$, up to this point. Especially, the expansion \eqref{eq:asymptoticproductsum} associates the $\varepsilon$-dependence with the ``momentum variables'' $P_{X},\ X\in\fg,$ in \eqref{eq:quantcomm}. But, in view of applications in Born-Oppenheimer reduction schemes and adiabatic perturbation theory it seems to be useful to be able to shift the $\varepsilon$-dependence to the variables dual to the ``momenta'', thereby changing from the microscopic scale to the macroscopic scale ($X'_{g} = \varepsilon X_{g},\ g'=\exp(\varepsilon X_{g})$). Furthermore, it can be advantageous to switch the roles of ``momenta'' and ``positions'' by means of a suitable integral transform on $L^{2}(G)$, as seen form \cite{TeufelAdiabaticPerturbationTheory} (Section 5).\\[0.1cm]
To exemplify this point, we recall how the Fourier transform, and its $\varepsilon$-scaled version, affect the case of pseudo-differential operators on $\R^{n}$. Namely, if we consider a symbol $\sigma\in S^{m}_{\rho,\delta}$ or $\cS'(\R^{2n})$, and its action on $\cS(\R^{n})$ via Weyl quantisation,
\begin{align}
\label{eq:fourierweylaction}
(A_{\sigma}\Psi)(q) & = \frac{1}{(2\pi\varepsilon)^{n}}\int_{\R^{2n}}dx\ d\xi\ \sigma(\tfrac{1}{2}(q+x),\xi)e^{\frac{i}{\varepsilon}\xi\cdot(q-x)}\Psi(x),
\end{align}
which is adapted to the representation of the commutation relations \eqref{eq:canonicalcomm} by $Q = q\!\ \cdot$ and \mbox{$P=-i\varepsilon\nabla_{q}$,} we may interchanges the roles of $Q$ and $P$ w.r.t. to the Weyl quantisation by applying the $\varepsilon$-scaled Fourier transform,
\begin{align}
\label{eq:stscaledft}
\mathcal{F}_{\varepsilon}[\Psi](p) = \hat{\Psi}_{(\varepsilon)}(p) & := \int_{\R^{n}}dq\ e^{-\frac{i}{\varepsilon}p\cdot q}\Psi(q), \\ \nonumber
\mathcal{F}^{-1}_{\varepsilon}[\Phi](q) = \check{\Phi}^{(\varepsilon)}(q) & := \int_{\R^{n}}\frac{dp}{(2\pi\varepsilon)^{n}}\ e^{\frac{i}{\varepsilon}p\cdot q}\Phi(p),
\end{align}
to obtain
\begin{align}
\label{eq:inversefourierweylaction}
(\hat{A}_{\sigma}\hat{\Psi}_{(\varepsilon)})(p) & = \frac{1}{(2\pi\varepsilon)^{n}}\int_{\R^{2n}}dx\ d\xi\ \sigma(x,\tfrac{1}{2}(p+\xi))e^{-\frac{i}{\varepsilon}(p-\xi)\cdot x}\hat{\Psi}_{(\varepsilon)}(\xi),
\end{align}
which is adapted to the representation of \eqref{eq:canonicalcomm} by $Q=-i\varepsilon\nabla_{p}$ and $P=p\cdot\ $. Consider, for example, a standard Born-Oppenheimer type Hamiltonian
\begin{align}
\label{eq:sthamiltonian}
H & = -\frac{\varepsilon^{2}}{2}(\nabla_{q}+ iA(q))^{2}\otimes\mathds{1}_{\fH_{f}} + V(q)
\end{align}
acting on the coupled quantum system $\fH=L^{2}(\R^{n})\otimes\fH_{f}$. \eqref{eq:sthamiltonian} is obtained as Weyl quantisation \eqref{eq:fourierweylaction} of the $\varepsilon$-dependent symbol
\begin{align}
\label{eq:sthamiltonianfouriersymbol}
\sigma_{H}(q,p) & = \frac{1}{2}(p-\varepsilon A(q))^{2}\otimes\mathds{1}_{\fH_{f}} + V(q).
\end{align}
On the Fourier transformed side, the Hamiltonian takes the form \eqref{eq:inversefourierweylaction}:
\begin{align}
\label{eq:stfourierhamiltonian}
\hat{H} & = \frac{\varepsilon^{2}}{2}(p-A(i\nabla_{p}))^{2}\otimes\mathds{1}_{\fH_{f}}+V(i\nabla_{p})\ \ \ \textup{(unscaled)},\\ \nonumber
\hat{H}_{\varepsilon} & = \frac{1}{2}(p-\varepsilon A(i\varepsilon\nabla_{p}))^{2}\otimes\mathds{1}_{\fH_{f}}+V(i\varepsilon\nabla_{p})\ \ \ \textup{(}\varepsilon\textup{-scaled)}. 
\end{align}
While these observation are almost trivial at this level, equation \eqref{eq:sthamiltonian} and \eqref{eq:stfourierhamiltonian} illustrate the fact, that it is beneficial to have suitable integral transforms, possibly with $\varepsilon$-scaling, at hand to decide whether a given operator can be written as Weyl quantisation of a, possibly operator valued, symbol to apply Born-Oppenheimer reduction or space adiabatic perturbation theory. A less trivial example is given by a Hamiltonian with periodic potential $V_{\Gamma}$ ($\Gamma\subset\R^{n}$ is the periodicity lattice of $V_{\Gamma}$) and slowly varying external electromagnetic fields $A, \phi$ considered by Panati, Teufel and Spohn \cite{PanatiEffectiveDynamicsFor, TeufelAdiabaticPerturbationTheory}:
\begin{align}
\label{eq:latticehamiltonian}
H & = \frac{1}{2}(-i\nabla_{q}-A(\varepsilon q))^{2}+V_{\Gamma}(q)+\phi(\varepsilon q),
\end{align}
which can be rewritten as Weyl quantisation of an operator valued symbol by means of the Bloch-Floquet transform $U:L^{2}(\R^{n})\rightarrow\fH_{\Gamma^{*}}$ (cf. \cite{FollandHarmonicAnalysisIn}, Chapter 1.10):
\begin{align}
\label{eq:blochfloquet}
U[\Psi](p,q) & := \sum_{\gamma\in\Gamma}e^{-ip\cdot(q+\gamma)}\Psi(q+\gamma),\ (q,p)\in\R^{n},\ \Psi\in\cS(\R^{n}),
\end{align}
where $\fH_{\Gamma^{*}}=\{\Phi\in L^{2}_{\textup{loc}}(\R^{n},L^{2}(\mathds{T}^{n}))\ |\ \Phi(p+\gamma^{*},q)=e^{-i\gamma^{*}\cdot q}\Phi(p,q)\}$ and $\Gamma^{*}$ is the lattice dual to $\Gamma$.
Applying $U$ to \eqref{eq:latticehamiltonian} gives:
\begin{align}
\label{eq:blochfloquethamiltonian}
UHU^{*} & = \frac{1}{2}(-i\nabla^{\textup{per}}_{q} + p + A(i\varepsilon\nabla_{p}))^{2} + V_{\Gamma}(q) + \phi(i\varepsilon\nabla_{p}),
\end{align}
which can be understood as the Weyl quantisation of a $\Gamma^{*}$-equivariant symbol with values in bounded operators from $H^{2}(\mathds{T}^{n})$ to $L^{2}(\mathds{T}^{n})$, and thus makes \eqref{eq:latticehamiltonian} accessible to space adiabatic perturbation theory.\\[0.1cm]
This said, we return to the case of compact and simply connected Lie groups\footnote{General compact Lie groups can be dealt with by an appeal to the structure theorem for compact Lie groups (cf. \cite{BroeckerRepresentationsOfCompact}).}, where we apply the Stratonovich-Weyl-Fourier transform introduced by Figueroa, Gracia-Bond\'ia and V\'arilly \cite{FigueroaMoyalQuantizationWith, VarillyTheMoyalRepresentation}, based on ideas of Stratonovich \cite{StratonovichOnDistributionsIn}, to the pseudo-differential operators defined in the previous section, which gives rise to an alternative to the common Fourier transform from $L^{2}(G)$ to $L^{2}(\hat{G})$, and additionally makes the effect of (radial) scaling particularly transparent. Since this transform is somewhat non-standard, we recall the main steps of its construction in some detail:\\[0.1cm]
According to \cite{FigueroaMoyalQuantizationWith}, we construct the Stratonovich-Weyl operator $\Delta^{\pi}:\mathcal{O}_{\pi}\rightarrow\End(V_{\pi})$, which allows us to map functions on $\mathcal{O}_{\pi}$ to operator in $V_{\pi}$, via the coherent state formalism for compact semisimple Lie groups \cite{PerelomovGeneralizedCoherentStates, SimonTheClassicalLimit, LandsmanMathematicalTopicsBetween}:
\begin{itemize}
	\item[1.] For a unitary irreducible representation, $\pi\in\hat{G}$, we choose the corresponding (real) highest weight $\lambda_{\pi}\in\ft^{*}\subset\fg^{*}$. Let $\mathcal{O}_{\pi}=\{\theta\in\fg^{*}\ |\ \exists g\in G: \theta = Ad^{*}_{g}(\lambda_{\pi})\}\subset\fg^{*}$ be the coadjoint orbit of $G$ through $\lambda_{\pi}$. $\mathcal{O}_{\pi}\cong G/G_{\lambda_{\pi}}$ is homogeneous space, where $G_{\lambda_{\pi}}$ is the stabiliser of $\lambda_{\pi}$. It is the content of the Borel-Weil theorem that the correspondence $\pi\leftrightarrow\lambda_{\pi}$ is one-to-one up to unitary equivalence\footnote{It is even onto, if we restrict to integral coadjoint orbits}.
	\item[2.] Next, we choose a normalised weight vector $v_{\pi}\in V_{\pi}$ of $\lambda_{\pi}$. Moreover, we define the equivariant momentum map
\begin{align}
\label{eq:coadjointmomentum}
J_{\pi}(v)(X) & := \frac{1}{2\pi i}(v,d\pi(X)v)_{V_{\pi}},\ v\in V_{\pi},X\in\fg,
\end{align}
which satisfies $J_{\pi}(\pi(g)v)=Ad^{*}_{g}(J_{\pi}(v)),\ g\in G,$, $J_{\pi}(v_{\pi})=\lambda_{\pi}$ and $J_{\pi}^{-1}(\{\lambda_{\pi}\})=\{z v_{\pi}\ |\ z\in\C, |z|=1\}$ because the weight space of $\lambda_{\pi}$ is one-dimensional. Then, we have $J_{\pi}^{-1}(\mathcal{O}_{\pi})=\{\pi(g)v_{\pi}\ |\ g\in G\}$.
	\item[3.] For $\theta\in\mathcal{O}_{\pi}$ we choose $g_{\theta}\in G$ s.t. $Ad^{*}_{g_{\theta}}(\lambda_{\pi})=\theta$ and $g_{\lambda_{\pi}}=e$. $\theta\mapsto g_{\theta}$ is a measurable section w.r.t. to the Liouville measure $d\mu_{\pi}$ on $\mathcal{O}_{\pi}$ ($\mu_{\pi}(\mathcal{O}_{\pi})=\dim V_{\pi}=d_{\pi}$) induced by the natural invariant symplectic form $\omega_{\pi}$.
	\item[4.] We define the coherent state $v_{\theta}:=\pi(g_{\theta})v_{\pi}\in V_{\pi}$ for $\theta\in\mathcal{O}_{\pi}$, which is uniquely determined by $\theta$ up to a phase, since $J_{\pi}(v_{\theta})=\theta$. 
	\item[5.] For an operator $A\in\End(V_{\pi})$, we have the \textit{covariant} or \textit{lower symbol}
\begin{align}
\label{eq:lowersymbol}
L^{\pi}_{A}(\theta) & :=(v_{\theta},Av_{\theta})_{V_{\pi}},
\end{align}
which uniquely determines $A$, since the coherent states $\{v_{\theta}\}_{\theta\in\mathcal{O}_{\pi}}$ are complete by means of the natural Kähler structure on $\mathcal{O}_{\pi}$ (cf. \cite{SimonTheClassicalLimit, FigueroaMoyalQuantizationWith, WildbergerOnTheFourier}). The lower symbol is covariant w.r.t. $G$:
\begin{align}
\label{eq:lowersymbolcov}
L^{\pi}_{\pi(g)A\pi(g)^{*}}(\theta) & = L^{\pi}_{A}(Ad^{*}_{g^{-1}}(\theta)),\ g\in G.
\end{align}
	\item[6.] By duality  and the Riesz-Fr\'echet theorem ($d_{\pi}<\infty$\footnote{For complete systems of coherent states, the upper symbol still exist for a (strongly) dense set of operators in $\cB(\fH)$ in the case of infinite dimensional Hilbert spaces $\fH$, and it is still dual to the lower symbol (cf. \cite{SimonTheClassicalLimit}).}), we obtain the \textit{contravariant} or \textit{upper symbol} $A\mapsto U^{\pi}_{A}$:
\begin{align}
\label{eq:uppersymbol}
(A,B)_{HS} = \tr(A^{*}B) & = \int_{\mathcal{O}_{\pi}}d\mu_{\pi}(\theta)\ \overline{U^{\pi}_{A}(\theta)}L^{\pi}_{A}(\theta).
\end{align}
The normalisation of $d\mu_{\lambda}$ ensures that $U^{\pi}_{\mathds{1}_{V_{\pi}}}=1$. The upper symbol is covariant, as well:
\begin{align}
\label{eq:uppersymbolcov}
U^{\pi}_{\pi(g)A\pi(g)^{*}}(\theta) & = U^{\pi}_{A}(Ad^{*}_{g^{-1}}(\theta)),\ g\in G.
\end{align}
	\item[8.] The map $U^{\pi}_{A}\mapsto L^{\pi}_{A}$ defines a positive $G$-invariant invertible operator $K_{\pi}$\footnote{We have chosen $K_{\pi}$ to be the inverse of the operator defined in \cite{FigueroaMoyalQuantizationWith}, because in this way $K_{\pi}$ is a smoothing operator, as we will see in the next subsection \ref{subsec:csquant}.} on the finite dimensional space of functions $S_{\pi}:=\textup{span}_{\C}\{U^{\pi}_{A}\ |\ A\in\End(V_{\pi})\}\subset L^{2}(\mathcal{O}_{\pi})$. $G$-invariance is to be understood w.r.t. the quasiregular representation $(\rho(g)f)(\theta)=f(Ad^{*}_{g^{-1}}(\theta)),\ f\in L^{2}(\mathcal{O}_{\pi})$. Form the definition of lower and upper symbols, we infer that the kernel of $K_{\pi}$ is determined by the overlap function of the coherent states:
\begin{align}
\label{eq:overlapkernel}
L^{\pi}_{A}(\theta) & = (v_{\theta},A v_{\theta})_{V_{\pi}}=\int_{\mathcal{O}_{\pi}}d\mu_{\pi}\ (\theta') |(v_{\theta},v_{\theta'})_{V_{\pi}}|^{2}\ U^{\pi}_{A}(\theta').
\end{align}
By means of $K_{\pi}$, the \textit{Stratonovich-Weyl symbol} of $A$ is defined to be:
\begin{align}
\label{eq:swsymbol}
W^{\pi}_{A} & := K^{\frac{1}{2}}_{\pi}U^{\pi}_{A} = K^{-\frac{1}{2}}_{\pi}L^{\pi}_{A},
\end{align}
where $K^{\frac{1}{2}}_{\pi}$ is the positive square root of $K_{\pi}$. It has the following properties:
\begin{itemize}
\item[(a)] $\End(V_{\pi})\ni A\mapsto W^{\pi}_{A}\in S_{\pi}$ is linear and bijective,
\item[(b)] $W^{\pi}_{A^{*}} = \overline{W^{\pi}_{A}}$,
\item[(c)] $W^{\pi}_{\mathds{1}_{V_{\pi}}}=1$,
\item[(d)] $W^{\pi}_{\pi(g)A\pi(g)^{*}}  = \rho(g)(W^{\pi}_{A})$,
\item[(e)]
\begin{align}
\label{eq:swsymbolprop}
(A,B)_{HS} & = \tr(A^{*}B) \\ \nonumber
 & = \int_{\mathcal{O}_{\pi}}d\mu_{\pi}(\theta)\ \overline{W^{\pi}_{A}(\theta)}W^{\pi}_{B}(\theta).
\end{align}
\end{itemize}
	\item[9.] Finally, the \textit{Stratonovich-Weyl operator} $\Delta^{\pi}:\mathcal{O}_{\pi}\rightarrow\End(V_{\pi})$ is constructed in spirit of the (Fourier-)Weyl elements, already familiar from the previous subsection (see \eqref{eq:stweylquant}). Namely, we look for an operator valued function $\Delta^{\pi}=(\Delta^{\pi})^{*}$ s.t.:
\begin{align}
\label{eq:swoperator}
W^{\pi}_{A}(\theta) & = \tr(\Delta^{\pi}(\theta)A) = (\Delta^{\pi}(\theta),A)_{HS}, \\[0.25cm] \nonumber
A & = \int_{\mathcal{O}_{\pi}}d\mu_ {\pi}(\theta)\ W^{\pi}_{A}(\theta)\Delta^{\pi}(\theta) = (\Delta^{\pi},W^{\pi}_{A})_{L^{2}(\mathcal{O}_{\pi})}.
\end{align}
To construct $\Delta^{\pi}$, we decompose $S_{\pi}\cong\End(V_{\pi})\cong V_{\pi}\otimes\overline{V_{\pi}}\cong\bigoplus_{\eta}V_{\eta\in\hat{G}}^{\oplus N_{\pi}^{\eta}}\cong\bigoplus_{\eta\in\hat{G}}\bigoplus_{s=1}^{N^{\eta}_{\pi}}V_{\eta}$\footnote{The ambiguity in taking a square root of $K_{\pi}$ is reflected by the fact that $S_{\pi}$ is in general not irreducible as representation $V_{\rho}$ of $G$.}, where $N_{\pi}^{\eta}\in\N_{0}$ is the multiplicity of $\eta$ in $\pi\otimes\overline{\pi}$. Now, we can introduce the \textit{generalised spin-weighted spherical harmonics} as an orthonormal basis of $S_{\pi}$:
\begin{align}
\label{eq:generalisedswsh}
\left(Y^{\pi}_{\eta s k}(g_{\theta})_{l}\right)_{l=1,...,d_{\eta}} & := \left(\eta_{s}(g_{\theta})_{k l}\right)_{l=1,...,d_{\eta}}
\end{align}
for $s=1,...,N^{\eta}_{\pi}, k=1,...,d_{\eta}, \theta\in\mathcal{O}_{\pi}$, where the matrix elements are computed w.r.t. to a basis, $\{v^{\eta,s}_{i}\}_{i=1}^{d_{\eta}},\ v^{\eta,s}_{d_{\eta}}=v_{\eta,s}$, adapted to the weight decomposition $V_{\eta,s}=V_{\lambda_{\eta}}\oplus\bigoplus_{\lambda}V_{\lambda}$ of the s-th copy of $V_{\eta}$ in $V_{\pi}\otimes\overline{V_{\pi}}$. We use the same notation for any adapted basis for a unitary irreducible representation of $V$ of $G$. Here, $\lambda_{\eta}$ denotes the highest weight of $\eta$, and $v_{\eta}$ is a normalised weight vector of $\lambda_{\eta}$. The \textit{generalised spherical harmonics} $Y^{\pi}_{(\eta,l_{\eta}) s k}(\theta):=Y^{\pi}_{(\eta,l_{\eta}) s k}(g_{\theta})$ are obtained from the decomposition
\begin{align}
\label{eq:genralisedsh}
\pi(g)_{id_{\pi}}\overline{\pi(g)}_{jd_{\pi}} & = \sum_{\eta, l_{\eta}, s, k}C(\pi,i;\overline{\pi},j|\eta,k;s)\overline{C(\pi,d_{\pi};\overline{\pi},d_{\overline{\pi}}|\eta,0_{l_{\eta}};s)}Y^{\pi}_{(\eta,l_{\eta}) s k}(g),
\end{align}
since $v_{\pi}\otimes \overline{v}_{\pi}$ has (real) weight 0. The $C(\pi,m;\zeta,n|\eta,k;s)$'s denote the Clebsch-Gordan coefficients of the decomposition $\pi\otimes\zeta\cong\bigoplus_{\eta\in\hat{G}}\bigoplus_{s=1}^{N^{\eta}_{\pi,\zeta}}V_{\eta}$. These function on $\mathcal{O}_{\pi}$ (weight 0!) diagonalise the kernel $K_{\pi}$ (cf. \cite{FigueroaMoyalQuantizationWith}),
\begin{align}
\label{eq:diagonalkernel}
K_{\pi}(\theta,\theta') & = |(v_{\theta},v_{\theta'})|^{2} \\ \nonumber
 & = |(\pi(g_{\theta})v_{\pi},\pi(g_{\theta'})v_{\pi})|^{2} \\ \nonumber
 & = \Big|\sum_{i=1}^{d_{\pi}}\pi(g_{\theta'})_{id_{\pi}}\overline{\pi(g_{\theta})}_{jd_{\pi}}\Big|^{2} \\[-0.25cm] \nonumber 
 & = \sum_{\substack{ \eta,l_{\eta},s,k \\ \eta',l'_{\eta'},s',k' }}\sum_{i,j=1}^{d_{\pi}}C(\pi,i;\overline{\pi},j|\eta,k;s)\overline{C(\pi,i;\overline{\pi},j|\eta',k';s')} \\ \nonumber
 & \hspace{0.5cm}\times\overline{C(\pi,d_{\pi};\overline{\pi},d_{\overline{\pi}}|\eta,0_{l_{\eta}};s)}C(\pi,d_{\pi};\overline{\pi},d_{\overline{\pi}}|\eta',0_{l'_{\eta'}};s')Y^{\pi}_{(\eta,l_{\eta}) s k}(g_{\theta})\overline{Y}^{\pi}_{(\eta',l'_{\eta'}) s' k'}(g_{\theta'}) \\ \nonumber
 & = \sum_{\eta,l_{\eta},s,k}C(\pi,d_{\pi};\overline{\pi},d_{\overline{\pi}}|\eta,0_{l_{\eta}};s)^{2}Y^{\pi}_{(\eta,l_{\eta}) s k}(\theta)\overline{Y}^{\pi}_{(\eta,l_{\eta}) s k}(\theta'),\ \theta,\theta'\in\mathcal{O}_{\pi},
\end{align}
and thus determine the Stratonovich-Weyl operator $\Delta^{\pi}$,
\begin{align}
\label{eq:swop}
\Delta^{\pi}(\theta) & = \sum_{\eta,l_{\eta},s,k}C(\pi,d_{\pi};\overline{\pi},d_{\overline{\pi}}|\eta,0_{l_{\eta}};s)^{-1}Y^{\pi}_{(\eta,l_{\eta}) s k}(\theta)\int_{\mathcal{O}_{\pi}}d\mu_{\pi}(\theta')\overline{Y}^{\pi}_{(\eta,l_{\eta}) s k}(\theta')P_{\theta'} \\ \nonumber
 & = K_{\pi}^{-\frac{1}{2}}P_{\theta}.
\end{align}
Here, $P_{\theta}=v_{\theta}\otimes v_{\theta}^{*}$ denotes the projection onto the coherent state $v_{\theta}$. The phase convention for the Clebsch-Gordan coefficients is chosen s.t. $C(\pi,d_{\pi};\overline{\pi},d_{\overline{\pi}}|\eta,0_{l_{\eta}};s)>0$.
\end{itemize}
\begin{Remark}
\label{rem:swquant}
From \eqref{eq:swop}, we see that the operator norm of $\Delta^{\pi}(\theta)$ is uniformly bounded in $\theta\in\mathcal{O}_{\pi}$. Therefore, the quantisation formula,
\begin{align}
\label{eq:genswquant}
A_{f} & := \int_{\mathcal{O}_{\pi}}d\mu_{\pi}(\theta)\ f(\theta)\Delta^{\pi}(\theta),
\end{align}
defines an element of $\End(V_{\pi})$ for any $f\in L^{1}(\mathcal{O}_{\pi})$. Since the generalised spherical harmonics are smooth, \eqref{eq:genswquant} even makes sense for $f\in\cD'(\mathcal{O}_{\pi})$. By restricting to $f\in S_{\pi}\subset C^{\infty}(\mathcal{O}_{\pi})$ the quantisation, $f\mapsto A_{f}$, becomes nondegenerate, but in contrast to $C^{\infty}(\mathcal{O}_{\pi})$, which can be interpreted as the analog of the space $S^{\infty}_{\rho,\delta}$, $S_{\pi}$ is not closed under multiplication.
\end{Remark}
Let us state the properties of the Stratonovich-Weyl quantisation \eqref{eq:genswquant} as a
\begin{Theorem}
\label{thm:swquant}
The \textit{Stratonovich-Weyl quantisation}
\begin{align}
\label{eq:scaledswquant}
Q^{\SW}_{\varepsilon}(f) & := \int_{\mathcal{O}_{\pi}}d\mu_{\varepsilon^{-1}\pi}(\theta)\ f(\theta)\ \Delta^{\varepsilon^{-1}\pi}(\theta),\ Q^{\SW}_{0}(f):=f,\ \varepsilon^{-1}\in\N_{0},
\end{align}
is a degenerate strict deformation\footnote{$\textup{im}(Q^{\SW}_{\varepsilon})$ is closed.} quantisation of $C^{\infty}(\mathcal{O}_{\pi})$ into $\End(V_{\varepsilon^{-1}\pi})$ in the sense of theorem \ref{thm:localweylquant} (cf. \cite{LandsmanMathematicalTopicsBetween}, Definition II.1.1.1.). Here, $\varepsilon^{-1}\pi\in\hat{G}$ is determined by the highest weight \mbox{$\varepsilon^{-1}\lambda_{\pi}\in\overline{C}\cap I_{r}$,} the intersection of the closed fundamental Weyl chamber $\overline{C}$ and the lattice of (real) integral weights $I_{r}^{*}$. The Poisson structure on $\mathcal{O}_{\pi}$ is induced from the (minus) Lie-Poisson structure on $\fg^{*}$.
\begin{Proof}
Degeneracy of the quantisation follows, because $\dim(\End(V_{\varepsilon^{-1}\pi}))<\infty$. For $f\in C^{\infty}(\mathcal{O}_{\pi})$, we observe that the Stratonovich-Weyl quantisation $Q^{\SW}_{\varepsilon}(f)$ is related to Berezin quantisation $Q^{\textup{B}}_{\varepsilon}(f) = \int_{\mathcal{O}_{\pi}}d\mu_{\pi}(\theta)\ f(\theta)\ P_{\theta}$ (cf. \cite{LandsmanMathematicalTopicsBetween}, Section III.1.11) by the operator $K_{\varepsilon^{-1}\pi}$:
\begin{align}
\label{eq:swbrelquant}
Q^{\SW}_{\varepsilon}(f) & = Q^{\textup{B}}_{\varepsilon}(K^{\frac{1}{2}}_{\varepsilon^{-1}\pi}f).
\end{align}
But, Landsman proves in \cite{LandsmanMathematicalTopicsBetween}, section II.1.11, that $Q^{\textup{B}}_{\varepsilon}$ is a strict quantisation of $C^{\infty}(\mathcal{O}_{\pi})$. Although, we need to slightly correct the $\varepsilon$-expansion of $Q^{\textup{B}}_{\varepsilon}$, which is erroneous in \cite{LandsmanMathematicalTopicsBetween}. Thus, if we controlled the $\varepsilon$-expansion of $K^{\frac{1}{2}}_{\varepsilon^{-1}\pi}$ to order $\varepsilon$, we would be able to decide the strictness of $Q^{\SW}_{\varepsilon}$ from the strictness of $Q^{\textup{B}}_{\varepsilon}$. To find the required $\varepsilon$-expansion of $K^{\frac{1}{2}}_{\varepsilon^{-1}\pi}$, we compute the $\varepsilon$-expansion of $K_{\varepsilon^{-1}\pi}$,
\begin{align}
\label{eq:Kexpansion}
K_{\varepsilon^{-1}\pi} = K^{(0)}_{\pi} + \varepsilon K^{(1)}_{\pi} + O(\varepsilon^{2}),
\end{align} 
and apply functional calculus, i.e.
\begin{align}
\label{eq:fcinverse}
K^{\frac{1}{2}}_{\varepsilon^{-1}\pi} & = \left(\id_{C(\mathcal{O}_{\pi})}+\left(K_{\varepsilon^{-1}\pi}-\id_{C(\mathcal{O}_{\pi})}\right)\right)^{\frac{1}{2}} \\ \nonumber
&  = \id_{C(\mathcal{O}_{\pi})} + \frac{1}{2}\left(K_{\varepsilon^{-1}\pi}-1\right)+O(\varepsilon^{2}) = \id_{C(\mathcal{O}_{\pi})} + \frac{1}{2}\varepsilon K^{(1)}_{\pi} + O(\varepsilon^{2}),
\end{align}
since $\lim_{\varepsilon\rightarrow0}K_{\varepsilon^{-1}\pi}= K^{(0)}_{\pi} = \id_{C(\mathcal{O}_{\pi})}$ (cf. \cite{LandsmanMathematicalTopicsBetween}, Theorem III.1.11.1.), and $\forall f\in C(\mathcal{O}_{\pi}): ||K_{\varepsilon^{-1}\pi}f||_{\infty}\leq||f||_{\infty}$. Now, let us show that $K_{\varepsilon^{-1}\pi}$ actually has an expansion of the form \eqref{eq:Kexpansion}. To this end, we analyse $K_{\varepsilon^{-1}\pi}$ in the form of \eqref{eq:overlapkernel}:
\begin{align}
\label{eq:Kkernelexpansion1}
\left(K_{\varepsilon^{-1}\pi}f\right)(\theta) & = \int_{\mathcal{O}_{\pi}}d\mu_{\varepsilon^{-1}\pi}(\theta')\ |(v_{\theta},v_{\theta'})_{V_{\varepsilon^{-1}\pi}}|^{2}\ f(\theta') \\ \nonumber
 & = \int_{\mathcal{O}_{\pi}}d\mu_{\varepsilon^{-1}\pi}(\theta')\ |(v_{\varepsilon^{-1}\pi},(\varepsilon^{-1}\pi)(g_{\theta}^{-1}g_{\theta'})v_{\varepsilon^{-1}\pi})_{V_{\varepsilon^{-1}\pi}}|^{2}\ f(\theta') \\ \nonumber
 & = d_{\varepsilon^{-1}\pi} \int_{G_{\lambda_{\pi}}}dh\ \int_{G/G_{\lambda_{\pi}}}dg_{\theta'}\ |(v_{\varepsilon^{-1}\pi},(\varepsilon^{-1}\pi)(g_{\theta}^{-1}g_{\theta'})v_{\varepsilon^{-1}\pi})_{V_{\varepsilon^{-1}\pi}}|^{2}\ F(g_{\theta'}h) \\ \nonumber
 & = d_{\varepsilon^{-1}\pi}\int_{G}dg\ |(v_{\varepsilon^{-1}\pi},\pi(g^{-1}_{\theta}g)v_{\varepsilon^{-1}\pi})_{V_{\varepsilon^{-1}\pi}}|^{2}\ F(g) \\ \nonumber
 & = d_{\varepsilon^{-1}\pi}\int_{G}dg\ |(v_{\varepsilon^{-1}\pi},\pi(g)v_{\varepsilon^{-1}\pi})_{V_{\varepsilon^{-1}\pi}}|^{2}\ F(g_{\theta}g),
\end{align}
where we have used the (right) $G_{\lambda_{\pi}}$-invariance of $g\mapsto|(v_{\varepsilon^{-1}\pi},\pi(g^{-1}_{\theta}g)v_{\varepsilon^{-1}\pi})_{V_{\varepsilon^{-1}\pi}}|^{2}$, as $v_{\varepsilon^{-1}\pi}$ is a highest weight vector. Here, $F=f\circ p\in C^{\infty}(G)$ is the (right) $G_{\lambda_{\pi}}$-invariant functions corresponding to $f\in C^{\infty}(\mathcal{O}_{\pi})$ via $p:G\rightarrow G/G_{\lambda_{\pi}}\cong\mathcal{O}_{\pi}$. Again, exploiting the fact that $v_{\varepsilon^{-1}\pi}$ is a highest weight vector, we find:
\begin{align}
\label{eq:cartancompositerelation}
\forall g\in G:\ (v_{\varepsilon^{-1}\pi},\pi(g)v_{\varepsilon^{-1}\pi})_{V_{\varepsilon^{-1}\pi}} & = (v_{\pi},\pi(g)v_{\pi})_{V_{\pi}}^{\varepsilon^{-1}},
\end{align}
since the Cartan composite $V_{\lambda_{1}+\lambda_{2}}$ of two highest weights $\lambda_{1},\lambda_{2}\in\overline{C}\cap I^{*}_{r}$ has multiplicity 1 in $V_{\lambda_{1}}\otimes V_{\lambda_{2}}$ (cf. \cite{BroeckerRepresentationsOfCompact}, VI.2.8). This allows us to write \eqref{eq:Kkernelexpansion1} in the form
\begin{align}
\label{eq:Kkernelexpansion2}
\left(K_{\varepsilon^{-1}\pi}f\right)(\theta) & = d_{\varepsilon^{-1}\pi}\int_{G}dg\ |(v_{\pi},\pi(g)v_{\pi})_{V_{\pi}}|^{2\varepsilon^{-1}}\ F(g_{\theta}g) \\ \nonumber
 & = d_{\varepsilon^{-1}\pi}\int_{G}dg\ e^{-2\varepsilon^{-1}S(g)}\ f(p(g_{\theta}g)) \\ \nonumber
 & = d_{\varepsilon^{-1}\pi}\int_{G/G_{\lambda_{\pi}}}dg_{\theta'}\ e^{-2\varepsilon^{-1}S(g_{\theta'})}\ f(p(g_{\theta}g_{\theta'})) \\ \nonumber
 & = d_{\varepsilon^{-1}\pi}\int_{\mathcal{O}_{\pi}}d\mu_{\pi}(\theta')\ e^{-2\varepsilon^{-1}S_{\pi}(\theta')}\ f(g_{\theta}\cdot\theta') \\ \nonumber
 & = d_{\varepsilon^{-1}\pi}\int_{U_{\lambda_{\pi}}}d\mu_{\pi}(\theta')\ e^{-2\varepsilon^{-1}S_{\pi}(\theta')}\ f(g_{\theta}\cdot\theta') + O(\varepsilon^{\infty}),
\end{align}
where $S(g):=-\log|(v_{\pi},\pi(g)v_{\pi})_{V_{\pi}}|$, which descends to $S_{\pi}$ on $\mathcal{O}_{\pi}$ by (right) $G_{\lambda_{\pi}}$-invariance. The restriction to an arbitrary open neighbourhood $U_{\lambda_{\pi}}$ of $\lambda_{\pi}$ is justified by the fact that the positive function $S_{\pi}$ assumes its sole absolute minimum at $\lambda_{\pi}$, $S_{\pi}(\lambda_{\pi})=0$, and the simple estimate:
\begin{align}
\label{eq:exponentialerror}
\bigg|d_{\varepsilon^{-1}\pi}\int_{\mathcal{O}_{\pi}\setminus U_{\lambda_{\pi}}}\!\!\!\!\!\!d\mu_{\pi}(\theta')\ e^{-2\varepsilon^{-1}S_{\pi}(\theta')}\ f(g_{\theta}\cdot\theta')\bigg| & \leq d_{\varepsilon^{-1}\pi}\int_{\mathcal{O}_{\pi}\setminus U_{\lambda_{\pi}}}\!\!\!\!\!\!d\mu_{\pi}(\theta')\ e^{-2\varepsilon^{-1}S_{\pi}(\theta')}\ |f(g_{\theta}\cdot\theta')| \\ \nonumber
 & \leq d_{\varepsilon^{-1}\pi} ||f(g_{\theta}\cdot(\!\ .\!\ ))||_{\infty}e^{-2\varepsilon\inf_{\theta'\in\mathcal{O}_{\pi}\setminus U_{\lambda_{\pi}}}\!\!\!\!S_{\pi}(\theta')}\in O(\varepsilon^{\infty}),
\end{align}
as $d_{\varepsilon^{-1}\pi}\in O(\varepsilon^{-\frac{1}{2}\dim\mathcal{O}_{\pi}})$ by Weyl's formula (cf. \cite{BroeckerRepresentationsOfCompact}, VI.1.7). Next, we choose $U_{\lambda_{\pi}}$ small enough to identify it with a neighbourhood $W_{\lambda_{\pi}}\subset T_{\lambda_{\pi}}\mathcal{O}_{\pi}\cong\fg/\fg_{\lambda_{\pi}}$ of $0$ via the exponential map. Due to the decomposition of $\fg_{\C} = \ft_{\C}\oplus\bigoplus_{\alpha\in R^{+}}\fg_{\alpha}\oplus\fg_{-\alpha}$ in to root spaces $\fg_{\pm\alpha},\ \alpha\in R^{+}$, we have $\fg_{\lambda_{\pi}}=\ft\oplus\bigoplus_{\substack{\!\!\!\!\!\!\alpha\in R^{+} \\ \langle\lambda_{\pi},\alpha\rangle_{\fg^{*}}=0}}(\fg\cap(\fg_{\alpha}\oplus\fg_{-\alpha}))$, leading to complex coordinates $(z,\overline{z})=\{(z_{\alpha},\overline{z}_{\alpha})\}_{\alpha\in R^{+}_{\lambda_{\pi}}}$ via
\begin{align}
\label{eq:exponentialcoordinatescoad}
g_{\theta(z,\overline{z})} & = \exp\Bigg(\sum_{\alpha\in R^{+}_{\lambda_{\pi}}}(z_{\alpha}E_{\alpha}-\overline{z}_{\alpha}E_{-\alpha})\Bigg),
\end{align}
where we defined $R^{+}_{\lambda_{\pi}}:=\{\alpha\in R^{+}\ |\ \langle\lambda_{\pi},\alpha\rangle_{\fg^{*}}\neq0\}$, and introduced a Cartan-Weyl basis \\ $\{H_{i}\}_{i=1}^{r}\subset\ft,\ \{E_{\alpha},E_{-\alpha}\}_{\alpha\in R^{+}}$, $E_{\pm\alpha}\in\fg_{\alpha}$, for $\fg_{\C}$ (cf. \cite{HumphreysIntroductionToLie}):
\begin{align}
\label{eq:cartanweylbasis}
[H_{i},H_{j}] & = 0,\ i,j=1,...,r\ ,\\ \nonumber
[H_{i}, E_{\pm\alpha}] & = \pm2\pi i\alpha(H_{i})E_{\pm\alpha},\ i=1,...,r\ ,\ \alpha\in R^{+}\ ,\\ \nonumber
[E_{\alpha},E_{-\alpha}] & = H_{\alpha},\ \alpha\in R^{+}, \\ \nonumber
[E_{\alpha},E_{\beta}] & = N_{\alpha,\beta}E_{\alpha+\beta},\ \alpha,\beta\in R^{+}\cup(-R^{+}),\ \alpha+\beta\neq0,\ N_{\alpha,\beta}\neq0\ \textup{iff}\ \alpha+\beta\in R^{+}\cup(-R^{+}).
\end{align}
Here, $H_{\alpha}=\frac{2T_{\alpha}}{\langle T_{\alpha}, T_{\alpha}\rangle_{\fg}}$ is the co-root associated with $\alpha = \frac{1}{2\pi i}\langle T_{\alpha},\ .\ \rangle_{\fg}\in R^{+}$\footnote{Here, $<\ .\ ,\ .\ >_{\fg}$ extends to a non-degenerate bilinear form on $\fg_{\C}$.}.\\
Thus, we arrive at
\begin{align}
\label{eq:cxKintegral}
\left(K_{\varepsilon^{-1}\pi}f\right)(\theta) & = d_{\varepsilon^{-1}\pi}\int_{W_{\lambda_{\pi}}\subset\fg/\fg_{\lambda}}\underbrace{\Big(\prod_{\alpha\in R^{+}_{\lambda_{\pi}}}\frac{dz_{\alpha}d\overline{z}_{\alpha}}{2\pi}\Big)}_{:=\sqrt{2\pi}^{-\dim(\mathcal{O}_{\pi})}dz d\overline{z}}J(z,\overline{z})e^{-2\varepsilon^{-1}S_{\pi}(z,\overline{z})}f(g_{\theta}\cdot(z,\overline{z})) \\[-0.5cm] \nonumber
 &\hspace{0.5cm} + O(\varepsilon^{\infty}),
\end{align}
where $J$ is the Jacobian associated with the exponential map. With \eqref{eq:cxKintegral} at hand, we are in a position to determine the $\varepsilon$-expansion of $K_{\varepsilon^{-1}\pi}$ to order $\varepsilon$ by an appeal to Laplace's method (cf. \cite{ErdelyiAsymptoticExpansions}), i.e. we insert the Taylor expansion of $S_{\pi}(z,\overline{z})=-\frac{1}{2}\log|(v_{\pi},\pi(g_{\theta(z,\overline{z})})v_{\pi})_{V_{\pi}}|^{2}$ around $(z_{\lambda_{\pi}},\overline{z}_{\lambda_{\pi}})=0$ up to fourth(!) order into \eqref{eq:cxKintegral}, and invoke the unique extension of the unitary representation $\pi$ to a holomorphic representation of $G_{\C}$:
\begin{align}
\label{eq:Sexpansion1}
& (v_{\pi},\pi(g_{\theta(z,\overline{z})})v_{\pi})_{V_{\pi}} \\ \nonumber
 & = \sum_{n=0}^{\infty}\frac{1}{n!}\big(v_{\pi},\Big(\sum_{\alpha\in R^{+}_{\lambda_{\pi}}}(z_{\alpha}d\pi(E_{\alpha})-\overline{z}_{\alpha}d\pi(E_{-\alpha}))\Big)^{n}v_{\pi}\big)_{V_{\pi}} \\ \nonumber
 & = 1 + \sum_{\alpha\in R^{+}_{\lambda_{\pi}}}(z_{\alpha}\underbrace{(v_{\pi},d\pi(E_{\alpha})v_{\pi})_{V_{\pi}}}_{=0}-\overline{z}_{\alpha}\underbrace{(v_{\pi},d\pi(E_{-\alpha})v_{\pi})_{V_{\pi}}}_{=0}) \\ \nonumber
 &\hspace{0.65cm}+\frac{1}{2}\sum_{\alpha,\beta\in R^{+}_{\lambda_{\pi}}}\Big(z_{\alpha}z_{\beta}\underbrace{(v_{\pi},d\pi(E_{\alpha})d\pi(E_{\beta})v_{\pi})_{V_{\pi}}}_{=0}+\overline{z}_{\alpha}\overline{z}_{\beta}
\underbrace{(v_{\pi},d\pi(E_{-\alpha})d\pi(E_{-\beta})v_{\pi})_{V_{\pi}}}_{=0} \\[-0.25cm] \nonumber
 &\hspace{2.6cm}-z_{\alpha}\overline{z}_{\beta}\underbrace{(v_{\pi},d\pi(E_{\alpha})d\pi(E_{-\beta})v_{\pi})_{V_{\pi}}}_{\mathclap{=\delta_{\alpha,\beta}(v_{\pi},d\pi(H_{\alpha})v_{\pi})_{V_{\pi}}=\delta_{\alpha,\beta}\langle\lambda_{\pi},\alpha\rangle_{\fg^{*}}(\langle\alpha,\alpha\rangle_{\fg^{*}})^{-1}}}-\overline{z}_{\alpha}z_{\beta}\underbrace{(v_{\pi},d\pi(E_{-\alpha})d\pi(E_{\beta})v_{\pi})_{V_{\pi}}}_{=0}\Big) \\ \nonumber
\label{eq:Sexpansion2}
 &+\frac{1}{6}\!\!\!\!\sum_{\alpha,\beta,\gamma\in R^{+}_{\lambda_{\pi}}}\!\!\!\!\Big(z_{\alpha}z_{\beta}z_{\gamma}\underbrace{(v_{\pi},d\pi(E_{\alpha})d\pi(E_{\beta})d\pi(E_{\gamma})v_{\pi})_{V_{\pi}}}_{=0}-z_{\alpha}z_{\beta}\overline{z}_{\gamma}\underbrace{(v_{\pi},d\pi(E_{\alpha})d\pi(E_{\beta})d\pi(E_{-\gamma})v_{\pi})_{V_{\pi}}}_{=\delta_{\alpha,\gamma-\beta}N_{\beta,-\gamma}\langle\lambda_{\pi},\alpha\rangle_{\fg^{*}}(\langle\alpha,\alpha\rangle_{\fg^{*}})^{-1}} \\ \nonumber
 &\hspace{0.5cm}-z_{\alpha}\overline{z}_{\beta}z_{\gamma}\underbrace{(v_{\pi},d\pi(E_{\alpha})d\pi(E_{-\beta})d\pi(E_{\gamma})v_{\pi})_{V_{\pi}}}_{=0}
+z_{\alpha}\overline{z}_{\beta}\overline{z}_{\gamma}\underbrace{(v_{\pi},d\pi(E_{\alpha})d\pi(E_{-\beta})d\pi(E_{-\gamma})v_{\pi})_{V_{\pi}}}_{=\delta_{\alpha-\beta,\gamma}N_{\alpha,-\beta}\langle\lambda_{\pi},\gamma\rangle_{\fg^{*}}(\langle\gamma,\gamma\rangle_{\fg^{*}})^{-1}} \\ \nonumber
 &\hspace{0.5cm}-\overline{z}_{\alpha}z_{\beta}z_{\gamma}\underbrace{(v_{\pi},d\pi(E_{-\alpha})d\pi(E_{\beta})d\pi(E_{\gamma})v_{\pi})_{V_{\pi}}}_{=0}
+\overline{z}_{\alpha}z_{\beta}\overline{z}_{\gamma}\underbrace{(v_{\pi},d\pi(E_{-\alpha})d\pi(E_{\beta})d\pi(E_{-\gamma})v_{\pi})_{V_{\pi}}}_{=0} \\[-0.25cm] \nonumber
 &\hspace{0.5cm}+\overline{z}_{\alpha}\overline{z}_{\beta}z_{\gamma}\underbrace{(v_{\pi},d\pi(E_{-\alpha})d\pi(E_{-\beta})d\pi(E_{\gamma})v_{\pi})_{V_{\pi}}}_{=0}
-\overline{z}_{\alpha}\overline{z}_{\beta}\overline{z}_{\gamma}\underbrace{(v_{\pi},d\pi(E_{-\alpha})d\pi(E_{-\beta})d\pi(E_{-\gamma})v_{\pi})_{V_{\pi}}}_{=0}\Big)\\ \nonumber
 &+\frac{1}{24}\!\!\!\!\sum_{\substack{\alpha,\beta, \\ \gamma,\zeta}\in R^{+}_{\lambda_{\pi}}}\!\!\!\!\Big(\!z_{\alpha}z_{\beta}z_{\gamma}z_{\zeta}\!\underbrace{(v_{\pi},d\pi(E_{\alpha})d\pi(E_{\beta})d\pi(E_{\gamma})d\pi(E_{\zeta})v_{\pi})_{V_{\pi}}}_{=0} \\[-0.25cm] \nonumber
 &\hspace{2cm}-z_{\alpha}\overline{z}_{\beta}z_{\gamma}z_{\zeta}\!\underbrace{(v_{\pi},d\pi(E_{\alpha})d\pi(E_{-\beta})d\pi(E_{\gamma})d\pi(E_{\zeta})v_{\pi})_{V_{\pi}}}_{=0} \\ \nonumber
 &\hspace{2cm}+z_{\alpha}\overline{z}_{\beta}\overline{z}_{\gamma}z_{\zeta}\underbrace{(v_{\pi},d\pi(E_{\alpha})d\pi(E_{-\beta})d\pi(E_{-\gamma})d\pi(E_{\zeta})v_{\pi})_{V_{\pi}}}_{=0}
 \\ \nonumber
 &\hspace{2cm}-z_{\alpha}\overline{z}_{\beta}\overline{z}_{\gamma}\overline{z}_{\zeta}\underbrace{(v_{\pi},d\pi(E_{\alpha})d\pi(E_{-\beta})d\pi(E_{-\gamma})d\pi(E_{-\zeta})
v_{\pi})_{V_{\pi}}}_{\neq0} \\ \nonumber
 &\hspace{2cm}+z_{\alpha}\overline{z}_{\beta}z_{\gamma}\overline{z}_{\zeta}\underbrace{(v_{\pi},d\pi(E_{\alpha})d\pi(E_{-\beta})d\pi(E_{\gamma})d\pi(E_{-\zeta})v_{\pi})_{V_{\pi}}}_{\neq0}
 \\ \nonumber
 &\hspace{2cm}+z_{\alpha}z_{\beta}\overline{z}_{\gamma}\overline{z}_{\zeta}\underbrace{(v_{\pi},d\pi(E_{\alpha})d\pi(E_{\beta})d\pi(E_{-\gamma})d\pi(E_{-\zeta})v_{\pi})_{V_{\pi}}}_{\neq0} \\ \nonumber
 &\hspace{2cm}-z_{\alpha}z_{\beta}\overline{z}_{\gamma}z_{\zeta}\underbrace{(v_{\pi},d\pi(E_{\alpha})d\pi(E_{\beta})d\pi(E_{-\gamma})d\pi(E_{\zeta})v_{\pi})_{V_{\pi}}}_{=0}
 \\ \nonumber
 &\hspace{2cm}-z_{\alpha}z_{\beta}z_{\gamma}\overline{z}_{\zeta}\underbrace{(v_{\pi},d\pi(E_{\alpha})d\pi(E_{\beta})d\pi(E_{\gamma})d\pi(E_{-\zeta})v_{\pi})_{V_{\pi}}}_{\neq0}\\ \nonumber
&\hspace{2cm}+\overline{z}_{\alpha}z_{\beta}z_{\gamma}z_{\zeta}\underbrace{(v_{\pi},d\pi(E_{-\alpha})d\pi(E_{\beta})d\pi(E_{\gamma})d\pi(E_{\zeta})v_{\pi})_{V_{\pi}}}_{=0} \\ \nonumber
 &\hspace{2cm}-\overline{z}_{-\alpha}
\overline{z}_{\beta}z_{\gamma}z_{\zeta}\underbrace{(v_{\pi},d\pi(E_{\alpha})d\pi(E_{-\beta})d\pi(E_{\gamma})d\pi(E_{\zeta})v_{\pi})_{V_{\pi}}}_{=0} \\ \nonumber
 &\hspace{2cm}+\overline{z}_{\alpha}\overline{z}_{\beta}\overline{z}_{\gamma}z_{\zeta}\!\underbrace{(\!v_{\pi},d\pi(E_{-\alpha})d\pi(E_{-\beta})d\pi(E_{-\gamma})d\pi(E_{\zeta})v_{\pi}\!)_{V_{\pi}}}_{=0}
 \\ \nonumber
 &\hspace{2cm}-\overline{z}_{\alpha}\overline{z}_{\beta}\overline{z}_{\gamma}\overline{z}_{\zeta}\!\underbrace{(\!v_{\pi},d\pi(E_{-\alpha})d\pi(E_{-\beta})d\pi(E_{-\gamma})
d\pi(E_{-\zeta})v_{\pi}\!)_{V_{\pi}}}_{=0} \\ \nonumber
 &\hspace{2cm}+\overline{z}_{\alpha}\overline{z}_{\beta}z_{\gamma}\overline{z}_{\zeta}\!\underbrace{(v_{\pi},d\pi(E_{-\alpha})d\pi(E_{-\beta})d\pi(E_{\gamma})d\pi(E_{-\zeta})v_{\pi})_{V_{\pi}}}_{=0} \\ \nonumber
 &\hspace{2cm}+\overline{z}_{\alpha}z_{\beta}\overline{z}_{\gamma}\overline{z}_{\zeta}\!\underbrace{(v_{\pi},d\pi(E_{-\alpha})d\pi(E_{\beta})d\pi(E_{-\gamma})d\pi(E_{-\zeta})
v_{\pi})_{V_{\pi}}}_{=0} \\ \nonumber
 &\hspace{2cm}-\overline{z}_{\alpha}z_{\beta}\overline{z}_{\gamma}z_{\zeta}\underbrace{(v_{\pi},d\pi(E_{-\alpha})d\pi(E_{\beta})d\pi(E_{-\gamma})d\pi(E_{\zeta})
v_{\pi})_{V_{\pi}}}_{=0}
 \\ \nonumber
 &\hspace{2cm}-\overline{z}_{\alpha}z_{\beta}z_{\gamma}\overline{z}_{\zeta}\underbrace{(v_{\pi},d\pi(E_{-\alpha})d\pi(E_{\beta})d\pi(E_{\gamma})d\pi(E_{-\zeta})v_{\pi})_{V_{\pi}}}_{=0}\Big)\\ \nonumber
 &+O((z,\overline{z})^{5})
\end{align}
where we repeatedly exploited the fact that $v_{\pi}$ is a highest weight vector, the commutation relations \eqref{eq:cartanweylbasis} and the (unitary) representation $\pi$ implements the adjointness relations $d\pi(E_{\alpha})^{*}=d\pi(E_{-\alpha}),\ \alpha\in R^{+}$\footnote{This can be seen from the fact that $\{H_{\alpha}E_{\alpha},E_{-\alpha}\}$ generates an $\mathfrak{sl}_{2}$-subalgebra, and the Verma module construction (cf. \cite{HallLieGroupsLie}).}. At this point it is important to note that non-zero contributions at order $(z,\overline{z})^{3}$ only arise for terms with $\alpha\neq\beta, \alpha\neq\gamma, \beta\neq\gamma$ due to the geometry of root systems, i.e. the only multiples of a root $\alpha$ occurring in the decomposition of $\fg_{\C}$ are $\pm\alpha$.  \eqref{eq:Sexpansion1} and \eqref{eq:Sexpansion2} imply:
\begin{align}
\label{eq:Sexpansion3}
 |(v_{\pi},\pi(g_{\theta(z,\overline{z})})v_{\pi})_{V_{\pi}}|^{2} & = 1 -\!\!\!\!\sum_{\alpha\in R^{+}_{\lambda_{\pi}}}\!\!\frac{\langle\lambda_{\pi},\alpha\rangle_{\fg^{*}}}{\langle\alpha,\alpha\rangle_{\fg^{*}}}z_{\alpha}\overline{z}_{\alpha} \\ \nonumber
 &\hspace{0.5cm}+\frac{1}{6}\Bigg(\sum_{\substack{\alpha,\beta,\gamma\in R^{+}_{\lambda_{\pi}} \\ \alpha\neq\beta,\alpha\neq\gamma,\beta\neq\gamma}}\big(-z_{\alpha}z_{\beta}\overline{z}_{\gamma}(v_{\pi},d\pi(E_{\alpha})d\pi(E_{\beta})d\pi(E_{-\gamma})v_{\pi})_{V_{\pi}}\\[-0.75cm] \nonumber
 &\hspace{3.6cm}+z_{\alpha}\overline{z}_{\beta}\overline{z}_{\gamma}(v_{\pi},d\pi(E_{\alpha})d\pi(E_{-\beta})d\pi(E_{-\gamma})v_{\pi})_{V_{\pi}}\big)\\ \nonumber
 &\hspace{1.1cm}\underbrace{\left.+\sum_{\substack{\alpha,\beta,\gamma\in R^{+}_{\lambda_{\pi}} \\ \alpha\neq\beta,\alpha\neq\gamma,\beta\neq\gamma}}\big(-\overline{z}_{\alpha}\overline{z}_{\beta}z_{\gamma}\overline{(v_{\pi},d\pi(E_{\alpha})d\pi(E_{\beta})d\pi(E_{-\gamma})v_{\pi})_{V_{\pi}}}\right.\ \ \Bigg)}_{=0,\ \textup{due to}\ d\pi(E_{\alpha})^{*}=d\pi(E_{-\alpha}),\ \alpha\in R^{+}}\\[-1.25cm] \nonumber
 &\hspace{3.6cm}+\overline{z}_{\alpha}z_{\beta}z_{\gamma}\overline{(v_{\pi},d\pi(E_{\alpha})d\pi(E_{-\beta})d\pi(E_{-\gamma})v_{\pi})_{V_{\pi}}}\big) \\[0.75cm] \nonumber
 & + \frac{1}{12}\Bigg(3\sum_{\alpha,\beta\in R^{+}_{\lambda_{\pi}}}\!\!\!\!\frac{\langle\lambda_{\pi},\alpha\rangle_{\fg^{*}}}{\langle\alpha,\alpha\rangle_{\fg^{*}}}\frac{\langle\lambda_{\pi},\beta\rangle_{\fg^{*}}}{\langle\beta,\beta\rangle_{\fg^{*}}}z_{\alpha}\overline{z}_{\alpha}z_{\beta}\overline{z}_{\beta} \\ \nonumber
 & + \sum_{\alpha,\beta,\gamma,\zeta\in R^{+}_{\lambda_{\pi}}}\!\!\left(-z_{\alpha}\overline{z}_{\beta}\overline{z}_{\gamma}\overline{z}_{\zeta}(v_{\pi},d\pi(E_{\alpha})d\pi(E_{-\beta})d\pi(E_{-\gamma})d\pi(E_{-\zeta})v_{\pi})_{V_{\pi}}\right) \\[-0.5cm] \nonumber
 &\hspace{2.1cm}+z_{\alpha}\overline{z}_{\beta}z_{\gamma}\overline{z}_{\zeta}(v_{\pi},d\pi(E_{\alpha})d\pi(E_{-\beta})d\pi(E_{\gamma})d\pi(E_{-\zeta})v_{\pi})_{V_{\pi}} \\ \nonumber
 &\hspace{2.1cm}+z_{\alpha}z_{\beta}\overline{z}_{\gamma}\overline{z}_{\zeta}(v_{\pi},d\pi(E_{\alpha})d\pi(E_{\beta})d\pi(E_{-\gamma})d\pi(E_{-\zeta})v_{\pi})_{V_{\pi}} \\[-0.25cm] \nonumber
 &\hspace{2.1cm}-z_{\alpha}z_{\beta}z_{\gamma}\overline{z}_{\zeta}(v_{\pi},d\pi(E_{\alpha})d\pi(E_{\beta})d\pi(E_{\gamma})d\pi(E_{-\zeta})v_{\pi})_{V_{\pi}}\ \ \ \ \Bigg) \\
 & + O((z,\overline{z})^{5}).
\end{align}
Thus, we find by expanding the logarithm in $S_{\pi}$:
\begin{align}
\label{eq:Sexpansion5}
S_{\pi}(z,\overline{z}) & = \frac{1}{2}\sum_{\alpha\in R^{+}_{\lambda_{\pi}}}\frac{\langle\lambda_{\pi},\alpha\rangle_{\fg^{*}}}{\langle\alpha,\alpha\rangle_{\fg^{*}}}z_{\alpha}\overline{z}_{\alpha} \\ \nonumber
 &\ \ \ + \frac{1}{24}\Bigg(3\sum_{\alpha,\beta\in R^{+}_{\lambda_{\pi}}}\frac{\langle\lambda_{\pi},\alpha\rangle_{\fg^{*}}}{\langle\alpha,\alpha\rangle_{\fg^{*}}}\frac{\langle\lambda_{\pi},\beta\rangle_{\fg^{*}}}{\langle\beta,\beta\rangle_{\fg^{*}}}z_{\alpha}\overline{z}_{\alpha}z_{\beta}\overline{z}_{\beta} \\[-0.25cm] \nonumber
 &\ \ \ \hspace{1cm} -\!\!\!\!\!\!\sum_{\alpha,\beta,\gamma,\zeta\in R^{+}_{\lambda_{\pi}}}\big(-z_{\alpha}\overline{z}_{\beta}\overline{z}_{\gamma}\overline{z}_{\zeta}(v_{\pi},d\pi(E_{\alpha})d\pi(E_{-\beta})d\pi(E_{-\gamma})d\pi(E_{-\zeta})v_{\pi})_{V_{\pi}}\Bigg) \\[-0.5cm] \nonumber
 &\hspace{3.3cm}+z_{\alpha}\overline{z}_{\beta}z_{\gamma}\overline{z}_{\zeta}(v_{\pi},d\pi(E_{\alpha})d\pi(E_{-\beta})d\pi(E_{\gamma})d\pi(E_{-\zeta})v_{\pi})_{V_{\pi}} \\ \nonumber
 &\hspace{3.3cm}+z_{\alpha}z_{\beta}\overline{z}_{\gamma}\overline{z}_{\zeta}(v_{\pi},d\pi(E_{\alpha})d\pi(E_{\beta})d\pi(E_{-\gamma})d\pi(E_{-\zeta})v_{\pi})_{V_{\pi}} \\ \nonumber
 &\hspace{3.3cm}-z_{\alpha}z_{\beta}z_{\gamma}\overline{z}_{\zeta}(v_{\pi},d\pi(E_{\alpha})d\pi(E_{\beta})d\pi(E_{\gamma})d\pi(E_{-\zeta})v_{\pi})_{V_{\pi}}\big) \\ \nonumber
 &\ \ \ + O((z,\overline{z})^{5}) \\ \nonumber
 & = \frac{1}{2}\sum_{\alpha\in R^{+}_{\lambda_{\pi}}}\frac{\langle\lambda_{\pi},\alpha\rangle_{\fg^{*}}}{\langle\alpha,\alpha\rangle_{\fg^{*}}}z_{\alpha}\overline{z}_{\alpha} + P_{4}(z,\overline{z}) + O((z,\overline{z})^{5}).
\end{align}
The integration in \eqref{eq:cxKintegral} can be extended from $W_{\lambda_{\pi}}$ to $\fg/\fg_{\lambda_{\pi}}$ at the expense of a term of order $\varepsilon^{\infty}$ by an estimate similar to \eqref{eq:exponentialerror}, if the integrands are suitably continued beyond $W_{\lambda_{\pi}}$, which leads to ($\langle\lambda_{\pi},\alpha\rangle_{\fg^{*}}>0$ for $\alpha\in R^{+}_{\lambda_{\pi}}$):
\begin{align}
\label{eq:laplaceapproxK}
 & \left(K_{\varepsilon^{-1}\pi}f\right)(\theta) \\ \nonumber
 & = \sqrt{2\pi}^{\ -\dim(\mathcal{O}_{\pi})}\!d_{\varepsilon^{-1}\pi}\!\!\!\int_{\fg/\fg_{\lambda}}\!\!\!\!\!\!\!dz\ d\overline{z}\ J(z,\overline{z})e^{-2\varepsilon^{-1}S_{\pi}(z,\overline{z})}f(g_{\theta}\cdot(z,\overline{z}))+O(\varepsilon^{\infty}) \\ \nonumber
 & = \sqrt{2\pi}^{\ -\dim(\mathcal{O}_{\pi})}\!d_{\varepsilon^{-1}\pi}\!\!\!\int_{\fg/\fg_{\lambda}}\!\!\!\!\!\!\!\!\!\!\!dz\ \!d\overline{z}\ \!J(z,\overline{z})e^{-\varepsilon^{-1}(\sum_{\alpha\in R^{+}_{\lambda_{\pi}}}\!\!\!\!\frac{\langle\lambda_{\pi},\alpha\rangle_{\fg^{*}}}{\langle\alpha,\alpha\rangle_{\fg^{*}}}z_{\alpha}\overline{z}_{\alpha} + 2P_{4}(z,\overline{z}) + O((z,\overline{z})^{5}))}  f(g_{\theta}\cdot(z,\overline{z}))\\ \nonumber
 &\hspace{0.5cm}+O(\varepsilon^{\infty}) \\ \nonumber
 & = \sqrt{\frac{\varepsilon}{2\pi}}^{\ \dim(\mathcal{O}_{\pi})}\!\!\!d_{\varepsilon^{-1}\pi}\int_{\fg/\fg_{\lambda}}\!\!\!\!\!\!\!\!\!\!\!dz\ \!d\overline{z}\ \!J(\sqrt{\varepsilon}z,\sqrt{\varepsilon}\overline{z})e^{-(\sum_{\alpha\in R^{+}_{\lambda_{\pi}}}\!\!\!\!\frac{\langle\lambda_{\pi},\alpha\rangle_{\fg^{*}}}{\langle\alpha,\alpha\rangle_{\fg^{*}}}z_{\alpha}\overline{z}_{\alpha} + 2\varepsilon P_{4}(z,\overline{z}))} f(g_{\theta}\cdot(\sqrt{\varepsilon}z,\sqrt{\varepsilon}\overline{z}))\\ \nonumber
 &\hspace{0.5cm}+O(\varepsilon^{3}) \\ \nonumber
 & =  \sqrt{\frac{\varepsilon}{2}}^{\ \dim(\mathcal{O}_{\pi})}\!\!\!\!\!\!d_{\varepsilon^{-1}\pi}\bigg(\prod_{\alpha\in R^{+}_{\lambda_{\pi}}}\frac{\langle\alpha,\alpha\rangle_{\fg^{*}}}{\langle\lambda_{\pi},\alpha\rangle_{\fg^{*}}}\bigg)\Bigg(J(0,0)f(g_{\theta}\cdot(0,0)) \\ \nonumber
  &\ \ \ +\varepsilon\bigg(\sum_{\alpha\in R^{+}_{\lambda_{\pi}}}\frac{\langle\alpha,\alpha\rangle_{\fg^{*}}}{\langle\lambda_{\pi},\alpha\rangle_{\fg^{*}}}\partial_{z_{\alpha}}\partial_{\overline{z}_{\alpha}}(Jf(g_{\theta}\cdot(\ .\ )))(0,0) + J(0,0)f(g_{\theta}\cdot(0,0))C_{4}(\lambda_{\pi})\bigg)\Bigg)+O(\varepsilon^{2}) \\ \nonumber
  & =  \bigg(\prod_{\alpha\in R^{+}_{\lambda_{\pi}}}\frac{\langle\alpha,\alpha\rangle_{\fg^{*}}}{2\langle\delta,\alpha\rangle_{\fg^{*}}}\bigg)\Bigg(J(0,0)f(g_{\theta}\cdot(0,0))+\varepsilon\bigg(\sum_{\alpha\in R^{+}_{\lambda_{\pi}}}\frac{\langle\alpha,\alpha\rangle_{\fg^{*}}}{\langle\lambda_{\pi},\alpha\rangle_{\fg^{*}}}\partial_{z_{\alpha}}\partial_{\overline{z}_{\alpha}}(Jf(g_{\theta}\cdot(\ .\ )))(0,0) \\ \nonumber
  &\hspace{4cm}+ J(0,0)f(g_{\theta}\cdot(0,0))\bigg(C_{4}(\lambda_{\pi})+\sum_{\alpha\in R^{+}_{\lambda_{\pi}}}\frac{\langle\delta,\alpha\rangle_{\fg^{*}}}{\langle\lambda_{\pi},\alpha\rangle_{\fg^{*}}}\bigg)\bigg)\Bigg)+O(\varepsilon^{2}).
\end{align}
Here, $C_{4}(\lambda_{\pi})$ results from contribution of $e^{-2\varepsilon P_{4}(z,\overline{z})}$ to the order $\varepsilon$. The half-integer orders of $\varepsilon$ coming from the Taylor expansions of $J$ and $f(g_{\theta}\cdot(\ .\ ))$ vanish, because
\begin{align}
\label{eq:cxgaussianintegrals}
\int_{\fg/\fg_{\lambda_{\pi}}}dz\ d\overline{z}\ e^{-\sum_{\alpha\in R^{+}_{\lambda_{\pi}}}z_{\alpha}\overline{z}_{\alpha}}z^{\beta}\overline{z}^{\gamma} & =\sqrt{\pi}^{\dim(\mathcal{O}_{\pi})}\beta!\delta_{\beta,\gamma},\ \ \ \beta,\gamma\in\N_{0}^{\frac{1}{2}\dim(\mathcal{O}_{\pi})}.
\end{align}
Furthermore, we have
\begin{align}
\label{eq:coadexpdet}
J(0,0) & = \bigg(\prod_{\alpha\in R^{+}_{\lambda_{\pi}}}\frac{\langle\alpha,\alpha\rangle_{\fg^{*}}}{2\langle\delta,\alpha\rangle_{\fg^{*}}}\bigg)^{-1},
\end{align}
as we know that $\lim_{\varepsilon\rightarrow0}K_{\varepsilon^{-1}\pi}=\id_{C(\mathcal{O}_{\pi})}$. This allows us to \eqref{eq:laplaceapproxK} into a slightly simpler form:
\begin{align}
\label{eq:simplelaplaceapproxK}
\left(K_{\varepsilon^{-1}\pi}f\right)(\theta) & = f(\theta)+\varepsilon\Bigg(\bigg(\prod_{\alpha\in R^{+}_{\lambda_{\pi}}}\frac{\langle\alpha,\alpha\rangle_{\fg^{*}}}{2\langle\delta,\alpha\rangle_{\fg^{*}}}\bigg)\sum_{\alpha\in R^{+}_{\lambda_{\pi}}}\frac{\langle\alpha,\alpha\rangle_{\fg^{*}}}{\langle\lambda_{\pi},\alpha\rangle_{\fg^{*}}}\partial_{z_{\alpha}}\partial_{\overline{z}_{\alpha}}(Jf(g_{\theta}\cdot(\ .\ )))(0,0) \\[-0.25cm] \nonumber
  &\hspace{4cm}+ f(\theta)\bigg(C_{4}(\lambda_{\pi})+\sum_{\alpha\in R^{+}_{\lambda_{\pi}}}\frac{\langle\delta,\alpha\rangle_{\fg^{*}}}{\langle\lambda_{\pi},\alpha\rangle_{\fg^{*}}}\bigg)\Bigg)+O(\varepsilon^{2}) \\[-0.5cm] \nonumber
  & = f(\theta) + \varepsilon\left(K^{(1)}_{\pi}f\right)(\theta) + O(\varepsilon^{2}).
\end{align}
Now, following Landsman (cf. \cite{LandsmanMathematicalTopicsBetween}, Theorem III.1.11.4.), given any $\Phi_{\varepsilon}\in V_{\varepsilon^{-1}\pi},\ ||\Phi_{\varepsilon}||=1,$ we have:
\begin{align}
\label{eq:berezinswexpansion}
 &(\Phi_{\varepsilon},(Q^{\SW}_{\varepsilon}(f)Q^{\SW}_{\varepsilon}(f')-Q^{\SW}_{\varepsilon}(ff'))\Phi_{\varepsilon})_{V_{\varepsilon^{-1}\pi}} \\ \nonumber
 & = (\Phi_{\varepsilon},(Q^{\textup{B}}_{\varepsilon}(K^{\frac{1}{2}}_{\varepsilon^{-1}\pi}f)Q^{\textup{B}}_{\varepsilon}(K^{\frac{1}{2}}_{\varepsilon^{-1}\pi}f')-Q^{\textup{B}}_{\varepsilon}(K^{\frac{1}{2}}_{\varepsilon^{-1}\pi}(ff')))\Phi_{\varepsilon})_{V_{\varepsilon^{-1}\pi}} \\ \nonumber
 & = (\Phi_{\varepsilon},(Q^{\textup{B}}_{\varepsilon}(f)Q^{\textup{B}}_{\varepsilon}(f')-Q^{\textup{B}}_{\varepsilon}(ff'))\Phi_{\varepsilon})_{V_{\varepsilon^{-1}\pi}} \\ \nonumber
 &\ \ \ +\frac{\varepsilon}{2} (\Phi_{\varepsilon},(Q^{\textup{B}}_{\varepsilon}(K^{(1)}_{\pi}f)Q^{\textup{B}}_{\varepsilon}(f')+Q^{\textup{B}}_{\varepsilon}(f)Q^{\textup{B}}_{\varepsilon}(K^{(1)}_{\pi}f')-Q^{\textup{B}}_{\varepsilon}(K^{(1)}_{\pi}(ff')))\Phi_{\varepsilon})_{V_{\varepsilon^{-1}\pi}} + O(\varepsilon^{2})
\end{align}
for all $f,f'\in C^{\infty}(\mathcal{O}_{\pi},\R)$.\\
At this point, it is important to recall the equality $||A||=\sup_{||\Phi_{\varepsilon}||=1}|(\Phi_{\varepsilon},A\Phi_{\varepsilon})_{V_{\varepsilon^{-1}\pi}}|$ for $A\in\End(V_{\varepsilon^{-1}\pi}),\ A^{*}=A$. The terms in $O(\varepsilon)$ satisfy bounds of the form $C\varepsilon^{k}||f||_{\infty,m}||f'||_{\infty,n}||\Phi_{\varepsilon}||^{2}$ for $k\in\N,\ m,n\in\N_{0},\ C>0$, which will imply Dirac's condition (see theorem \ref{thm:localweylquant}) in the proof of strictness of $Q^{\SW}_{\varepsilon}$, if $Q^{\textup{B}}_{\varepsilon}$ is strict. Similarly, Rieffel's and von Neumann's conditions will follow from the strictness of $Q^{\textup{B}}_{\varepsilon}$ due to:
\begin{align}
\label{eq:rnvcondexpansion}
& (\Phi_{\varepsilon},Q^{\SW}_{\varepsilon}(f)\Phi_{\varepsilon})_{V_{\varepsilon^{-1}\pi}} \\ \nonumber
& = (\Phi_{\varepsilon},Q^{\textup{B}}_{\varepsilon}(f)\Phi_{\varepsilon})_{V_{\varepsilon^{-1}\pi}} + \varepsilon(\Phi_{\varepsilon},Q^{\textup{B}}_{\varepsilon}(K^{(1)}_{\pi}f)\Phi_{\varepsilon})_{V_{\varepsilon^{-1}\pi}} + O(\varepsilon^{2}), \\[0.25cm] \nonumber
& (\Phi_{\varepsilon},(\tfrac{i}{\varepsilon}[Q^{\SW}_{\varepsilon}(f),Q^{\SW}_{\varepsilon}(f')]-Q^{\SW}_{\varepsilon}(\{f,f'\}_{-}))\Phi_{\varepsilon})_{V_{\varepsilon^{-1}\pi}} \\ \nonumber
& = (\Phi_{\varepsilon},(\tfrac{i}{\varepsilon}[Q^{\textup{B}}_{\varepsilon}(f),Q^{\textup{B}}_{\varepsilon}(f')]-Q^{\textup{B}}_{\varepsilon}(\{f,f'\}_{-}))\Phi_{\varepsilon})_{V_{\varepsilon^{-1}\pi}} + O(\varepsilon).
\end{align}
Finally, the strictness of $Q^{\textup{B}}_{\varepsilon}$ can be concluded from Landsman's argument subject to some minor modifications. More precisely, Landsman considers the first term in the last line of \eqref{eq:berezinswexpansion} in the form:
\begin{align}
\label{eq:landsmanslaplace}
& (\Phi_{\varepsilon},(Q^{\textup{B}}_{\varepsilon}(f)Q^{\textup{B}}_{\varepsilon}(f')-Q^{\textup{B}}_{\varepsilon}(ff'))\Phi_{\varepsilon})_{V_{\varepsilon^{-1}\pi}} \\ \nonumber
& = d_{\varepsilon^{-1}\pi}\int_{G}dg\ F(g)(\Phi_{\varepsilon},(\varepsilon^{-1}\pi)(g)v_{\varepsilon^{-1}\pi})_{V_{\varepsilon^{-1}\pi}}I_{\varepsilon}(g), \\[0.25cm] \nonumber
I_{\varepsilon}(g) & := d_{\varepsilon^{-1}\pi}\int_{G}dh\ (v_{\pi},\pi(h)v_{\pi})_{V_{\pi}}^{\varepsilon^{-1}}F_{\varepsilon^{-1}\pi}(g,h), \\ \nonumber
F'_{\varepsilon^{-1}\pi}(g,h) & := ((\varepsilon^{-1}\pi)(gh)v_{\varepsilon^{-1}\pi},\Phi_{\varepsilon})_{V_{\varepsilon^{-1}\pi}}(F'(gh)-F'(g)),
\end{align}
and then subjects $I_{\varepsilon}(g)$ to an asymptotic expansion by Laplace's method analogous to that of $\left(K_{\varepsilon^{-1}\pi}f\right)(\theta)$. In contrast to the previous calculation, neither $I_{\varepsilon}$ nor $F'_{\varepsilon^{-1}\pi}$ are (right) $G_{\lambda_{\pi}}$-invariant, but are only (right) $G_{\lambda_{\pi}}$-equivariant:
\begin{align}
\label{eq:landsmankernelequivariance}
F'_{\varepsilon^{-1}\pi}(g,hg_{\lambda_{\pi}}) & = e^{-\frac{i}{\varepsilon}\phi(g_{\lambda_{\pi}})}F'_{\varepsilon^{-1}\pi}(g,h), & F'_{\varepsilon^{-1}\pi}(gg_{\lambda_{\pi}},h) & = e^{-\frac{i}{\varepsilon}\phi(g_{\lambda_{\pi}})}F'_{\varepsilon^{-1}\pi}(g,g_{\lambda_{\pi}}hg_{\lambda_{\pi}}^{-1}), \\ \nonumber
I_{\varepsilon}(gg_{\lambda_{\pi}}) & = e^{-\frac{i}{\varepsilon}\phi(g_{\lambda_{\pi}})}I_{\varepsilon}(g), & g_{\lambda_{\pi}}\in G_{\lambda_{\pi}},&\ \phi:G_{\lambda_{\pi}}\rightarrow\R
\end{align}
which ensures the invariance of \eqref{eq:landsmanslaplace}. Nonetheless, the functions $h\mapsto(v_{\pi},\pi(h)v_{\pi})_{V_{\pi}}^{\varepsilon^{-1}}F_{\varepsilon^{-1}\pi}(g,h)$ are (right) $G_{\lambda_{\pi}}$-invariant for every $g\in G$. By the same arguments used in \eqref{eq:Kkernelexpansion2}, \eqref{eq:exponentialerror} and \eqref{eq:cxKintegral}, as $|F'_{\varepsilon^{-1}\pi}(g,h)|\leq2||f'||_{\infty}$, we have
\begin{align}
\label{eq:landsmanskernel}
I_{\varepsilon}(g) & = \sqrt{2\pi}^{\ -\dim(\mathcal{O}_{\pi})}d_{\varepsilon^{-1}\pi}\int_{\fg/\fg_{\lambda_{\pi}}}\!\!\!\!\!\!dz\ d\overline{z}\ J(z,\overline{z})\ e^{-\varepsilon(-\log(v_{\pi},\pi(g_{\theta(z,\overline{z})})v_{\pi})_{V_{\pi}})}F'_{\varepsilon^{-1}\pi}(g,g_{\theta(z,\overline{z})}) \\ \nonumber
&\hspace{0.5cm} + O(\varepsilon^{\infty}).
\end{align}
Employing \eqref{eq:Sexpansion1} and \eqref{eq:Sexpansion2}, we find:
\begin{align}
\label{eq:landsmanskernelexpansion1}
 & -\log(v_{\pi},\pi(g_{\theta(z,\overline{z})})v_{\pi})_{V_{\pi}} \\ \nonumber
 & = -(\left(v_{\pi},\pi(g_{\theta(z,\overline{z})})v_{\pi})_{V_{\pi}}-1\right)-\frac{1}{2}\left((v_{\pi},\pi(g_{\theta(z,\overline{z})})v_{\pi})_{V_{\pi}}-1\right)^{2}+O((z,\overline{z})^{5}) \\ \nonumber
 & = \frac{1}{2}\sum_{\alpha\in R^{+}_{\lambda_{\pi}}}\frac{\langle\lambda_{\pi},\alpha\rangle_{\fg^{*}}}{\langle\alpha,\alpha\rangle_{\fg^{*}}}z_{\alpha}\overline{z}_{\alpha}-\frac{1}{6}\sum_{\substack{\alpha,\beta,\gamma\in R^{+}_{\lambda_{\pi}} \\ \alpha\neq\beta,\alpha\neq\gamma,\beta\neq\gamma}}\left(-z_{\alpha}z_{\beta}\overline{z}_{\gamma}\delta_{\alpha,\gamma-\beta}N_{\beta,-\gamma}\langle\lambda_{\pi},\alpha\rangle_{\fg^{*}}(\langle\alpha,\alpha\rangle_{\fg^{*}})^{-1}\right.\\[-0.75cm] \nonumber
 &\hspace{6.6cm}\left.+z_{\alpha}\overline{z}_{\beta}\overline{z}_{\gamma}\delta_{\alpha-\beta,\gamma}N_{\alpha,-\beta}\langle\lambda_{\pi},\gamma\rangle_{\fg^{*}}(\langle\gamma,\gamma\rangle_{\fg^{*}})^{-1}\right)\\ \nonumber
 &\ \ \ + \frac{1}{24}\Bigg(3\sum_{\alpha,\beta\in R^{+}_{\lambda_{\pi}}}\frac{\langle\lambda_{\pi},\alpha\rangle_{\fg^{*}}}{\langle\alpha,\alpha\rangle_{\fg^{*}}}\frac{\langle\lambda_{\pi},\beta\rangle_{\fg^{*}}}{\langle\beta,\beta\rangle_{\fg^{*}}}z_{\alpha}\overline{z}_{\alpha}z_{\beta}\overline{z}_{\beta} \\ \nonumber
 &\ \ \ -\!\!\!\!\!\!\sum_{\alpha,\beta,\gamma,\zeta\in R^{+}_{\lambda_{\pi}}}\!\!\big(-z_{\alpha}\overline{z}_{\beta}\overline{z}_{\gamma}\overline{z}_{\zeta}(v_{\pi},d\pi(E_{\alpha})d\pi(E_{-\beta})d\pi(E_{-\gamma})d\pi(E_{-\zeta})v_{\pi})_{V_{\pi}}\Bigg) \\[-0.5cm] \nonumber
 &\hspace{2.1cm}+z_{\alpha}\overline{z}_{\beta}z_{\gamma}\overline{z}_{\zeta}(v_{\pi},d\pi(E_{\alpha})d\pi(E_{-\beta})d\pi(E_{\gamma})d\pi(E_{-\zeta})v_{\pi})_{V_{\pi}} \\ \nonumber
 &\hspace{2.1cm}+z_{\alpha}z_{\beta}\overline{z}_{\gamma}\overline{z}_{\zeta}(v_{\pi},d\pi(E_{\alpha})d\pi(E_{\beta})d\pi(E_{-\gamma})d\pi(E_{-\zeta})v_{\pi})_{V_{\pi}} \\ \nonumber
 &\hspace{2.1cm}-z_{\alpha}z_{\beta}z_{\gamma}\overline{z}_{\zeta}(v_{\pi},d\pi(E_{\alpha})d\pi(E_{\beta})d\pi(E_{\gamma})d\pi(E_{-\zeta})v_{\pi})_{V_{\pi}}\big) \\ \nonumber
 &\ \ \ + O((z,\overline{z})^{5}),
\end{align}
which yields the corrected expansion of $I_{\varepsilon}(g)$ to order $\varepsilon$:
\begin{align}
\label{eq:landsmankernelexpansion2}
I_{\varepsilon}(g) & = \bigg(\prod_{\alpha\in R^{+}_{\lambda_{\pi}}}\frac{\langle\alpha,\alpha\rangle_{\fg^{*}}}{\langle\delta,\alpha\rangle_{\fg^{*}}}\bigg)\Bigg(J(0,0)\underbrace{F'_{\varepsilon^{-1}\pi}(g,e)}_{=0} \\ \nonumber
 &\hspace{3cm}+\varepsilon\bigg(\sum_{\alpha\in R^{+}_{\lambda_{\pi}}}2\frac{\langle\alpha,\alpha\rangle_{\fg^{*}}}{\langle\lambda_{\pi},\alpha\rangle_{\fg^{*}}}\partial_{z_{\alpha}}\partial_{\overline{z}_{\alpha}}(JF'_{\varepsilon^{-1}\pi}(g,g_{\theta(\!\ .\!\ ,\!\ .\!\ )}))(0,0) \\ \nonumber
 &\hspace{3cm}+ J(0,0)\underbrace{F'_{\varepsilon^{-1}\pi}(g,e)}_{=0}\bigg(C_{3}(\lambda_{\pi})+2C_{4}(\lambda_{\pi})+\sum_{\alpha\in R^{+}_{\lambda_{\pi}}}\frac{\langle\delta,\alpha\rangle_{\fg^{*}}}{\langle\lambda_{\pi},\alpha\rangle_{\fg^{*}}}\bigg)\bigg)\Bigg)+O(\varepsilon^{2}) \\ \nonumber
 & = \varepsilon\sqrt{2}^{\ \dim(\mathcal{O}_{\pi})}J(0,0)^{-1}\Bigg(\sum_{\alpha\in R^{+}_{\lambda_{\pi}}}2\frac{\langle\alpha,\alpha\rangle_{\fg^{*}}}{\langle\lambda_{\pi},\alpha\rangle_{\fg^{*}}}\partial_{z_{\alpha}}\partial_{\overline{z}_{\alpha}}(JF'_{\varepsilon^{-1}\pi}(g,g_{\theta(\ .\ ,\ .\ )}))(0,0)\Bigg) + O(\varepsilon^{2}) \\ \nonumber
 & \underset{\mathclap{\substack{ \\ \\ (\partial_{z_{\alpha}}J)(0,0)=0 \\ (\partial_{\overline{z}_{\alpha}}J)(0,0)=0}}}{=}\varepsilon\sqrt{2}^{\ \dim(\mathcal{O}_{\pi})}\Bigg(\sum_{\alpha\in R^{+}_{\lambda_{\pi}}}2\frac{\langle\alpha,\alpha\rangle_{\fg^{*}}}{\langle\lambda_{\pi},\alpha\rangle_{\fg^{*}}}\partial_{z_{\alpha}}\partial_{\overline{z}_{\alpha}}(F'_{\varepsilon^{-1}\pi}(g,g_{\theta(\ .\ ,\ .\ )}))(0,0)\Bigg) + O(\varepsilon^{2})
\end{align}
The terms of order $(z,\overline{z})^{3}$ do not yield contributions containing first derivatives of $F'_{\varepsilon^{-1}\pi}$, because of the constraints $\alpha\neq\beta,\ \alpha\neq\gamma,\ \beta\neq\gamma$ and \eqref{eq:cxgaussianintegrals}. Clearly, the expansion is compatible with (right) $G_{\lambda_{\pi}}$-equivariance, as can be seen from \eqref{eq:landsmankernelequivariance} and the $G_{\lambda_{\pi}}$-invariance of the differential operator:
\begin{align}
\label{eq:invlaplaceop}
\Delta_{\lambda_{\pi}} & := \bigg(\sum_{\alpha\in R^{+}_{\lambda_{\pi}}}2\frac{\langle\alpha,\alpha\rangle_{\fg^{*}}}{\langle\lambda_{\pi},\alpha\rangle_{\fg^{*}}}\partial_{z_{\alpha}}\partial_{\overline{z}_{\alpha}}\bigg)_{|z=0,\overline{z}=0}.
\end{align}
Now, strictness of $Q^{\textup{B}}_{\varepsilon}$ follows from Landsman's argument.
\end{Proof}
\end{Theorem}
\begin{Remark}
\label{rem:swquantspecial}
The Stratonovich-Weyl quantisation on coadjoint orbits can be interpreted as the analog of Weyl quantisation on $\R^{n}$. In view of \eqref{eq:swsymbolprop} it is distinguished from Berezin quantisation by the ``tracial property'' leading to the Stratonovich-Weyl-Fourier transform as pointed out by Figueroa, Gracia-Bond{\'i}a and V{\'a}rilly \cite{FigueroaMoyalQuantizationWith}.
\end{Remark}
Now, we are in a position to define the Stratonovich-Weyl-Fourier transform:
\begin{Definition}[cf. \cite{FigueroaMoyalQuantizationWith}]
\label{def:swftransform}
The \textup{Stratonovich-Weyl-Fourier transform} is the composition of the Fourier transform $\mathcal{F}:L^{2}(G)\subset L^{1}(G)\rightarrow L^{2}(\hat{G})$ with the Stratonovich-Weyl symbol map $W:\hat{G}\rightarrow L^{2}(\bigcup_{\pi\in\hat{G}}\mathcal{O}_{\pi})=:L^{2}(\mathcal{O}_{G})$, i.e.:
\begin{align}
\label{eq:swftransform}
\mathcal{F}_{\SW}[\Psi](\pi,\theta) & = \hat{\Psi}_{\SW}(\pi,\theta) \\ \nonumber
 & = W^{\pi}_{\mathcal{F}[\Psi]}(\theta) \\ \nonumber
 & = \int_{G}dg\ \Psi(g)\tr(\Delta^{\pi}(\theta)\pi(g)) \\ \nonumber
 & = \int_{G}dg\ \Psi(g)E(g;\pi,\theta),\ \Psi\in L^{1}(G),\ \pi\in\hat{G},\ \theta\in\mathcal{O}_{\pi},
\end{align}
where we introduced the integral kernel $E(g;\pi,\theta)=\tr(\Delta^{\pi}(\theta)\pi(g)) = W^{\pi}_{\pi(g)}(\theta)$. The \textup{space of integral coadjoint orbits}, $\mathcal{O}_{G}$, is endowed with the integral:
\begin{align}
\label{eq:coadjointint}
\int_{\mathcal{O}_{G}}d\mu_{\mathcal{O}_{G}}(\pi,\theta)\Phi(\pi,\theta) & := \sum_{\pi\in\hat{G}}d_{\pi}\int_{\mathcal{O}_{\pi}}d\mu_{\pi}(\theta)\ \Phi(\pi,\theta),\ \Phi\in L^{1}(\mathcal{O}_{G}).
\end{align}
\end{Definition}
It follows from the next theorem that the Stratonovich-Weyl-Fourier transform is well-defined and isometric from $L^{2}(G)$ to $L^{2}(\mathcal{O}_{G})$ (the range is $\bigoplus_{\pi\in\hat{G}}S_{\pi}$). Moreover, it intertwines the convolution product and the \textit{twisted product} coming from:
\begin{align}
\label{eq:swtwistedproduct}
W^{\pi}_{AB}(\theta) = (W^{\pi}_{A}\star W^{\pi}_{B})(\theta) = \int_{\mathcal{O}_{\pi}}d\mu_{\pi}(\theta')\int_{\mathcal{O}_{\pi}}d\mu_{\pi}(\theta'')\tr(\Delta^{\pi}(\theta)\Delta^{\pi}(\theta')\Delta^{\pi}(\theta''))W^{\pi}_{A}(\theta')W^{\pi}_{B}(\theta'').
\end{align}
\begin{Theorem}[cf \cite{FigueroaMoyalQuantizationWith}, Theorem 5]
\label{thm:swftransform}
The Stratonovich-Weyl-Fourier transform satisfies the inversion formula
\begin{align}
\label{eq:swfinversion}
\Psi(g) & = \int_{\mathcal{O}_{G}}d\mu_{\mathcal{O}_{G}}(\pi,\theta)\overline{E}(g;\pi,\theta)\mathcal{F}_{\SW}[\Psi](\pi,\theta),
\end{align}
and the Parseval-Plancherel identity
\begin{align}
\label{eq:ppid}
\int_{G}dg\ |\Psi(g)|^{2} & = \int_{\mathcal{O}_{G}}d\mu_{\mathcal{O}_{G}}(\pi,\theta)\ |\mathcal{F}_{\SW}[\Psi](\pi,\theta)|^{2}.
\end{align}
Furthermore, the convolution product $\ast$ on $L^{1}(G)$ is intertwined with the twisted product $\star$:
\begin{align}
\label{eq:convolutiontwist}
\mathcal{F}_{\SW}[\Psi\ast\Phi] & = \mathcal{F}_{\SW}[\Psi]\star\mathcal{F}_{\SW}[\Phi].
\end{align}
\end{Theorem}
The integral kernel $E:G\times\mathcal{O}_{G}\rightarrow\C$ has important properties that entail its independence of the representative in $\pi\in\hat{G}$ (property 2).
\begin{Proposition}[cf. \cite{FigueroaMoyalQuantizationWith}, Theorem 4]
\label{prop:swfkernel}
The integral kernel $E$ satisfies:
\begin{itemize}
	\item[1.] $\overline{E}(g;\pi,\theta) = E(g^{-1};\pi,\theta)$,
	\item[2.] $E(\alpha_{h}(g);\pi,\theta) = E(g;\pi,Ad^{*}_{h^{-1}}(\theta))$, 
	\item[3.] $\int_{\mathcal{O}_{\pi}}d\mu_{\pi}(\theta)\ E(g;\pi,\theta) = \tr(\pi(g)) = \chi_{\pi}(g)$,
	\item[4.] $\int_{G}dg\ E(g;\pi,\theta)\overline{E}(g;\pi,\theta') = d_{\pi}^{-1}I_{\pi}(\theta,\theta')$,
	\item[5.] $(E(g;\ .\ )\star\mathcal{F}_{\SW}[\Psi])(\pi,\theta) = \mathcal{F}_{\SW}[U_{g}\Psi](\pi,\theta),\ \Psi\in L^{2}(G)$,
	\item[6.] $E(g;\ .\ )\star E(h;\ .\ ) = E(gh;\ .\ )$.
\end{itemize}
\end{Proposition}
\begin{Remark}
\label{rem:swfkernel}
Property 5 \& 6 of proposition \ref{prop:swfkernel} tell us that the action of the exponentials $U_{g}, g\in G,$ of the ``momenta'' $P_{X}, X\in\fg,$ in \eqref{eq:quantcomm} is turned in to (non-commutative) multiplication with the functions $E(g;\ .\ ),\ g\in G$ by the Stratonovich-Weyl-Fourier transform. Thus, the range of the latter qualifies as (non-commutative) flux representation for loop quantum gravity in the terminology of \cite{BaratinNonCommutativeFlux}.
\end{Remark}
As mentioned in the beginning of this subsection, the Stratonovich-Weyl-Fourier transform allows us to study the effect of scaling in the global Fourier correspondence for compact Lie groups. To this end, we distinguish to types of $\varepsilon$-scaled Stratonovich-Weyl-Fourier transforms ($\varepsilon=\frac{1}{k},\ k\in\N$):
\begin{itemize}
	\item[1.](``position'' scaling): $\mathcal{F}^{\varepsilon}_{\SW}[\Phi](\pi,\theta) := \int_{G}dg\ E(g^{\frac{1}{\varepsilon}};\pi,\theta)\Psi(g)$,
	\item[2.](``momentum'' scaling): $\mathcal{F}_{\SW,\varepsilon}[\Phi] (\pi,\theta):= \int_{G}dg\ E(g;\varepsilon^{-1}\pi,\theta)\Psi(g)$,\\[0.1cm] where $\varepsilon^{-1}\pi$ is the unitary irreducible representation of $G$ with highest weight $\varepsilon^{-1}\lambda_{\pi}$.
\end{itemize}
As above, the restriction to discrete scalings, $\varepsilon=\frac{1}{k},\ k\in\N,$ is necessary as otherwise the powers $g^{\frac{1}{\varepsilon}}$ would not be uniquely defined, and $\varepsilon^{-1}\lambda_{\pi}$ would not belong to $\overline{C}\cap I_{r}^{*}$. Clearly, these transformations coincide for commutative Lie groups, where the Stratonovich-Weyl-Fourier transform coincides with usual Fourier transform. Structurally, the ``momentum'' scaled transform seems to be favoured, as it can be expressed in terms of the unscaled transform:
\begin{align}
\label{eq:momentumscaling}
\mathcal{F}_{\SW,\varepsilon}[\Phi] (\pi,\theta) & = \int_{G}dg\ E(g;\varepsilon^{-1}\pi,\theta)\Psi(g) = \mathcal{F}_{\SW}[\Psi](\varepsilon^{-1}\pi,\theta).
\end{align}
In contrast, the ``position'' scaled transform is not related to unscaled transform, because the $\varepsilon$-th power is not a diffeomorphism of $G$ (unless $\varepsilon=1$). Therefore, we stick to the ``momentum'' scaled transform in the following. Unfortunately, we immediately recognize, that the scaled transform no longer defines an invertible map, as it restricts the Stratonovich-Weyl-Fourier transform to those integral coadjoint orbits associated with the sub-lattice $\varepsilon^{-1}I_{r}^{*}\subset I_{r}^{*}$. Again, this is a feature forced upon us by the rigidity of compact Lie groups. In subsection \ref{subsec:u1bohr}, we will discuss a possible way to remove this restriction for $G=U(1)$.\\[0.1cm]
Nonetheless, the Stratonovich-Weyl-Fourier transform provides us with a $\varepsilon$-scaled transform for systems modelled on integral coadjoint orbits $\mathcal{O}_{\pi}$, which was already exploited in \cite{StottmeisterCoherentStatesQuantumI} for $G=SU(2)$. To see how this works in the general case, we note that $\lambda_{\pi}$ and $\varepsilon^{-1}\lambda_{\pi}$ have the same, possibly degenerate, orbit type $\mathcal{O}_{\pi}$, since $G_{\lambda_{\pi}}=G_{\varepsilon^{-1}\lambda_{\pi}}$. Thus, we may study the ``semiclassical'' limit, $\varepsilon\rightarrow0$, of sequences of operators $\{A_{\varepsilon}\}_{\varepsilon^{-1}\in\N},\ A_{\varepsilon}\in\End(V_{\varepsilon^{-1}\pi}),$ in terms of their Stratonovich-Weyl symbols $W^{\varepsilon^{-1}\pi}_{A_{\varepsilon}}\in S_{\varepsilon^{-1}\pi}\subset C^{\infty}(\mathcal{O}_{\pi})\subset L^{2}(\mathcal{O}_{\pi})$ and the twisted products $\star_{\varepsilon}$, e.g.
\begin{align}
\label{eq:generalisedpauli}
A_{\varepsilon}=i\varepsilon\sum_{i=1}^{n}B_{i}d(\varepsilon^{-1}\pi)(\tau_{i})\ \Leftrightarrow\ W^{\varepsilon^{-1}\pi}_{A_{\varepsilon}}(\theta)=i\varepsilon\sum_{i=1}^{n}B_{i}\frac{d}{dt}_{|t=0}E(e^{t\tau_{i}};\varepsilon^{-1}\pi,\theta),
\end{align}
which is a generalisation of the magnetic part of the Pauli Hamiltonian to $G$. Here, $\{\tau_{i}\}_{i=1}^{n}$ is a basis of $\fg$ and $B=(B_{i})_{i=1,..,n}\in\R^{n}$. Another physical application, apart from spin-orbit coupling discussed in \cite{StottmeisterCoherentStatesQuantumI, FaureTopologicalPropertiesOf}, where these structures feature prominently, is the description of classical and quantum particles with internal symmetry in external gauge fields, which are governed by the classical respectively quantum \textit{Wong equations} (cf. \cite{LandsmanStrictdeformationQuantization, LandsmanMathematicalTopicsBetween}), and its relation to \textit{quantum chaos} (cf. \cite{SchraderSemiclassicalAsymptoticsGauge, TaylorSemiclassicalSpectraOf, GuilleminClusteringTheoremsWith, ZelditchOnAQuantum}).\\[0.1cm]
If we wanted to avoid working with degenerate coadjoint orbits, i.e. those corresponding to singular integral weights $\lambda,\ \langle\alpha,\lambda\rangle_{\fg^{*}}=0$ for some $\alpha\in R^{+}$, we could incorporate the usual shift by half the sum of the positive roots, $\delta=\tfrac{1}{2}\sum_{\alpha\in R^{+}}\alpha$, in to the correspondence between dominant integral weights and integral coadjoint orbits,
\begin{align}
\label{eq:deltashiftedorbit}
\lambda_{\pi}\rightarrow\mathcal{O}^{\delta}_{\pi}:=\{\theta\in\fg^{*}\ |\ \exists g\in G: \theta = Ad^{*}_{g}(\lambda_{\pi}+\delta)\},
\end{align}
which allows us to work solely with coadjoint orbits of strongly dominant integral weights, since
\begin{align}
\label{eq:deltashiftedweights}
\lambda\in\overline{C}\cap I^{*}_{r}\Leftrightarrow\lambda+\delta\in C\cap I^{*}_{r}
\end{align}
(cf. \cite{BroeckerRepresentationsOfCompact, HumphreysIntroductionToLie}). The coadjoint orbits of strongly dominant integral weights are isomorphic to the simply connected generalised flag variety $G/T\cong G_{\C}/B^{+}$, where $B^{+}$ is the standard Borel subgroup associated with the positive roots $R^{+}$ (cf. \cite{BroeckerRepresentationsOfCompact, PerelomovGeneralizedCoherentStates}).
\\[0.1cm]
Another advantage of the Stratonovich-Weyl-Fourier transform is that it enables us to pass from matrix valued symbols on $G\times\hat{G}$ in the definition of the global pseudo-differential calculus of subsection \eqref{subsubsec:globalcalc} to genuine functions on $G\times\mathcal{O}_{G}$. Namely, the Stratonovich-Weyl symbol of a symbol $\sigma\in\hat{\cD'}(\hat{G}\times G)$ is just the Stratonovich-Weyl-Fourier transform of its left convolution kernel $F_{\sigma}$:
\begin{align}
\label{eq:swsymbolpseudo}
\sigma_{\SW}(\pi,\theta;g) & = W^{\pi}_{\sigma(\pi,g)}(\theta) = \tr(\Delta^{\pi}(\theta)\sigma(\pi,g)) \\ \nonumber
 &\ = \int_{G}dh\ E(h;\pi,\theta)F_{\sigma}(h,g) = \hat{F}_{\SW}(\pi,\theta;g).
\end{align}
In this way, the action of the symbol $\sigma$ on $C^{\infty}(G)$ is related to the twisted product $\star$:
\begin{align}
\label{eq:swfpseudodifferential}
\left(\rho_{\uL}(F_{\sigma})\Psi\right)(g) & = \int_{\mathcal{O}_{G}}d\mu_{\mathcal{O}_{G}}\ \left(\overline{E}(g;\ .\ )\star\sigma_{\SW}(\ .\ ;g)\right)(\pi,\theta)\hat{\Psi}_{\SW}(\pi,\theta).
\end{align}
Also, the composition of two symbols $\sigma,\tau$ becomes expressible in terms of the twisted product (cp. proposition \ref{prop:knquant}, 6.):
\begin{align}
\label{eq:swfsymbolproduct}
\mathcal{F}_{\SW}[F_{\sigma}\ast_{\uL}F_{\tau}](\pi,\theta;g) & = \int_{G}dh\ F_{\sigma}(h,g)\left(E(h;\ .\ )\star\tau_{\SW}(\ .\ ;h^{-1}g)\right)(\pi,\theta).
\end{align}
\subsection{Coherent states \& quantisation}
\label{subsec:csquant}
In the previous subsection, we have encountered the concept of Berezin quantisation w.r.t. a (complete) system of coherent states (cf. \cite{BerezinQuantization, BerezinGeneralConceptOf}), and its close relation to Stratonovich-Weyl quantisation on coadjoint orbits. As explained in section \ref{sec:covcs}, we have a (complete) system of coherent states for any compact Lie group $G$ at our disposal, which naturally leads to the question how Berezin quantisation on $T^{*}G$ in terms of these coherent states relates to the Kohn-Nirenberg and Weyl quantisation discussed in subsection \ref{sec:compact}. Moreover, it is important to understand, whether this Berezin quantisation of $T^{*}G$ gives a suitable framework for the Born-Oppenheimer approximation and adiabatic perturbation theory. As we will argue, the equivalent roles played by Berezin and (Stratonovich-)Weyl quantisation on coadjoint orbits with regard to such a framework is quite special to the case of compact phase spaces, and therefore finite dimensional Hilbert spaces, while the existence of an analog of the (smoothing) operator $K^{\frac{1}{2}}_{\pi}$ \eqref{eq:swsymbol} will be the reason, why Weyl quantisation is favoured, if the construction of a twisted product for phase space functions is intended.\\[0.1cm]
Let us start by defining Berezin quantisation on $T^{*}G$ in terms of the coherent states from section \ref{sec:covcs}. To this end, we note that the coherent states $\Psi^{t}_{z}\in L^{2}(G),\ t\in\R,\ z\in G_{\C},$ \eqref{eq:cs} provides us with a \textit{coherent pure state quantisation} of $T^{*}G$ (in the sense of \cite{LandsmanMathematicalTopicsBetween}). Therefore, the following definition makes sense for small enough $t>0$ (see \eqref{eq:measureequiv}, cf. also \cite{SimonTheClassicalLimit}):
\begin{Definition}
\label{def:berezinquantG}
Given $f\in L^{p}(T^{*}G,dg\ dX)$, $1\leq p\leq\infty$, its \textit{Berezin quantisation} is
\begin{align}
\label{eq:berezinquantG}
Q^{\uB}_{t}(f) & := \int_{G_{\C}}dz\ \nu_{t}(z)\ f(\Phi^{-1}(z))\ |\Psi^{t}_{z}\rangle\langle\Psi^{t}_{z}|,
\end{align}
which defines an operator in $\mathcal{S}_{p}(L^{2}(G))$, the $p$th Schatten class in $\cB(L^{2}(G))$, since $\tr\left(|Q^{\uB}_{t}|^{p}\right)\leq||f||^{p}_{p})$ for $1\leq p<\infty$ and $||Q^{\uB}_{t}(f)||\leq||f||_{\infty}$. $f\circ\Phi^{-1}$ is called the \textup{upper} or \textup{contravariant symbol} of the operator $Q^{\uB}_{t}(f)$. The \textup{lower} or \textup{covariant symbol} of an operator $A\in \cB(L^{2}(G))$ is
\begin{align}
\label{eq:lowersymbolG}
L^{t}_{A}(z,\overline{z}) & := \frac{\langle\Psi^{t}_{z}|A|\Psi^{t}_{z}\rangle}{\langle\Psi^{t}_{z}|\Psi^{t}_{z}\rangle},
\end{align}
which satisfies $||L^{t}_{A}||_{\infty}\leq||A||$, and $||L^{t}_{A}|^{p}||^{p}_{p}\leq\tr\left(|A|^{p}\right)$ for $A\in\mathcal{S}_{p}(L^{2}(G))$.
\end{Definition}
\begin{Remark}
\label{rem:berezinquantG}
Berezin quantisation would take a more natural form if the conjecture \ref{con:csres} turned out to be true, as in that case the normalised projection $P_{t}(z,\overline{z})$ onto the coherent state vectors $\Psi^{t}_{z}$ would provide a resolution of unity w.r.t. to the Liouville measure on $T^{*}G$, and \eqref{eq:berezinquantG} would take the form:
\begin{align}
\label{eq:conberezinquantG}
Q^{\uB}_{t}(f) & = C_{t} \int_{G}\int_{\fg}dg\ dX\ f(g,X)\ P_{t}(ge^{iX},ge^{-iX}).
\end{align}
\end{Remark}
\begin{Corollary}[cf. \cite{SimonTheClassicalLimit}, Appendix 1]
\label{cor:berezinquantG}
Berezin quantisation is real, $Q^{\uB}_{t}(f)^{*}=Q^{\uB}_{t}(\overline{f})$, and positive, $Q^{\uB}_{t}(f)\geq0$ if $f\geq0$ a.e.. The upper and lower symbols associated with Berezin quantisation are dual to one another\footnote{This shows that Berezin quantisation lacks the ``tracial property'' of Kohn-Nirenberg and (Stratonovich)-Weyl quantisations (cp. proposition \ref{prop:knquant}, property 5 and \eqref{eq:swsymbolprop}).}, i.e.
\begin{align}
\label{eq:upperlowerdualityG}
\tr(AQ^{\uB}_{t}(f)) & = \int_{G_{\C}}\frac{dz\ \nu_{t}(z)}{\langle\Psi^{t}_{z}|\Psi^{t}_{z}\rangle}L^{t}_{A}(z,\overline{z})f(\Phi^{-1}(z)),\ \ \ A\in\mathcal{S}_{q}(L^{2}(G)),\ f\in L^{p}(T^{*}G,dg\ dX),
\end{align}
for $\frac{1}{q}+\frac{1}{p}=1$. Moreover, the normalised coherent state projections $P_{t}(z,\overline{z})$ are complete, i.e. $\img(Q^{\uB}_{t})$ is sequentially strongly dense in $\cB(L^{2}(G))$, due to the analyticity of the coherent states $\Psi^{t}_{z}$, as the latter implies $\forall A\in\mathcal{S}_{1}(L^{2}(G))$: $L^{t}_{A}=0$ if and only if $A=0$.
\end{Corollary}
While the corollary tells us, that the upper symbol exists for a sequentially strongly dense subspace of operators in $\cB(L^{2}(G))$, it does not ensure that all operators in $\cB(L^{2}(G))$ can be obtained from Berezin quantisation, as $L^{2}(G)$ is infinite dimensional in contrast to the representation spaces $V_{\pi}$ considered in the previous subsection. Therefore, the existence of an upper symbol for a product of Berezin quantisations $Q^{\uB}_{t}(f)Q^{\uB}_{t}(f')$ cannot be concluded. Nevertheless, we might wonder whether there exists a set of functions $S^{t}_{\uB}$ on $T^{*}G$, such that on the one hand $Q^{\uB}_{t}:S^{t}_{\uB}\rightarrow \cB(L^{2}(G))$ is nondegenerate, and $\forall f,f'\in S^{t}_{\uB}:\exists f''\in S^{t}_{\uB}: Q^{\uB}_{t}(f)Q^{\uB}_{t}(f')=Q^{\uB}_{t}(f'')=:Q^{\uB}_{t}(f\star_{t}f')$, while on the other hand Berezin quantisation of $S^{t}_{\uB}$ encompasses sufficiently many ``interesting'' operators, and $\star_{t}$ can be (asymptotically) expanded w.r.t. the Poisson bracket on $T^{*}G$. To analyse this question in some detail, we first consider the relation between upper and lower symbol, if both exist, of a given Berezin quantisation. Similar to \eqref{eq:overlapkernel}, the lower symbol of $Q^{\uB}_{t}(f)$ can be expressed in terms of the upper symbol and the overlap function of the coherent states:
\begin{align}
\label{eq:overlapkernelG}
L^{t}_{Q^{\uB}_{t}(f)}(z,\overline{z}) & = (\langle\Psi^{t}_{z}|\Psi^{t}_{z}\rangle)^{-1}\int_{G_{\C}}dz'\ \nu_{t}(z')\ |\langle\Psi^{t}_{z}|\Psi^{t}_{z'}\rangle|^{2}\ f(\Phi^{-1}(z')).
\end{align}
This expression is also valid in the case of standard coherent states for $\R^{n}$, where it takes the well-known explicit form (see the discussion at the beginning of subsection \ref{sec:newsbh}, cp. \cite{FollandHarmonicAnalysisIn, LandsmanMathematicalTopicsBetween}):
\begin{align}
\label{eq:standardoverlapkernel}
L^{t}_{Q^{\uB}_{t}(f)}(z,\overline{z}) & = \int_{\C^{n}}\tfrac{d\Re(z)\ d\Im(z)}{(2\pi t)^{n}}\ e^{-\frac{1}{2t}(z-z')(\overline{z}-\overline{z}')}\ f(\Phi^{-1}(z)) = \left(e^{2t\partial_{z}\partial_{\overline{z}}}f\circ\Phi^{-1}\right)(z),
\end{align}
from which it can be inferred that $L^{t}_{Q^{\uB}_{t}}$ is the restriction of an entire function on $\C^{2n}$ even if $f\in\cS'(\R^{2n})$ due to the smoothing nature of $e^{2t\partial_{z}\partial_{\overline{z}}} = e^{\frac{t}{2}\Delta_{(q,p)}},\ z=q+ip=\Phi(q,p)$. Moreover, Weyl quantisation (see \eqref{eq:stweylquant}), $Q^{\uB}_{t}(f)=A_{\sigma}$, fits into the picture in the same way as for the coadjoint orbits \eqref{eq:swsymbol} (cp. \cite{FollandHarmonicAnalysisIn}):
\begin{align}
\label{eq:standardweylsmoothing}
L^{t}_{Q^{\uB}_{t}(f)}(q,p) & = \left(e^{\frac{t}{4}\Delta_{(q,p)}}\sigma\right)(q,p), & \sigma(q,p) & = \left(e^{\frac{t}{4}\Delta_{(q,p)}}f\right)(q,p),
\end{align}
which tells us that Weyl symbols are typically better behaved than upper symbols. Especially, since we would expect the (asymptotic) expansion of a twisted product $\star_{t}$ to be determined by local products of derivatives of the factors, which would be problematic in case of distributional symbols.\\[0.1cm]
A similar situation is to be expected for a compact Lie group $G$, and the Berezin quantisation \eqref{eq:berezinquantG} of $T^{*}G$, as the latter is non-compact. We will show this explicitly for $G=U(1)$\footnote{A generalisation to $G=U(1)^{n}$ is straightforward, and will be omitted.}, or more precisely for $U(1)$-equivariant Berezin quantisation. The coherent states from section \ref{sec:covcs} take the following explicit form for $G=U(1)$ (cp. \cite{KowalskiCoherentStatesFor}):\\[0.1cm]
Considering the Hilbert spaces
\begin{align}
\label{eq:u1equivarianths}
\fH_{j_{0}} & :=\{\Psi\in L^{2}_{\loc}(\R)\ |\ \forall j\in\Z: \Psi(\varphi+2\pi j) = e^{2\pi i j_{0}j}\Psi(\varphi)\},\ j_{0}\in[0,1),
\end{align}
equipped with the scalar product
\begin{align}
\label{eq:u1equivariantsp}
(\Psi_{1},\Psi_{2})_{j_{0}} & := \int_{[0,2\pi)}\frac{d\varphi}{2\pi}\ \overline{\Psi_{1}(\varphi)}\Psi_{2}(\varphi),
\end{align}
we have the following orthonormal bases of eigenfunction of $J := -i\partial_{\varphi},\ D(J):=\fH_{j_{0}}\cap H^{1}_{\loc}(\R)$ (subject to the boundary conditions associated with \eqref{eq:u1equivarianths}):
\begin{align}
\label{eq:u1equivariantb}
\Psi^{j_{0}}_{j}\in\fH_{j_{0}}: \Psi^{j_{0}}_{j}(\varphi) & := e^{i(j+j_{0})\varphi} = \langle\varphi|j+j_{0}\rangle,\ j\in\Z.
\end{align}
Moreover, the Hilbert spaces $\fH_{j_{0}}$ are representation spaces of the $U(1)$-Weyl algebra (cp. \eqref{eq:quantcommexp})
\begin{align}
\label{eq:u1weylalgebra}
& (V_{t}(\beta)\Psi)(\varphi):=\Psi(\varphi+t\beta), (U_{t}(m)\Psi)(\varphi):=e^{i m\varphi}\Psi(\varphi),\ \Psi\in\fH_{j_{0}}, \\ \nonumber
& V_{t}(\beta)U_{t}(m) = e^{i t m\beta}U_{t}(m)V_{t}(\beta),\\ \nonumber
& U_{t}(m)U_{t}(n) = U_{t}(m+n),\ V_{t}(\beta)V_{t}(\gamma) = V_{t}(\beta+\gamma), \\ \nonumber
& U_{t}(0) = \mathds{1} = V_{t}(0),\ U_{t}(m)^{*} = U_{t}(-m),\ V_{t}(\beta)^{*} = V_{t}(-\beta),\ m,n\in\Z,\ \beta,\gamma\in\R.
\end{align}
In terms of the bases the (equivariant) coherent states are:
\begin{align}
\label{eq:u1equivariantcs}
\Psi^{t,j_{0}}_{\xi}\in\fH_{j_{0}}: \Psi^{t,j_{0}}_{\xi}(\varphi) & = \sum_{j\in\Z}(\xi e^{tj_{0}})^{-j}e^{-\frac{t}{2}j^{2}}\Psi^{j_{0}}_{j}(\varphi) \\ \nonumber
 & = \langle\varphi|\xi,j_{0}\rangle_{t}
\end{align}
for $\xi\in\C^{*}=U(1)_{\C}\cong T^{*}U(1)$. These vectors are eigenfunctions of the \textit{annihilation operators} $X_{t}:=e^{-\frac{t}{2}}U_{t}(1)e^{-tJ}$,
\begin{align}
\label{eq:u1equivariantaop}
X_{t}|\xi,j_{0}\rangle_{t} & = \xi|\xi,j_{0}\rangle_{t},
\end{align}
where the former can be obtained from the complexifier method, $X_{t}:=e^{-\frac{t}{2}J^{2}}U e^{\frac{t}{2}J^{2}}$ (cf. \cite{ThiemannGaugeFieldTheory1, ThiemannGaugeFieldTheory2, ThiemannGaugeFieldTheory3, HallCoherentStatesOn}). Now, given a Berezin quantisation $Q^{\uB}_{t}(f)$ of $f\in C^{\infty}_{b}(T^{*}U(1))$, the formula connecting the upper symbol $f\circ\Phi^{-1},\ \Phi(e^{i\varphi},l)=e^{i\varphi}e^{-l} = e^{-l+i\varphi},$ and the lower symbol $L^{t}_{Q^{\uB}_{t}(f)}$ is:
\begin{align}
\label{eq:u1overlapkernel1}
L^{t}_{Q^{\uB}_{t}(f)}(\xi,\overline{\xi}) & = \int_{\C^{*}}\frac{d\xi'\wedge d\overline{\xi}'}{4\pi i \sqrt{\pi t}}f(\Phi^{-1}(\xi'))\ |\xi'|^{-2}\ e^{-\frac{(\log|\xi'|-j_{0}t)^{2}}{t}}\frac{|_{t}\langle\xi,j_{0}|\xi',j_{0}\rangle_{t}|^{2}}{_{t}\langle\xi,j_{0}|\xi,j_{0}\rangle_{t}} \\ \nonumber
 & = \int_{[0,2\pi)}\int_{\R}\frac{d\varphi'\ dl'}{2\pi\sqrt{\pi t}}f(\varphi',l')\ e^{-\frac{(l'-j_{0}t)^{2}}{t}}\frac{|_{t}\langle\xi,j_{0}|\Phi(e^{i\varphi'},l'),j_{0}\rangle_{t}|^{2}}{_{t}\langle\xi,j_{0}|\xi,j_{0}\rangle_{t}},
\end{align}
where we chose to base the Berezin quantisation on the resolution of unity \eqref{eq:csresmnemonic}. Putting $\xi=e^{-l+i\varphi}$ and using
\begin{align}
\label{eq:u1equivariantcssp}
_{t}\langle\Phi(e^{i\varphi},l),j_{0}|\Phi(e^{i\varphi'},l'),j_{0}\rangle_{t} & = \sum_{j\in\Z}e^{-tj^{2}}e^{j((l+l')+i(\varphi-\varphi')-2j_{0}t)} \\ \nonumber
 & = \vartheta_{3}(\tfrac{i}{2\pi}(-(l+l')-i(\varphi-\varphi')+2j_{0}t)|\tfrac{it}{\pi}) \\ \nonumber
 & = \sqrt{\frac{\pi}{t}}e^{\frac{1}{4t}(-(l+l')-i(\varphi-\varphi')+2j_{0}t)^{2}}\vartheta_{3}(\tfrac{1}{2t}(-(l+l')-i(\varphi-\varphi'))+j_{0}|\tfrac{i\pi}{t}) \\ \nonumber
 &  = \sqrt{\frac{\pi}{t}}\sum_{j\in\Z}e^{\frac{1}{4t}(((l-j_{0}t)+(l'-j_{0}t))+i(\varphi-\varphi')-2\pi i j)^{2}},
\end{align}
we have:
\begin{align}
\label{eq:u1overlapkernel2}
L^{t}_{Q^{\uB}_{t}(f)}(\Phi(e^{i\varphi},l)) & = \frac{e^{-\frac{(l-j_{0}t)^{2}}{t}}}{\vartheta_{3}(\tfrac{l}{t}-j_{0}|\tfrac{i\pi}{t})}\int_{[0,2\pi)}\int_{\R}\frac{d\varphi'\ dl'}{2\pi t}f(\varphi',l')\ e^{-\frac{(l'-j_{0}t)^{2}}{t}} \\ \nonumber
 & \hspace{0.5cm}\times\sum_{j,k\in\Z}e^{\frac{1}{4t}(((l-j_{0}t)+(l'-j_{0}t))+i(\varphi-\varphi')-2\pi i j)^{2}}e^{\frac{1}{4t}(((l-j_{0}t)+(l'-j_{0}t))-i(\varphi-\varphi')+2\pi i k)^{2}}\\ \nonumber
 & = \int_{[0,2\pi)}\int_{\R}\frac{d\varphi'\ dl'}{2\pi t\ \vartheta_{3}(\tfrac{l}{t}-j_{0}|\tfrac{i\pi}{t})}\ f(\varphi',l')\ e^{-\frac{1}{2t}((l-j_{0}t)-(l'-j_{0}t))^{2}} \\ \nonumber
 &\hspace{0.5cm}\times\sum_{j,k\in\Z}e^{-\frac{1}{2t}(\varphi-\varphi'-2\pi j)^{2}}e^{-\frac{1}{4t}(2\pi(j-k))}e^{-\frac{1}{2t}(2\pi(j-k))(\varphi-\varphi'-2\pi j+i((l-j_{0}t)+(l'-j_{0}t)))} \\ \nonumber
 &\hspace{-1.1cm}\underset{\substack{n=j-k \\ (SL_{2}(\Z)\ \textup{transform})}}{=} \int_{[0,2\pi)}\int_{\R}\frac{d\varphi'\ dl'}{2\pi t\ \vartheta_{3}(\tfrac{l}{t}-j_{0}|\tfrac{i\pi}{t})}\ f(\varphi',l')\ e^{-\frac{1}{2t}((l-j_{0}t)-(l'-j_{0}t))^{2}} \\ \nonumber
 &\hspace{1.5cm}\times\sum_{j,n\in\Z}e^{-\frac{1}{2t}(\varphi-\varphi'-2\pi j)^{2}}e^{-\frac{\pi^{2}}{t}n^{2}}e^{-\frac{\pi}{t}n(\varphi-\varphi'-2\pi j+i((l-j_{0}t)+(l'-j_{0}t)))} \\ \nonumber
 & =\sum_{n\in\Z}\!e^{-\frac{\pi^{2}}{t}n^{2}}\!e^{-2\pi i n(\frac{l}{t}-j_{0})}\!\!\int_{\R}\!\frac{dl'}{2\pi t\ \!\vartheta_{3}(\tfrac{l}{t}\!-\!j_{0}|\tfrac{i\pi}{t})}\!\!\int_{[0,2\pi)}\!\!\!\!\!\!\!\!\!\!d\varphi'\ \!f(\varphi',l') \\ \nonumber
 &\hspace{0.5cm}\times\sum_{j\in\Z}e^{-\frac{1}{2t}((l-l'-i\pi n)^{2}+(\varphi-\varphi'+\pi n - 2\pi j)^{2})} \\ \nonumber
 & =\underbrace{\sum_{n\in\Z}e^{-\frac{\pi^{2}}{t}n^{2}}e^{-2\pi i n(\frac{l}{t}-j_{0})}}_{=\vartheta_{3}(\frac{l}{t}-j_{0}|\frac{i\pi}{t})}\!\!\int_{\R}\frac{dl'}{2\pi t\ \vartheta_{3}(\tfrac{l}{t}-j_{0}|\tfrac{i\pi}{t})}\int_{[0,2\pi)}\!\!\!\!\!\!\!\!\!\!d\varphi'\ \!f(\varphi',l') \\ \nonumber
 &\hspace{0.5cm}\times\sum_{j\in\Z}e^{-\frac{1}{2t}((l-l')^{2}+(\varphi-\varphi'+2\pi j)^{2})} \\ \nonumber
 & =\frac{1}{2\pi t}\int_{\R}dl'\sum_{j\in\Z}\int_{[2\pi j,2\pi(j+1)}d\varphi'\ \underbrace{f(\varphi'-2\pi j,l')}_{=f(\varphi',l')}\ e^{-\frac{1}{2t}((l-l')^{2}+(\varphi-\varphi')^{2})} \\ \nonumber
 & =\frac{1}{2\pi t}\int_{\R}dl'\ \int_{\R}d\varphi'\ f(\varphi',l')\ e^{-\frac{1}{2t}((l-l')^{2}+(\varphi-\varphi')^{2})} \\ \nonumber
 & =\left(e^{\frac{t}{2}\Delta_{(\varphi,l)}}f\right)(\varphi,l),
\end{align}
where we used the invariance of the measure $dl\wedge d\varphi = (2\pi i |\xi|^{2})^{-1}d\xi\wedge d\overline{\xi}$ under the substitution $\varphi'\mapsto\varphi+\pi n,\ l'\mapsto l'-i\pi n$, because $\Phi(e^{i\varphi},l) = e^{-l+\varphi} = e^{-(l-i\pi n)+i(\varphi+\pi n)} = \Phi(e^{(-1)^{n}i\varphi},l-i\pi n)$, and identified $f$ with its periodic extension to $T^{*}\R$.
\begin{Remark}
\label{rem:u1overlapkernel}
The relation $L^{t}_{Q^{\uB}_{t}(f)}(\Phi(e^{i\varphi},l)) = (e^{\frac{t}{2}\Delta_{(\varphi,l)}}f)(\varphi,l)$ is in accordance with the covering $\R\rightarrow\R/2\pi\Z\cong U(1)$ and the observation that $\vartheta_{3}$ arises from a $2\pi\Z$-periodisation of the Euclidean heat kernel, $\tfrac{1}{2\pi}\vartheta_{3}(\tfrac{l}{2\pi}|\tfrac{it}{2\pi}) = \sum_{j\in\Z}\tfrac{1}{\sqrt{2\pi t}}e^{-\frac{(l-2\pi j)^{2}}{2t}}$. Furthermore, an analogous calculation as in \eqref{eq:u1overlapkernel1} shows that the lower symbol of a product of two Berezin quantisations $Q^{\uB}_{t}(f),\ Q^{\uB}_{t}(f')$ satisfies
\begin{align}
\label{eq:u1productkernel}
& L^{t}_{Q^{\uB}_{t}(f)Q^{\uB}_{t}(f')}(\Phi(e^{i\varphi},l)) \\[0.25cm] \nonumber
 & = \int_{\R^{2}}\frac{d\varphi' dl'}{2\pi t}\ e^{-\frac{1}{2t}(\varphi'+il')(\varphi'-il')}f(\varphi+\sqrt{2t}\varphi',l+\sqrt{2t}l') \\ \nonumber
 & \hspace{0.5cm}\times\int_{\R^{2}}\frac{d\varphi'' dl''}{2\pi t}e^{-\frac{1}{2t}(\varphi''+il'')(\varphi''-il'')}e^{-\frac{1}{2t}(\varphi'-il')(\varphi''+il'')}f'(\varphi+\sqrt{2t}\varphi'',l+\sqrt{2t}l''),
\end{align}
which is familiar from the $\R^{n}$-case, as well.
\end{Remark}
If we were to base the Berezin quantisation on the resolution of unity \eqref{eq:newcsres},
\begin{align}
\label{eq:newu1overlapkernel1}
L^{t}_{Q^{\uB}_{t}(f)}(\xi,\overline{\xi}) & = C_{t}\int_{\C^{*}}\frac{d\xi'\wedge d\overline{\xi}'}{4\pi i}|\xi'|^{-2}f(\Phi^{-1}(\xi'))\underbrace{\frac{|_{t}\langle\xi,j_{0}|\xi',j_{0}\rangle_{t}|^{2}}{_{t}\langle\xi,j_{0}|\xi,j_{0}\rangle_{t}\ _{t}\langle\xi',j_{0}|\xi',j_{0}\rangle_{t}}}_{= \tr\left(P^{j_{0}}_{t}(\xi,\overline{\xi})P^{j_{0}}_{t}(\xi',\overline{\xi}')\right)},
\end{align}
which was already proven for $G=U(1)$, we would obtain a similar result:
\begin{align}
\label{eq:newu1overlapkernel2}
 & L^{t}_{Q^{\uB}_{t}(f)}(\Phi(e^{i\varphi},l)) \\[0.25cm] \nonumber
 & = \sum_{n\in\Z}\frac{e^{-\frac{\pi^{2}}{t}n^{2}}e^{-2\pi i n(\frac{l}{t}-j_{0})}}{\vartheta_{3}(\tfrac{l}{t}-j_{0}|\tfrac{i\pi}{t})}\!\!\int_{\R}\frac{C_{t}dl'}{2\pi\ \vartheta_{3}(\tfrac{l'}{t}-j_{0}|\tfrac{i\pi}{t})}\int_{[0,2\pi)}\!\!\!\!\!\!\!\!\!\!d\varphi'\ f(\varphi',l')\sum_{j\in\Z}e^{-\frac{1}{2t}((l-l'-i\pi n)^{2}+(\varphi-\varphi'+\pi n - 2\pi j)^{2})} \\ \nonumber
 & = \frac{C_{t}}{2\pi}\int_{\R}dl'\ \frac{\vartheta_{3}((\tfrac{l}{t}-j_{0})+(\tfrac{l'}{t}-j_{0})|\tfrac{i2\pi}{t})}{\vartheta_{3}(\tfrac{l}{t}-j_{0}|\tfrac{i\pi}{t})\vartheta_{3}(\tfrac{l'}{t}-j_{0}|\tfrac{i\pi}{t})}\int_{\R}d\varphi'\ f(\varphi',l')\ e^{-\frac{1}{2t}((l-l')^{2}+(\varphi-\varphi')^{2})}.
\end{align}
There is yet another interesting way to obtain the relation between the upper and lower symbols, namely via applying the commutation relation between creation and annihilation operators,
\begin{align}
\label{eq:ancrccr}
X_{t}X_{t}^{*} & = e^{2t}X_{t}^{*}X_{t},
\end{align}
to any operator $A\in \cB(L^{2}(U(1)))$ in (anti-)Wick-ordered form:
\begin{align}
\label{eq:u1wickop}
A & = \sum_{m,n\in\Z}\left(A^{W}_{t}\right)_{mn}(X_{t}^{*})^{m}X_{t}^{n} \\ \nonumber
 & = \sum_{m,n\in\Z}\underbrace{\left(A^{W}_{t}\right)_{mn}e^{-2tmn}}_{=:\left(A^{aW}_{t}\right)_{mn}}X_{t}^{n}(X_{t}^{*})^{m}.
\end{align} 
From the resolution of unity \eqref{eq:csresmnemonic} and \eqref{eq:u1equivariantaop}, we infer that the upper and lower symbol of $A$ are given by the Laurent series:
\begin{align}
\label{eq:u1laurent}
f_{A}(\Phi^{-1}(\xi)) & = A^{aW}_{t}(\xi,\overline{\xi}) & L^{t}_{A}(\xi,\overline{\xi}) & = A^{W}_{t}(\xi,\overline{\xi}) \\ \nonumber
 & = \sum_{m,n\in\Z}\left(A^{aW}_{t}\right)_{mn}\ \xi^{m}\ \overline{\xi}^{n}, &  & = \sum_{m,n\in\Z}\left(A^{W}_{t}\right)_{mn}\ \xi^{m}\ \overline{\xi}^{n}, \\ \nonumber
& = \sum_{m,n\in\Z}e^{-2tmn}\left(A^{W}_{t}\right)_{mn}\ \xi^{m}\ \overline{\xi}^{n} & &
\end{align}
which evidently satisfy
\begin{align}
\label{eq:u1laurentrel}
L^{t}_{A} & = e^{\frac{t}{2}\Delta}f_{A}.
\end{align}
Unfortunately, such simple reasoning is not available for general compact Lie groups, because the commutation relations of the creation and annihilation operators do not close among themselves.\\[0.1cm]
To conclude this section, we also compute the analogs of the relations \eqref{eq:standardweylsmoothing} for $U(1)$-equivariant Kohn-Nirenberg\footnote{We stick to the Kohn-Nirenberg quantisation in this subsection, because it avoids the use of square roots in the group, and leads to similar results regarding the regularity of symbols as Weyl quantisation.} symbols (cp. \eqref{eq:knfourier} and \eqref{eq:knpseudo}). In this case, the Weyl elements are obtained from the Weyl algebra \eqref{eq:u1weylalgebra}:
\begin{align}
\label{eq:u1weylelements}
W^{j_{0}}_{t}(\varphi,k) & := \frac{t}{2\pi}\sum_{m\in\Z}\int_{[0,2\pi/t)}d\beta\ e^{-i(m\varphi+\beta(k+j_{0}))}U_{t}(m)V_{t}(\beta),
\end{align}
which satisfy the relation
\begin{align}
\label{eq:u1weylorthogonality}
\tr_{\fH_{j_{0}}}(W^{j_{0}}_{t}(\varphi,k)^{*}W^{j_{0}}_{t}(\varphi',k')) & = 2\pi\sum_{m\in\Z}\delta(\varphi-\varphi'-2\pi m)\delta_{k,k'}.
\end{align}
Now, Kohn-Nirenberg quantisation and dequantisation takes the form:
\begin{align}
\label{eq:u1knquant}
A_{\sigma,t} & :=\frac{1}{2\pi}\sum_{k\in\Z}\int_{[0,2\pi)}d\varphi\ \sigma_{t}(\varphi,k)W^{j_{0}}_{t}(\varphi,k), & \sigma_{A,t}(\varphi,k) & := \tr_{\fH_{j_{0}}}(W^{j_{0}}_{t}(\varphi,k)^{*}A_{t}),
\end{align}
\begin{align} \nonumber
\left(A_{\sigma,t}\Psi\right)(\varphi) & = \frac{1}{2\pi}\sum_{k\in\Z}\int_{[0,2\pi)}d\varphi'\ e^{ik(\varphi-\varphi')}\sigma_{t}(\varphi,k)\Psi(\varphi') = \sum_{k\in\Z}e^{ik\varphi}\sigma_{t}(\varphi,k)\hat{\Psi}(k),\ \Psi\in C^{\infty}_{b}(\R)\cap\fH_{j_{0}}.
\end{align}
There are natural symbol spaces associated with \eqref{eq:u1knquant} (cf. \cite{RuzhanskyPseudoDifferentialOperators}), which are in close analogy with those familiar from pseudo-differential operators on $\R^{n}$ ($m\in\R, 0\leq\delta\leq\rho\leq1$):
\begin{align}
\label{eq:u1knquantsymbols}
\sigma\in S^{m}_{\rho,\delta}(U(1)\times\Z)\ :\Leftrightarrow\ &\ \forall k\in\Z:\sigma(\ .\ ,k)\in C^{\infty}(U(1))\!\!\!\!\!\! & & \\ \nonumber
&\ \&\ \forall\alpha,\beta\in\N_{0}:\forall(\varphi,k)\in U(1)\!\!\!\!\!\! &\times\Z:\exists C_{\alpha\beta}>0: & \\ \nonumber
& & |(\partial^{\alpha}_{\varphi}\Delta^{\beta}_{k}\sigma)(\varphi,k)| & \leq C_{\alpha\beta}\langle k\rangle^{m-\rho\beta+\delta\alpha}, \\[0.25cm] \nonumber
S^{-\infty}(U(1)\times\Z) & :=\bigcap_{m\in\R}S^{m}_{\rho,\delta}(U(1)\times\Z),\!\!\!\!\!\! & S^{\infty}_{\rho,\delta}(U(1)\times\Z) & :=\bigcup_{m\in\R}S^{m}_{\rho,\delta}(U(1)\times\Z).
\end{align}
Here, $(\Delta_{k}f)(k):=f(k+1)-f(k)$ for $f:\Z\rightarrow\C$ is the forward difference. If we allow for $U(1)$-equivariant symbols, i.e. $\sigma(\varphi+2\pi j,k)=e^{2\pi i j j_{0}}\sigma(\varphi,k)e^{-2\pi i j j'_{0}}$, we can encompass operators $A_{\sigma}:\fH_{j'_{0}}\rightarrow\fH_{j_{0}}$, as well.\\
Clearly, we could have invoked $U(1)$-equivariant symbols,
\begin{align}
\label{eq:u1equivsym}\nonumber
S^{m}_{\rho,\delta,(j_{0},j'_{0})}(\R^{2}) & =\{\sigma\in S^{m}_{\rho,\delta}(\R^{2})\ |\ \forall j\in\Z:\sigma(q+2\pi j,p)=\sigma(\varphi+2\pi j,k)=e^{2\pi i j j_{0}}\sigma(q,p)e^{-2\pi i j j'_{0}}\} \\ 
 & \subset S^{m}_{\rho,\delta}(\R^{2}),
\end{align}
from the usual symbol classes instead of \eqref{eq:u1knquantsymbols}:
\begin{align}
\label{eq:equivariantknquant}
(A_{\sigma,t}\Psi)(q) & = \frac{1}{2\pi t}\int_{\R}dp\int_{[0,2\pi)}dq\ e^{\frac{i}{t}p(q-q')}\sigma(q,p)\Psi(q') = \frac{1}{2\pi t}\int_{\R}dp\ e^{\frac{i}{t}pq}\sigma(q,p)\cF_{t}[\Psi](p)
\end{align}
for $\Psi\in C^{\infty}_{b}(\R)\cap\fH_{j_{0}}$. We will comment on a similar dichotomy for almost-periodic pseudo-differential operators in the following subsection \ref{subsec:u1bohr}. In the present case the distinction is only apparent, because symbols in $S^{m}_{\rho,\delta}(U(1)\times\Z)$ can be interpolated by those in $S^{m}_{\rho,\delta}(\R^{2})$ (cf. \cite{RuzhanskyPseudoDifferentialOperators}, Corollary 4.6.13.).\\[0.1cm]
To conclude the present subsection, we provide the relations between upper, lower and Kohn-Nirenberg symbols, which display a smoothing from upper to Kohn-Nirenberg to lower symbols similar to \eqref{eq:standardweylsmoothing}:
\begin{align}
\label{eq:u1knsmoothing}
L^{t}_{A_{\sigma,t}}(\Phi(e^{i\varphi},l)) & = \frac{\sqrt{2}}{2\pi\vartheta_{3}(\tfrac{l}{t}-j_{0}|\tfrac{i\pi}{t})}\sum_{k\in\Z}\int_{[0,2\pi)}d\varphi'\sigma_{t}(\varphi',k)e^{-\frac{1}{t}(l-t(k+j_{0}))^{2}} \\ \nonumber
 &\hspace{4cm}\times\sum_{j\in\Z}e^{-\frac{1}{2t}((\varphi-\varphi'-2\pi j)-i(l-t(k+j_{0})))^{2}} \\ \nonumber
 & = \frac{\sqrt{2}}{2\pi\vartheta_{3}(\tfrac{l}{t}-j_{0}|\tfrac{i\pi}{t})}\sum_{k\in\Z}\int_{\R}d\varphi'\sigma_{t}(\varphi',k)e^{-\frac{1}{t}(l-t(k+j_{0}))^{2}}e^{-\frac{1}{2t}((\varphi-\varphi')-i(l-t(k+j_{0})))^{2}} \\ \nonumber
 & = \frac{\sqrt{2}}{2\pi\vartheta_{3}(\tfrac{l}{t}-j_{0}|\tfrac{i\pi}{t})}\sum_{k\in\Z}\int_{\R}d\varphi'\sigma_{t}(\varphi',k)e^{-\frac{1}{2t}((l-t(k+j_{0}))^{2}+(\varphi-\varphi')^{2})}e^{\frac{i}{t}(\varphi-\varphi')(l-t(k+j_{0}))} \\[0.25cm] \nonumber
\sigma_{t}(\varphi,k) & = \frac{\sqrt{2}}{2\pi t}\int_{\R}dl'\int_{\R}d\varphi'f_{A_{\sigma,t}}(\varphi',l')e^{-\frac{1}{2t}((t(k+j_{0})-l)^{2}+(\varphi-\varphi')^{2})}e^{-\frac{i}{t}(\varphi-\varphi')(t(k+j_{0})-l)}.
\end{align}
These formulas are obtained by a calculation completely analogous to \eqref{eq:u1overlapkernel2}.
\subsection{Scaled Fourier transforms for $G=U(1)$ and an extension to $\R_{\Bohr}$}
\label{subsec:u1bohr}
In subsection \ref{subsec:swt}, we have discussed the issue of defining a $\varepsilon$-scaled integral transform for a compact (simply connected) Lie group $G$ by means of the Stratonovich-Weyl-Fourier transform. While the resulting transform \eqref{eq:momentumscaling} on $L^{2}(G)$ is has its merits, when applied to systems modelled on coadjoint orbits of $G$, its use is limited in the analysis of pseudo-differential operators on $C^{\infty}(G)$ as it is not invertible due to the fundamental discreteness inherent to representation theory of $G$, i.e. any irreducible representation of $G$ is uniquely (up to isomorphism) determined by an integral dominant (real) weight $\lambda_{\pi}\in\overline{C}\cap I^{*}_{r}$.\\[0.1cm]
In the following, first restricting to semisimple $G$, we will pursue the question, whether it is possible to lift this discreteness by associating a representation $\pi_{\lambda}$ to any dominant weight $\lambda\in\overline{C}$ ($\overline{C}\subset\ft^{*}$ admits a natural $\R_{+}$-action as it is a convex cone). To be a bit more precise, we will consider (complex linear) representations $d\pi_{\lambda}$ of $\fg_{\C}$ resp. $U(\fg_{\C})$\footnote{$U(\fg_{\C})$ denotes the universal enveloping algebra of $\fg_{\C}$.} instead of $G$ to lift the integrality condition. A natural way to achieve this is to exploit the construction of irreducible representations of $G$ by means of Verma modules (cf. \cite{HumphreysIntroductionToLie, HumphreysRepresentationsOfSemisimple, HallLieGroupsLie}). Since the representation theory of $\fg_{\C}$ is typically formulated with respect to (infinitesimal) integral weights $\lambda\in2\pi i I^{*}_{r}=I^{*}$, and their complex linear extensions to $\ft^{*}_{\C}$, instead of integral real weights, we will switch to using the former for the reminder of the section. The same applies to the roots of $\fg$, and the subset of positive roots $R^{+}$.
\begin{Definition}[cf. \cite{HumphreysRepresentationsOfSemisimple}, I.1.3.]
\label{def:verma}
Given $\lambda\in\ft^{*}$, the $U(\fg_{\C})$-module
\begin{align}
\label{eq:verma}
M(\lambda) & := U(\fg_{\C})\otimes_{U(\fb^{+})}\C_{\lambda}
\end{align}
is called the \textup{Verma module of highest weight} $\lambda$. Here,
\begin{align}
\label{eq:stborel}
\fb^{+}:=\ft_{\C}\oplus\underbrace{\bigoplus_{\alpha\in R^{+}}\fg_{\alpha}}_{=:\fn^{+}}
\end{align}
is the \textup{standard Borel subalgebra} of $\fg_{\C}$ associated with $R^{+}$, and $\C_{\lambda}$ is the 1-dimensional $U(\fb^{+})$-module defined by
\begin{align}
\label{eq:vermalborelrep}
d\pi_{\lambda}(H+N)1 & := \lambda(H)1,\ H\in\ft_{\C},\ N\in\fn^{+}.
\end{align}
We denote by $L(\lambda):=M(\lambda)/N(\lambda)$ the unique irreducible quotient module w.r.t. to the unique maximal submodule $N(\lambda)\subset M(\lambda)$ (cf. \cite{HumphreysRepresentationsOfSemisimple}, Theorem I.1.2.).
\end{Definition}
\begin{Remark}
\label{rem:verma}
The Verma module construction \eqref{eq:verma} also works for $G=U(1)^{n}$. Since $R^{+}=\emptyset$ and $\fg_{\C}=\ft_{\C}\cong\C^{n}$ in this case, we have $M(\lambda)=\C_{\lambda}$, which is irreducible.\\
For semisimple $G$, the irreducible representation $L(\lambda)$ of $\fg_{\C}$ is finite dimensional if and only if $\lambda$ is dominant integral. $M(\lambda)$ is freely generated by $U(\fn^{-})(1\otimes1),\ \fn^{-}:=\bigoplus_{\alpha\in R^{+}}\fg_{-\alpha}$.
\end{Remark}
Unfortunately, the reason, why $L(\lambda)$ does not integrate to a (unitary) representation of $G$, when $\lambda$ is not dominant integral, can be traced back to the fact, that there is no Hilbert space structure on $L(\lambda)$ compatible with the adjointness relations $d\pi_{\lambda}(E_{\alpha})^{*}=d\pi_{\lambda}(E_{-\alpha}),\ \alpha\in R^{+},$ and $d\pi_{\lambda}(H_{i})^{*}=d\pi_{\lambda}(H_{i}),\ i=1,...,r,$ for a choice of Cartan-Weyl basis \eqref{eq:cartanweylbasis} (cf. \cite{FuchsAffineLieAlgebras}) but only a Krein space structure. This is easily seen, in the case of $\fg_{\C}=\mathfrak{sl}_{2} (\overline{C}\cap I^{*}\cong\N,\ R^{+} = \{2\})$:
\begin{align}
\label{eq:sl2innerproduct}
(d\pi_{\lambda}(E_{-2})^{k}v_{\lambda},d\pi_{\lambda}(E_{-2})^{k}v_{\lambda})_{\lambda} & = (v_{\lambda},d\pi_{\lambda}(E_{2})^{k}d\pi_{\lambda}(E_{-2})^{k}v_{\lambda})_{\lambda} \\ \nonumber
 & = (v_{\lambda},d\pi_{\lambda}(E_{2})^{k-1}(d\pi_{\lambda}(E_{-2})d\pi_{\lambda}(E_{2}) + d\pi_{\lambda}(H_{2}))d\pi_{\lambda}(E_{-2})^{k-1}v_{\lambda})_{\lambda} \\ \nonumber
 & = k(\lambda(H_{2})+1-k)(d\pi_{\lambda}(E_{-2})^{k-1}v_{\lambda},d\pi_{\lambda}(E_{-2})^{k-1}v_{\lambda})_{\lambda} \\ \nonumber
 & = \left(\prod_{l=1}^{k}l(\lambda(H_{2})+1-l)\right)(v_{\lambda},v_{\lambda})_{\lambda},\ k\in\N_{0}
\end{align}
where we denoted the highest weight vector $1\otimes 1\in M(\lambda)$ by $v_{\lambda}$, and invoked the commutation relations \eqref{eq:cartanweylbasis}. The last line in \eqref{eq:sl2innerproduct} shows that the inner product of $v^{k}_{\lambda}=d\pi_{\lambda}(E_{-2})^{k}v_{\lambda},\ k\in\N_{0},$ becomes negative for certain $k>\lambda(H_{2})+1$, if $\lambda(H_ {2})\notin\N$. Moreover, the modules $M(\lambda),\ \lambda(H_{2})\notin\N,$ are irreducible implying that there is no compatible Hilbert space structure. For $\lambda(H_{2})\in\N$, we obtain the submodule $N(\lambda)=\textup{span}_{\C}\{v_{\lambda}^{k}\ |\ k\geq\lambda(H_{2})+1\}$ of null vectors, which can be factored out. A similar argument applies to general semisimple $\fg_{\C}$ by an appeal to the $\mathfrak{sl}_{2}$-submodules generated by $\{E_{\alpha},E_{-\alpha},H_{\alpha}\},\ \alpha\in R^{+}$.\\[0.1cm]
In contrast, if $G=U(1)^{n}$ the set of roots is empty, and the standard inner product on $\C_{\lambda}\cong\C$ is compatible with the adjointness relations. Furthermore, the representation $\C_{\lambda}$ integrates to a (unitary) representation of $\R^{n}$:
\begin{align}
\label{eq:u1integratedverma}
\pi(x\cdot H)_{\lambda}1 & = e^{i\sum_{i=1}^{n}x_{i}\lambda(H_{i})}1,\ x\cdot H = \sum_{i=1}^{n}x_{i}H_{i},\ x\in\R^{n}.
\end{align}
Thus, for $G=U(1)^{n}$ we may form the direct sums of Hilbert spaces
\begin{align}
\label{eq:u1vermasum}
\fH_{n} & := \bigoplus_{\{\lambda(H_{i})\}_{i=1}^{n}\in\R^{n}}\C_{\lambda},\ n\in\N, & \fH_{n}\cong\fH_{1}^{\otimes n}.
\end{align}
The Hilbert space $\fH_{1}=l^{2}(\R)$ can be identified with $L^{2}(\R_{\textup{Bohr}})$, the $L^{2}$-space on the Bohr compactification of $\R$, which can be defined as the $L^{2}$-closure of $\textup{span}_{\C}\{e_{\lambda}:\R\rightarrow\mathds{T}\ |\ \lambda\in\R\},\ e_{\lambda}(x):=e^{i\lambda x},$ w.r.t. the ergodic mean (the Haar measure on $\R_{\Bohr}$)
\begin{align}
\label{eq:bohrmean}
(f,f')_{\Bohr} = \int_{\R_{\Bohr}}d\mu_{\Bohr}(x)\ \overline{f(x)}f'(x)& := \lim_{T\rightarrow\infty}\frac{1}{2T}\int_{[-T,T]}dx\ \overline{f(x)}f'(x).
\end{align}
$L^{2}(\R_{\Bohr})$ is also naturally isomorphic with the space of Besicovitch almost periodic functions $B^{2}(\R)$ on $\R$ (cf. \cite{ShubinAlmostPeriodicFunctions, BesicovitchAlmostPeriodicFunctions}). In analogy with the Bloch-Floquet transform of $L^{2}(\R)$, we have a direct sum (instead of direct integral) decomposition over the elementary cell $[0,1)$ (Brillouin zone) w.r.t. the Hilbert spaces $\fH_{j_{0}}$ (see \eqref{eq:u1equivarianths}):
\begin{align}
\label{eq:l2bohrdecomposition}
L^{2}(\R_{\Bohr}) & \cong\bigoplus_{j_{0}\in[0,1)}\fH_{j_{0}}.
\end{align}
Interestingly, a similar function space realization can be obtained for the simple quotients of the Verma modules of $\mathfrak{sl}_2$. Namely, we realise $L(\lambda),\ \lambda=\lambda(H_{2}),$ as a subspace of $L^{2}_{\loc}(\R^{2})$ by
\begin{align}
\label{eq:sl2vermafsp}
v_{\lambda}(x,y) & :=e_{\lambda}(x), & d\pi_{\lambda}(H_{2}) & :=-i(\partial_{x}-\partial_{y})\\ \nonumber
d\pi_ {\lambda}(E_{2}) & :=-ie_{1}(x-y)\partial_{y}, & d\pi_{\lambda}(E_{-2}) & :=-ie_{-1}(x-y)\partial_{x}.
\end{align}
Then, $L(\lambda)$ is spanned by the weight vectors:
\begin{align}
\label{eq:sl2vermaspan}
v^{k}_{\lambda}(x,y) & := \left(d\pi_{\lambda}(E_{-2})^{k}v_{\lambda}\right)(x,y) =  \left(\prod_{l=1}^{k}(\lambda+1-l)\right)e_{\lambda-k}(x)e_{k}(y),\ k\in\N_{0}, \\ \nonumber
v^{0}_{\lambda}(x,y) & := v_{\lambda}(x,y),
\end{align}
which satisfy $v^{k}_{\lambda}(x+2\pi m,y+2\pi n) = e^{2\pi i j_{\lambda}m}v^{k}_{\lambda}(x,y),\ j_{\lambda}=\lambda\mod 1\in[0,1),\ m,n\in\Z$. Therefore, $L(\lambda)$ constitutes a (diagonal) subspace of $\fH_{j_{\lambda}}\otimes\fH_{0}\cong\{\Psi\in L^{2}_{\loc}(\R^{2})\ |\ \forall m,n\in\Z: \Psi(x+2\pi m,y+2\pi n) = e^{2\pi i j_{\lambda}m}\Psi(x,y)\}=:L^{2}_{(j_{\lambda},0)}(\R^{2})$. If $L(\lambda)$ is endowed with the inner product $(\ .\ ,\ .\ )_{(j_{\lambda},0)}$ coming from $\fH_{j_{\lambda}}\otimes\fH_{0}$, we will have:
\begin{align}
(v^{k}_{\lambda},v^{k'}_{\lambda})_{(j_{\lambda},0)} & = \left(k!\binom{\lambda}{k}\right)^{2}\delta_{k,k'},\ k,k'\in\N_{0},
\end{align}
implying the (modified) adjointness relations
\begin{align}
\label{eq:sl2modifiedadjoint}
d\pi_{\lambda}(H_{2})^{*} & = d\pi_{\lambda}(H_{2}), \\ \nonumber
d\pi_{\lambda}(E_{2})^{*}\left(\lambda+d\pi_{\lambda}(H_{2})\right)v^{k}_{\lambda} & = \left(\lambda-d\pi_{\lambda}(H_{2})\right)d\pi_{\lambda}(E_{-2})v^{k}_{\lambda},\ k\in\N_ {0}.
\end{align}
Here, we consistently set $d\pi_{\lambda}(E_{2})^{*}v^{\lambda}_{\lambda}=0$ for $\lambda\in\N_{0}$. Thus, as in \eqref{eq:sl2innerproduct}, we find that the tension between Hilbert space structure and adjointness relations is due to the root vectors $E_{2},\ E_{-2}$, while the adjointness relation for the generator of the Cartan subalgebra $H_{2}$ is the expected one. The completion $\hat{L}(\lambda)$ of $L(\lambda)$ w.r.t. the Hilbert space structure induced by $(\ .\ ,\ .\ )_{(j_{\lambda},0)}$ is given by (possibly finite) $l^{2}$-series w.r.t. the vectors $v^{k}_{\lambda},\ k\in\N_{0}$:
\begin{align}
\label{eq:sl2vermacompletion}
\Psi\in\hat{L}(\lambda) &\ :\Leftrightarrow\ \Psi=\sum_{k\in\N_{0}}\Psi_{k}\left(k!\binom{\lambda}{k}\right)^{-1}v^{k}_{\lambda},\ \sum_{k\in\N_{0}} |\Psi_{k}|^{2}<\infty.
\end{align}
The realisation \eqref{eq:sl2vermafsp} of $L(\lambda)$ is closely related to the Schwinger representation of $\mathfrak{sl}_{2}$ on $\C[X,Y]$:
\begin{align}
\label{eq:sl2schwinger}
a^{*}_{+}=X & := e_{1}(x), & a_{+}=\partial_{X} & := -ie_{-1}(x)\partial_{x}, \\ \nonumber
a^{*}_{-}=Y & := e_{1}(y), & a_{-}=\partial_{Y} & := -ie_{-1}(y)\partial_{y}, \\[0.25cm] \nonumber
E_{\pm2} & \doteq a^{*}_{\pm}a_{\mp}, & H_{2} & \doteq a^{*}_{+}a_{+}-a^{*}_{-}a_{-},
\end{align}
which can be readily generalised to abtritrary $\fg_{\C}$. Namely, given a (complex) linear, finite dimensional representation $d\pi:\fg_{\C}\rightarrow\End(V_{\pi})$ and $d_{\pi}$ copies of the CCR-algebra, e.g. creation and annihilation operators on the (bosonic) Fock space $\mathcal{F}_{s}(\C^{d_{\pi}})=\bigoplus_{n=0}^{\infty}\textup{Sym}^{n}\C^{d_{\pi}}$, we define
\begin{align}
\label{eq:schwingerrep}
S_{\pi}^{g_{1},g_{2}}(X) & := \sum_{m,n=1}^{d_{\pi}}d\pi(X)_{mn}\left(a^{*}_{m}a_{n} + g_{1}a^{*}_{m} + g_{2}a_{n}+g_{1}g_{2}\right),\ X\in\fg_{\C},g_{1},g_{2}\in\C,
\end{align}
which satisfies $[S_{\pi}^{g_{1},g_{2}}(X),S^{g_{1},g_{2}}(Y)]=S_{\pi}^{g_{1},g_{2}}([X,Y]),\ X,Y\in\fg_{\C},$ and coincides with the Schwinger representation of $\mathfrak{sl}_{2}$ for $g_{1}=0=g_{2}$ and $\pi=\pi_{\frac{1}{2}}$, the fundamental representation\footnote{Setting $g=g_{1}=\overline{g_{2}}$ enforces the adjointness relations of $\pi$ onto $S^{g}_{\pi}$.}.\\[0.1cm]
For the remainder of the subsection, we return to $G=U(1)^{n}$, and further restrict to $n=1$, in view of \eqref{eq:u1vermasum} and $(\R^{n})_{\Bohr}\cong(\R_{\Bohr})^{n}$, since in this case the extension from integral weights to arbitrary weights works to full extent. Moreover, the ``decompactification'' from the spaces $\fH_{j_{0}}$, as representations of the $U(1)$-Weyl algebra \eqref{eq:u1weylalgebra}, to the space $\fH_{1}=l^{2}(\R)\cong B^{2}(\R)\cong L^{2}(\R_{\Bohr})$, which is a representation of the standard 1-particle Weyl algebra (obtained from the algebraic state $\omega_{0}(W(\alpha,\beta))=\delta_{\alpha,0}$), not only lifts the problem of $\varepsilon$-scaling, but additionally allows to handle the square root necessary for replacing the Kohn-Nirenberg quantisation by a genuine Weyl quantisation. There are essentially two of the latter, the first is natural in view of the isomorphism $l^{2}(\R)\cong B^{2}(\R)$ (cf. \cite{ShubinAlmostPeriodicFunctions, ShubinDifferentialAndPseudodifferential}), while the second is tied to the identification $l^{2}(\R)\cong L^{2}(\R_{\Bohr})$, and extends the one proposed by Fewster and Sahlmann (cf. \cite{FewsterPhaseSpaceQuantization}):\\[0.1cm]
The Weyl elements are as usual (cp. \eqref{eq:u1weylalgebra}):
\begin{align}
\label{eq:bohrweylelements}
& W_{\varepsilon}(\alpha,\beta) := e^{\frac{i\varepsilon}{2}\alpha\beta}U_{\varepsilon}(\alpha)V_{\varepsilon}(\beta), \\ \nonumber
& W_{\varepsilon}(\alpha,\beta)^{*} = W_{\varepsilon}(-\alpha,-\beta),\ W_{\varepsilon}(0,0) = \mathds{1}, \\ \nonumber
& W_{\varepsilon}(\alpha,\beta)W_{\varepsilon}(\gamma,\delta) = e^{-\frac{i\varepsilon}{2}(\alpha\delta-\gamma\beta)}W_{\varepsilon}(\alpha+\gamma,\beta+\delta),\ \alpha,\beta,\gamma,\delta\in\R.
\end{align}
Let us also introduce some frequently used test function and Sobolev spaces on $\R_{\Bohr}$ and its (topological) dual group $\R_{\disc}$ ($\R$ with the discrete topology, cf. \cite{ShubinAlmostPeriodicFunctions}):
\begin{align}
\label{eq:bohrfsp}
d(\R) & := \{\hat{\Psi}:\R\rightarrow\C\ \!|\ \!\supp(\hat{\Psi})\ \textup{finite}\}, & \check{d}(\R) & := \textup{span}_{\C}\{e_{\lambda}\ \!|\ \!\lambda\!\in\!\R\} \\ \nonumber
& & & = \Trig(\R), \\ \nonumber
d'(\R) & := \C^{\R}, & \check{d}'(\R) & := \{T\!=\!\sum_{\lambda\in\R}\hat{T}(\lambda)e_{\lambda}\ \!|\ \!\hat{T}\!\in\!\C^{\R}\} \\ \nonumber
& & & = \Trig'(\R), \\ \nonumber
H^{s}_{p}(\R_{\Bohr}) & := \overline{\Trig(\R)}^{||\ .\ ||_{(s,p)}}, & \bigg|\bigg|\sum_{\lambda\in\R}\hat{\Psi}(\lambda)
e_{\lambda}(x)\bigg|\bigg|_{(s,p)}\!\!\!\! & := \!\bigg(\sum_{\lambda\in\R}(\langle\lambda\rangle^{s}|\hat{\Psi}(\lambda)|)^{p}\!\bigg)^{\frac{1}{p}}, \\ \nonumber
H^{s}_{\infty}(\R_{\Bohr}) & := \{T\in\Trig'(\R)\ \!|\ \!||T||_{s,\infty}\!<\infty\}, & ||T||_{(s,\infty)} & :=\sup_{\lambda\in\R}|\langle\lambda\rangle^{s}\hat{T}(\lambda)| \\ \nonumber
H^{\infty}_{p}(\R_{\Bohr}) & := \bigcap_{s\in\R}H^{s}_{p}(\R_{\Bohr}), & H^{-\infty}_{p}(\R_{\Bohr}) & := \bigcup_{s\in\R}H^{s}_{p}(\R_{\Bohr}).
\end{align}
for $s\in\R,p\in[0,\infty)$. The space $d(\R)$ bears some similarities with the space of test functions $\cD(\R)=C^{\infty}_{0}(\R)$, especially we may endowed it with the strict inductive limit topology coming from
\begin{align}
\label{eq:bohrtestfunctions}
d(\R) & = \bigcup_{\substack{F\subset\R \\ |F|<\infty}}d_{F}(\R), & d_{F}(\R) & := \{\hat{\Psi}:\R\rightarrow\C\ |\ \textup{supp}(\hat{\Psi})\subset F\}\cong\C^{|F|}, \\ \nonumber
 & & ||\hat{\Psi}||_{F,k} & := \max_{\lambda\in F}|\langle\lambda\rangle^{k}\hat{\Psi}(\lambda)|,\ \ \ k\in\N_{0},
\end{align}
although the limit is uncountable in this case. There is a natural isomorphism, $\overline{\Trig(\R)}^{||\ .\ ||_{\infty}}=CAP(\R)\cong C(\R_{\Bohr})$, between the uniformly almost periodic functions on $\R$ and the continuous function on $\R_{\Bohr}$, which allows to define the space of smooth functions on $\R_{\Bohr}$:
\begin{align}
\label{eq:bohrsmooth}
C^{\infty}(\R_{\Bohr}) = CAP^{\infty}(\R) & := CAP(\R)\cap C^{\infty}_{b}(\R),
\end{align}
carrying the natural Fr\'echet space topology induced by $C^{\infty}_{b}(\R)$. Furthermore, we have $H^{0}_{2}(\R_{\Bohr})=L^{2}(\R_{\Bohr})\cong B^{2}(\R_{\Bohr})$, and $H^{0}_{1}(\R_{\Bohr})\subset CAP(\R)\subset H^{0}_{2}(\R_{\Bohr}),$ $H^{\infty}_{1}(\R_{\Bohr})\subset CAP^{\infty}(\R)\subset H^{\infty}_{2}(\R_{\Bohr}),$ $H^{s}_{p}(\R_{\Bohr})\subset H^{s'}_{p'}(\R_{\Bohr}),\ p\leq p',\ s'\leq s$\footnote{The space $H^{0}_{1}(\R_{\Bohr})=:CAP_{a}(\R)$ of uniformly almost periodic functions with absolutely convergent Fourier series is a generalisation of the Wiener algebra.}. But $H^{\infty}_{p}(\R_{\Bohr})$ is not embedded in $CAP(\R)$ for $p>1$ (in contrast with the usual situation for Sobolev spaces on $\R$ or $\T$), which can be related to $\R_{\disc}$ not being $\sigma$-finite w.r.t. the counting measure, e.g. the function,
\begin{align}
\label{eq:counterexamplesobolev}
\hat{\Psi}(\lambda) & :=\left\{\begin{matrix}\frac{1}{n}& : & \lambda=\frac{1}{n},\ n\in\N \\[0.1cm] 0 & : & \textup{else}\end{matrix}\right. ,
\end{align}
belongs to $\bigcap_{p>1}H^{\infty}_{p}(\R_{\Bohr})$, since $\forall n\in\N:\ 2^{-|p s|}\leq\langle\frac{1}{n}\rangle^{p s}\leq2^{|p s|}$, while it is not in $H^{\infty}_{1}(\R_{\Bohr})$, because the harmonic series is divergent. The inner products of the various realizations of $\fH_{1}$ induce dualities between $H^{s}_{p}(\R_{\Bohr}),\ \!d(\R),\ \!\Trig(\R)$ and $H^{-s}_{q}(\R_{\Bohr}),\ \!d'(\R),\ \!\Trig'(\R)$ for $\tfrac{1}{p}+\tfrac{1}{q}=1$, and $\widehat{\ }$ (or $\check{\ }$) in \eqref{eq:bohrfsp} indicates that the corresponding vectors or spaces are related via the (inverse) Fourier transform on $\R_{\Bohr}$ (or $\R_{\disc}$):
\begin{align}
\label{eq:bohrfourier}
\hat{\Psi}(\lambda) & := (e_{-\lambda},\Psi)_{\Bohr},\ \Psi\in H^{0}_{1}(\R_{\Bohr}), & \check{\Phi}(x) & := (e_{x},\Phi)_{l^{2}(\R)},\ \Phi\in l^{1}(\R).
\end{align}
The images of the spaces $H^{s}_{p}(\R_{\Bohr})$ under the Fourier transform are denoted by $h^{s}_{p}(\R)$. Before we introduce the aforementioned Weyl quantisations, we add a short
\begin{Remark}
\label{rem:bohrschwartz}
We might ask, whether there is a analogue $s(\R)\subset l^{2}(\R)$ of the Schwartz space $\cS(\R)\subset L^{2}(\R)$. Since $\cS(\R)$ is typically defined in terms of the uniform boundedness of expression of the form $(\langle x\rangle^{m}\partial_{x}^{n}f)(x),\ m,n\in\N_{0}$, we need a viable substitute for the partial derivatives as elements of $l^{2}(\R)$ are never differentiable. Additionally, we expect $\check{s}(\R)\subset H^{\infty}_{p}(\R_{\Bohr})\ \forall p\in[1,\infty]$ to hold, which requires a summability condition on the elements of $s(\R)$, because decay conditions at infinity do not suffice in view of \eqref{eq:counterexamplesobolev}. A natural discretised replacement of the partial derivatives, already familiar from the context of $s(\Z)\subset l^{2}(\Z)$, is the (scaled) forward difference $(\Delta_{x,x_{0}} f)(x):=f(x+x_{0})-f(x)$, and because of the relative nesting of the Sobolev spaces, we could propose the following definition:
\begin{align}
\label{eq:bohrschwartz}
s(\R) & := \{\hat{\Psi}:\R\rightarrow\C\ |\ \sum_{\lambda\in\R}\langle\lambda\rangle^{m}|(\Delta^{n}_{\lambda,\lambda_{0}}\hat{\Psi})(\lambda)|<\infty\},\ \lambda_{0}\in\R,\ m,n\in\N_{0}.
\end{align}
But, as $(\Delta^{n}_{\lambda,\lambda_{0}}\hat{\Psi})(\lambda) = \sum_{k=0}^{n}(-1)^{n-k}\binom{n}{k}\hat{\Psi}(\lambda+k\lambda_{0})$, it is easy to see that $\check{s}(\R)=H^{\infty}_{1}(\R_{\Bohr})$ in this case. Similar observations hold if we replace the forward difference by the backward or central difference. Interestingly, there is a useful duality between $\cS(\R)\subset L^{2}(\R)$ and $s(\R)\subset l^{2}(\R)$ respectively $H^{\infty}_{1}(\R_{\Bohr})\subset L^{2}(\R_{\Bohr})$, which is compatible with inner products, Fourier transforms and multiplier actions (cp. \cite{FewsterPhaseSpaceQuantization}). Namely, for $f\in\cS(\R)$ and $\hat{\Psi}\in s(\R)\ (\Psi\in H^{\infty}_{1}(\R_{\Bohr}))$, we have
\begin{align}
\label{eq:schwartzsobolevduality}\nonumber
(\cF[f],\hat{\Psi})_{l^{2}(\R)} & = \sum_{\lambda\in\R}\overline{\cF[f](\lambda)}\hat{\Psi}(\lambda), &  |(\cF[f],\hat{\Psi})_{l^{2}(\R)}| & \leq||\Psi||_{(s,1)}\sup_{\lambda\in\R}\langle\lambda\rangle^{-s}|\cF[f](\lambda)| \\ \nonumber
 & & & <\infty, \\
(\cF[f],\hat{\Psi})_{l^{2}(\R)} & = (\check{\cF[f]},\Psi)_{\Bohr} & & \\ \nonumber
 & = (f,\Psi)_{L^{2}(\R)} & & \\ \nonumber
 &  = (\cF[f],\cF[\Psi])_{L^{2}(\R)}, & &
\end{align}
because $\Psi\cdot f\in\cS(\R),\ \ \cF[f]\cdot\hat{\Psi}\in s(\R)$. \\
Furthermore, the duality is compatible with the representations $\pi_{F}$ and $\pi_{0}$ of the Weyl algebra on $L^{2}(\R)$ and $l^{2}(\R)$, respectively:
\begin{align}
\label{eq:weylalgebraduality}
(\pi_{F}(W_{\varepsilon}(\alpha,\beta)^{*})f,\Psi)_{L^{2}(\R)} & = (\pi_{F}(W_{\varepsilon}(\beta,-\alpha)^{*})\cF_{\varepsilon}[f],\hat{\Psi}_{(\varepsilon)})_{l^{2}(\R)} \\ \nonumber
 & = (\cF_{\varepsilon}[f],\pi_{0}(W_{\varepsilon}(\beta,-\alpha))\hat{\Psi}_{(\varepsilon)})_{l^{2}(\R)} \\ \nonumber
 & = (f,\pi_{0}(W_{\varepsilon}(\alpha,\beta))\Psi)_{L^{2}(\R)}.
\end{align}
Here, we used the $\varepsilon$-scaled version of \eqref{eq:bohrfourier}: $\hat{\Psi}_{(\varepsilon)}(\lambda) = (e_{-\frac{\lambda}{\varepsilon}},\Psi)_{\Bohr}$ (cp. also \eqref{eq:stscaledft}, although we switched the sign of the Fourier exponential, as to fit with the use of left convolution kernels in the case of compact Lie groups).
\end{Remark}
The two choices of Weyl quantisation associated with $B^{2}(\R)$ and $l^{2}(\R)$, respectively, $L^{2}(\R_{\Bohr})$ resemble the dichotomy already mentioned in the previous subsection \ref{subsec:csquant} (see the comment below \eqref{eq:u1knquantsymbols}), and arise from (so far formal) expressions:
\begin{align}
\label{eq:apweylquant}
(A_{\sigma}\Psi)(x) & = \frac{1}{2\pi\varepsilon}\int_{\R}d\lambda\int_{\R}dx'\ \sigma(\tfrac{1}{2}(x+x'),\lambda)e_{-\lambda}(\tfrac{x-x'}{\varepsilon})\Psi(x'),\ \Psi\in\Trig(\R)\ \textup{or}\ CAP^{\infty}(\R),
\end{align}
and
\begin{align}
\label{eq:bohrweylquant}
(A_{\sigma}\Phi)(\lambda) & = \sum_{\lambda'\in\R}\int_{\R_{\Bohr}}d\mu_{\Bohr}(x)\ \sigma(x,\tfrac{1}{2}(\lambda+\lambda'))e_{\frac{\lambda-\lambda'}{\varepsilon}}(x)\Phi(\lambda'),\ \Phi\in d(\R)\ \textup{or}\ \check{\Phi}\in H^{\infty}_{1}(\R_{\Bohr}), \\[0.25cm] \nonumber
(A_{\sigma}\Psi)(x) & = \sum_{\lambda\in\R}\int_{\R_{\Bohr}}d\mu_{\Bohr}(x')\ \sigma(\tfrac{1}{2}(x+x'),\lambda)e_{-\lambda}(\tfrac{x-x'}{\varepsilon})\Psi(x'),\ \Psi\in\Trig(\R)\ \textup{or}\ H^{\infty}_{1}(\R_{\Bohr}).
\end{align}
Fundamentals on the theory of operators defined by \eqref{eq:apweylquant} in Kohn-Nirenberg form can be found in the works of Shubin \cite{ShubinAlmostPeriodicFunctions, ShubinDifferentialAndPseudodifferential} (cf. also \cite{CoburnCAlgebrasOf}), which we recall to some extent, as to allow for immediate comparison with operators of the form \eqref{eq:bohrweylquant}. Following this, we will make precise the definition of the latter.\\[0.1cm]
Pseudo-differential operators like \eqref{eq:apweylquant} with $\sigma\in S^{m}_{\rho,\delta}(\R^{2})$ (Hörmander's symbol classes) can be defined on $C^{\infty}_{b}(\R)$, which contains $\Trig(\R)$ and $CAP^{\infty}(\R)$, by the usual means of oscillatory integrals (cf. \cite{ShubinDifferentialAndPseudodifferential}). In the context of almost periodic functions, the admissible symbols\footnote{We only consider the largest of the admissible symbol classes introduced by Shubin.} $APS^{m}_{\rho,\delta}(\R^{2})\subset S^{m}_{\rho,\delta}(\R^{2})$ are adapted to preserve the property of almost periodicity, i.e.
\begin{align}
\label{eq:apsymbols}
\sigma\in APS^{m}_{\rho,\delta}(\R^{2})\subset C^{\infty}(\R^{2})\ :\Leftrightarrow\ & \R\ni\lambda\mapsto\sigma(\ .\ ,\lambda)\in CAP(\R)&\!\!\!\!\!\!\!\! \textup{is continuous}\ \ \ \ & \\ \nonumber
 & \& \forall\alpha,\beta\in\N_{0}:\forall(x,\lambda)\in\R^{2}: &\!\!\!\!\!\!\!\! \exists C_{\alpha\beta}>0:\ \ \ \ \ \ \ \ \ & \\ \nonumber
 & &\!\!\!\!\!\!\!\! |(\partial^{\alpha}_{x}\partial_{\lambda}^{\beta}\sigma)(x,\lambda)| & \leq C_{\alpha\beta}\langle\lambda\rangle^{m-\rho\beta+\delta\alpha}, \\[0.25cm] \nonumber
APS^{-\infty}(\R^{2}) & :=\bigcap_{m\in\R}APS^{m}_{\rho,\delta}(\R^{2}), &\!\!\!\!\!\!\!\!  APS^{\infty}_{\rho,\delta}(\R^{2})\ \ \ \ \ & \!\!\!\!\!\!\!\!\!\!:=\bigcup_{m\in\R}APS^{m}_{\rho,\delta}(\R^{2})
\end{align}
for $m\in\R, 0\leq\delta\leq\rho\leq1$
\begin{Definition}
\label{def:apweylquant}
An operator $A_{\sigma}$ on $CAP^{\infty}(\R)$ defined by \eqref{eq:apweylquant} with $\sigma\in APS^{m}_{\rho,\delta}(\R^{2})$ is called an almost-periodic pseudo-differential operator.
\end{Definition}
Statements familiar form the theory of pseudo-differential operators on $\R^{n}$ about composition, adjoints, asymptotic expansions and continuity w.r.t. to $CAP^{\infty}(\R)$ and the scales of Sobolev spaces $H^{s}_{p}(\R_{\Bohr})$ remain valid, as expected, because all necessary operations on the symbols preserve almost-periodicity. An especially interesting and useful property of almost-periodic pseudo-differential operators is
\begin{Proposition}[cf. \cite{ShubinDifferentialAndPseudodifferential}]
\label{prop:apadjoint}
Given $A_{\sigma}$ with $\sigma\in APS^{m}_{\rho,\delta}(\R^{2}),\ \delta<\rho,$ there exist the formal adjoint $A^{*}_{\sigma}$ w.r.t. to $(\ ,\ )_{L^{2}(\R)}$ and $(\ ,\ )_{\Bohr}$, i.e.
\begin{align}
\label{eq:apadjoint1}
(A^{*}_{\sigma}\Psi,\Phi)_{L^{2}(\R)} & = (\Psi,A_{\sigma}\Phi)_{L^{2}(\R)},\ \Psi,\Phi\in\cS(\R), \\[0.25cm]
\label{eq:apadjoint2}
(A^{*}_{\sigma}\Psi,\Phi)_{\Bohr} & = (\Psi,A_{\sigma}\Phi)_{\Bohr},\ \Psi\in H^{\infty}_{p}(\R_{\Bohr}),\ \Phi\in H^{\infty}_{q}(\R_{\Bohr})\ (\tfrac{1}{p}+\tfrac{1}{q}=1).
\end{align}
The symbol of $A^{*}_{\sigma}$ is $\overline{\sigma}$.
\end{Proposition}
Furthermore, Shubin \cite{ShubinAlmostPeriodicFunctions} proves the equality $||A_{\sigma}||_{\cB(L^{2}(\R))} = ||A_{\sigma}||_{\cB(B^{2}(\R))}$ for bounded $A_{\sigma}$, which entails the equality of spectra $\textup{spec}_{\cB(L^{2}(\R))}(A_{\sigma}) = \textup{spec}_{\cB(B^{2}(\R))}(A_{\sigma})$ by the preceding proposition (this continues to hold for (hypo)elliptic $A_{\sigma}$). Nevertheless, the quality of the spectra w.r.t. $L^{2}(\R)$ and $B^{2}(\R)$ can differ significantly, e.g. the spectrum of the Laplace operator $\Delta$, which is essentially self-adjoint on $\cS(\R)$ and $\Trig(\R)$, respectively, is absolutely continuous in the first and pure point in the second case.\\[0.1cm]
Now, we come to the definition of pseudo-differential operators in terms of \eqref{eq:bohrweylquant}, which we will call \textit{Bohrian pseudo-differential operators}. Let us first remark that the formulas \eqref{eq:bohrweylquant} are closer in structure to those applied in the definition of pseudo-differential operators on compact Lie groups (see subsection \ref{sec:compact}), while the almost-periodic pseudo-differential operators heavily exploit the special relation between $\R$ and $\R_{\Bohr}$. The same reasoning applies to $U(1)$-equivariant pseudo-differential operators (see subsection \ref{subsec:csquant}), where the special relation between $\R$ and $U(1)$ via the covering morphism $\R\rightarrow\R/2\pi\Z\cong U(1)$ is of avail. But, in contrast with $U(1)$-equivariant operators, where the analogues, \eqref{eq:u1knquant} and \eqref{eq:equivariantknquant}, of \eqref{eq:apweylquant} and \eqref{eq:bohrweylquant} are essentially equivalent due to the sparseness of $\Z\subset\R$ and the possibility of smooth interpolation, the situation will be different in the present case.
\begin{Definition}
\label{def:bohrsymbol}
A function $\sigma:\R_{\disc}\times\R_{\Bohr}\rightarrow\C$ is called a \textup{symbol in} $S^{m}_{\rho,\delta}(\R_{\Bohr}\times\R_{\disc})=s^{m}_{\rho,\delta}$, where $m\in\R,\ 0\leq\delta<\rho\leq1$, if $\forall\lambda\in\R: \sigma(\ .\ ,\lambda)\in H^{\infty}_{1}(\R_{\Bohr})$ and $\forall\alpha,\beta\in\N_{0}:\exists C_{\alpha,\beta}>0$ s.t.:
\begin{align}
\label{eq:bohrsymbol}
\forall(x,\lambda)\in\R_{\Bohr}\times\R_{\disc}: |(\partial^{\alpha}_{x}\Delta_{\lambda}^{\beta}\sigma)(x,\lambda)| & \leq C_{\alpha\beta}\langle\lambda\rangle^{m-\rho\beta+\delta\alpha}.
\end{align}
We set:
\begin{align}
\label{eq:bohrsymbolsp}
s^{-\infty} & = \bigcap_{m\in\R}s^{m}_{\rho,\delta}, & s^{\infty}_{\rho,\delta} & = \bigcup_{m\in\R}s^{m}_{\rho,\delta}.
\end{align}
\end{Definition}
A simple application of the (discrete) Leibniz formula and Peetre's inequality, $\forall r,\lambda,\lambda'\in\R:$ \mbox{$\langle\lambda+\lambda'\rangle^{r}\leq 2^{|r|}\langle\lambda\rangle^{|r|}\langle\lambda'\rangle^{r}$,} gives:
\begin{Corollary}
\label{cor:bohrsymbolprod}
Let $\sigma\in s^{m}_{\rho,\delta}$ and $\tau\in s^{m'}_{\rho',\delta'}$. Then $\forall\alpha,\beta\in\N_{0}:\partial^{\alpha}_{x}\Delta^{\beta}_{\lambda}\sigma\in s^{m-\rho\beta+\delta\alpha}$ and $\sigma\tau\in s^{m+m'}_{\min(\rho,\rho'),\max(\delta,\delta')}$.
\end{Corollary}
In view of remark \ref{rem:bohrschwartz}, the definition of symbol classes $s^{m}_{\rho,\delta}$ for \eqref{eq:bohrweylquant} requires summability conditions or restrictions on the Fourier spectrum supplementing the usual decay conditions.
\begin{Definition}
\label{def:polysymbol}
A symbol $\sigma\in s^{m}_{\rho,\delta}$ is said to have \textup{polynomially bounded spectral growth of order $\gamma,\gamma'\in\R$}, if $|\textup{supp}(\hat{\sigma}^{1}(\ .\ ,\lambda))\cap K_{\lambda'}|=:N_{\lambda'}(\lambda)\leq C_{\gamma,\gamma'}\langle\lambda\rangle^{\gamma}\langle\lambda'\rangle^{\gamma'}$ for $K_{\lambda'}=[-\lambda',\lambda']\subset\R,\ \lambda'\in\R$, where
\begin{align}
\label{eq:partialbohrfourier}
\hat{\sigma}^{1}(\lambda',\lambda) & := \int_{\R_{\Bohr}}d\mu_{\Bohr}(x)e_{\lambda'}(x)\sigma(x,\lambda).
\end{align}
Clearly, symbols of polynomially bounded spectral growth of order $\gamma$ form a subspace $s^{m,(\gamma,\gamma')}_{\rho,\delta}\subset s^{m}_{\rho,\delta}$. Furthermore, the property to be of polynomially bounded spectral growth is, on the one hand, preserved under multiplication in the Fourier domain and, on the other hand, not preserved under convolution in the Fourier domain. 
\end{Definition}
An important subclass of symbols is given by
\begin{Definition}
\label{def:equivariantbohrsymbols}
A symbol $\sigma$ in $s^{m,(\gamma,\gamma'=1)}_{\rho,\delta}$ is called $U(1)_{\lambda_{0}}$-\textup{equivariant}, $\lambda_{0}\in\R$, if $\ \forall\lambda\in\R:\textup{supp}(\hat{\sigma}^{1}(\ .\ ,\lambda))\subset\Z^{j_{0}}_{\lambda_{0}}:=\lambda_{0}(\Z+j_{0})$ for $j_{0}\in[0,1)$.
Similarly, $T\in\Trig'(\R)$ is $U(1)_{\lambda_{0}}$-equivariant, if $\textup{supp}(\hat{T})\subset\Z^{j_{0}}_{\lambda_{0}}$. The various Fourier images of the $U(1)_{\lambda_{0}}$-equivariant versions of the function spaces on $\R_{\Bohr}$ (see\eqref{eq:bohrfsp} and below) are denoted by $d(\Z^{j_{0}}_{\lambda_{0}}),\ d'(\Z^{j_{0}}_{\lambda_{0}}),\ h^{s}_{p}(\Z^{j_{0}}_{\lambda_{0}}),$ etc.
\end{Definition}
Next, we give meaning to \eqref{eq:bohrweylquant} for $\sigma\in s^{m,(\gamma,\gamma')}_{\rho,\delta},\ \Phi\in d(\R)$ in a similar fashion as one does for \eqref{eq:apweylquant} with $\sigma\in S^{m}_{\rho,\delta},\ \Psi\in C^{\infty}_{b}(\R)$. To this end, we will work primarily with the first formula in \eqref{eq:bohrweylquant}, and interpret the second formula as a mnemonic for the dual operator defined by the $\varepsilon$-scaled Fourier transform. On the contrary, Fewster and Sahlmann \cite{FewsterPhaseSpaceQuantization} construct operators by means of the second formula. It is apparent that the first formula is natural, when working with the ``volume representation'' in loop quantum cosmological models.
\begin{Proposition}
\label{prop:polyoperator}
Let $\sigma\in s^{m,(\gamma,\gamma')}_{\rho,\delta}$, then $A_{\sigma}:d(\R)\rightarrow h^{\infty}_{1}(\R)$. If $\sigma$ is $U(1)_{\lambda_{0}}$-equivariant and $\lambda'_{0}\in\R$ satisfies $\tfrac{\lambda'_{0}}{\lambda_{0}}=\tfrac{p}{q}\in\mathds{Q}$, we have $A_{\sigma}:h^{\infty}_{1}(\Z^{j'_{0}}_{\lambda'_{0}})\rightarrow h^{\infty}_{1}(\Z^{j''_{0}}_{\lambda''_{0}})$ for some $\lambda''_{0}\in\R$ and $j''_{0}\in[0,1)$ ($\varepsilon=1$).\\
In general, if $\sigma\in\Trig'(\R)\otimes d'(\R)$, $A_{\sigma}$ defines a quadratic form $Q_{\sigma}$ on $d(\R)$:
\begin{align}
\label{eq:bohrquadraticform}
Q_{\sigma}(\Phi_{1},\Phi_{2}) & := (\Phi_{1},A_{\sigma}\Phi_{2})_{l^{2}(\R)},\ \Phi_{1},\Phi_{2}\in d(\R).
\end{align}
The \textup{formal adjoint} $A^{*}_{\sigma}$, defined by
\begin{align}
\label{eq:bohrformaladjoint}
(A^{*}_{\sigma}\Phi_{1},\Phi_{2})_{l^{2}(\R)} & = (\Phi_{1},A_{\sigma}\Phi_{2})_{l^{2}(\R)},\ \Phi_{1},\Phi_{2}\in d(\R),
\end{align}
has symbol $\overline{\sigma}$.
\begin{Proof}
For simplicity, we restrict to $\varepsilon=1$, since the general case only leads to some trivial rescaling in the formula to follow. First, let $\sigma\in s^{m,\gamma}_{\rho,\delta}$ and $\Phi\in d(\R)$, then we use $e_{\lambda}(x)=\langle\lambda\rangle^{-2M}(1-\Delta_{x})^{M}e_{\lambda}(x),\ M\in\N_{0},$ and the translation invariance of $\mu_{\Bohr}$ to regularize \eqref{eq:bohrweylquant}:
\begin{align}
\label{eq:polyreg}
(A_{\sigma}\Phi)(\lambda) & = \sum_{\lambda'\in\textup{supp}(\Phi)}\!\!\!\!\Phi(\lambda')\int_{\R_{\Bohr}}\!\!\!\!\!\!d\mu_{\Bohr}(x')\langle\lambda-\lambda'\rangle^{-2M}((1-\Delta_{x'})^{M}\sigma)(x',\tfrac{1}{2}(\lambda+\lambda')))e_{\lambda-\lambda'}(x').
\end{align}
By assumption, i.e. $|\textup{supp}(\Phi)|<\infty$ and $\sigma\in s^{m,(\gamma,\gamma')}_{\rho,\delta}$, $|\textup{supp}(A_{\sigma}\Phi)\cap K_{\lambda}|\leq C_{\gamma''}\langle\lambda\rangle^{\gamma''}$ for some $\gamma''\in\R$. Therefore, for arbitrary $s\in\R$ and large enough $M\in\N_{0}$
\begin{align}
\label{eq:polybound}
 & ||A_{\sigma}\Phi||_{(s,1)} \\ \nonumber
 & \leq \sum_{\lambda\in\textup{supp}(A_{\sigma}\Phi)}\langle\lambda\rangle^{s}\sum_{\lambda'\in\textup{supp}(\Phi)}\langle\lambda-\lambda'\rangle^{-2M}\sum_{\alpha=0}^{M}\binom{M}{\alpha}\sup_{x'\in\R}|(\partial^{2\alpha}_{x'}\sigma)(x',\tfrac{1}{2}(\lambda+\lambda'))||\Phi(\lambda')| \\ \nonumber
 & \leq \sum_{\alpha=0}^{M}\binom{M}{\alpha}C_{2\alpha0}\sum_{\lambda'\in\textup{supp}(\Phi)}|\Phi(\lambda')|\sum_{\lambda\in\textup{supp}(A_{\sigma}\Phi)}\langle\lambda\rangle^{s}\langle\lambda-\lambda'\rangle^{-2M}\langle\tfrac{1}{2}(\lambda+\lambda')\rangle^{m+2\delta\alpha} \\ \nonumber
 & \leq 2^{|s|+2M}\sum_{\alpha=0}^{M}\binom{M}{\alpha}C_{2\alpha0}2^{2|m+2\delta\alpha|}\sum_{\lambda'\in\textup{supp}(\Phi)}\langle\lambda'\rangle^{2M+|m+2\delta\alpha|}|\Phi(\lambda')|\sum_{\lambda\in\textup{supp}(A_{\sigma}\Phi)}\langle\lambda\rangle^{-2(M-\delta\alpha)+s+m} \\ \nonumber
 & <\infty.
\end{align}
Here, we used Peetre's inequality to arrive at the next to last line, and concluded finiteness of the last sum from, using the polynomial growth bound on $|\textup{supp}(A_{\sigma}\Phi)\cap K_{\lambda}|$,
\begin{align}
\label{eq:polysum}
\sum_{\lambda\in\textup{supp}(A_{\sigma}\Phi)}\langle\lambda\rangle^{-2(M-\delta\alpha)+s+m} & \leq \sum_{\lambda\in\textup{supp}(A_{\sigma}\Phi)}\langle\lambda\rangle^{-2M(1-\delta)+s+m} \\ \nonumber
 & = \lim_{n\rightarrow\infty}\sum_{m=0}^{n}\sum_{\lambda\in\textup{supp}(A_{\sigma}\Phi)\cap K_{m}\setminus K_{m-1}}\langle\lambda\rangle^{-2M(1-\delta)+s+m},\ K_{-1}=\emptyset \\ \nonumber
 & \leq \lim_{n\rightarrow\infty}\sum_{m=0}^{n}\langle m-1\rangle^{-2M(1-\delta)+s+m}(N_{m}-N_{m-1}) \\ \nonumber
 & \leq C'_{\gamma''}\lim_{n\rightarrow\infty}\sum_{m=0}^{1}\langle m-1\rangle^{-2M(1-\delta)+s+m+\gamma''},
\end{align}
which is finite for large enough $M$, because $\delta<1$. To prove the second assertion, we observe that for $\Phi\in h^{\infty}_{1}(\Z^{j'_{0}}_{\lambda'_{0}})$ and $U(1)_{\lambda_{0}}$-equivariant $\sigma$:
\begin{align}
\label{eq:fractionallatticesupport}
(A_{\sigma}\Phi)(\lambda) & \neq0\hspace{2cm}\Leftrightarrow\ &\ \lambda & = \underbrace{\lambda-\lambda'}_{\in\Z^{j_{0}}_{\lambda_{0}}} + \underbrace{\lambda'}_{\in\Z^{j'_{0}}_{\lambda'_{0}}}\in\Z^{j_{0}}_{\lambda_{0}}+\Z^{j'_{0}}_{\lambda'_{0}}.
\end{align}
But, $Z^{j_{0}}_{\lambda_{0}}+\Z^{j'_{0}}_{\lambda'_{0}}=\{(\lambda_{0}m+\lambda'_{0}n)+(\lambda_{0}j_{0}+\lambda'_{0}j'_{0})\ |\ m,n\in\Z\}\subset\tfrac{\lambda_{0}}{q}\Z+\tfrac{\lambda_{0}}{q}(q j_{0}+p j'_{0}).$
Thus, setting $\lambda''_{0}:=\tfrac{\lambda_{0}}{q}=\tfrac{\lambda'_{0}}{p}$ and $j''_{0} := (q j_{0}+p j'_{0}) \mod 1$, we have $A_{\sigma}\Phi\in d'(\Z^{j''_{0}}_{\lambda''_{0}})$. Now, we regularise the expression for $A_{\sigma}\Phi$ in the same way as above to show that we even have $A_{\sigma}\Phi\in h^{\infty}_{1}(\Z^{j''_{0}}_{\lambda''_{0}})$ ($s\in\R$):
\begin{align}
\label{eq:equivariantbound}
 & ||A_{\sigma}\Phi||_{(s,1)} \\ \nonumber
 & \leq \sum_{\lambda\in\Z^{j''_{0}}_{\lambda''_{0}}}\langle\lambda\rangle^{s}\sum_{\lambda'\in\Z^{j'_{0}}_{\lambda'_{0}}}\langle\lambda-\lambda'\rangle^{-2M}\sum_{\alpha=0}^{M}\binom{M}{\alpha}\sup_{x'\in\R}|(\partial^{2\alpha}_{x'}\sigma)(x',\tfrac{1}{2}(\lambda+\lambda'))||\Phi(\lambda')| \\ \nonumber
 & \leq 2^{|s|+2M}\sum_{\alpha=0}^{M}\binom{M}{\alpha}C_{2\alpha0}2^{2|m+2\delta\alpha|}\sum_{\lambda'\in\Z^{j'_{0}}_{\lambda'_{0}}}\langle\lambda'\rangle^{2M+|m+2\delta\alpha|}|\Phi(\lambda')|\sum_{\lambda\in\Z^{j''_{0}}_{\lambda''_{0}}}\langle\lambda\rangle^{-2(M-\delta\alpha)+s+m} \\ \nonumber
 & < \infty,
\end{align}
where, again, we employed Peetre's inequality, $\Phi\in h^{\infty}_{1}(\Z^{j'_{0}}_{\lambda'_{0}})$ and the finiteness of the last sum for large enough $M$ ($\delta<1$). \\
The statements concerning the quadratic form defined by general $A_{\sigma}$ and the formal adjoint $A^{*}_{\sigma}$ are obvious from the finiteness of all sums and the behaviour of the Fourier transform w.r.t. complex conjugation.
\end{Proof} 
\end{Proposition}
\begin{Remark}
\label{rem:equivariantbohrsymbols}
Taking a look at \eqref{eq:fractionallatticesupport}, it becomes evident that the condition $\tfrac{\lambda'_{0}}{\lambda_{0}}\in\mathds{Q}$ in proposition \ref{prop:polyoperator} cannot be relaxed easily, because two relatively irrational lattices may generated arbitrary dense support for $A_{\sigma}\Phi$ in the bounded intervals $K_{\lambda}$. 
\end{Remark}
\begin{Definition}
\label{def:fractionalbohrsymbols}
Two operators $A_{\sigma},\ A_{\tau}$ with $U(1)_{\lambda^{\sigma}_{0}}$-, respectively, $U(1)_{\lambda^{\tau}_{0}}$-equivariant symbols $\sigma,\ \tau$, defined on the (rational scales of) spaces
\begin{align}
\label{eq:fractionalscalesspaces}
h^{\infty}_{1}(\lambda^{\sigma}_{0}) & := \bigcup_{\substack{\lambda'_{0}\in\R \\ \frac{\lambda'_{0}}{\lambda^{\sigma}_{0}}\in\mathds{Q}}}\bigcup_{j'_{0}\in[0,1)}h^{\infty}_{1}(\Z^{j'_{0}}_{\lambda'_{0}}) , & h^{\infty}_{1}(\lambda^{\tau}_{0}) & := \bigcup_{\substack{\lambda''_{0}\in\R \\ \frac{\lambda''_{0}}{\lambda^{\tau}_{0}}\in\mathds{Q}}}\bigcup_{j''_{0}\in[0,1)}h^{\infty}_{1}(\Z^{j''_{0}}_{\lambda''_{0}}) 
\end{align}
are called \textup{relatively rational}, if $\tfrac{\lambda^{\sigma}_{0}}{\lambda^{\tau}_{0}}\in\mathds{Q}$ ($\varepsilon =1$). Clearly, in this case $h^{\infty}_{1}(\lambda^{\sigma}_{0})=h^{\infty}_{1}(\lambda^{\tau}_{0})$ (other function space are defined similarly\footnote{In contrast with $d(\R)$, the test function space $d(\lambda_{0})$ can be given a \textit{countable} strict inductive limit topology, because $\mathds{Q}$ is countable.}).
\end{Definition}
\begin{Corollary}
\label{cor:fractionalbohrops}
Relatively rational operators generate an algebra with common domain $h^{\infty}_{1}(\lambda_{0})$ for some $\lambda_{0}\in\R$.
\end{Corollary}
\begin{Corollary}
\label{cor:fractionalscaling}
Let $\sigma$ be a $U(1)_{\lambda_{0}}$-equivariant symbol, then $A^{(\varepsilon)}_{\sigma}$ and $A^{(\varepsilon')}_{\sigma}$ are relatively rational if and only if $\tfrac{\varepsilon'}{\varepsilon}\in\mathds{Q}$. Here, we made the $\varepsilon$-dependence of \eqref{eq:bohrweylquant} explicit.
\end{Corollary}
If we want to use symbolic calculus in the analysis of Bohrian pseudo-differential operators, we will need a statement concerning the asymptotic summation of symbols in $s^{m}_{\rho,\delta}$.
\begin{Proposition}
\label{prop:bohrasymptoticsum}
Let $\{m_{j}\}^{\infty}_{j=1}\subset\R$ be such that $\lim_{j\rightarrow\infty}m_{j}=-\infty,\ m:=\max_{j\in\N}m_{j}$, and $\sigma_{j}\in s^{m_{j}}_{\rho,\delta}$ for all $j\in\N$. Then, there exists a symbol $\sigma\in s^{m}_{\rho,\delta}$, unique up to $s^{-\infty}$, such that $a\sim\sum_{j=1}^{\infty}\sigma_{j}$, i.e.:
\begin{align}
\label{eq:bohrasymptoticsum}
\forall n\in\N:\exists k_{n}\in\N:\forall k\geq k_{n}:\ \sigma-\sum_{j=1}^{k}\sigma_{j}\in s^{m_{n}}_{\rho,\delta}.
\end{align}
If the symbols $\sigma_{j},\ j\in\N,$ are of polynomially bounded spectral growth of order $(\gamma,\gamma')$ or $U(1)_{\lambda_{0}}$-equivariant, then so is $\sigma$.
\begin{Proof}
Using standard excision function techniques and literally repeating the argument in \cite{RuzhanskyPseudoDifferentialOperators}, Theorem 4.4.1, accomplishes the proof. The statement concerning polynomially bounded spectral growth and equivariance follows, because the excision functions only touch the second argument of the symbols.
\end{Proof}
\end{Proposition} 
We close this section with a discussion of the differences between almost-periodic pseudo-differential operators and Bohrian pseudo-differential operators:\\[0.1cm]
Our first observation is, as already mentioned above, that almost-periodic pseudo-differential operators are not equivalent to Bohrian pseudo-differential operators, at least not without extending the symbol classes for almost-periodic pseudo-differential operators to include non-smooth elements, in contrast with the analogous situation in the $U(1)$-equivariant case. This is most easily inferred from an explicit example (important in loop quantum cosmology):
\begin{align}
\label{eq:nonexample}
A\ :\ & h^{s}_{p}(\R)\rightarrow h^{s-1}_{p}(\R),\ s\in\R, & (A\Phi)(\lambda) & := |\lambda|\Phi(\lambda),\ \Phi\in h^{s+1}_{p}(\R),
\end{align}
which has the symbol $\sigma_{A}(x,\lambda)=|\lambda|$, which makes sense for \eqref{eq:apweylquant} as well as \eqref{eq:bohrweylquant}. Clearly, $\sigma_{A}\in s^{1,(0,0)}_{1,0}$, but $\sigma\notin APS^{m}_{\rho,\delta}$ for any $m,\rho,\delta$, because this require smoothness in the second argument. Since an element $\Phi\in h^{s}_{p}(\R)$ can have support at any point $\lambda\in\R$, there is no smooth interpolation of $\sigma_{A}$ (the same observation holds for the spaces $h^{\infty}_{1}(\lambda_{0})$).\\
Second, since symbols of almost-periodic pseudo-differential operators form a subclass of H\"ormander's symbols, it is possible to transfer much of the usual symbolic calculus to their setting. This remains partially true for Bohrian pseudo-differential operator, if we replace the symbolic calculus with a discrete version familiar from the $U(1)$-equivariant case (cf. \cite{RuzhanskyPseudoDifferentialOperators}). From a conceptional point of view it is useful to introduce Fourier-Weyl elements, and the associated (de-)quantisation formulas (cp. \eqref{eq:stweylquant}, and below):
\begin{align}
\label{eq:fourierweylbohr}
\hat{W}_{\varepsilon}(\lambda,x) & = \sum_{\beta\in\R}\int_{\R_{\Bohr}}\!\!\!\!\!\!d\mu_{\Bohr}(\alpha)\ e_{-\lambda}(\alpha)e_{-\beta}(x)\ W_{\varepsilon}(\alpha,\beta),  \\ \nonumber
\tr_{l^{2}(\R)}(\hat{W}_{\varepsilon}(\lambda,x)\hat{W}_{\varepsilon}(\lambda',x')) & = \delta_{\lambda,\lambda'}\delta_{\Bohr}(x-x'),\\ \nonumber
A_{\sigma} & = \sum_{\lambda\in\R}\int_{\R_{\Bohr}}\!\!\!\!\!\!d\mu_{\Bohr}(x)\ \sigma(x,\lambda)\ \hat{W}_{\varepsilon}(\lambda,x), \\ \nonumber
\sigma_{A}(x,\lambda) & =\tr_{l^{2}(\R)}(\hat{W}_{\varepsilon}(\lambda,x)A),
\end{align}
which are to interpreted in a distributional sense with $\delta_{\Bohr}=\sum_{\lambda\in\R}e_{-\lambda}\in H^{0}_{\infty}(\R_{\Bohr})$. These formulas can be used to derive the product formula for symbols, $\rho=\sigma\star_{\varepsilon}\tau$, corresponding to the operator product $A_{\rho}=A_{\sigma}A_{\tau}$ (if defined):
\begin{align}
\label{eq:bohrtwistprod}
\rho(x,\lambda) & = \sum_{\lambda'\in\R}\int_{\R_{\Bohr}}\!\!\!\!\!\!d\mu_{\Bohr}(x')e_{-\lambda'}(x)e_{\lambda}(x') \\ \nonumber
& \hspace{0.5cm}\times\sum_{\lambda''\in\R}\int_{\R_{\Bohr}}\!\!\!\!\!\!d\mu_{\Bohr}(x'')\hat{\sigma}(\lambda''\!,x'')\hat{\tau}(\lambda'\!\!-\!\lambda''\!,x'\!\!-\!x'')e_{-\frac{\varepsilon}{2}\lambda''}(x'\!\!-\!x'')e_{\frac{\varepsilon}{2}(\lambda'-\lambda'')}(x'') \\ \nonumber
& = \sum_{\lambda'\in\R}\sum_{\lambda''\in\R}e_{-\lambda'}(x)e_{-\lambda''}(x)\ \hat{\sigma}^{1}(\lambda'',\lambda+\tfrac{\varepsilon}{2}\lambda')\hat{\tau}^{1}(\lambda',\lambda-\tfrac{\varepsilon}{2}\lambda''), \\ \nonumber
\hat{\rho}^{1}(\lambda',\lambda) & = \sum_{\lambda''\in\R}\hat{\sigma}^{1}(\lambda'',\lambda+\tfrac{\varepsilon}{2}(\lambda'-\lambda''))\hat{\tau}^{1}(\lambda'-\lambda'',\lambda-\tfrac{\varepsilon}{2}\lambda''),
\end{align}
which is completely analogous with the formula for the standard Moyal product \eqref{eq:sttwistprod}. For $\sigma,\tau\in\Trig(\R)\otimes d(\R)$ the expression \eqref{eq:bohrtwistprod} is convergent, but in general, e.g. for $A_{\sigma}, A_{\tau}$ relatively rational, it has to be interpreted in an oscillatory sense, i.e. it should be regularised in the way we used to define $A_{\sigma}$.
\begin{Remark}
\label{rem:fractionalopcomp}
The formula \eqref{eq:bohrtwistprod} shows that the composition $A_{\rho}=A_{\sigma}A_{\tau}$ of relatively rational operators $A_{\sigma}, A_{\tau}$ is also relatively rational to $A_{\sigma}, A_{\tau}$.
\end{Remark}
Finally, we look into possible asymptotic expansions of \eqref{eq:bohrtwistprod}. Let us first assume that $\sigma$ and $\tau$ are smooth in the second argument, and belong to H\"ormander's symbol classes. Then, we apply a Taylor expansion of the product of $\hat{\sigma}^{1}$ and $\hat{\tau}^{1}$ in \eqref{eq:bohrtwistprod}:
\begin{align}
\label{eq:bohrsmoothexpansion}
\rho(x,\lambda) & = \sum_{\lambda'\in\R}e_{-\lambda'}(x)\sum_{\lambda''\in\R}e_{-\lambda''}(x) \\ \nonumber
 & \ \ \ \ \times\sum_{n=0}^{N}\frac{1}{n!}\left(\frac{-i\varepsilon}{2}\right)^{n}\sum_{k=0}^{n}\binom{n}{k}(-1)^{n-k}\left(\widehat{(\partial^{k}_{x}\partial^{n-k}_{\lambda}\sigma)}^{\!\!\!\!\!\!1}\ (\lambda'',\lambda)\widehat{(\partial^{n-k}_{x}\partial^{k}_{\lambda}\tau)}^{\!\!\!\!\!\!1}\ (\lambda',\lambda)\right) \\ \nonumber
 & \ \ \ \ + \left(\frac{-i\varepsilon}{2}\right)^{N+1}R^{(N)}_{\sigma,\tau}(x,\lambda), \\ \nonumber
 & = \sum_{n=0}^{N}\frac{1}{n!}\left(\frac{-i\varepsilon}{2}\right)^{n}\sum_{k=0}^{n}\binom{n}{k}(-1)^{n-k}\left((\partial^{k}_{x}\partial^{n-k}_{\lambda}\sigma)(\lambda'',\lambda)(\partial^{n-k}_{x}\partial^{k}_{\lambda}\tau)(\lambda',\lambda)\right) \\ \nonumber
 & \ \ \ \  + \left(\frac{-i\varepsilon}{2}\right)^{N+1}R^{(N)}_{\sigma,\tau}(x,\lambda), \\ \nonumber
R^{(N)}_{\sigma,\tau}(x,\lambda) & = \frac{1}{N!}\sum_{\lambda'\in\R}e_{-\lambda'}(x)\sum_{\lambda''\in\R}e_{-\lambda''}(x)\sum_{k=0}^{N+1}\binom{N+1}{k}(-1)^{N+1-k} \\ \nonumber
 & \ \ \ \ \ \ \ \ \ \times\int_{0}^{1}dt(1-t)^{N}\widehat{(\partial^{k}_{x}\partial^{n-k}_{\lambda}\sigma)}^{\!\!\!\!\!\!1}\ (\lambda'',\lambda+\tfrac{t\varepsilon}{2}\lambda')\widehat{(\partial^{n-k}_{x}\partial^{k}_{\lambda}\tau)}^{\!\!\!\!\!\!1}\ (\lambda',\lambda-\tfrac{t\varepsilon}{2}\lambda''),
\end{align}
which resembles the well-know formulas from the $\R^{n}$-case, besides the fact that we have to deal with the Fourier series on $\R_{\disc}$ instead of the Fourier transform, which leads to different convergence properties (see above). If we do not want to impose smoothness of the symbols and work with the classes $s^{m}_{\rho,\delta}$ (see definition \ref{def:bohrsymbol}), we will have to replace the Taylor expansion by some non-smooth analogue. A least, in the case of relatively rational operators, this is achieved by the \textit{discrete Taylor expansion} or \textit{Newton series} (cf. \cite{RuzhanskyPseudoDifferentialOperators}, Theorem 3.3.21), because this essentially reduces the situation to the $U(1)$-equivariant case. For $ \Phi:\Z^{n}\rightarrow\C$, we have (in multi-index notation):
\begin{align}
\label{eq:discretetaylor}
\Phi(\lambda+\lambda') & = \sum_{\substack{\alpha\in\N_{0} \\ |\alpha|\leq N}}\frac{1}{\alpha!}\lambda'^{(\alpha)}(\Delta^{\alpha}_{\lambda}\Phi)(\lambda) + r^{N}_{\Phi}(\lambda,\lambda'), & \lambda'^{(\alpha)} & =\prod_{i=1}^{n}\lambda'^{(\alpha_{i})}_{i},\ \lambda'^{(\alpha_{i})}_{i}:=\lambda'\cdot...\cdot(\lambda'-\alpha_{i}+1), \\ \nonumber
 & = \sum_{\substack{\alpha\in\N_{0} \\ |\alpha|\leq N}}\binom{\lambda'}{\alpha}(\Delta^{\alpha}_{\lambda}\Phi)(\lambda) + r^{N}_{\Phi}(\lambda,\lambda'), &  & = \alpha!\binom{\lambda'}{\alpha}
\end{align}
where the remainder $r^{(N)}_{\Phi}(\lambda,\lambda')$ satisfies:
\begin{align}
\label{eq:discretetaylorremainder}
|(\Delta^{\beta}_{\lambda}r^{(N)}_{\Phi})(\lambda,\lambda')| & \leq \max_{\substack{|\alpha|=N+1 \\ \lambda''\in Q(\lambda')}}|\lambda'^{(\alpha)}(\Delta^{\alpha+\beta}_{\lambda}\Phi)(\lambda+\lambda'')|, \\ \nonumber
Q(\lambda') & :=\{\lambda''\in\Z^{n}\ |\ |\lambda''_{i}|\leq|\lambda'_{i}|,\ i=1,...,n\}. 
\end{align}
Therefore, if $A_{\sigma}$ and $A_{\tau}$ are relatively rational operators with $\sigma\in s^{m_{\sigma}}_{\rho,\delta}$ and $\sigma\in s^{m_{\tau}}_{\rho,\delta}$, their composition $A_{\rho}=A_{\sigma}A_{\tau}$ is defined, and we find:
\begin{align}
\label{eq:rationalbohrtwistprod}
\rho(x,\lambda) & = \sum_{\lambda'\in\Z^{j^{\tau}_{0}}_{\lambda^{\tau}_{0}}}\sum_{\lambda''\in\Z^{j^{\sigma}_{0}}_{\lambda^{\sigma}_{0}}}e_{-\lambda'}(x)e_{-\lambda''}(x)\ \hat{\sigma}^{1}(\lambda'',\lambda+\tfrac{\varepsilon}{2}\lambda')\hat{\tau}^{1}(\lambda',\lambda-\tfrac{\varepsilon}{2}\lambda'') \\ \nonumber
 & = e_{-(\lambda^{\sigma}_{0}j^{\sigma}_{0}+\lambda^{\tau}_{0}j^{\tau}_{0})}(x)\sum_{m_{\tau}\in\Z}\sum_{m_{\sigma}\in\Z}e_{-(\lambda^{\sigma}_{0}m_{\sigma}+\lambda^{\tau}_{0}m_{\tau})}(x)\ \hat{\sigma}^{1}(\lambda^{\sigma}_{0}(m_{\sigma}+j^{\sigma}_{0}),\lambda+\tfrac{\varepsilon}{2}\lambda^{\tau}_{0}(m_{\tau}+j^{\tau}_{0}))\\[-0.4cm] \nonumber
 &\hspace{7.2cm}\times\hat{\tau}^{1}(\lambda^{\tau}_{0}(m_{\tau}+j^{\tau}_{0}),\lambda-\tfrac{\varepsilon}{2}\lambda^{\sigma}_{0}(m_{\sigma}+j^{\sigma}_{0})) \\ \nonumber
 & = e_{-(\lambda^{\sigma}_{0}j^{\sigma}_{0}+\lambda^{\tau}_{0}j^{\tau}_{0})}(x)\!\sum_{m_{\tau}\in\Z}\!\sum_{m_{\sigma}\in\Z}\!e_{(\lambda^{\sigma}_{0}m_{\sigma}-\lambda^{\tau}_{0}m_{\tau})}(x) \\ \nonumber
 &\hspace{0.5cm}\times\sum^{N}_{n=0}\!\frac{1}{n!}\!\sum^{n}_{k=0}\!\binom{n}{k}m^{(k)}_{\tau}\!\left(\Delta^{k}_{\lambda,\frac{\varepsilon}{2}\lambda^{\tau}_{0}}\hat{\sigma}^{1}\right)(\lambda^{\sigma}_{0}(-m_{\sigma}\!+\!j^{\sigma}_{0}),\lambda\!+\!\tfrac{\varepsilon}{2}\lambda^{\tau}_{0}j^{\tau}_{0}) \\[-0.4cm] \nonumber
 &\hspace{5.75cm}\times m^{(n-k)}_{\sigma}\!\left(\Delta^{n-k}_{\lambda,\frac{\varepsilon}{2}\lambda^{\sigma}_{0}}\hat{\tau}^{1}\right)(\lambda^{\tau}_{0}(m_{\tau}\!+\!j^{\tau}_{0}),\lambda\!-\!\tfrac{\varepsilon}{2}\lambda^{\sigma}_{0}j^{\sigma}_{0}) \\ \nonumber
 &\hspace{1cm} + r^{(N)}_{\sigma,\tau}(x,\lambda;\tfrac{\varepsilon}{2}) \\ \nonumber
 & = \sum^{N}_{n=0}\frac{1}{n!}\sum^{n}_{k=0}\binom{n}{k}\bigg(\Big(\overline{D}^{j^{\sigma}_{0}}_{x,\lambda^{\sigma}_{0}}\Big)^{(n-k)}\Delta^{k}_{\lambda,\frac{\varepsilon}{2}\lambda^{\tau}_{0}}\sigma\bigg)(x,\lambda+\tfrac{\varepsilon}{2}\lambda^{\tau}_{0}j^{\tau}_{0}) \\ \nonumber
 &\hspace{5.75cm}\times\bigg(\Big(D^{j^{\tau}_{0}}_{x,\lambda^{\tau}_{0}}\Big)^{(k)}\Delta^{n-k}_{\lambda,\frac{\varepsilon}{2}\lambda^{\sigma}_{0}}\tau\bigg)(x,\lambda-\tfrac{\varepsilon}{2}\lambda^{\sigma}_{0}j^{\sigma}_{0}) \\ \nonumber
 &\hspace{1cm} + r^{(N)}_{\sigma,\tau}(x,\lambda;\tfrac{\varepsilon}{2}),
\end{align}
where $D^{j_{0}}_{x,\lambda_{0}}:=-\tfrac{i}{\lambda_{0}}\partial_{x}-j_{0}$, and we employed the scaled forward difference $(\Delta_{\lambda,\lambda_{0}}f)(\lambda)=f(\lambda+\lambda_{0})-f(\lambda)$. To arrive at the last line, we applied the inverse Fourier transform (the series are in $h^{\infty}_{1}(\R)$) together with the identity:
\begin{align}
\label{eq:bohrfourierderivative}
\lambda^{(k)}\hat{\Psi}(\lambda) & = \int_{\R_{\Bohr}}d\mu_{\Bohr}(x)\ e_{\lambda}(x)(i\partial_{x})^{(k)}\Psi(x),\ \Psi\in H^{\infty}_{1}(\R_{\Bohr}).
\end{align}
The formula \eqref{eq:rationalbohrtwistprod} constitutes a discrete analogue of the usual asymptotic Weyl product formula. Although, the expansion is not manifestly given in orders of $\varepsilon$, it still has the crucial property that the contribution at order $n$ belong to the symbol classes $s^{m_{\sigma}+m_{\tau}-n(\rho-\delta)}_{\rho,\delta}$ leading to contributions by strictly smaller operators (assuming a suitable version of the Calder\'on-Vaillancourt theorem holds, cf. \cite{CalderonAClassOf, FollandHarmonicAnalysisIn, ShubinDifferentialAndPseudodifferential}) with every increase in the order of the expansion for $\delta<\rho$ (use corollary \ref{cor:bohrsymbolprod}). Finally, we would like to conclude that \eqref{eq:rationalbohrtwistprod} qualifies as an asymptotic expansion, which would hold true, if we were to show that $r^{(N)}_{\sigma,\tau}\in s^{m_{\sigma}+m_{\tau}-(N+1)(\rho-\delta)}_{\rho,\delta}$. In view of the positive results of Ruzhansky and Turunen for the $U(1)$-Kohn-Nirenberg calculus (cf. \cite{RuzhanskyPseudoDifferentialOperators}, Theorem 4.7.10), we fully expect this to be the case.\\[0.1cm]
A simple boundedness theorem of Sobolev type for $U(1)_{\lambda_{0}}$-equivariant operators $A_{\sigma}$ with $\sigma\in s^{m}_{0,0}$ (this encompasses the important case $\sigma\in s^{m}_{\rho,0}\subset s^{m}_{0,0}$) can be proved by means of the (discrete) Young's inequality (cf. \cite{RuzhanskyPseudoDifferentialOperators}, where a similar reasoning is applied to $U(1)$-Kohn-Nirenberg operators, Proposition 4.2.3):
\begin{Lemma}
\label{lem:youngineq}
Given a function $h:\Z^{j_{0}}_{\lambda_{0}}\times\Z^{j'_{0}}_{\lambda'_{0}}\rightarrow\C$, s.t.
\begin{align}
\label{eq:youngineqbounds}
C_{1}:=\sup_{\lambda\in\Z^{j_{0}}_{\lambda_{0}}}\sum_{\lambda'\in\Z^{j'_{0}}_{\lambda'_{0}}}|h(\lambda,\lambda')| & <\infty, & C_{2}:=\sup_{\lambda'\in\Z^{j'_{0}}_{\lambda'_{0}}}\sum_{\lambda\in\Z^{j_{0}}_{\lambda_{0}}}|h(\lambda,\lambda')| & <\infty,
\end{align}
we may define $\left(K_{h}
\Phi\right)(\lambda):=\sum_{\lambda'\in\Z^{j'_{0}}_{\lambda'_{0}}}h(\lambda,\lambda')\Phi(\lambda')$ for all $\Phi\in h^{0}_{p}(\Z^{j'_{0}}_{\lambda'_{0}})$, and have:
\begin{align}
\label{eq:youngineq}
||K_{h}\Phi||_{(0,p)} & \leq C_{1}^{\frac{1}{p}}C_{2}^{\frac{1}{q}}||\Phi||_{(0,p)},\ \ \ 1\leq p,q\leq\infty,\ \tfrac{1}{p}+\tfrac{1}{q}=1.
\end{align}
\begin{Proof}
The results is a simple repetition of the argument presented in \cite{RuzhanskyPseudoDifferentialOperators} for scales affine $\Z$-lattices.
\end{Proof}
\end{Lemma}
Another useful result presented in \cite{RuzhanskyPseudoDifferentialOperators}, which generalises to our case, is
\begin{Lemma}[cp. \cite{RuzhanskyPseudoDifferentialOperators}, Lemma 4.2.1]
\label{lem:bohrfouriersymbolbounds}
Given $\sigma\in s^{m}_{\rho,\delta}$, its Fourier transform w.r.t. the first variable satisfies:
\begin{align}
\label{eq:bohrfouriersymbolbounds}
\forall r\in\R_{\geq0},\ \beta\in\N_{0}:\ \left|\left(\Delta^{\beta}_{\lambda}\hat{\sigma}^{1}\right)(\lambda',\lambda)\right| & \leq C_{r,\beta}\langle\lambda'\rangle^{-r}\langle\lambda\rangle^{m-\rho\beta+\delta r}.
\end{align}
\begin{Proof}
In analogy with the usual regularisation arguments, we have for $M\in\N_{0}$
\begin{align}
\label{eq:bohrfourierboundsestimate}
\left|\left(\Delta^{\beta}_{\lambda}\hat{\sigma}^{1}\right)(\lambda',\lambda)\right|  & = \left|\int_{\R_{\Bohr}}d\mu_{\Bohr}(x)\ e_{\lambda'}(x)\sigma(x,\lambda)\right| \\ \nonumber
 & = \left|\int_{\R_{\Bohr}}d\mu_{\Bohr}(x)\ \langle\lambda'\rangle^{-2M}\left((1-\partial_{x}^{2})^{M}e_{\lambda'}(x)\right)\sigma(x,\lambda)\right| \\ \nonumber
 & = \left|\int_{\R_{\Bohr}}d\mu_{\Bohr}(x)\ \langle\lambda'\rangle^{-2M}e_{\lambda'}(x)(1-\partial_{x}^{2})^{M}\sigma(x,\lambda)\right| \\ \nonumber
 & \leq \langle\lambda'\rangle^{-2M}\sum_{\alpha=0}^{M}\binom{M}{\alpha}C_{2\alpha\beta}\langle\lambda\rangle^{m-\rho\beta+\delta2\alpha} \\ \nonumber
 & \leq \underbrace{\left(\sum_{\alpha=0}^{M}\binom{M}{\alpha}C_{2\alpha\beta}\right)}_{=:C_{2M,\beta}}\langle\lambda'\rangle^{-2M}\langle\lambda\rangle^{m-\rho\beta+\delta2M}.
\end{align}
The boundary term arising from the partial integrations vanishes, because $\forall\lambda\in\R: \sigma(\ .\ ,\lambda)\in C^{\infty}_{b}(\R)$. The result follows for $M=\tfrac{p}{q}\in\Q_{\geq0}$ from
\begin{align}
\label{eq:qinterpolation}
\left|\left(\Delta^{\beta}_{\lambda}\hat{\sigma}^{1}\right)(\lambda',\lambda)\right| & = \left(\left|\left(\Delta^{\beta}_{\lambda}\hat{\sigma}^{1}\right)(\lambda',\lambda)\right|^{2q}\right)^{\frac{1}{2q}} \\ \nonumber
 & \leq \left(C_{2p,\beta}\langle\lambda'\rangle^{-2p}\langle\lambda\rangle^{m-\rho\beta+\delta2p}\right)^{\frac{1}{2q}}\left(C_{0,\beta}\langle\lambda\rangle^{m-\rho\beta}\right)^{\frac{2q-1}{2q}} \\ \nonumber
 & = C_{\frac{p}{q},\beta}\langle\lambda'\rangle^{-\frac{p}{q}}\langle\lambda\rangle^{m-\rho\beta+\delta\frac{p}{q}},
\end{align}
and for $M=r\in\R_{\geq0}$ by continuity.
\end{Proof}
\end{Lemma}
Now, a Calder\'on-Vaillancourt type result can be proved:
\begin{Theorem}
\label{thm:bohrpseudosobolevbound}
A $U(1)_{\lambda_{0}}$-equivariant operator $A_{\sigma}:h^{\infty}_{1}(\Z^{j'_{0}}_{\lambda'_{0}})\rightarrow h^{\infty}_{1}(\Z^{j''_{0}}_{\lambda''_{0}})$ as in proposition \ref{prop:polyoperator} with $\sigma\in s^{m}_{\rho,\delta}$ extends uniquely to bounded operator from $h^{s}_{p}(\Z^{j'_{0}}_{\lambda'_{0}})$ to $h^{s-t}_{p}(\Z^{j''_{0}}_{\lambda''_{0}})$ for $p\in[1,\infty)$ and any $s,t\in\R$, s.t.
\begin{align}
\label{eq:bohrpseudosobolevineq}
\exists r\in\R_{\geq0}: \delta r\leq t-m\ \ \ \&\ \ \ (1-\delta)r>(|m|-1+|t|+|s-t|).
\end{align}
In particular, this justifies to call operators with $\sigma\in s^{-\infty}$ \textup{(infinitely) smoothing}. Moreover, if $\delta=0$, we can choose $t=m$.
\begin{Proof}
For any $t\in\R$, we estimate by Peetre's inequality:
\begin{align}
\label{eq:bohrpseudosobolevbound1}
||A_{\sigma}\Phi||^{p}_{(s-t,p)} & = \sum_{\lambda''\in\Z^{j''_{0}}_{\lambda''_{0}}}\langle\lambda''\rangle^{p(s-t)}\bigg|\sum_{\lambda'\in\Z^{j'_{0}}_{\lambda'_{0}}}\hat{\sigma}^{1}(\lambda''-\lambda',\tfrac{1}{2}(\lambda''+\lambda'))\Phi(\lambda')\bigg|^{p} \\ \nonumber
 & \leq 2^{p|s-t|}\sum_{\lambda''\in\Z^{j''_{0}}_{\lambda''_{0}}}\bigg|\sum_{\lambda'\in\Z^{j'_{0}}_{\lambda'_{0}}}\langle\lambda''-\lambda'\rangle^{|s-t|}\langle\lambda'\rangle^{-t}\hat{\sigma}^{1}(\lambda''-\lambda',\tfrac{1}{2}(\lambda''+\lambda'))\langle\lambda'\rangle^{s}\Phi(\lambda')\bigg|^{p}.
\end{align}
Now, setting $h_{\sigma}(\lambda'',\lambda') := \langle\lambda''-\lambda'\rangle^{|s-t|}\langle\lambda'\rangle^{-t}\hat{\sigma}^{1}(\lambda''-\lambda',\tfrac{1}{2}(\lambda''+\lambda'))$, we would like to employ Young's inequality (see lemma \ref{lem:youngineq}), which will be possible since for any $r\in\R_{\geq0}$:
\begin{align}
\label{eq:bohrsymbolyoungconditions}\nonumber
C^{\sigma}_{1} & := \sup_{\lambda''\in\Z^{j''_{0}}_{\lambda''_{0}}}\sum_{\lambda'\in\Z^{j'_{0}}_{\lambda'_{0}}}\!\!\!\!|h_{\sigma}(\lambda'',\lambda')| \leq C_{r,0}\!\!\sup_{\lambda''\in\Z^{j''_{0}}_{\lambda''_{0}}}\sum_{\lambda'\in\Z^{j'_{0}}_{\lambda'_{0}}}\!\!\langle\lambda''-\lambda'\rangle^{|s-t|}\langle\lambda''\rangle^{-t}\langle\lambda''-\lambda'\rangle^{-r}\langle\tfrac{1}{2}(\lambda''+\lambda')\rangle^{m+\delta r} \\ 
 & \leq 2^{3|m+\delta r|+|t|}C_{r,0}\sup_{\lambda''\in\Z^{j''_{0}}_{\lambda''_{0}}}\sum_{\lambda'\in\Z^{j'_{0}}_{\lambda'_{0}}}\langle\lambda''-\lambda'\rangle^{-r(1-\delta)+|m|+|t|+|s-t|}\langle\lambda''\rangle^{-t+m+\delta r} \\ \nonumber
 & \leq 2^{3|m+\delta r|+|t|}C_{r,0}\Big(\sup_{\lambda''\in\Z^{j''_{0}}_{\lambda''_{0}}}\langle\lambda''\rangle^{-t+m+\delta r}\Big)\sum_{\lambda'\in\Z^{j''_{0}}_{\lambda''_{0}}}\langle\lambda'\rangle^{-r(1-\delta)+|m|+|t|+|s-t|},\\[0.25cm] \nonumber
C^{\sigma}_{2} & := \sup_{\lambda'\in\Z^{j'_{0}}_{\lambda'_{0}}}\sum_{\lambda''\in\Z^{j''_{0}}_{\lambda''_{0}}}\!\!\!\!|h_{\sigma}(\lambda'',\lambda')| \leq C_{r,0}\!\!\sup_{\lambda'\in\Z^{j'_{0}}_{\lambda'_{0}}}\sum_{\lambda''\in\Z^{j''_{0}}_{\lambda''_{0}}}\!\!\langle\lambda''-\lambda'\rangle^{|s-t|}\langle\lambda''\rangle^{-t}\langle\lambda''-\lambda'\rangle^{-r}\langle\tfrac{1}{2}(\lambda''+\lambda')\rangle^{m+\delta r} \\ \nonumber
 & \leq 2^{3|m+\delta r|}C_{r,0}\sup_{\lambda'\in\Z^{j'_{0}}_{\lambda'_{0}}}\sum_{\lambda''\in\Z^{j''_{0}}_{\lambda''_{0}}}\langle\lambda''-\lambda'\rangle^{-r(1-\delta)+|m|+|s-t|}\langle\lambda'\rangle^{-t+m+\delta r} \\ \nonumber
 & \leq 2^{3|m+\delta r|}C_{r,0}\Big(\sup_{\lambda'\in\Z^{j'_{0}}_{\lambda'_{0}}}\langle\lambda'\rangle^{-t+m+\delta r}\Big)\sum_{\lambda''\in\Z^{j''_{0}}_{\lambda''_{0}}}\langle\lambda''\rangle^{-r(1-\delta)+|m|+|s-t|},
\end{align}
where we used the estimate on $\hat{\sigma}^{1}$ from lemma \ref{lem:bohrfouriersymbolbounds} and, repeatedly, Peetre's inequality. In each case, the last line follows from the rationality of $\tfrac{\lambda'_{0}}{\lambda''_{0}}$. Thus, if \eqref{eq:bohrpseudosobolevineq} holds, we will have $C^{\sigma}_{1}<\infty$ and $C^{\sigma}_{2}<\infty$ ($0\leq\delta<1$), from which we conclude by Young's inequality:
\begin{align}
\label{eq:bohrpseudosobolevbound2}
||A_{\sigma}\Phi||^{p}_{(s-m,p)} & \leq 2^{p|s-m|}||K_{h_{\sigma}}(\langle\ .\ \rangle^{s}\Phi)||^{p}_{0,p} \leq 2^{p|s-m|}C^{\sigma}_{1}(C^{\sigma}_{2})^{\frac{p}{q}}||\Phi||^{p}_{s,p}.
\end{align}
Finally, because $h^{\infty}_{1}(\R)\subset h^{s}_{p}(\R)$ is dense for $s\in\R,\ p\in[1,\infty)$, the existence of a unique bounded extension of $A_{\sigma}$ follows. The remaining statements are obvious.
\end{Proof}
\end{Theorem}
\section{Conclusions and perspectives}
\label{sec:con}
In the main part of this article, we have introduced and analysed a Weyl quantisation for loop quantum gravity-type models aiming at the implementation of the program space adiabatic perturbation theory in the latter. As we have argued, a complete  implementation requires the Weyl quantisation to be scalable with the quantisation parameter $\varepsilon$ (also called adiabatic parameter) for a perturbation theory in orders of $\varepsilon$ to be possible. Unfortunately, due to the fact, that models {\`a} la loop quantum gravity are constructed via projective limits of function algebras living on the co-tangent bundles of compact Lie groups (in the phase space approach), there results a fundamental asymmetry in the treatment of configuration and momentum space degrees of freedom in the quantum theory. This asymmetry entails the scalability (with $\varepsilon$) of the local Weyl quantisation w.r.t. to Ashtekar-Isham-Lewandowski representation, if and only if the adiabatic parameter $\varepsilon$ is associated with the momentum variables: An effect that its easily understood from the observation that the momentum space degrees of freedom are modelled by the co-tangent space directions, which possess a (local) vector space structure. In contrast, the configuration space degrees of freedom, which are modelled by the compact Lie group underlying the co-tangent bundle, do not admit suitable, i.e. compatible with the (local) commutation relations defining the transformation group algebra, $\varepsilon$-scale transformations, because the existence of the latter would require exponential map to be a homomorphism of Lie groups (existence of arbitrary (real) powers of group element in a homomorphic way).
In the global Weyl quantisation, where the co-tangent bundle spaces are replaced by the representation theoretical dual of the Lie group, the problem of scalability also shows up for the momentum space degrees of freedom (unitary equivalence classes of irreducible representations), and manifest itself in the rigid structure of the lattice of integral highest weights. Since the representation theoretical dual and the group itself are in a one-to-one correspondence (Tannaka-Krein duality \cite{BroeckerRepresentationsOfCompact}, Doplicher-Roberts theorem \cite{DoplicherANewDuality}), the question of scalability in the local and global setting can be unified under the theme of $\varepsilon$-scalable Fourier transforms, which we discussed in subsection \ref{subsec:swt} \& \ref{subsec:u1bohr}).\\
One reason, why we are not satisfied with the local Weyl quantisation, and its scalability w.r.t. to the momentum variables, is exemplified by toy models of loop quantum cosmology-type. In those, we find that it is quite natural for the adiabatic parameter $\varepsilon$ to be associated with the configuration space degrees of freedom (holonomies of $U(1)$): A feature that may well persist in full loop quantum gravity-type models.\\
Another reason is implicit in the construction of the (local) calculus Paley-Wiener-Schwartz symbols, which requires an analytic momentum dependence of the quantisable functions, because, after a fibre-wise Fourier transform from the co-tangent bundle to the tangent bundle, the dual distribution is required to be of compact support. Thus, again the non-global nature of the exponential map of a compact Lie group is of some disadvantage, as it severely restricts (analyticity!) the class of quantisable function, and therefore the class of dequantisable operators.\\
Such analyticity condition are not necessary in the global calculus, but the problems due to compactness show up in the way mentioned before.\\
An almost satisfactory solution to the difficulties introduced by the compactness of the (truncated) configuration spaces, can, up to this point, only be achieved in $U(1)^{n}$-models, where its possible to proceed from $U(1)$ to the still compact, but representation theoretically less rigid, Bohr compactification of the reals $\R_{\Bohr}$. Regrettably, this appears not to be a viable strategy in non-Abelian models\cite{ShubinAlmostPeriodicFunctions}.\\
Since we also discussed the use of coherent state/Berezin quantisations for compact Lie groups (section \ref{sec:cst} and subsection \ref{subsec:csquant}) in our research on the implementability of the ideas of Born and Oppenheimer in loop quantum gravity-type models, it should be once more pointed out, that, unless a suitable $\star$-product is constructed from these quantisations, the existence of which is highly questionable for non-compact (truncated) phase spaces, it seems to be very difficult to construct a systematic perturbation theory. Thus, while the use of coherent states is conceptually tempting, it seems to be technically and computationally disfavoured. 
\begin{acknowledgments}
\label{sec:ack}
AS gratefully acknowledges financial support by the Ev. Studienwerk e.V.. This work was supported in parts by funds from the Friedrich-Alexander-University, in the context of its Emerging Field Initiative, to the Emerging Field Project ``Quantum Geometry’’.
\end{acknowledgments}
\bibliography{boa2a.bbl, boa2aNotes.bib}
\bibliographystyle{aipnum4-1}
\label{sec:ref}
\end{document}